\documentstyle[amsfonts,epsfig,12pt]{article}

\def\mypagenumber{1}

\def\myend{\end{document}}

\normalsize

\newcounter{sxn}

\newcounter{axn}

\date{}

\newdimen\mybaselineskip
\newcommand{\beeq}{\begin{equation}}
\newcommand{\eneq}{\end{equation}}
\newcommand{\be}{\begin{eqnarray}}
\newcommand{\ee}{\end{eqnarray}}
\newcommand{\bpic}{\begin{picture}}
\newcommand{\epic}{\end{picture}}

\def\dd{\partial}
\def\la{\raise.16ex\hbox{$\langle$} \, }
\def\ra{\, \raise.16ex\hbox{$\rangle$} }

\def\psibar{ \psi \kern-.65em\raise.6em\hbox{$-$} }
\def\mbar{ m \kern-.78em\raise.4em\hbox{$-$}\lower.4em\hbox{} }

\def\n@space{\nulldelimiterspace=0pt \mathsurround=0pt }
\def\huge#1{{\hbox{$\left#1\vbox to 20.5pt{}\right.\n@space$}}}

\def\myskip{\noalign{\kern 8pt}}
\def\myeqspace{\noalign{\kern 10pt}}

\def\boxit#1{$\vcenter{\hrule\hbox{\vrule\kern3pt
    \vbox{\kern3pt\hbox{#1}\kern3pt}\kern3pt\vrule}\hrule}$}
\def\bigbox#1{$\vcenter{\hrule\hbox{\vrule\kern5pt
     \vbox{\kern5pt\hbox{#1}\kern5pt}\kern5pt\vrule}\hrule}$}

\def\ignore#1{{}}

\begin{document}
\bibliographystyle{unsrt}
\footskip 1.0cm

\thispagestyle{empty}
\setcounter{page}{\mypagenumber}


\begin{flushright}{
UFIFT-HEP-03-32\\} {LPT-ORSAY 03-107\\}

\end{flushright}

\vspace{2.5cm}
\begin{center}
{\LARGE \bf {Gravitational Lensing by Dark Matter Caustics }}\\
\vskip 1 cm {\large{Vakif K. Onemli  }}\footnote{e-mail:~
onemli@zephyr.th.u-psud.fr}\\
\vspace{.5cm} {\it Department of Physics, University of Florida,
Gainesville, FL 32611 USA}\\
\begin{abstract}
There are compelling reasons to believe that the dark matter of
the universe is constituted, in large part, by non-baryonic
collisionless particles with very small primordial velocity
dispersion. Such particles are called cold dark matter (CDM). The
leading candidates are axions and weakly interacting massive
particles (WIMPs). The word ``collisionless'' indicates that the
particles are so weakly interacting that they move purely under
the influence of gravity. Galaxies are surrounded by CDM and
hence, because of gravity, CDM keeps falling onto galaxies from
all directions. CDM infall produces a discrete number of flows and
caustics in the halos of galaxies. There are two types of caustics
in the halos of galaxies: inner and outer. An outer caustic is a
simple fold ($A_2$) catastrophe located on a topological sphere
enveloping the galaxy. An inner caustic is a closed tube whose
cross-section is an elliptic umbilic ($D_{-4}$) catastrophe, with
three cusps.

In space, caustics are the boundaries that separate regions with
differing numbers of flows. One side of a caustic surface has two
more flows than the other. The density of CDM particles becomes
very large as one approaches the caustic from the side with the
extra flows. Dark matter caustics have specific density profiles
and therefore precisely calculable gravitational lensing
properties. This dissertation presents a formalism that simplifies
the relevant calculations, and apply it to four specific cases. In
the first three cases, the line of sight is tangent to a smooth
caustic surface. The curvature of the surface at the tangent point
is positive, negative, or zero. In the fourth case, the line of
sight passes near a cusp.  For each case we derive the map between
the image and source planes.  In some cases, a point source has
multiple images and experiences infinite magnification when the
images merge. A promising approach to reveal the locations of
caustics is to observe the distortions in the images of extended
sources (such as radio jets) passing by the caustics.
\end{abstract}

\end{center}

\vspace*{1.5cm}

\newpage

\normalsize
\baselineskip=23pt plus 1pt minus 1pt
\parindent=27pt
\vskip 4 cm
\begin{center}
{\bf
Acknowledgments}
\end{center}

This dissertation is based on a research paper on gravitational
lensing by dark matter caustics. Caustic rings of dark matter were
discovered by Dr. Pierre Sikivie who graciously allowed me to draw
extensively on his work as he served as my guide during the course
of the research that yielded this dissertation. I am deeply
indebted to him for accepting me as one of his students. It has
been a great privilege to work under his supervision in this new
area of research in particle astrophysics. This attempt to widen
the range of the discussion and to make the material more
pedagogical would not have been possible without his help and
encouragement. I would like to thank him also for his patience and
his willingness to share both his thoughts and his time
generously.

It is a pleasure to express my thanks to the other members of my
supervisory committee: Dr. James Fry, Dr. David Grossier, Dr.
Zongan Qiu, Dr. Pierre Ramond, Dr. Richard P. Woodard, and Dr.
John Yelton. I have been honored by their presence on my
committee; and benefitted from their contributions as teachers,
collaborators, and coordinators throughout the years of my stay at
the Physics Department.

I would like to thank Dr. Richard P. Woodard for his unfailing
interest, encouragement and support. He was always available to
assist me with every kind of problem I had, from physics to
personal ones. His enthusiasm and dedication influenced me deeply.
I was very fortunate that I had the chance to attend his lectures
and collaborate with him. For the ``cydonies,'' I am forever
grateful. Dear Richard, thank you for caring about my career. You
will be missed. As in the good old days ``all the brothers were
valiant.''

My thanks also go to the other members of the Department of
Physics of the University of Florida for making my studies
enjoyable as well as educational. I would also like to thank my
friends, whose support I have relied on; and my collaborators: Dr.
Christos Charmousis, Dr. Zongan Qiu, Dr. Pierre Sikivie, Dr.
Bayram Tekin, Dr. Murat Tas, and Dr. Richard Woodard, for the
insights they shared with me. I thank Carol Lauriault for
carefully editing the manuscript. I am particularly grateful to my
grandmother, my parents and my brother for their love and support.
Finally, the Gibson Dissertation Fellowship I received is most
gratefully acknowledged.

\newpage
\section{Introduction}

A large amount of astronomical evidence indicates the existence in
the universe of more gravitationally interacting matter than
luminous matter. Most of the matter in the universe is dark. The
stuff that is responsible for holding galaxies (and clusters of
galaxies) together is a peculiar kind of matter that we neither
see nor detect by any means. This cosmic mystery began puzzling
astronomers in the 1930s and the case still remains open. For a
review of the sources and distributions of dark matter, see
Sikivie \cite{SikivieSourcesDistributions} and the references
therein.

In 1932, Oort \cite{Oort} studied the motions of galactic disk
stars in the direction perpendicular to the disk. He applied the
virial theorem to the distribution of vertical star velocities,
and estimated the density of the galactic disk as $\rho_{\rm
disk}\simeq 1.2\cdot 10^{-23}{\frac{\rm gr}{\rm cm^3}}$. When the
density due to the matter seen in stars, interstellar gas, and
stellar remnants like white dwarfs is calculated, however, one
finds an estimate considerably less than the above value. This
implies the existence of dark matter in the disk. Because this
dark matter is in the disk rather than the halo, it is expected to
be dissipative, hence baryonic in nature.

In 1933, Zwicky \cite{Zwicky} used measurements of the line of
sight velocities of galaxies in the Coma cluster to estimate the
total mass of the cluster using the virial theorem: \beeq \langle
T\rangle =-\frac{1}{2}\langle U\rangle\, ,\label{virial}\eneq
where $T$ and $U$ are the kinetic and potential energy of the
system respectively. Bracket $\langle \, \rangle$ denotes time
average. In these estimates, however, it is important to remember
that the virial theorem is valid for a closed system in mechanical
equilibrium and that it applies to the time average of the system.
Whether the observed clusters satisfy these requirements is
questionable. Nevertheless, assuming the applicability of the
theorem, and using for the kinetic energy of $N$ galaxies in the
cluster: \beeq \langle T\rangle =\frac{1}{2} N\langle mv^2\rangle
\ ,\eneq the potential energy of the $\frac{1}{2}N(N-1)$
independent galaxy pairs: \beeq \langle U\rangle =-\frac{1}{2}G
N(N-1)\frac{\langle m^2\rangle}{\langle r\rangle}\ ,\eneq and
taking $N\simeq N-1$, one can estimate the total mass: \beeq
M=N\langle m\rangle \simeq \frac{2 r \langle v^2\rangle}{G}\, .
\eneq By measuring the quantities $r$ and $v$, it is possible to
estimate $M$. The result obtained was about 400 times the mass in
the luminous parts of the cluster galaxies.

During the seventies, the rotation curves (velocity as a function
of radial distance from the center) of spiral galaxies began to be
measured over much larger distances than before. Spiral galaxies
are large scale structures containing billions of stars, arranged
in the form of a rotating disk with a central bulge. Assuming that
the stars have a circular orbit around the galactic center, the
rotation velocity of a star can be calculated by balancing the
centrifugal and the gravitational forces:\beeq \frac{G M(r)
m}{r^2}=\frac{v_{\rm rot}^2 m}{r}\, ,\eneq where $M(r)$ is the
mass interior to the orbit radius $r$, $m$ is the mass of the star
and $v_{\rm rot}(r)$ is the rotation velocity at $r$. Here, we
used the fact that for cylindrically or spherically symmetric
distributions, the forces due to the mass lying outside the orbit
compensate exactly. We have, therefore  \beeq v_{\rm
rot}(r)=\sqrt{\frac{GM(r)}{r}}\ .\eneq If we assume that the bulge
is spherically symmetric with a constant density of $\rho$, then
$M(r)=\frac{4}{3}\pi r^3\rho$. Hence, for the innermost part of a
galaxy, a rotation curve where $v_{\rm rot}\sim r$ is expected.
For the outside of the galaxy, where $r>$ disc-radius, $M(r)$ is
equal to the total mass of the galaxy $M(r)=M_{\rm galaxy}$, which
is a constant. In this case \beeq v_{\rm
rot}(r)=\sqrt{\frac{GM_{\rm galaxy}}{r}}\sim\frac{1}{\sqrt{r}}\,
.\eneq The data show, however, that $v_{\rm rot}\sim {\rm
constant}$ for large $r$. This means that $M(r)\sim r$, which
implies the existence of a huge unseen mass extending far beyond
the visible region. The distribution of globular star clusters
suggests a spherical distribution. Theoretical model calculations
show that pure disk galaxies have a tendency to become bars (i.e.,
within the central nucleus, a bar-like structure forms)
\cite{BinneyTremaine}. Barred galaxies do exist, but they are
relatively rare. A spherical halo of dark matter increases the
stability of the pure disk structure, producing a ratio of barred
spirals consistent with observations \cite{OstrikerPeebles}.
Considerations of the stellar dynamics in elliptical galaxies also
imply that they contain a significant fraction of dark
matter\cite{Saglia}.

These observations lead to the hypothesis of a halo of dark matter
whose density is $d_{\rm dm}(r)\sim\frac{1}{r^2}$ at large $r$.
The halo distribution is usually modelled by the function\beeq
\rho_{\rm dm}(r)=\frac{\rho_{\rm
dm}(0)}{1+\left(\frac{r}{r_0}\right)^2}\, ,\eneq where $r_0$ is
called the core radius. For our own galaxy, $v_{\rm rot}\simeq
220\; \frac{\rm km}{\rm s}$, $r_0$ is a few kpc, and $\rho_{\rm
dm}(0)\sim 10^{-23}\frac{\rm gr}{\rm cm^3}$.

According to Wilkinson Microwave Anisotropy Probe (WMAP) results
\cite{Bennett,Spergel,Peiris}, the density parameter \be
\Omega=\frac{\rho}{\rho_c}=\frac{8\pi G\rho}{3H_0^2}\,
\label{8piGr}\ee of the universe has the numerical value
$\Omega_{\rm univ}=1.02 \pm 0.02$. In Eq. \ref{8piGr} $\rho_c$ is
the critical density for closing the universe and $H_0$ is the
present Hubble expansion rate. Most of the energy density of the
universe is in the form of vacuum energy $\Omega_{\Lambda}=0.73\pm
0.04$. The density due to matter is $\Omega_m=0.27\pm 0.04$ and
$0.17\pm 0.01$ of $\Omega_m$ is baryonic: $\Omega_b=0.044\pm
0.004$. Since the density of luminous matter is $\Omega_{\rm
lum.}<0.006$, we conclude that some baryons are dark. Recall that
there is dark matter associated with the disk of a galaxy. Because
it is in the disk rather than in a halo, this dark matter must be
dissipative which presumably means that it is baryonic (disk dark
matter must have sufficiently strong interactions to be
concentrated in a disk by dissipating its energy while conserving
its angular momentum). Likely hiding places for these dark baryons
may be black holes, cold white dwarfs, and brown dwarfs (i.e.,
stars too low in mass to burn by nuclear fusion). Objects of this
kind, generically called MACHOs (for massive compact halo
objects), have been searched for by looking for the gravitational
lensing of background stars by MACHOs that happen to pass close to
the line of sight. Many of them have been discovered
\cite{m1,m2,m3,m4}. We still do not know what makes up most of the
missing mass in the universe.

The success of nucleosynthesis in producing the primordial
abundances of light elements \cite{KT}, also requires that
$0.011\leq \Omega_b\leq 0.12$. Studies of large scale structure
formation support the view that most of the dark matter consists
of non-baryonic, collisionless but gravitationally interacting
particles such as axions and/or Weakly Interacting Massive
Particles (WIMPs) (like neutralinos, photinos, higgsinos, etc.)
\cite{KT}. The particles must have small primordial velocity
dispersion; for this reason, they also are called ``Cold Dark
Matter (CDM)'' candidates. We discuss these CDM candidates in the
next section. Here, and throughout the dissertation, we assume
that they exist; and study interesting consequences of their
motion in the potential well of a galactic halo. We see that these
candidates produce surfaces (which we call caustics) in physical
space where the dark matter density is large, because folds exist
at the corresponding locations of the sheet on which the particles
lie in phase space. The CDM caustics may move about, but they are
stable. They also have specific density profiles, and hence
precisely calculable gravitational lensing properties. We present
a formalism that simplifies the relevant calculations; and apply
it to different cases to obtain their lensing signatures
\cite{lensingbycaustics}.

Let us start with a discussion of the phase space
(velocity-position space) structure of CDM halos. Since the
particles are in three-dimensional (3D) space, the phase space is
six-dimensional. The primordial velocity dispersion of the cold
dark matter candidates is of a very small order \cite{sing} \beeq
\delta v_a (t) \sim 1.5\cdot 10^{-17} \left( {10^{-5} {\rm
eV}\over m_a}\right)^{4.7/5.7}~\left({t_0\over t}\right)^{2/3}\,
,\label{delvax}\eneq for axions and \beeq \delta v_\chi (t) \sim
10^{-11} \left({{\rm GeV}\over
m_\chi}\right)^{1/2}~\left({t_0\over t}\right)^{2/3}\, ,
\label{delvWIMP}\eneq for WIMPs, where $t_0$ is the present age of
the universe, and $m_a$ and $m_\chi$ are respectively the masses
of the axion and the WIMP. Calculations for $\delta v_a (t)$ and
$\delta v_\chi (t)$ are given in Sections \ref{sect:Axions} and
\ref{sect:WIMPs}, respectively. These estimates of primordial
velocity dispersions are approximate, but in the context of this
dissertation and of galaxy formation in general, $\delta v_a$ and
$\delta v_\chi$ are entirely negligible. Therefore, CDM particles
lie on a very thin 3D sheet in 6D phase space. The thickness of
the sheet is the primordial velocity dispersion $\delta v$ of the
particles. Moreover, the present average number density $n$ of the
particles can be expressed in terms of their energy density in
units of the critical density. For axions \beeq n_a(t_0) =
{1.5~10^{64} \over {\rm pc}^3}\, ~\Omega_a~ \left({h \over
0.7}\right)^2~\left({10^{-5}{\rm eV} \over m_a}\right)\,
,\label{numdenaxions}\eneq see Section \ref{sect:Axions}.
Likewise, for WIMPs \beeq n_\chi(t_0) = {1.5~10^{50} \over {\rm
pc}^3}\, ~\Omega_\chi~ \left({h \over 0.7}\right)^2~\left({{\rm
GeV} \over m_ \chi}\right)\, ,\label{numdenWIMPs}\eneq see Section
\ref{sect:WIMPs}. The large exponents indicate that the density of
cold dark matter particles is enormous in terms of astronomical
length scales. Since they are effectively collisionless, the
particles experience only gravitational forces. These are
universal and vary only on huge distances compared to the
interparticle distance. Hence the sheet on which the particles lie
in phase space is continuous.  It cannot break and its evolution
is constrained by topology. If each of the aforementioned CDM
candidates is present, the phase space sheet has a layer for each
species, with a thickness proportional to the corresponding
velocity dispersion.

\begin{figure}[ht] \centering
\includegraphics[height=9cm,width=13cm]{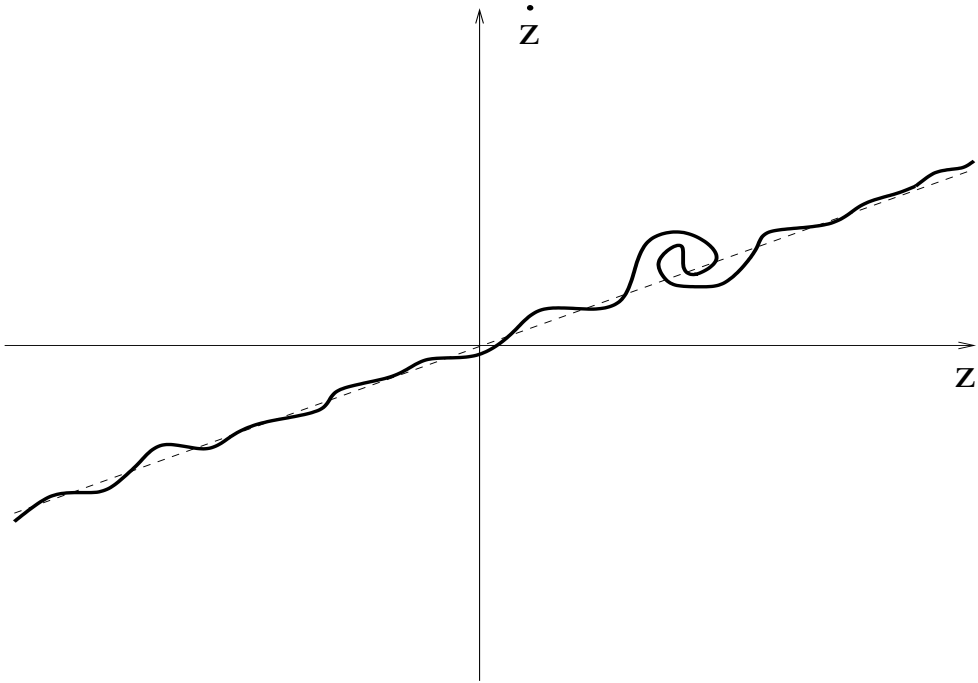}
\caption{The wiggly line is the intersection of the $(z,\dot{z})$
plane with the 3D sheet on which the collisionless dark matter
particles lie in phase space.  The thickness of the line is the
primordial velocity dispersion.  The amplitude of the wiggles in
the $\dot{z}$ direction is the velocity dispersion associated with
density perturbations.  Where an overdensity grows in the
nonlinear regime, the line winds up in clockwise fashion.  One
such overdensity is shown.} \label{fig:sheet}
\end{figure}

\begin{figure}[ht] \centering
\includegraphics[height=9cm,width=13cm]{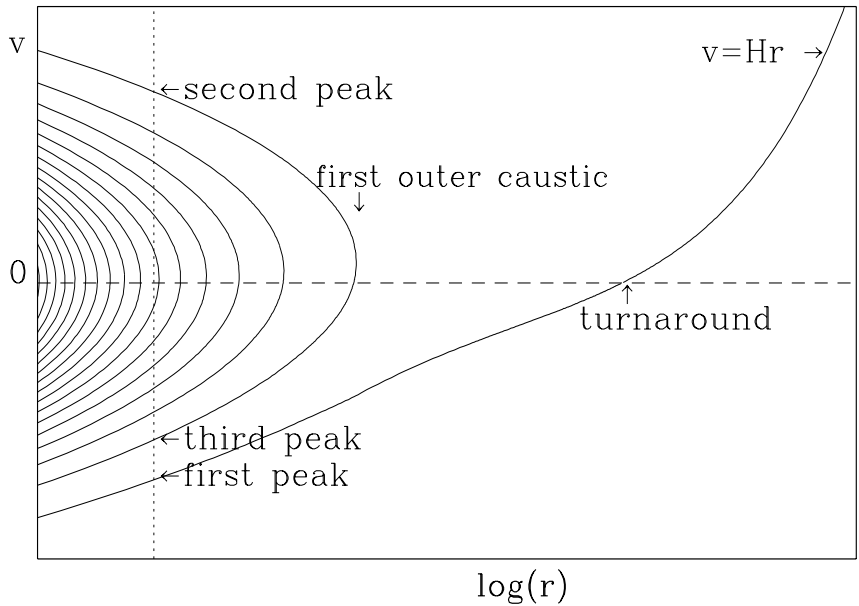}
\caption{Snapshot of the phase space distribution of CDM particles
in a galactic halo.  For simplicity, spherical symmetry is
assumed: r is galactocentric distance and v is radial velocity.
The solid line indicates the location of the particles.  The
dotted line corresponds to observer position. Each intersection of
the solid and dotted lines corresponds to a CDM flow at the
observer's location.  ``Turnaround'' refers to the moments in a
particle's history when it has zero radial velocity with respect
to the galactic center. A ``caustic'' appears wherever the phase
space line folds back. Particles pile up and hence the density
diverges there.} \label{fig:phase}
\end{figure}

The phase space sheet is located on the 3D hypersurface of points
$(\vec{r}, \vec{v})~:~\vec{v} = H(t)\vec{r} + \Delta \vec{v}
(\vec{r},t)$, where $H(t) = {2\over 3t}$ is the Hubble expansion
rate and $\Delta \vec{v} (\vec{r},t)$ is the peculiar velocity
field.  Figure \ref{fig:sheet} shows a 2D section of phase space
along the $(z,\dot{z})$ plane.  The wiggly line is the
intersection of the 3D sheet on which the particles lie in phase
space with the plane of the figure.  The thickness of the line is
the velocity dispersion $\delta v$, whereas the amplitude of the
wiggles in the line is the peculiar velocity $\Delta v$. If there
were no peculiar velocities, the line would be straight, since in
that case $\dot{z} = H(t) z$. The peculiar velocities are associated with density perturbations
and grow by gravitational instability as $\Delta v\sim t^{2/3}$.
On the other hand the primordial velocity dispersion decreases on
average as $\delta v \sim t^{-2/3}$, consistent with Liouville's
theorem.  When a large overdensity enters the nonlinear regime,
the particles in the vicinity of the overdensity fall back onto
it.  This implies that the phase space sheet ``winds up'' in
clockwise fashion wherever an overdensity grows in the nonlinear
regime.  One such overdensity is shown in Fig. \ref{fig:sheet}.
Before density perturbations enter the nonlinear regime, there is
only one value of velocity (i.e., a single flow) at a typical
location in physical space, because the phase space sheet covers
physical space only once.  On the other hand, inside an
overdensity in the nonlinear regime, the sheet wraps up in phase
space, turning clockwise in any two dimensional cut $(z, \dot{z})$
of that space. The physical space coordinate axis $z$ is in an
arbitrary direction, and $\dot{z}$ is its associated velocity. The
outcome of this process in a galactic halo is an odd number of
flows at any physical point in the halo \cite{ips}. One flow is
associated with particles falling through the galaxy for the first
time ($n=1$); another is associated with particles falling through
the galaxy for the second time ($n=2$), and so on (Fig.
\ref{fig:phase}).

At the boundary between two regions, one of which has $n$ flows
and the other $n + 2$ flows, the physical space density is very
large because the phase space sheet has a fold there.  At the
fold, the phase space sheet is tangent to velocity space and
hence, in the limit of zero velocity dispersion $(\delta v = 0)$,
the physical space density diverges, since it is the integral of
the phase space density over velocity space.  The structure
associated with such a phase space fold is called a ``caustic''
\cite{cr,sing,Tre}. Generically, caustics are surfaces in physical
space, since they separate regions with differing number of flows.
One side of a caustic surface has two more flows than the other.
Because caustic surfaces occur wherever the number of flows
changes, they are topologically stable, in direct analogy with
domain walls. It is easy to show (Section \ref{sec:caustic
surfaces}) that,  in the limit of zero velocity dispersion, the
density diverges as $d \sim {1\over\sqrt{\sigma}}$ when the
caustic is approached from the side with $n+2$ flows, where
$\sigma$ is the distance to the caustic.  If the velocity
dispersion is small, but nonzero, the divergence is cut off so
that the density at the caustic is no longer infinite, but merely
very large.

Zel'dovich \cite{Zel} emphasized the importance of caustics in
large scale structure formation, and suggested using the name
``pancakes'' for them.  The reason galaxies tend to lie on
surfaces \cite{Huchra} such as ``the Great Wall'' is undoubtedly
that the 3D sheet on which the dark matter particles and baryons
lie in phase space acquires folds on very large scales, producing
caustics appropriately called Zel'dovich pancakes. Sikivie
\cite{sing} derived the minimal CDM caustic structure that must
occur in galactic halos \cite{sing}. There are two types of
caustics in the halos of galaxies: inner and outer. The outer
caustics are simple fold ($A_2$) catastrophes located on
topological spheres surrounding the galaxy.

We saw above that where a localized overdensity is growing in the
nonlinear regime, the line at the intersection of the phase space
sheet with the $(z,\dot{z})$ plane winds up in a clockwise
fashion. The onset of this process is shown in Fig.
\ref{fig:sheet}. Of course, the picture is qualitatively the same
in the $(x,\dot{x})$ and $(y,\dot{y})$ planes.  In this view, the
process of galactic halo formation is the winding up of the phase
space sheet of collisionless dark matter particles.  When the
galactic center is approached from any direction, the local number
of flows increases.  First, there is one flow, then three flows,
then five, seven. . . . The boundary between the region with one
(three, five, . . .) and the region with three (five, seven,
. .
.) flows is the location of a caustic, which is topologically a
sphere surrounding the galaxy (Fig. \ref{fig:phase}). When these
caustic spheres are approached from the inside, the density
diverges as $d\sim {1\over\sqrt{\sigma}}$ in the zero velocity
dispersion limit. Outer caustics occur where a given outflow
reaches its furthest distance from the galactic center before
falling back in.

\begin{figure}[ht] \centering
\includegraphics[height=9cm,width=44cm]{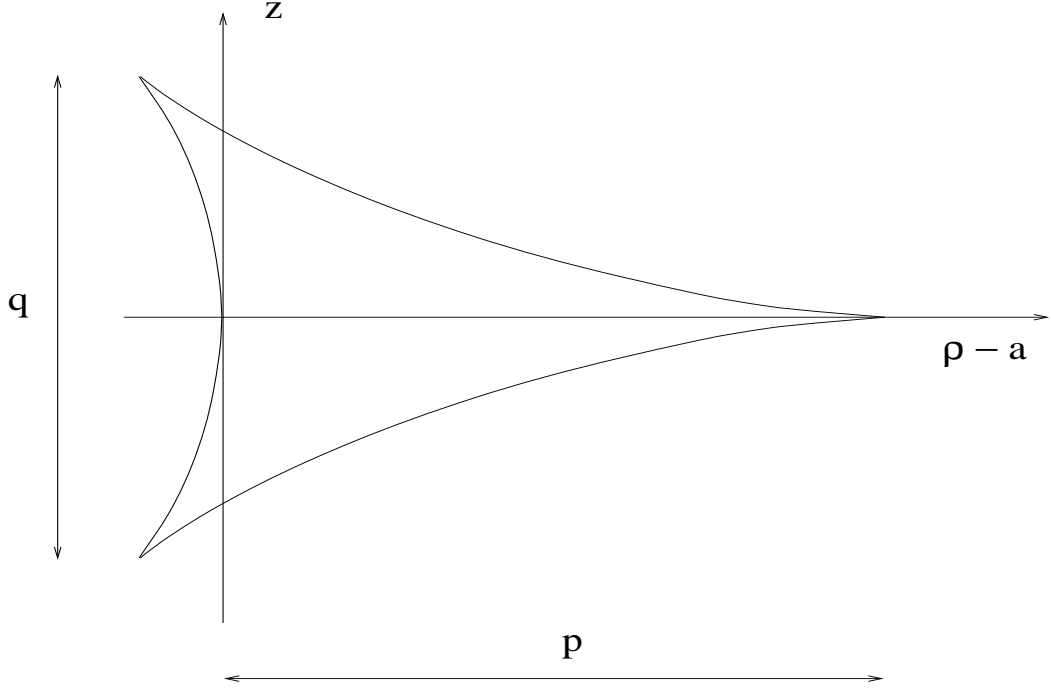}
\caption{Cross-section of a caustic ring in the case of axial and
reflection symmetry, and $p,q \ll a$.} \label{fig:fig6}
\end{figure}

The inner caustics are rings \cite{cr}.  They are located near
where the particles with the most angular momentum in a given
inflow reach their distance of closest approach to the galactic
center before going back out. In the absence of angular momentum,
the caustic rings collapse to a caustic point at the center
whereas the outer caustics are unaffected. A caustic ring is more
precisely a closed tube with a special structure. Its transverse
cross-section is a closed line with three cusps, one of which
points away from the galactic center (Fig. \ref{fig:fig6}).
In the language of Catastrophe Theory, such a singularity is called
a $D_{-4}$ (or {\it elliptic umbilic}) catastrophe \cite{sing}. We
call it a ``tricusp.'' The existence of these caustics and their
topological properties are independent of any assumptions of
symmetry. Their derivations are discussed in Section \ref{Flow}.

Dark matter caustics have very well-defined density profiles, and
hence calculable gravitational lensing signatures \cite{Hogan}. In
this dissertation we derive these signatures in a number of
specific cases. In the limit of zero velocity dispersion ($\delta
v = 0$), the density diverges when one approaches a caustic
surface on the side that has two extra flows, as the inverse
square root of the distance to the surface. This divergence is cut
off if there is velocity dispersion, because the location of the
caustic surface gets smeared over some distance $\delta x$.  For
the dark matter caustics in galactic halos, $\delta x$ and $\delta
v$ are related \cite{cr} by
\begin{equation}
\delta x \sim {R~\delta v \over v} \label{sme}\ ,
\end{equation}
where $v$ is the order of magnitude of the velocity of the
particles in the flow and $R$ is the distance scale over which
that flow turns around (i.e., changes its direction).  For a
galaxy like our own, $v = 500$ km/s and $R = 200$ kpc are typical
orders of magnitude.

As mentioned earlier, the primordial velocity dispersion of the
leading cold dark matter candidates is very small.  Using the
estimates of $\delta v$ given above, one finds that axion caustics
in galactic halos are typically smeared over
\begin{equation}
\delta x_a \sim 5\cdot 10^{9}\, {\rm cm} \left({10^{-5} {\rm eV}
\over m_a} \right)^{4.7/5.7} \label{dxa}
\end{equation}
as a result of their primordial velocity dispersion; whereas WIMP
caustics are smeared over
\begin{equation}
\delta x_\chi
 \sim 3 \cdot 10^{15}\, {\rm cm} \left({{\rm GeV}
\over m_\chi}\right)^{1/2} \ . \label{dxW}
\end{equation}
It should be kept in mind, however, that a cold dark matter flow
may have an effective velocity dispersion that is larger than its
primordial velocity dispersion.  Effective velocity dispersion
occurs when the sheet on which the dark matter particles lie in
phase space is wrapped up on scales that are small compared to the
galaxy as a whole.  It is associated with the clumpiness of the
dark matter falling onto the galaxy.  The effective velocity
dispersion of a flow may vary from point to point, taking larger
values where more small scale structure has formed; and taking the
minimum primordial value where no small scale structure has
formed.  For a coarse-grained observer, the dark matter caustic is
smeared over $\delta x$ given by Eq. \ref{sme} where $\delta v$ is
the effective velocity dispersion of the flow.

Little is known about the size of the effective velocity
dispersion of dark matter flows in galactic halos.  Sikivie
\cite{milk}, however, interpreted a triangular feature in the IRAS
map of the Milky Way as the imprint on baryonic matter of the
caustic ring of dark matter nearest to us.  The sharpness of the
feature's edges implies an upper limit of 20 pc on the distance
$\delta x$ over which that caustic is smeared; and hence an upper
limit of order 50 m/s on the effective velocity dispersion of the
corresponding flow.

The gravitational lensing effects of a caustic surface are largest
when the line of sight is near tangent to the surface, because the
contrast in column density is largest there.  The effects depend
on the curvature of the caustic surface at the tangent point in
the direction of the line of sight: the smaller the curvature, the
larger the effects.  A caustic is an oriented surface because one
side has two more flows than the other.  We considered three cases
of gravitational lensing by a smooth caustic surface.  In the
first case, the line of sight is near tangent to a caustic surface
that curves toward the side with two extra flows. We call such a
surface ``concave.'' In the second case, the surface is
``convex,'' (i.e., it curves away from the side with two extra
flows). In the third case, the caustic surface has zero curvature
at the tangent point (the radius of curvature is infinite), but
the tangent line is entirely outside the side with two extra
flows. Caustic surfaces may have cusps.  The outer dark matter
caustics of galactic halos are smooth topological spheres, which
have no singularities, but the inner caustics of galactic halos
are closed tubes whose cross-section has three cusps. Therefore,
the fourth case we consider has a line of sight near a cusp, and
parallel to the plane of the cusp.

Gravitational lensing produces a map of an object surface onto an
image surface.  The magnification is the inverse of the Jacobian
of this map.  Because dark matter caustics have well-defined
density profiles, it is a neat mathematical exercise to calculate
their gravitational lensing characteristics.  The images of
extended sources may show distortions that can be unambiguously
attributed to lensing by dark matter caustics in the limit of
perfect observations.  We see that in three of the cases
considered, a point source can have multiple images.  In those
cases when two images merge, the Jacobian of the map vanishes and
the magnification diverges.  So, at least in theory, it seems that
gravitational lensing is a very attractive tool for investigating
dark matter caustics.  Observation of the calculated lensing
signatures would give direct evidence for caustics and CDM.

We have been particularly motivated by the possibility
\cite{Hogan} that the observer might be able to distinguish
between axions and WIMPs by determining the distance over which
the caustics are smeared.  The nearby caustic, whose position is
revealed \cite{milk} by a triangular feature in the IRAS map of
the Milky Way plane, is only 1 kpc away from us in the direction
of observation.  By observing the gravitational lensing due to
that caustic, one may be able to measure $\delta x$ as small as
$10^{13}$ cm, assuming an angular resolution of $3 \cdot 10^{-9}$
radians.  If $\delta x$ turned out to be that small, the WIMP dark
matter hypothesis would be severely challenged (Eq. \ref{dxW}).

Unfortunately, as shown below, the gravitational lensing due to a
caustic only a kpc away from us is too weak to be observed with
current instruments.  It is well known that gravitational lensing
effects are proportional to ${D_L D_{LS} \over D_S}$ where $D_S,
D_L$ and $D_{LS}$ are respectively the distances from the observer
to the source, from the observer to the lens, and from the lens to
the source. We see below that, for the gravitational lensing
effects of dark matter caustics to be observable using the current
technology, the lenses and sources must be as far away as possible
(at the cosmological distances of order Gpc). Even then, the
observation of such effects will be difficult. Unfortunately, at
Gpc distances it is not possible to measure $\delta x$ as small as
Eqs. \ref{dxa} and \ref{dxW} with foreseeable technology. So, it
seems unlikely that one will be able to distinguish between dark
matter candidates on the basis of the gravitational lensing
characteristics of the caustics they form. Henceforth, unless
otherwise stated, the velocity dispersion is set equal to zero.

The outline of this dissertation is as follows. Section\
\ref{chap:CDMcandidates } is devoted to the review of the main CDM
candidates. In Sections \ref{sect:WIMPs} and \ref{sect:Axions} we
study the WIMPs and axions, respectively. Each candidate's status
in particle physics and production mechanism in the early universe
are discussed. Calculations show that their velocity dispersions
are negligible. In Section \ref{Flow}, we study the caustics
associated with the infall of collisionless dark matter particles
onto a galaxy. In Section \ref{sect:causticsingeneral} a general
discussion of caustics is given. Caustic surfaces are introduced
in Section \ref{sec:caustic surfaces}. We prove the existence of
CDM caustics and classify them as $outer$ and $inner$. We show
that the outer caustics are topological spheres and the inner
caustics are rings with three cusps. We find $n+2$ flows (where
$n$ is always odd) inside an outer caustic; and $n$-flows outside.
Estimates of the radii and fold coefficients of the outer caustics
(based on the self similar infall model) are given in Section
\ref{sect:denprofofoutercaustics}. We find $n+4$ flows inside the
caustic rings; and $n+2$ outside. In Section
\ref{sect:axiallysymmetricrings}, we give a detailed analysis of
the caustic rings under the additional assumptions that the flow
is axially and reflection symmetric; and that the transverse
dimensions of the ring, $p$ and $q$, are small compared to the
ring radius $a$.  In Section \ref{subsect:flowatthecaustic}, we
show that under these assumptions, the flow near the ring is
described in terms of five parameters: $a$, $b$, $\tau_0$, $u$ and
$s$. We study the differential geometry of the ring surfaces
produced by this flow in Section \ref{difgeo}. We conclude the
section by estimating the density profiles of the caustic rings,
using the self-similar infall model.

In Section \ref{sec:gf}, the general formalism of gravitational
lensing is reviewed. We show how the calculations can be
streamlined for the case of gravitational lensing by dark matter
caustics.  In Section \ref{sec:fc}, we calculate the gravitational
lensing properties of dark matter caustics in the four cases
mentioned above.  In Section \ref{Conclusions}, we summarize our
conclusions.

\newpage
\section{Cold Dark Matter Candidates}
\label{chap:CDMcandidates }
\subsection{Weakly Interacting Massive
Particles (WIMPs)} \label{sect:WIMPs}

WIMPs is an acronym for weakly interacting massive particles. In
this section we derive some properties of WIMPs, in particular
their velocity dispersion, $\delta_\chi\equiv \sqrt{\langle
v^2_\chi\rangle}$, where $\langle \; \rangle$ indicates averaging,
and $\chi$ denotes a WIMP. First, we briefly introduce some
relevant concepts and calculational
tools\cite{KT,Padmanabhan,Bergstrom,leeweinberg}.

The key to understanding thermodynamics in an expanding universe
is the comparison of the particle interaction rates and the
expansion rate of the universe \cite{KT}. The interaction rate per
particle of a species is defined as \beeq \Gamma_{\rm
int}(t)\equiv n\,\langle \sigma_{\rm int} |\vec{v}|\rangle (t)
\label{annr} \eneq where $n(t)$ is the number density of target
particles, $\vec{v}$ is the relative velocity, and $\sigma_{\rm
int}$ is the interaction cross-section. Since $\sigma_{\rm int}$
is a function of energy and interacting particles have random
thermal velocities, averaging of the combination $\sigma_{\rm int}
|\vec{v}|$ is necessary. The expansion rate of the universe is
given by $H(t)\equiv \dot R(t)/R(t)$, where $R(t)$ is the scale
factor of the comoving coordinates. Since in the relativistic
regime, $n(t)$ scales as $R^{-3}(t)$, and $\langle \sigma_{\rm
int}\vec{v}\rangle$ is typically a declining function of energy,
$\Gamma_{\rm int}(t)$ decreases at least as fast as $R^{-3}\sim
t^{-3/2}$. In the non-relativistic regime, $n(t)$ and hence
$\Gamma_{\rm int}(t)$ decreases exponentially. The $H(t)$, on the
other hand, changes in time as $t^{-1}$ in both regimes. Thus,
$\Gamma_{\rm int}(t)/H(t)\rightarrow\infty$, as $t\rightarrow 0$.
In time, crossover between $\Gamma_{\rm int}(t)$ and $H(t)$ is
unavoidable. This means that there is a certain time $t_D$, called
decoupling time, such that $\Gamma_{\rm int}(t_D)\simeq H(t_D)$.
After $t_D$, a particle species effectively stops interacting with
the rest of the matter; and its distribution evolves
independently.

Next let us consider WIMPs in particular. We assume that they
carry a quantum number, like R parity in the case of the lightest
supersymmetric partner (LSP), that keeps them stable. Thus, the
number of WIMPs per comoving volume can only change via
annihilation processes mediated by $Z^0$, such as the two-body
final state annihilations: \beeq \chi\bar\chi \leftrightarrow
X\bar X  \label{reac} \eneq where $X$ denotes particle species
(like $\nu$, $e^-$, $\mu^-$, $\pi^-$, $u$, $d$, $s$, etc.). A bar
indicates the anti-particle. We represent the average annihilation
cross-section summed over all annihilation channels (not only the
two-body final state annihilations) by $\langle \sigma_{\rm
ann}|\vec{v}|\rangle$. Let us also assume that there is no
particle number asymmetry between $\chi$'s and $\bar\chi$'s, or
that it is negligible. When $T$ drops below $m_\chi$, the $\chi$
particles become nonrelativistic and their energy density
$\rho_\chi\simeq m_\chi n_\chi$ is Boltzmann suppressed as long as
it is in thermal equilibrium. Their number density is given by
\beeq n_\chi\simeq g_{\chi}\left(\frac{m_\chi
T}{2\pi}\right)^{3/2} \exp{(-m_\chi/T)} \label{nchiBoltz} \eneq
where $T(t)$ is the temperature and $g_\chi$ is the number of
internal degrees of freedom of the $\chi$'s (such as spin, color,
etc.). The number of internal degrees of freedom $g_i$ of a
species, in general, is an important quantity in the calculations
(because each adds independently to the number and energy
densities, pressure, etc.). For example, photons have two
polarization states, hence $g_\gamma=2$; neutrinos have only one
polarization state, therefore $g_\nu=1$; and electrons and muons
have $g_{e^- ,\;\mu}=2$, since they have two spin states. Internal
degrees of freedom of the antiparticles are counted independently;
and their number is the same as for the particles. It is clear
from Eq. \ref{nchiBoltz} that if a massive species remained in
thermal equilibrium until the present, its abundance would be
negligible. On the other hand, if the annihilations of the species
stop at a temperature $T_F=T(t_F)$, called the freeze-out
temperature, $n_\chi$ would decrease only as $R^{-3}$ after $t_F$,
exceeding its equilibrium value. Such particles could have a
significant relic abundance even today.

Using Eqs. \ref{annr} and \ref{nchiBoltz}, the annihilation rate
of $\chi$ can be written as \beeq \Gamma_{\rm ann}\simeq
g_{\chi}\left(\frac{m_\chi T}{2\pi}\right)^{3/2}
\exp{(-m_\chi/T)}\langle\sigma_{\rm ann}|\vec{v}|\rangle\; .
\label{Gam} \eneq The average of $\sigma_{\rm ann} |\vec{v}|$ for
the annihilation reactions is given \cite{KT} as \beeq
\langle\sigma_{ann}|\vec{v}|\rangle\equiv
\sigma_0\left(\frac{T}{m_\chi}\right)^{n} \; , \label{sigann}
\eneq (where $n=0$ for s-wave annihilation, $n=1$ for p-wave
annihilation, etc.). The $\sigma_0$ for a WIMP with mass in the
range $T\leq m_\chi\leq m_{Z^0}$ is \beeq \sigma_0\simeq
\frac{c}{2\pi} G_F^2 m_\chi^2\; , \label{sigma0} \eneq where $G_F$
is the Fermi constant. The value $c$ depends on the number of
annihilation channels and whether $\bar \chi$ is distinct from
$\chi$ (Dirac-type) or not (Majorana-type). Eqs.
\ref{Gam}-\ref{sigma0} yield \beeq \Gamma_{\rm ann}(T)\simeq
\frac{c\; G_F}{(2\pi)^{5/2}} g_{\chi}
\left(\frac{m_\chi}{T}\right)^{1/2 -n} T^2 m_\chi^3
\exp{(-m_\chi/T)}\; . \label{G} \eneq

The Hubble parameter $H(T)$ can be determined from the Friedmann
equation for the early universe: \beeq H^2(t)=\frac{8\pi
G}{3}\rho(t)\; , \label{Friedmann} \eneq where $G$ is the Newton's
constant and $\rho(t)$ is the total energy density. Since the
energy density of a nonrelativistic species in thermal equilibrium
is exponentially suppressed compared to that of a relativistic
species (assuming for the moment that no other species is out of
equilibrium), summing only the energy densities of the
relativistic particles in thermal equilibrium at a given
temperature is sufficient. Hence, the energy density has the form
of the Stefan-Boltzmann law: \beeq
\rho(T)\simeq\frac{\pi^2}{30}g_{\rm eff}(T)\; T^4\; . \label{SB}
\eneq

Here we introduce the effective degeneracy factor $g_{\rm
eff}(T)$, which counts the total number of internal degrees of
freedom of the particles that are relativistic and in thermal
equilibrium at temperature $T$ (particle species whose mass
$m_i\ll T$). The expression for $g_{\rm eff}(T)$ also contains the
factor $7/8$ for fermions relative to bosons. It is useful to
calculate $g_{\rm eff}(T)$ for a temperature (say, 1 TeV) at which
all the particles of the Standard Model were relativistic and in
thermal equilibrium. The total number of internal degrees of
freedom of the gauge and Higgs bosons is $28$ (the Higgs boson is
spinless, hence it has just one spin state. The $W^\pm$ and $Z^0$
bosons are massive spin $s=1$ particles, hence each have $2s+1=3$
spin states. The photon and each of the eight gluons are massless
spin 1 particles, therefore each has 2 helicity states: $+1$ and
$-1$. If we add them up, the total number of bosonic degrees of
freedom is $9\cdot 2+ 3\cdot 3+ 1=28$) and for fermions it is
$90$. (Each of the six quarks comes in 3 colors and 2 spins. Each
of the three charged leptons has two spin states. Each of the
three neutrinos has one helicity state. Therefore, the total
fermionic particle degeneracy is $6\cdot 6+ 3\cdot 2+ 3\cdot 1=
45$. Because of the anti-particles, the total fermionic degeneracy
doubles, hence $2\cdot 45=90$). Thus, \beeq g_{\rm eff}(T=1
\rm{TeV})= 28+\frac{7}{8}\cdot 90=106.75\;\;\; .\nonumber\eneq

For some species of particles, as happens for neutrinos, the
interaction rate becomes smaller than the expansion rate. Those
particles will have lower equilibrium temperatures than the
photons (see below), but will still remain relativistic. To allow
this possibility, we introduce a specific temperature $T_i$ for
each kind of relativistic particle and define the total number of
effective degrees of freedom as \beeq g_{\rm
eff}(T)\equiv\sum_{i={\rm boson}} g_i\left(\frac{T_i}{T}\right)^4
+\frac{7}{8}\sum_{j={\rm
fermion}}g_j\left(\frac{T_j}{T}\right)^4\, .\eneq A useful formula
can immediately be obtained by inserting Eq. \ref{SB} into the
Friedmann equation (Eq. \ref{Friedmann})\beeq H^2=\frac{8\pi
G}{3}\frac{\pi^2}{30}g_{\rm eff} T^4=2.76 \frac{g_{\rm eff}
T^4}{m^2_{\rm Pl}}\nonumber\eneq where $G$ is given as the inverse
square of the Planck mass $\equiv m_{\rm Pl}$, or \beeq
H=\frac{1.66}{m_{\rm Pl}}\sqrt{g_{\rm eff}}\; T^2\;
.\label{H1.66T2}\eneq  Since the scale factor $R(t)\sim\sqrt{t}$
in the era of radiation domination, $H(t)=\dot{R}/R=1/(2t)$. Thus,
we find the time-temperature relation in the radiation
domination:\beeq t=0.3\frac{m_{\rm Pl}}{\sqrt{g_{\rm eff}}\;
T^2}\; .\label{timetemperature}\eneq This formula is valid at
temperatures around 1 MeV, where most of nucleosythesis and
neutrino decoupling occurred.

Annihilations effectively stop at temperature $T_F=T(t_F)$ when
the ratio of the rates $\Gamma(T_F)/H(T_F)\sim 1$. Then, Eqs.
\ref{G}-\ref{SB} yield \beeq \frac{\Gamma(T_F)}{H(T_F)}\simeq
\frac{c\sqrt{90}}{(2\pi)^4}\frac{G_F^2}{\sqrt{G}}\frac{g_\chi}{\sqrt{g_{\rm
eff}}}\left(\frac{m_\chi T}{2\pi}\right)^{1/2 -n} m_\chi^3
\exp{(-m_\chi /T)}\sim 1\; . \label{dec} \eneq If the $\chi$'s are
Dirac-type particles, $c\simeq 5$ and $n=0$. Taking $g_\chi=2$ and
$g_{\rm eff}\simeq 60$, the logarithm of Eq. \ref{dec} yields
\beeq \frac{m_\chi}{T_F}\simeq
16.4+\frac{1}{2}\ln{\left(\frac{m_\chi}{T_F}\right)}
+3\ln{\left(\frac{m_\chi}{\rm GeV}\right)}\; . \eneq In this
equation, to leading order, $m_\chi/T_F \sim 16.4$. Iterating in
this order, the logarithmic term corrects the result as \beeq
\frac{m_\chi}{T_F} \simeq 17.8 + 3\ln{\left(\frac{m_\chi}{\rm
GeV}\right)}\; .\eneq Although $n_\chi R^3$ freezes out at
$T_F\sim m_\chi/20$, the energy distribution of the WIMPs is kept
thermalized by collisions with $\nu, e, {\rm etc.}$ (e.g. , $\nu
+\chi\rightarrow \nu +\chi$), down to the decoupling temperature
$T_D\sim {\rm MeV}$ where all the WIMP interactions effectively
stop and WIMPs stream freely. Neutrino decoupling takes place also
at this temperature. To see that the decoupling temperature of
neutrinos is indeed around an MeV, consider, for example, the
interactions between relativistic charged leptons and neutrinos. A
typical process maintaining thermal equilibrium, such as $\nu_e^-
+e^+\rightarrow \nu_\mu +\mu^+$ mediated by $W^+$, has a weak
interaction cross-section: \beeq \sigma_{\rm weak}\sim
\frac{\alpha^2 s}{(s-m^2_W)^2}\sim\frac{\alpha^2 s}{m^4_W}\;
,\eneq where $s$ is the square of the total four momenta of the
incoming or outgoing particles (which is also equal to the square
of the total energy in the center of momentum frame), $m_W$ is the
mass of the $W^+$ boson and $\alpha=1/137$. Since $s$ is of order
the energy squared of the reacting particles, and average energy
is proportional to $T$, the cross-section can be expressed as
\beeq \sigma_{\rm weak}\sim \frac{\alpha^2 T^2}{m_W^4}\; .\eneq
The interaction rate $\Gamma_{weak}=n\langle\sigma_{\rm
weak}|\vec{v}|\rangle$ is thus\beeq \Gamma_{\rm weak}\sim
\frac{\alpha^2 T^5}{m^4_W}\; ,\eneq where we have used
$|\vec{v}|=c=1$ and the number density for a relativistic fermion
species
$n_{i}=\frac{3}{4}\left(\frac{\zeta(3)}{\pi^2}g_iT^3\right)\sim
T^3$. Comparing the above rate with the Hubble expansion rate
$H\sim T^2/m_{\rm Pl}$ (using Eq. \ref{H1.66T2}), we have\beeq
\frac{\Gamma_{\rm weak}}{H}\sim \frac{\alpha^2 T^3 m_{\rm
Pl}}{m^4_W}\; . \eneq The decoupling occurs when the above ratio
drops below unity, implying\beeq
T_D\sim\left(\frac{m^4_W}{\alpha^2 m_{\rm
Pl}}\right)^{\frac{1}{3}}\sim 4 \; {\rm MeV}\; .\eneq

Turning back to the case of WIMPs, being non-relativistic at
$t_D$, the $\chi$s satisfy \beeq \frac{1}{2}m_\chi\langle
v_\chi^2\rangle_{T_D}=\frac{3}{2} T_D \; . \eneq Therefore the
velocity dispersion of the WIMPs at $t_D$ is \beeq \delta
v_{\chi}(T_D)=\sqrt{\langle
v_\chi^2\rangle_{T_D}}=\sqrt{\frac{3T_D}{m_\chi}}\; . \eneq For
any free particle moving in an expanding universe, momentum
decreases in inverse proportion to the scale factor:
$p(T)=p(T_D)\left(R(T_D)/ R(T)\right)$. The velocity dispersion of
the WIMPs as a function of time is therefore \beeq \delta
v_{\chi}(t)=\sqrt{\frac{3T_D}{m_\chi}}\;\frac{R(t_D)}{R(t)}
\label{vd} \; .\eneq The ratio $R(t_D)/R(t)$ can be determined
using conservation of entropy, which implies $T\sim 1/R$ (see
below). For a numerical estimate of the ratio today, depending on
the decoupling era of $\chi$s, we can either use the present
temperature of the CMBR photons $T_{0\gamma}$ or the temperature
of the relic neutrinos $T_{0\nu}$. The latter can be calculated
from $T_{0\gamma}$. The difference between $T_{0\gamma}$ and
$T_{0\nu}$ is due to the early decoupling of neutrinos and the
subsequent annihilation of electron-positron pairs: $e^-
e^+\rightarrow\gamma\gamma$ near $T=1\;{\rm MeV}$. To understand
this so called ``reheating'' effect of photons, we recall that the
expansion of the universe is adiabatic, and hence the entropy of
particles in thermal equilibrium
$S(T)=R^3(T)\left(\rho(T)+p(T)\right)/T$ is conserved. Using the
relativistic expressions for the energy density $\rho(T)=\pi^2
g_{\rm eff} T^4/30$ and the pressure $p(T)=\rho(T)/3$, the entropy
can be written as $S(T)=2\pi^2g^*_{\rm eff}(R(T)T)^3/45$. Here we
define \beeq g^*_{\rm eff}\equiv\sum_{i={\rm bos}}
g_i\left(\frac{T_i}{T}\right)^3 +\frac{7}{8}\sum_{j={\rm
fer}}g_j\left(\frac{T_j}{T}\right)^3\; .\eneq Thus, \beeq g^*_{\rm
eff}(R(T)T)^3=C \label{g} \eneq is a conserved quantity. After
decoupling, neutrinos move freely. They remain in a thermal
Fermi-Dirac distribution with a temperature $T_\nu= T_D
R(T_D)/R(T)=K R^{-1}(T)$ provided $T_\nu\gg m_\nu$, and their
entropy $S_\nu (T)$ is separately conserved. If we consider the
rest of the particles in thermal equilibrium, using Eq. \ref{g},
we calculate their temperature as $T=(C/g_{\rm eff}^*)^{1/3}
R^{-1}(T)$. Since at the time of decoupling, photons, neutrinos,
and the rest of the matter had the same temperature, $(C/g_{\rm
eff}^*(T_D))^{1/3}=K$. Thus, in spite of the decoupling of
neutrinos, as long as $g_{\rm eff}$ does not change they all
continue to have the same temperature. However $g_{\rm eff}$
changes when the temperature of the universe falls below $T\simeq
m_e$. Below this temperature, the mean energy of photons is not
sufficient to create $e^- e^+$ pairs. Hence annihilations $e^-
e^+\rightarrow\gamma\gamma$ exhaust the $e^- e^+$ pairs.
Therefore, for $T_D>T\geq m_e$, photons $(g_\gamma =2)$ are in
equilibrium with electrons $(g_{e^-}=2)$ and positrons
$(g_{e^+}=2)$ giving $g_{\rm eff}^*=2+(7/8)\cdot 4=11/2$. For
$T\ll m_e$, however, only photons are relativistic, giving $g_{\rm
eff}^*=2$. If we apply the conservation of entropy Eq. \ref{g} to
the particles which are in equilibrium with radiation before and
after $e^- e^+$ annihilation, we find \beeq \frac{\left(
R(T)T_\gamma\right)^3_{T\ll m_e}}{\left(
R(T)T_\gamma\right)^3_{T\geq m_e}}=\frac{(g_{\rm eff}^*)_{T\geq
m_e}}{(g_{\rm eff}^*)_{T\ll m_e}}=\frac{11}{4}\; . \label{114}
\eneq Just before $e^- e^+$ annihilation, neutrinos (although they
were decoupled) had the same temperature as photons: \beeq
(R(T)T_\nu)_{T\geq m_e}=(R(T)T_\gamma)_{T\geq m_e}=K\; . \eneq
Using the above equation in Eq. \ref{114}, we obtain \be \left(
R(T)T_\gamma\right)_{T\ll
m_e}&=&\left(\frac{11}{4}\right)^{1/3}\left(
R(T)T_\nu\right)_{T\geq m_e}\nonumber\\
&=&\left(\frac{11}{4}\right)^{1/3}(R(T)T_\nu)_{T\ll m_e}\; . \ee
In the last equation the constancy of $(R(T)T_\nu)$ is used. After
$e^- e^+$ annihilation, $g_{\rm eff}^*$ does not change any more.
Therefore, the relation \beeq T_\nu
=\left(\frac{4}{11}\right)^{1/3}T_\gamma
\label{temperatureneutrino}\eneq remains valid even today.

Now, let's turn back to the estimation of the velocity dispersion
given in Eq. \ref{vd}. If $\chi$s decouple before $T\geq m_e$,
they are not affected by the reheating. Therefore, \beeq \delta
v_\chi=\sqrt{\frac{3T_D}{m_\chi}}\; \frac{T_\nu}{T_D}\; .
\label{velocitydispersionneutrino}\eneq Let us estimate the
present value of the above $\delta v_\chi$ numerically. We take
$T_D\sim 1$MeV and $m_\chi\sim 100$GeV. Since the present value of
$T_\gamma=2.725{\rm K}= 2.348\cdot 10^{-13}$GeV, using Eq.
\ref{temperatureneutrino} we find $T_\nu=1.945{\rm K}= 1.676\cdot
10^{-13}$GeV. Thus, from Eq. \ref{velocitydispersionneutrino}, we
obtain the present value of the velocity dispersion as\beeq \delta
v_\chi(t_0)=9.180\cdot 10^{-12}\sqrt{\frac{1{\rm
GeV}}{m_\chi}}\sqrt{\frac{1{\rm MeV}}{T_D}}\;
,\label{presentveldispneutrino}\eneq where $t_0$ denotes the
present age of the universe. To insert the time dependence of
$\delta_\chi$ back, using Eq. \ref{vd}, we write\beeq \delta
v_\chi(t)=\sqrt{\frac{3T_D}{m_\chi}}\frac{R(t_D)}{R(t)}=
\sqrt{\frac{3T_D}{m_\chi}}\frac{R(t_D)}{R(t_0)}\frac{R(t_0)}{R(t)}\;
. \eneq Because $\sqrt{\frac{3T_D}{m_\chi}}\frac{R(t_D)}{R(t_0)}$
is the present value of the velocity dispersion $\delta
v_\chi(t_0)$ estimated in Eq. \ref{presentveldispneutrino} and
$R(t)\sim t^{\frac{2}{3}}$ in the era of matter domination, we
have\be \delta v_\chi(t)&=&\delta
v_\chi(t_0)\frac{R(t_0)}{R(t)}=\delta
v_\chi(t_0)\left(\frac{t_0}{t}\right)^{\frac{2}{3}}=9.180\cdot
10^{-12}\sqrt{\frac{1{\rm GeV}}{m_\chi}}\sqrt{\frac{1{\rm
MeV}}{T_D}}\left(\frac{t_0}{t}\right)^{\frac{2}{3}}\nonumber\\&\simeq&
10^{-11}\left(\frac{{\rm
GeV}}{m_\chi}\right)^{\frac{1}{2}}\left(\frac{{\rm
MeV}}{T_D}\right)^{\frac{1}{2}}\left(\frac{t_0}{t}\right)^{\frac{2}{3}}\;
\label{deltavW}.\ee

If on the other hand, $\chi$s decouple after $T\simeq m_e$, they
benefit from the reheating caused by $e^+ e^-$ annihilation, and
\beeq \delta v_\chi=\sqrt{\frac{3T_D}{m_\chi}}\;
\frac{T_\gamma}{T_D}\; . \label{veldispphoton}\eneq Then, in this
case, \be\delta v_\chi(t)&=&1.286\cdot 10^{-11}\sqrt{\frac{1{\rm
GeV}}{m_\chi}}\sqrt{\frac{1{\rm MeV}}{T_D}}
\left(\frac{t_0}{t}\right)^{\frac{2}{3}}\; , \ee which is not very
different from the value obtained in Eq. \ref{deltavW}.

The present average WIMP number density $n_\chi(t_0)$ can be
expressed in terms of $\Omega_\chi\equiv
\frac{\rho_\chi(t_0)}{\rho_{\rm c}(t_0)}$, which is the present
WIMP energy density in units of the critical density $\rho_{\rm
c}$:\beeq
n_\chi(t_0)=\frac{\rho_\chi(t_0)}{m_\chi}=\frac{\rho_c(t_0)
\Omega_\chi}{m_\chi}\, . \label{nden0}\eneq Recall that the
Friedmann equation for a flat universe is \beeq
H^2(t_0)=\frac{8\pi G}{3}\rho_c (t_0)\ ,\nonumber\eneq where
$H(t_0)\equiv \frac{\dot{R}(t_0)}{R(t_0)}$ is the present Hubble
expansion rate of the universe. Therefore the present critical
density is\beeq \rho_c(t_0)=8.1\cdot 10^{-11} h^2 ({\rm
eV})^4=4.85\cdot 10^{82} \frac{h^2}{({\rm pc})^4}\,
,\label{rho_c}\eneq where we used $H(t_0)=h\, 100\frac{{\rm
km}}{{\rm s}\cdot {\rm Mpc}}=h\, 3.34\cdot 10^{-10} \frac{1}{{\rm
pc}}$. Using Eqs. \ref{nden0}, \ref{rho_c} and the conversion
${\rm eV}=1.564 \cdot 10^{23}\frac{1}{\rm pc}$ we find
\begin{equation}
n_\chi(t_0) = {1.5~10^{50} \over {\rm pc}^3}\, ~\Omega_\chi~
\left({h \over 0.7}\right)^2~\left({{\rm GeV} \over
m_\chi}\right)\; . \label{nW}
\end{equation}

We next discuss another leading species of cold dark matter
candidates, namely axions.

\subsection{Axions} \label{sect:Axions} Axions have been
postulated to solve the strong CP problem of Quantum
Chromodynamics (QCD). QCD is the theory of strong interactions. It
is a gauge theory on the gauge group SU(3) of ``color.'' Its
dynamical variables are a color octet of gauge fields $A_\mu^a$,
called gluons, and a family of color triplet spinor fields $q_f$,
called quarks. The flavour index $f$, labels the various triplets.
$A_\mu^a$ is analogous to the vector potential of electrodynamics.
Here, in addition to the spacetime index $\mu$, the gauge field
carries the $SU(3)$ group index $a$. In an analogy with Quantum
Electrodynamics (QED), the action for the gauge sector is at first
taken to be\beeq {{\mathcal{L}}_{\rm gauge}} =-\frac{1}{4}\int
d^4x G^{a\mu\nu}G_{\mu\nu}^a \eneq where \beeq
G_{\mu\nu}^a=\partial_\mu A_\nu^a-\partial_\nu A_\mu^a + g f^{abc}
A_\mu^b A_\nu^c \eneq is the gluon field strength tensor, $g$ is
the gauge coupling, and $f^{abc}$ are the structure constants of
the $SU(3)$ algebra: $[T^a , T^b]=if^{abc}T^c$. The analogy
between QCD and QED, however, fails where the vacuum structure of
the theories is concerned. In both theories, the vacuum
configuration of the theory is characterized by the vanishing of
the field strength tensor $F_{\mu\nu}$. In QED, however, all the
vector potentials which yield $F_{\mu\nu}=0$, are related to each
other by a gauge transformation. For any vacuum configuration,
$A_\mu$ can be made to vanish by a gauge transformation; see the
Appendix. Thus the QED vacuum is unique. In QCD, on the other
hand, not all solutions of $G^a_{\mu\nu}=0$ are related to each
other by a gauge transformation. These physically distinct vacuum
states can be classified by the set of integers, $n$, and hence
the most general ground state $|\Theta\rangle$, called
$\Theta$-vacuum, can be expressed as a superposition of all the
degenerate states $|n\rangle$: \beeq |\Theta\rangle
=\sum_{n=-\infty}^{\infty} \exp{(-in\Theta)}|n\rangle \; ,\eneq
where the extra parameter $\Theta$ has period $2\pi$, and needs to
be measured; see the Appendix. The effects of the $\Theta$-vacuum
can be included in the Lagrangian density of QCD by the addition
of a topological term \beeq
{{\mathcal{L}}}_{\Theta}=-\frac{g^2\Theta}{32\pi^2}
G^{a\mu\nu}{}^*{G}^a_{\mu\nu}=\partial_\mu K_\mu\;
,\label{LTK}\eneq where (Eq. \ref{totder}) \beeq
K_\mu\equiv-\epsilon_{\mu\nu\rho\sigma}{\rm Tr}(
A_\nu\partial_\rho A_\sigma -i\frac{2g}{3}A_\nu A_\rho A_\sigma)\;
.\nonumber\eneq Such a term affects neither the equations of
motion nor the perturbative aspects of the theory (Feynman rule
for the vertex associated with a four-divergence would necessarily
have a multiplicative factor of the sum of the momenta at the
vertex which vanishes automatically due to momentum conservation).
Therefore, the existence of the $\Theta$-term in a Lagrangian does
not necessarily mean that physics depends on it. As remarked in
the above, Classical Electrodynamics and perturbative QED are
unaffected by the presence of such a term in their Lagrangians. In
QED, if there are non-perturbative effects, they are absolutely
negligible. This is because, the essential structure of a
non-perturbative effect can be represented by
$e^{-\frac{1}{\alpha_{e}}}$, which vanishes as the coupling
constant $\alpha_{e}\rightarrow 0$ (when the interaction is
turned-off) and can not be expanded in a Taylor series in
$\alpha_{e}$. Since $\alpha_{e}=1/137$, the $\Theta$-dependence
would be highly suppressed. Non-perturbative effects, however,
produce $\Theta$-dependence in QCD. These effects must be present
in the theory, otherwise QCD would not produce the phenomenology
at low energies.

To illustrate the claim given in Eq. \ref{LTK}, we evaluate the
vacuum to vacuum transition amplitude in the $SU(3)$ gauge theory:
\beeq
\langle\Theta'|e^{-iHt}|\Theta\rangle=\sum_{m,s}e^{im\Theta'}e^{is\Theta}
\langle m|e^{-iHt}|s\rangle\; . \label{vakvak} \eneq Recall that
the expectation value $\langle m|e^{-iHt}|s\rangle$ can be written
as the path integral over all $\left[A^a_\mu\right]_{s-m}$ that
connects the $m$th vacuum with the $s$th vacuum: \beeq \langle
m|e^{-iHt}|s\rangle =\int \left[DA_\mu\right]_{s-m} \exp{(-i\int
{\mathcal{L_{\rm gauge}}} d^4x)}\; . \eneq Then Eq. \ref{vakvak}
becomes \be\langle\Theta'|e^{-iHt}|\Theta\rangle &=&
\sum_{s,m}e^{-i(s-m) \Theta}e^{im(\Theta'-\Theta)}
\int\left[DA_\mu\right]_{s-m}
\exp{(-i\int {{\mathcal{L}}}_{\rm gauge} d^4x)}\nonumber\\
&=&\delta(\Theta -\Theta')\sum_n e^{in\Theta}\int
\left[DA_\mu\right]_{n}e^{-i\int d^4x {{\mathcal{L}}_{\rm
gauge}}}\; , \ee where the delta function is obtained by summing
over $m$, after $s$ was replaced by $n=s-m$. The phase factor
$e^{in\Theta}$ can be absorbed into the action using \beeq
n=\frac{g^2}{32\pi^2}\int d^4x G^{a\mu\nu}{}^*{G}^a_{\mu\nu}\; ,
\eneq where $g$ is the strong coupling constant and
${}^*{G}^{a\mu\nu}=\frac{1}{2}\epsilon^ {\mu\nu}\,_{\lambda\rho}
G^{a\lambda\rho}$ is the dual field strength tensor. Therefore,
the effective QCD Lagrange density is the sum of \beeq
{{\mathcal{L}}}_{\Theta}=-n\Theta=-\frac{\Theta g^2}{32\pi^2}\int
d^4x G^{a\mu\nu}{}^*{G}^a_{\mu\nu}, \label{LT} \eneq
${\mathcal{L}}_{\rm gauge}$ and the Dirac Lagrangian density for
the quark fields: \beeq {{\mathcal{L}}}_{ QCD}=\sum_f\left[\bar
q_f i {D\!\!\!\!\!\!\not} \;\;\;
q_{f}-(m_f{q}^\dagger_{Lf}q_{Rf}+h.c.)\right] -\frac{1}{4}
G^{a\mu\nu} G^a_{\mu\nu}-\frac{\Theta g^2}{32\pi^2}
G^{a\mu\nu}{}^*{G}^a_{\mu\nu}\; , \eneq where we define
${D\!\!\!\!\!\!\not}\;\,\equiv\gamma^\mu D_\mu =\gamma^\mu\left(
\partial_\mu -igA_\mu^a
T^a\right)$, $q_{Lf}\equiv(1/2)(1-\gamma_5)q_f$ and
$q_{Rf}\equiv(1/2)(1+\gamma_5)q_f$. The scale dependence of the
strong coupling constant $g$ is given at one loop by
\beeq\alpha_s(\mu)\equiv
\frac{g^2(\mu)}{4\pi}=\frac{2\pi}{(11-\frac{2}{3}n_f)\ln{(\frac{\mu}{\Lambda_f})}}\,
, \label{alpha_s}\eneq where $n_f$ is the number of quark flavors
with mass less $\mu$ and $\Lambda_f$ is the appropriate QCD scale.
The $m_f$ denote the quark masses. They originate in the
electroweak sector of the Standard Model which must violate CP to
explain $K_L\rightarrow 2\pi$. Hence, $m_f$ are complex numbers in
general.

In the Standard Model, the $\Theta$ term receives a contribution
due to electroweak effects involving the $m_f$. To see the reason,
consider the effect in the path integral \beeq Z[\eta , \bar\eta ,
j_\mu]=\int [Dq][D\bar q][DA]\exp{\left(i\int
d^4x\left[{{\mathcal{L}}}_{QCD}+\bar q \eta +\bar\eta q+j_\mu
A^\mu\right]\right)}\; , \label{Z} \eneq of a redefinition of all
the quark fields \beeq q_f\rightarrow
q'_f=\exp{(i\beta_f\gamma_5)}q_f\; , \label{kiral} \eneq where
$\alpha_f$ are a set of real phases. Recall that $\{\gamma_\mu ,
\gamma_5\}=0$, $\gamma^\dagger_5=\gamma_5=\gamma^{-1}_5$ and
$\gamma_5\gamma_\mu=-\gamma_\mu\gamma_5$. Therefore, \be
\exp{(i\beta_f(x)\gamma_5)}\gamma_\mu&=&\sum_{n=0}^\infty\frac{(i\beta_f(x))^n}{n!}
\gamma_5^n\gamma_\mu=\gamma_\mu\sum_{n=0}^\infty\frac{(i\beta_f(x))^n}{n!}
(-1)^n\gamma_5^n\nonumber\\
&=&\gamma_\mu\exp{(-i\beta_f(x)\gamma_5)}\; , \ee hence, \be \bar
q'_f(x)&=&{q'}^\dagger_f(x)\gamma^0=\left({\rm
e}^{i\beta_f(x)\gamma_5}q_f(x)\right)
^\dagger\gamma^0=q^\dagger_f(x){\rm
e}^{-i\beta_f(x)\gamma_5^\dagger}\gamma^0\nonumber\\
&=&q^\dagger_f(x)\gamma^0{\rm e}^{i\beta_f(x)\gamma_5}
=\bar{q}_f(x) {\rm e}^{i\beta_f(x)\gamma_5} \; .\ee The
transformations of the terms in the exponent of Eq. \ref{Z} can
then be calculated easily: \beeq \bar q'_f i\gamma_\mu\partial^\mu
q'_f=\bar{q}_f{\rm
e}^{i\beta_f(x)\gamma_5}i\gamma_\mu\partial^\mu{\rm
e}^{i\beta_f(x)\gamma_5}q_f\nonumber\eneq \beeq =\bar{q}_f{\rm
e}^{i\beta_f(x)\gamma_5}i\gamma_\mu\left[(i\partial^\mu\beta_f(x)\gamma_5){\rm
e}^{i\beta_f(x)\gamma_5}q_f +{\rm
e}^{i\beta_f(x)\gamma_5}\partial^\mu q_f\right]\nonumber\eneq\beeq
=-(\partial^\mu\beta_f(x))\bar{q}_f{\rm
e}^{i\beta_f(x)\gamma_5}\gamma_\mu\gamma_5{\rm
e}^{i\beta_f(x)\gamma_5}q_f+\bar{q}_fi
{\partial\!\!\!\!\!\!\not}\;\;\; q_f
=-(\partial^\mu\beta_f(x))\bar{q}_f\gamma_\mu\gamma_5
q_f+\bar{q}_fi {\partial\!\!\!\!\!\!\not}\;\;\; q_f\;
,\nonumber\eneq and \beeq\bar q'_f \gamma_\mu A^\mu q'_f=\bar q_f
\gamma_\mu {\rm e}^{-i\beta_f(x)\gamma_5}A^\mu {\rm
e}^{i\beta_f(x)\gamma_5} q_f=\bar q_f \gamma_\mu A^\mu q_f \;
.\nonumber\eneq The mass term in ${{\mathcal{L}}}_{QCD}$ is \beeq
{{\mathcal{L}}}_{\rm mass}=-\frac{1}{2}\sum_f m_f \bar q_f
(1+\gamma_5)q_{f} -\frac{1}{2}\sum_f m^\ast_f \bar q_f
(1-\gamma_5)q_f\; . \eneq Since the chiral transformation gives
\beeq m_f\bar q'_f q'_f=m_f\bar q_f {\rm e}^{i2\beta_f(x)\gamma_5}
q_f=m_f\left[ \cos{(2\beta_f)}\bar q_f q_f+i\sin{(2\beta_f)}\bar
q_f \gamma_5 q_f \right]\; ,\eneq \beeq m_f\bar q'_f \gamma_5
q'_f=m_f\bar q'_f \gamma_5 {\rm e}^{i2\beta_f(x)\gamma_5}
q_f=m_f\left[ \cos{(2\beta_f)}\bar q_f \gamma_5 q_f
+i\sin{(2\beta_f)}\bar q_f q_f \right]\; , \eneq the
${{\mathcal{L}}}_{\rm mass}$ transforms to \beeq
{{\mathcal{L}}}'_{\rm mass}=-\frac{1}{2}\sum_f  m'_f \bar q_f
(1+\gamma_5)q_{f} -\frac{1}{2}\sum_f {m'}^\ast_f \bar q_f
(1-\gamma_5)q_f\; , \eneq where \beeq m'_f=m_f{\rm
e}^{i2\beta_f(x)}\ ,\nonumber \eneq \beeq {m'}^\ast_f=m^\ast_f{\rm
e}^{-i2\beta_f(x)}\; . \eneq Finally, the measure for the path
integral over quark fields transforms \cite{Fujikawa} as \beeq
[dq][d\bar q]\rightarrow \exp{\left(\frac{-ig^2}{32\pi^2}\int d^4x
G^{a\mu\nu}{}^*{G}^a_{\mu\nu}\sum_f 2 \beta_f\right)}[dq] [d\bar
q]\; . \label{meas} \eneq Hence: \beeq Z=Z[\eta' , {\bar\eta}' ,
j_\mu]=\int [d q][d\bar q][DA]\exp{\left(i\int
d^4x\left[{{\mathcal{L}}}_{\beta(x)}+\bar q \eta' +{\bar\eta}'
q+j_\mu A^\mu\right]\right)}\; , \eneq where, in the source terms,
we shifted the chiral transformation from the fields to the
sources: \beeq \eta'_f (x)\equiv{\rm e}^{i\beta_f(x)\gamma_5}\eta
(x) \; ,\nonumber\eneq \beeq {\bar\eta}'_f (x)\equiv\bar\eta
(x){\rm e}^{i\beta_f(x)\gamma_5}\; , \eneq and \be
{{\mathcal{L}}}_{\beta(x)}=\sum_f\Big\{\bar q_f
i{D\!\!\!\!\!\!\not}\;\;\; q_{f}-\left({\rm
e}^{i2\beta_f}{m}_f{q}^\dagger_{Lf}q_{Rf}+h.c.\right)-(\partial^\mu\beta_f)\bar
q_f \gamma_\mu\gamma_5 q_f\Big\}\nonumber\\
 -\frac{1}{4} G^{a\mu\nu}
G^a_{\mu\nu} -\frac{(\Theta+2\sum_f\beta_f)g^2}{32\pi^2}\,
G^{a\mu\nu}{}^*{G}^a_{\mu\nu}\; . \label{L} \ee Using Noether's
theorem, which relates the divergence of the current associated
with an infinitesimal continuous transformation on the fields to
the change in the Lagrangian density under that transformation, we
calculate the current associated with the chiral transformations:
\beeq q'_f(x)=(1+i\beta_f(x)\gamma_5)q_f(x)+O(\beta_f^2(x))\;
,\nonumber \eneq \beeq
\bar{q}'_f(x)=\bar{q}(x)(1+i\beta_f(x)\gamma_5)+O(\beta_f^2(x))
\label{inf} \; .\eneq The infinitesimal form of Eq. \ref{L} is
\beeq
{{\mathcal{L}}}_{\beta(x)}={{\mathcal{L}}}_{QCD}-\sum_f[(\partial^\mu\beta_f)\bar
q_f \gamma_\mu\gamma_5
q_f+i2\beta_f({m}_f{q}^\dagger_{Lf}q_{Rf}+h.c.)+\frac{\beta_f
g^2}{16\pi^2} G^{a\mu\nu}{}^*{G}^a_{\mu\nu}]\nonumber\; . \eneq
The current associated with the transformation (Eq. \ref{inf}) is
\beeq J^5_\mu=-\sum_f\frac{\partial{{\mathcal{L}}}_\beta}
{\partial(\partial^\mu\beta_f)}=\sum_f \bar q_f \gamma_\mu\gamma_5
q_f \; .\eneq Noether's theorem states \beeq
\partial^\mu
J_\mu^5=-\sum_f\frac{\partial{{\mathcal{L}}}_\beta}{\partial\beta_f}=\sum_f
i2({m}_f{q}^\dagger_{Lf}q_{Rf}+h.c.)+\frac{g^2}{16\pi^2}
G^{a\mu\nu}{}^*{G}^a_{\mu\nu} \; .\eneq The explicit dependence of
${{\mathcal{L}}}_\beta$ on $\beta_f(x)$ in Eq. \ref{inf} yields
the non-conserved axial vector current \beeq
\partial_\mu J^\mu_5=\partial_\mu (\sum_f\bar q_f \gamma_\mu\gamma_5
q_f)=\sum_f[i2({m}_f
{q}^\dagger_{Lf}q_{Rf}+h.c.)+\frac{g^2}{16\pi^2}
G^{a\mu\nu}{}^*{G}^a_{\mu\nu}] \; .\eneq The first term on the
right hand side is due to the fermion mass, whereas the second is
due to the anomaly which is a quantum mechanical effect (in fact
the anomaly term is proportional to $g^2\hbar$).

Equation \ref L implies \cite{axionreview} that the physics of the
theory defined by ${{\mathcal{L}}}_{QCD}$ is unchanged under the
{\it global} (spacetime independent) transformations: \be
q_f&\rightarrow& {\rm e}^{i\beta_f\gamma_5}q_f\nonumber\\
m_f&\rightarrow& {\rm e}^{-i2\beta_f} m_f\nonumber\\
m_f^\ast&\rightarrow& {\rm e}^{i2\beta_f} m_f^\ast
\nonumber\\
\Theta&\rightarrow&\Theta - 2\sum_f\beta_f\; . \label{alp} \ee In
general, this set of transformations is not a symmetry of the
theory because, when the dynamical variables $q_f$ are
transformed, the parameters, $m_f$ and $\Theta$, are needed to be
transformed, to leave the Lagrangian invariant. If the physics
were $\Theta$ independent (transforming $\Theta$ would not make
any difference), then one would recover the classical result that
the theory has an axial $U_A(1)$ symmetry when $m_f=0$ ($m_f$
transformations would not exist). We know, however, that $U_A(1)$
is explicitly broken because otherwise there would be a fourth
pseudo Nambu-Goldstone boson $\eta'$, in addition to the three
pions, with $m_{\eta'}<\sqrt{3}m_\pi$. This can only be understood
if physics is $\Theta$-dependent. In this case, Eq. \ref{alp} is
not a symmetry even in the massless limit, because the parameter
$\Theta$ is transformed. Hence, $U_A(1)$ is broken even when the
quark masses are zero. (If there is $\Theta$-dependence
$\partial_\mu J_5^\mu\not= 0$, due to the
$G^{a\mu\nu}{}^*{G}^a_{\mu\nu}$-term, even if $m_f=0$. Hence the
anomaly explicitly breaks $U_A(1)$). Therefore, QCD must have
$\Theta$-dependence through non-perturbative effects (as noted
earlier, the $\Theta$-term is a four-divergence and hence does not
contribute in the perturbative quantum theory). These effects that
break $U_A(1)$ are known as QCD instanton effects; see the
Appendix. The instanton in QCD describes a classically forbidden,
but quantum mechanically allowed vacuum tunneling, in which the
charge $Q_f^5=\int d^3x q_f^\dagger \gamma_5 q_f$ associated with
$U_A(1)$ symmetry is not conserved. Therefore, the instantons
explicitly violate $U_A(1)$. The probability amplitude of an
instanton event (see the Appendix) \beeq
{{\mathcal{A}}}\sim\exp\left({-\frac{2\pi}{\alpha_s(\mu)}}\right)
\, ,\eneq where $\mu$ is the inverse of the instanton size and
$\alpha_s$ is defined in Eq. \ref{alpha_s}. This amplitude can not
be expanded in a Taylor series in $\alpha_s$. Thus, the instanton
is indeed a non-perturbative effect. At a finite $T$, using Eq.
\ref{alpha_s}, we have
\beeq{{\mathcal{A}}}\sim\exp\left({-\frac{2\pi}{\alpha_s(T)}}\right)
=\left(\frac{\Lambda_{QCD}}{T}\right)^{11-\frac{2}{3}n_f}\, .\eneq
At high temperatures non-perturbative (instanton) effects are
suppressed. Around $T\sim \Lambda_{QCD}$, instanton effects fully
turn-on and strongly break $U_A(1)$ symmetry of QCD.

The terms $i\bar{q}_f \gamma_5 q_f$ and
$G^{a\mu\nu}{}^*{G}^a_{\mu\nu}$ are P and CP odd. Hence their
presence in the theory (in the ${{\mathcal{L}}}_{\rm mass}$ and
${{\mathcal{L}}}_\Theta$, respectively) leads in general to P and
CP violation. Moreover, the transformations given in Eq. \ref{alp}
allow us to change the $m_f$ (for each flavor) and $\Theta$. The
apparent source of CP violation can be moved among the $m_f$ and
$\Theta$, by changing the phases of the quark masses. Since there
are six transformations on seven parameters, we can not find a
definition of quark fields that sets all the parameters real. We
might choose, for example, the $\Theta$ as zero, $m_u$, $m_d$, . .
., $m_b$ as real and $m_t$ as complex. One combination that can
not be removed from the theory is $\Theta-{\rm Arg}\prod_f
m_f=\Theta -2\sum_f\beta_f$. Hence, the $\Theta$ dependence of QCD
comes only through the combination of parameters
\beeq\bar\Theta\equiv\Theta-{\rm Arg}\prod_f m_f =\Theta
-2\sum_f\beta_f\ .\eneq Let us also note that the transformations
given in Eq. \ref{alp} leaves the $\bar\Theta$ invariant. For
example lets transform the up-quark field only, then for the
invariance, we have\be
q_u&\rightarrow& {\rm e}^{i\beta_u\gamma_5}q_u\nonumber\\
{\rm Arg} (m_u)&\rightarrow& {\rm Arg} (m_u) -2\beta_u\nonumber\\
\Theta&\rightarrow&\Theta - 2\beta_u\; , \ee and hence\beeq
\bar\Theta\rightarrow\bar\Theta\ . \eneq Let us suppose that
$\bar\Theta\not= 0$. Then, no matter how we shuffle the phases of
quark masses, it is impossible to set $\Theta=0$, and ${\rm Arg}
(m_f)=0$, $\forall f$. Therefore, it is inevitable to have some CP
violation, with an upper limit consistent with the experiments.

As we remarked, when $\bar\Theta\not= 0$, the $\bar\Theta$ term in
${{\mathcal{L}}}_{QCD}$ violates the P and T (and hence CP)
symmetries.  Since \beeq P: \;\; E^a_i\rightarrow -E^a_i\hskip 2cm
B^a_i\rightarrow B^a_i\eneq and\beeq T: \;\; E^a_i\rightarrow
E^a_i\hskip 2.3cm B^a_i\rightarrow -B^a_i \; ,\eneq
$G^{a\mu\nu}{}^*{G}^a_{\mu\nu}=-4E^a_i B^a_i$ is odd under $P$ and
$T$. Therefore, the $\bar\Theta$-term would induce a CP violating
term in the effective $\pi$-N coupling and produce an electric
dipole moment for the neutron. It is of order
\cite{Baluni,Crewther,Ramsey,Altarev1,Smith,Altarev2}$d_n\simeq
5\cdot 10^{-16}|\bar\Theta| \rm{e\cdot cm}$. Comparison with the
experimental bound $d_n\leq 10^{-25} \rm{e\cdot cm}$ constrains
\cite{Dress} $\bar\Theta\leq 10^{-10}$. This fine tuning
requirement on the parameter $\bar\Theta$ is known as ``the strong
CP problem.'' It is puzzling that the two independent
contributions in the $\bar\Theta$, $\Theta$ due to the vacuum
structure of QCD, and $2\sum_f\beta_f$ due to the electroweak
effects involving quark masses, seem to cancel each other.

The most elegant solution of the strong CP puzzle is to replace
the parameter $\bar \Theta$ by a dynamical field $\bar\Theta(x)$.
This is achieved by introducing a Higgs field whose expectation
value $v_{PQ}$ $spontaneously$ breaks a postulated global
$U(1)_{PQ}$ Peccei-Quinn (PQ) Symmetry
\cite{PecceiQuinn1,PecceiQuinn2}. The axion is the Nambu-Goldstone
boson associated with this $spontaneous$ symmetry breaking. The
defining properties of PQ symmetry
are\cite{axionreview}:\\
\hskip 1cm 1) it is an exact global symmetry of the classical
theory,
before anomalies,\\
\hskip 1cm 2) it is broken spontaneously,\\
\hskip 1cm 3) it is broken explicitly by the QCD instanton effects
which make the theory $\Theta$ dependent. To see how the PQ
mechanism works, consider the theory defined by \be
{{\mathcal{L}}}_{QCD+PQ}\!\!\!&=&\!\!\!-\frac{1}{4}
G^a_{\mu\nu}G^{a\mu\nu}+\frac{1}{2}\partial_\mu\phi^\ast\partial^\mu\phi-
\frac{\Theta}{32\pi^2}
G^a_{\mu\nu}{\tilde{G}}^{a\mu\nu}\nonumber\\
&&\!\!\!+\sum_f\left[\bar q_f {D\!\!\!\!\!\!\not}\;\;\;
q_f-(K_f{q}^\dagger_{Lf} q_{Rf}\phi +h.c.)\right]-V(\phi^\ast\phi)
\label{LPQ} \; ,\ee where $\phi$ is a complex scalar field. Under
$U(1)_{PQ}$, $\phi\rightarrow {\rm e}^{i\alpha}\phi$ and
$q_f\rightarrow {\rm e}^{i\alpha\gamma_5/2}q_f$. The potential $V$
is a ``Mexican hat'' potential; hence, $U(1)_{PQ}$ is
spontaneously broken: \beeq \langle \phi(x)\rangle =v_{PQ} {\rm
e}^{i\alpha(x)} \; .\eneq The quarks acquire masses: $m_f=K_f
v_{PQ} {\rm e}^{i\alpha(x)}$, and hence \beeq \bar\Theta(x)=\Theta
-{\rm Arg}(\prod_f m_f)=\Theta -{\rm Arg}(\prod _f
K_f)-\sum_f\alpha(x)=\Theta -{\rm Arg}(\prod _f K_f)-N\alpha(x)
\label{newalp} \; .\eneq In this toy model $N$ is the number of
quarks. Notice that here $\bar\Theta(x)$ is a function of the
dynamical field $\alpha(x)$, whereas in Eq. \ref{alp} $\bar\Theta$
is fixed. One can adopt a convention such that $\Theta -{\rm
Arg}(\prod_f m_f)=0$. In Eq. \ref{LPQ} the non-perturbative
effects which make the theory $\bar\Theta$-dependent produce an
effective potential for $\bar\Theta$. The field $\bar\Theta$
relaxes to the minimum of that potential. Therefore the axion is
not massless but acquires a small mass due to non-perturbative QCD
(instanton) effects. The zero temperature mass is given in terms
of up and down quark masses $m_u$ and $m_d$, and the pion mass
$m_\pi$ \cite{Weinberg,Wilczek} as\beeq m_a=\frac{\sqrt{m_u
m_d}}{m_u+m_d}\frac{m_\pi f_\pi}{f_a}\ ,\eneq where
$f_a=\frac{v_{PQ}}{N}$ and $f_\pi$ are the axion and pion decay
constants, and $N$ is a positive integer that describes the color
anomaly of $U_{PQ}(1)$. (Axion models have $N$ degenerate vacua
\cite{OfSikivie}). Numerically, \beeq m_a\simeq 6\cdot\; {\rm
eV}\left(\frac{10^{6}\; {\rm GeV}}{f_a}\right)\; .
\label{masszeroT}\eneq

We now study the cosmological implications of axion models
\cite{AbbottSikivie,PWW,DineF}. Let the potential for $\phi(x)$ in
Eq. \ref{LPQ} be \beeq
V(\phi)=\frac{\lambda}{4}(|\phi|^2-v^2_{PQ})^2\; . \eneq At
extremely high temperatures $U_{PQ}(1)$ symmetry is restored,
$\langle\phi(x)\rangle=0$. As the universe expands and cools below
$T_{PQ}\simeq v_{PQ}$, the potential looks like a ``Mexican hat''
and is minimized for \beeq \phi=v_{PQ}\exp{(i\alpha(x))}\; . \eneq
Notice that $V(\phi)$ is independent of $\alpha(x)$. This massless
degree of freedom is the axion: $a(x)\equiv v_{PQ}\alpha(x)$. The
system settles down in one of the continuously degenerate ground
states which are all equally likely. At much lower temperatures
$T\sim \Lambda_{QCD}$, axions acquire mass. This is because
$U(1)_{PQ}$ is {\it explicitly} broken by non-perturbative QCD
(instanton) effects at low temperatures. At high temperatures
these effects are suppressed. When $T$ approaches $\Lambda_{\rm
QCD}$, instanton effects turn on. They produce the effective
potential \beeq \tilde{V}(\bar\Theta)=m^2_a(T)
\frac{v^2_{PQ}}{N^2}(1-\cos{(\bar\Theta)})\; ,
\label{effectivepotential}\eneq where \cite{GrossPisarskiYaffe}
\beeq m_a(T)\simeq 0.1 m_a \left(\frac{\Lambda_{
QCD}}{T}\right)^{3.7}\; . \label{mass} \eneq The minimum of the
total potential is at the CP conserving value
$\bar\Theta(x)=N\alpha(x)=0$. The axion develops mass $m_a$, due
to the curvature of the potential at the minimum. The effective
Lagrangian density for the axion field is obtained by setting
$\alpha(x)=\frac{a(x)}{v_{PQ}}$: \beeq {{\mathcal{L}}}_{\rm
eff}=\frac{1}{2}\partial_\mu a \partial^\mu
a-m^2_a(T)\frac{v_{PQ}^2}{N^2} (1-\cos{(\frac{Na}{v_{\rm PQ}})})
\; , \eneq where we implicitly assumed that the signature of the
spacetime is $(+ - - -)$. Variation of the action $S=\int d^4x
\sqrt{-g} {{\mathcal{L}}} _{\rm eff}$ yields the sine-Gordon
equation for $\alpha(x)$: \beeq
\square\alpha(x)+\frac{1}{N}{m_a^2(T)}\sin{(N\alpha(x))}=0\; ,
\label{SG} \eneq where d'Alambertian $\square\equiv
(1/\sqrt{-g})\partial_\mu(\sqrt{-g} g^{\mu\nu}\partial_\nu)$. In a
Friedmann universe where \beeq ds^2=dt^2-R^2(t)d\vec{x}\cdot
d\vec{x} \label{met} \eneq Eq. \ref{SG} reads \beeq
\ddot\alpha+3H(t)\dot\alpha-\frac{1}{R^2(t)}\nabla^2\alpha
+\frac{1}{N}m^2_a\left(T(t)\right)\sin{(N\alpha)}=0\label{Dal}\;
.\eneq Near the minima, the potential (Eq.
\ref{effectivepotential}) $\tilde{V}(\alpha)\simeq
\frac{1}{2}m_a^2 v^2_{PQ}\alpha^2$. Hence, $\sin{(N\alpha)}\simeq
N\alpha$ in Eq. \ref{Dal}. In that case, the solution of Eq.
\ref{Dal} is a linear superposition of eigenmodes with definite
comoving wave vector $\vec{k}$: \beeq \alpha(\vec{x},t)=\int
d^3k\; \alpha(\vec{k}, t) e^{i\vec{k}\cdot \vec{x}} \; ,\eneq
where the $\alpha(\vec{k}, t)$ satisfy \beeq
\left(\partial_t^2+3\frac{\dot{R}(t)}{R(t)}\partial_t+\frac{k^2}{R^2(t)}
+m^2_a(t)\right)\alpha(\vec{k}, t)=0\; . \label{mod} \eneq We are
interested in the $k=0$ mode of vacuum misalignment contribution.
The other contributions (of similar magnitude) are discuss by
Chang, Hagmann and Sikivie \cite{SikivieColdAxPop}. Then Eq.
\ref{Dal} becomes \beeq \ddot\alpha +3H(t)\dot\alpha
+m^2_a(t)\alpha =0\; . \label{osc} \eneq This is the equation of a
damped harmonic oscillator with time dependent parameters. Since
the axion is massless when $T\simeq T_{PQ}$, no initial value
$\alpha_1$ of $\alpha$ is preferred dynamically. Although
$\alpha\simeq 0$ today, there is no reason to expect that was the
case initially. Therefore, $\alpha_1$ is chosen by some stochastic
process. For $m_a=0$ (as is the case for $T\gg\Lambda_{QCD}$), the
solution of Eq. \ref{osc} is obtained by inserting the ansatz
$\alpha=t^p$ to the differential equation and solving the
resulting algebraic equation $p^2+p/2=0$. The most general
solution is\beeq \alpha=\alpha_1 +\alpha_2 t^{-1/2}\; .\eneq
Quickly $\alpha$ approaches $\alpha_1$, which is a constant, and
starts to oscillate when $T$ approaches the critical temperature
$T_1$ where \cite{KT}\beeq m_a\left(T_1(t_1)\right)\sim
3H\left(T_1(t_1)\right)=\frac{3}{2t_1}\; . \label{m3H} \eneq To be
more specific let's define the critical temperature as \beeq
m_a\left(T_1(t_1)\right)t_1=\frac{3}{2}\ .\label{criticalt}\eneq

We will see below (Eq. \ref{omegasuba}) that $\Omega_a\sim 0.4$
when $m_a\sim 6\cdot 10^{-6} \rm{eV}$ (Eq. \ref{masszeroT}). In
that case $T_1$ is about $1{\rm GeV}$ (Eq. \ref{T_1ax}). The
typical wavelength of these modes is of the horizon size back
then. The corresponding momenta of the particle excitations at
birth are therefore: \beeq p_a(t_1)\sim
1/d_H=1/t_1=\frac{\sqrt{g_{\rm eff}}}{0.3}\frac{T_1^2}{m_{\rm
Pl}}\sim 1.6\cdot 10^{-9} {\rm eV}\left(\frac{10^{12}{\rm
GeV}}{f_a}\right)^{\frac{2}{5.7}}\; , \label{p_1} \eneq where we
used time-temperature relation Eq. \ref{timetemperature}. Since
their mass at that time $m_a(t_1)\sim 2.4\cdot 10^{-9} {\rm eV}
\left({10^{12}{\rm GeV}}/{f_a}\right)^{\frac{2}{5.7}}$, they were
semi-relativistic. Axions, however, gain mass rapidly. As the
temperature drops to $T\simeq \Lambda_{QCD}\sim 200 {\rm MeV}$,
their mass increases by a factor of $230\cdot\left( {10^{12}{\rm
GeV}}/{f_a}\right)^{\frac{3.7}{5.7}}$, and axions suddenly become
non-relativistic.\begin{figure}[ht] \centering
\includegraphics[height=6cm,width=13cm]{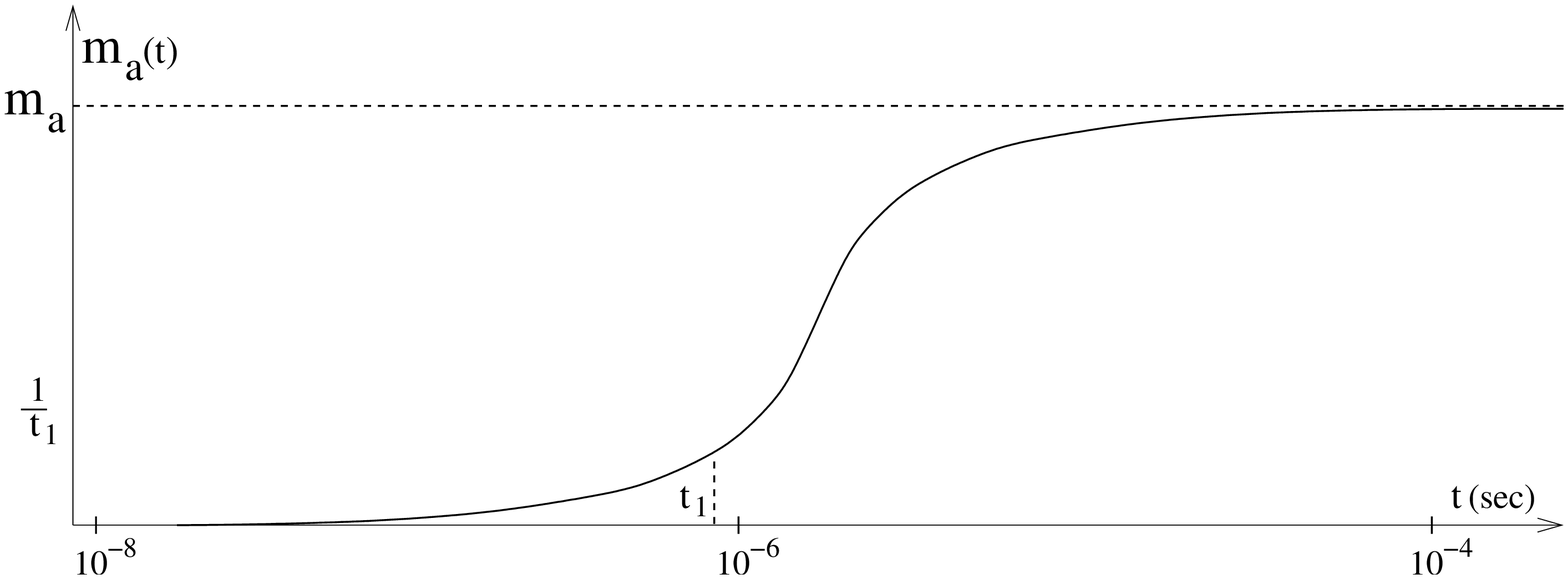}
\caption{Switch-on of the axion mass in the early universe.}
\label{fig:axionmass}
\end{figure}Because the axion couplings are small, energy dissipation of the
modes is negligible. The coherent oscillations produced by the
vacuum misalignment are good dark matter candidates.

The present velocity dispersion of these axions can be calculated
using Eq. \ref{p_1} and the conservation of entropy (Eq.
\ref{g}):\beeq g_{\rm eff}^*(T_1) T^3_1 R^3(T_1)=g_{\rm
eff}^*(T_{0\gamma}) T^3_{0\gamma} R^3(T_{0\gamma})\nonumber\
,\eneq where $g_{\rm eff}^*(T_{1})\sim 60$ and $g_{\rm
eff}^*(T_{0\gamma})=2$,\beeq \delta v_a(t_0)\equiv
v_a(t_0)=\frac{p_a(t_1)}{m_a}\frac{R(t_1)}{R(t_0)}\sim\frac{1}{m_a
t_1}\frac{R(t_1)}{R(t_0)}=\frac{1}{m_a t_1} \left(\frac{g_{\rm
eff}^*(T_{0\gamma})}{g_{\rm
eff}^*(T_{1})}\right)^{\frac{1}{3}}\frac{T_{0\gamma}}{T_{1}}\; .
\label{deltava}\eneq To estimate the velocity dispersion we need
to calculate $t_1$ and $T_1$. Using Eqs. \ref{masszeroT} and
\ref{mass} we find\beeq m_a(T)\sim 1.5\cdot 10^{-9}{\rm
eV}\left(\frac{10^{12}{\rm GeV}}{f_a}\right)\left(\frac{\rm
GeV}{T}\right)^{3.7} \ .\label{-9}\eneq Taking $g_{\rm eff}\sim
60$ in the time-temperature relation (Eq. \ref{timetemperature}),
we recast Eq. \ref{-9} to read\beeq m_a(T)\sim 3.1\cdot
10^{-51}({\rm GeV})^{\frac{5.7}{2}}\left(\frac{10^{12}{\rm
GeV}}{f_a}\right)t^{\frac{3.7}{2}} \ .\eneq Then definition
\ref{criticalt} yields\beeq t_1\sim 4\cdot 10^{-7}{\rm
sec}\left(\frac{f_a}{10^{12}{\rm GeV}}\right)^{\frac{2}{5.7}}\sim
6.1 \cdot 10^{17} ({\rm GeV})^{-1}\left(\frac{f_a}{10^{12}{\rm
GeV}}\right)^{\frac{2}{5.7}}\ .\label{t_1ax}\eneq The
corresponding temperature is \beeq T_1\sim 0.9\cdot {\rm
GeV}\left(\frac{10^{12}{\rm GeV}}{f_a}\right)^\frac{1}{5.7}\ .
\label{T_1ax}\eneq Inserting Eqs. \ref{mass}, \ref{t_1ax} and
\ref{T_1ax}, in Eq. \ref{deltava} we estimate the present value of
the axion velocity dispersion as \beeq\delta v_a(t_0)\sim 1.5
\cdot 10^{-17}\left( \frac{10^{-5}{\rm
eV}}{m_a}\right)^{\frac{4.7}{5.7}}\ .\eneq Since $\delta
v_a(t)=\delta v_a(t_0)\left(R(t_0)/R(t)\right)=\delta
v_a(t_0)\left(t_0/t\right)^\frac{2}{3}$ we have\beeq \delta
v_a(t)=1.5 \cdot 10^{-17}\left( \frac{10^{-5}{\rm
eV}}{m_a}\right)^{\frac{4.7}{5.7}}
\left(\frac{t_0}{t}\right)^\frac{2}{3}\label{deltavanumerik}\
.\eneq

Let us now calculate the energy density of the axion field due to
vacuum misalignment. The energy density $\rho$ of a homogeneous
scalar field around the minimum of its potential is \beeq
\rho=\frac{v^2_{PQ}}{2}\left[{\dot{\alpha}}^2+m_a^2(t)\alpha^2\right]
\label{rhoa} \; .\eneq The Virial Theorem implies \beeq \langle
{\dot{\alpha}}^2\rangle=m^2\langle\alpha^2\rangle=\frac{\rho}{v^2_{PQ}}
\label{vir} \; .\eneq Multiplying Eq. \ref{osc} by $\dot{\alpha}$
and using Eqs. \ref{rhoa}-\ref{vir} we obtain \beeq
\frac{\dot\rho}{\rho}=\frac{\dot m}{m}-3\frac{\dot R}{R}\; , \eneq
whose solution yields \beeq \rho ={\rm const.}
\frac{m_a(t)}{R^3(t)}\; . \eneq This equation means that as long
as the axion mass varies adiabatically, the number of axions per
comoving volume is conserved. Because the initial energy in the
coherent oscillations is $\rho_1=v^2_{PQ} m_a^2(t_1)\alpha_1^2/2$,
it is easy to estimate the present energy density $\rho_0$:\beeq
\rho_0=\rho_1\frac{m_a(t_0)}{m_a(t_1)}\frac{R^3(t_1)}{R^3(t_0)}=
\frac{1}{2}v^2_{PQ}m_a(t_1) m_a
\frac{R^3(t_1)}{R^3(t_0)}\alpha_1^2\ .\eneq Using the conservation
of entropy (Eq. \ref{g}), we find \beeq \rho_0\sim
\frac{1}{60}v^2_{PQ}m_a(t_1) m_a
\frac{T_{0\gamma}^3}{T_1^3}\alpha_1^2\ . \label{denks}\eneq If the
universe never inflated, our presently observable universe should
have $10^{30}$ causally disconnected domains when the $\alpha$
oscillations started. Therefore it is reasonable to assume that
each of the domains had arbitrarily chosen initial amplitudes and
take $\alpha_1$ as the $rms$ average of the uniform distribution
from $-\pi/N$ to $\pi/N$:\beeq
\left(\alpha_1\right)_{rms}=\left[\frac{N}{2\pi}\int_{-\frac{\pi}{N}}^{\frac{\pi}{N}}d\alpha_1
\alpha_1^2\right]^{1/2}=\frac{\pi}{\sqrt{3}N}\ . \eneq If, on the
other hand, the universe inflated after the PQ symmetry breaking,
then our observable universe originated from a single domain.
Although $\alpha_1$ had the same value everywhere in our universe,
it is arbitrary. Any value is equally likely. To determine
$\Theta_1$ in this case, measurement of the axion mass and density
are necessary. Assuming $\alpha\sim \frac{\pi}{\sqrt{3}N}$, and
using Eqs. \ref{masszeroT}, \ref{-9} and \ref{T_1ax} in Eq.
\ref{denks}, we find\beeq\rho_0\sim 3.5\cdot 10^{-30}\frac{\rm
gr}{\rm cm^3}\left(\frac{ f_a}{10^{12}{\rm
GeV}}\right)^{\frac{6.7}{5.7}}\, .\eneq Dividing by the critical
density $\rho_c=\frac{3H^2}{8\pi G}$, we obtain\beeq\Omega_a\sim
0.4\cdot\left(\frac{ f_a}{10^{12}{\rm
GeV}}\right)^{\frac{6.7}{5.7}}\left(\frac{0.7}{h}\right)^2\;
,\label{omegasuba}\eneq where $h$ is defined as usual by $H_0=h\;
100 {\rm km/sec}\cdot{\rm Mpc}$.

Finally, following the same steps between Eqs. \ref{nden0} and
\ref{nW}, we express the present axion number density $n_a$ in
terms of $\Omega_a$, the present axion energy density in units of
the critical density:
\begin{equation}
n_a(t_0) = {1.5~10^{64} \over {\rm pc}^3}\, ~\Omega_a~ \left({h
\over 0.7}\right)^2~\left({{10^{-5}\; {\rm eV}} \over
m_a}\right)~~~\ . \label{na}
\end{equation}

As can be seen from the large exponents in Eqs. \ref{nW} and
\ref{na}, WIMPs and axions exist in enormous numbers on
astronomical length scales. Moreover, they have negligible
velocity dispersion $\delta v_\chi$ and $\delta v_a$ implied by
the exponents in Eqs. \ref{deltavW} and \ref{deltavanumerik}.
Because of their small velocity dispersions, they are cold dark
matter (CDM) particles. CDM particles are so weakly interacting
(collisionless) that they have moved purely under the influence of
gravity since they decoupled. Galaxies are surrounded by unseen
CDM, and hence, because of gravity, CDM particles keep falling in
and out of galaxies from all directions. This continuous flow
produces a discrete number of flows anywhere in the halo of a
galaxy. This simple observation has rather interesting
consequences. Caustics occur where the flow number jumps from $n$
to $n+2$, with $n$ being an odd number. Generically, caustics are
the surfaces in space which separate these regions with differing
numbers of flows. The number density of CDM particles becomes very
large as one approaches the caustic from the side with the two
extra flows. The following section is devoted to the analysis of
the CDM caustic surfaces in galactic halos.

\newpage

\section{Cold Dark Matter Caustics} \label{Flow}

In this section, I make extensive use of the explanations and
derivations given in \cite{sing}. We have seen in the Introduction
that, before the onset of galaxy formation, the collisionless
(weakly interacting) CDM particles lie on a time dependent thin 3D
sheet in 6D phase space. The thickness of this sheet is the
primordial velocity dispersion $\delta v$. At the onset of galaxy
formation, the particles in a neighborhood of an overdensity fall
onto it. As a result, the phase space sheet winds up clockwise
wherever galaxies grow (Fig. \ref{fig:sheet}). Outside
overdensities, at a location in physical space, there is only one
value of velocity ( i.e., one single flow, the Hubble flow).
There, the phase space sheet covers physical space once. Hence, in
Fig. \ref{fig:sheet}, away from the overdensity the plot is one to
one. On the other hand, inside an overdensity, the phase space
sheet covers physical space multiple times, but always an odd
number of times, implying that there are an odd number of flows at
such locations in physical space (Figs. \ref{fig:sheet} and
\ref{fig:phase}). CDM caustics are associated with these flows
\cite{sing}. We examine them rigorously in the next section.

\subsection{Caustics in General} \label{sect:causticsingeneral}

Let us consider a flow of CDM particles with zero (or negligible)
velocity dispersion. To study the motion of the particles, we
adopt a parametrization of the 3D phase space sheet by labelling
each particle with an arbitrary 3-component parameter
$\vec{\alpha}=(\alpha_1, \alpha_2, \alpha_3)$. The phase space
sheet location at time $t$ is specified by the map $\vec{\alpha}
\rightarrow \vec{x} (\vec{\alpha},t)$ where $\vec{x}$ is the
position in physical space of particle $\vec{\alpha}$ at time $t$.
The velocity of particle $\vec{\alpha}$ is $\vec{v} = {\partial
\vec{x}\over
\partial t}~(\vec{\alpha}, t)$. Let $\vec\alpha_j (\vec r, t)$, with $j=1.\; .\; .n$,
be the solutions of \beeq \vec{r} = \vec{x} (\vec{\alpha},t)\
.\label{alpha(x)}\eneq Here $n$ is the number of distinct flows at
spatial location $\vec{r}$ and time $t$. In general, the number of
solutions $n$ jumps by $2$ on certain surfaces which are the
locations of caustics. The total number of particles is: \be
N&=&\int d^3\alpha ~{d^3N\over d\alpha_1 d\alpha_2 d\alpha_3}
~(\vec\alpha)\nonumber\\
&=&\int d^3r \sum_{j=1}^n ~{d^3N\over d\alpha_1 d\alpha_2
d\alpha_3} ~\left(\vec\alpha_j (\vec r,t)\right) {1\over \mid \det
\left({\partial \vec x\over
\partial \vec\alpha}\right)\mid_{\vec\alpha_j (\vec r, t)}} \ .
\label{2.1} \ee The density of particles in physical space can be
extracted from the above equation:
\begin{equation}
d(\vec r,t) = \sum_{j=1}^n~{d^3N\over d\alpha_1 d\alpha_2
d\alpha_3}~ (\vec\alpha_j (\vec r,t))~{1\over \mid D(\vec\alpha,t)
\mid_{\vec\alpha_j (\vec r, t)}}~~~~~~\ , \label{2.2}
\end{equation}
where
\begin{equation}
D(\vec\alpha, t) \equiv \det \left( {\partial \vec x\over \partial
\vec\alpha}\right)\equiv \det ({\mathcal{D}})\ ,\label{2.3}
\end{equation}
and ${\mathcal{D}}$ is the Jacobian. As expected, the density
$d(\vec r,t)$ is a reparametrization invariant: if we
reparametrize the particles $\vec{\alpha} \rightarrow
\vec{\beta}$, such that the n distinct flows $\beta_j(\vec{r}, t)$
are the solutions of $\vec{r}=\vec{x}(\vec{\beta}, t)$, then: \be
\frac{d^3N}{d\alpha_1 d\alpha_2
d\alpha_3}&\rightarrow&\frac{d^3N}{\left| \det
\left(\frac{\partial\vec{\alpha}}{\partial\vec{\beta}}\right)\right|d\beta_1
d\beta_2 d\beta_3}\; ,\nonumber\\\det \left( {\partial \vec x\over
\partial \vec\alpha}\right)&\rightarrow&\det \left(
\frac{\partial\vec{x}}{\partial\vec{\beta}}\;
\frac{\partial\vec{\beta}}{\partial\vec{\alpha}}\right)=\det
\left( \frac{\partial\vec{x}}{\partial\vec{\beta}} \right) \det
\left( \frac{\partial\vec{\beta}}{\partial\vec{\alpha}} \right)\;
, \ee and hence, the density retains its form given in Eq.
\ref{2.2} with $\vec\alpha$ replaced by $\vec\beta$.

Caustics are wherever the density diverges (i.e. wherever
$D(\vec{\alpha}, t)=0$). Thus, the map $\vec\alpha \rightarrow
\vec x$ is singular at the location of the caustics. Generically
the zeros of $D(\vec{\alpha}, t)$ are simple (i.e., the matrix
$\frac{\partial \vec x}{\partial \vec\alpha}$ has a single
vanishing eigenvalue).  The condition that one eigenvalue vanishes
imposes one constraint on the three parameters $\vec{\alpha}$.
Therefore, the location of a caustic is generically determined by
two independent labels, hence it is a 2D surface in physical
space.
\subsection{Caustic Surfaces} \label{sec:caustic surfaces}
\begin{figure}[ht]
\centering
\includegraphics[height=10.6cm,width=9cm]{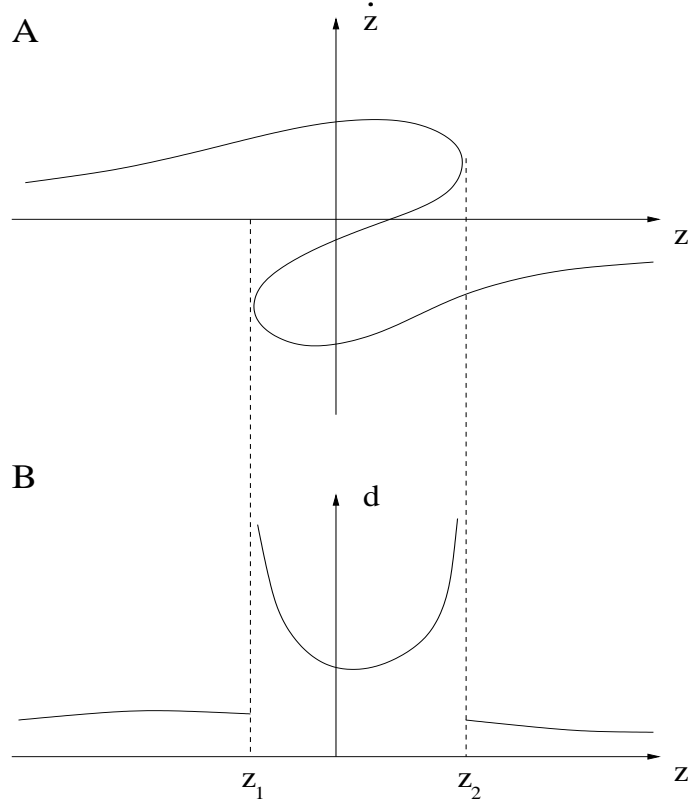}
\caption{A generic surface caustic. A) In phase space. B) In
physical space. The two dimensions ($x$ and $y$) into which the
caustic extends as a surface are not shown.  The physical space
density $d$ diverges at those locations ($z_1$ and $z_2$) where
the phase space sheet folds back. } \label{fig:fig2}
\end{figure}

Let's consider a generic caustic surface at a fixed time $t$.
Using the reparametrization invariance $\vec{\alpha} \rightarrow
\vec{\beta}(\vec{\alpha},t)$ of the flow, let's parametrize the
particles near the caustic such that the surface is at
$\beta_3=0$. Also let's choose a Cartesian coordinate system in a
neighborhood of point P on the caustic surface such that $\hat{z}$
is the perpendicular direction to the surface, whereas $\hat{x}$
and $\hat{y}$ are the parallel directions. Hence, in this
neighborhood $\partial z /\partial \beta_1 =\partial z /\partial
\beta_2=0$. Thus, we have
\begin{equation}
D = {\partial z\over \partial\beta_3} ~\det\left({\partial
(x,y)\over
\partial(\beta_1,\beta_2)}\right)\ ,
\label{2.4}
\end{equation}
around P.  The two dimensional matrix ${\partial (x,y)}/
{\partial(\beta_1,\beta_2)}$ is nonsingular, whereas $\partial z/
\partial\beta_3=0$ at $\beta_3 = 0$, hence, by construction $D$ vanishes at $\beta_3=0$, where
the caustic surface is. By Taylor expanding, we have up to the
second order in $\beta_3$:
\begin{equation}
z = z_0 + B\beta_3^2 \label{2.5}\; .
\end{equation}
We may choose the positive ${z}$-direction in such a way that
$B>0$. Then, since $\partial z/\partial \beta_3=2\sqrt{B(z-z_0)}$,
the determinant becomes
\begin{equation}
D = 2\sqrt{B(z-z_0)}~\det \left({\partial (x,y)\over \partial
(\beta_1, \beta_2)}\right)~~~~~{\rm for}~z> z_0 \ . \label{2.6}
\end{equation}
Hence, near a caustic surface located at $z = z_0$, the density
diverges as ${1\over \sqrt{z-z_0}}$ on the side of the surface
where $z>z_0$. Figure \ref{fig:fig2}.A below depicts a 2D cut of
phase space in the $(z, \dot{z})$ plane.The particles lie on the curve which is at the
intersection of the phase space sheet with the $(z, \dot{z})$
plane. The label $\beta_3$ gives the position of the particles on
this curve. The caustics, which are located at $z=z_1$ and
$z=z_2$, extend in the $x$ and $y$ directions. Figure
\ref{fig:fig2}.B depicts the density $d(z)$ which diverges as
$1/\sqrt{z-z_1}$ for $z\rightarrow z_1$ with $z>z_1$, and as
$1/\sqrt{z_2-z}$ for $z\rightarrow z_2$ with $z<z_2$. There is
only one flow for $z<z_1$ and for $z>z_2$, whereas there are three
distinct flows in the region $z_1<z<z_2$. The number of flows
changes by two at the locations of the caustics. The phase space
sheet has a fold at $z=z_1$ and $z=z_2$, where $\partial
z/\partial\beta_3=0$. Hence the velocity space is tangent to the
fold. Therefore, in physical space the particles pile up at $z_1$
and $z_2$ where the sheet folds back. More rigourously, if we
represent the 2D phase space density near a fold in the limit
$\delta v\rightarrow 0$ by \be \frac{d^2N}{dz d\dot
z}=N\delta(z-C(\dot{z})^2)\ ,\ee where $C$ and $N$ are numbers,
then, the density\be d=\int \frac{d^2N}{dz d\dot z} d\dot z &=&
N\int\delta(z-C(\dot{z})^2)d\dot z = \int\left\{ \frac{\delta(\dot
z
-\sqrt{\frac{z}{C}})}{\left|-2C\sqrt{\frac{z}{C}}\right|}+\frac{\delta(\dot
z +\sqrt{\frac{z}{C}})}{\left|2C\sqrt{\frac{z}{C}}\right|}\right\}
d\dot z\nonumber\ ,\\
&=&\frac{N}{\sqrt{Cz}}\equiv\frac{A}{\sqrt{z}}\, ,\ee where $A$
will be called the ``fold coefficient.'' Notice that $d$ diverges
as $z\rightarrow 0$ where the caustic is located.

\subsection{CDM Infall in Galactic Halos and Caustics} \label{CR}
In this section we will introduce the caustics associated with the
infall of CDM particles into the potential well of a galactic
halo. The existence and structure of the caustics will be studied
in Sections \ref{ExistenceOCDP} and \ref{sect:ExistenceCRDP} in
detail, using the discussion given here. We set the velocity
dispersion equal to zero $(\delta v=0)$ in our considerations. A
small non-vanishing velocity dispersion would provide a cutoff,
such that the density at the caustic would become very large
rather than infinite.

We have already seen in the introduction that the process of
galaxy formation involves winding up of the phase space sheet of
CDM particles which produces a discrete number of flows at a
location in space. If the galactic center is approached from an
arbitrary direction, the local number of flows increases. First
there is one flow, then there are three flows, then five etc.
(Fig. \ref{fig:phase}). The boundary between the region with one
(three, five,. . .) and the region with three (five, seven,. . .)
flows is the location of a caustic which is topologically a sphere
surrounding the galaxy. As will be shown in Section
\ref{sect:denprofofoutercaustics}, when these spheres are
approached from the side with the two extra flows the density
diverges as $\frac{1}{\sqrt{\sigma}}$ where $\sigma$ is the
distance to the surface. We called these spheres ``outer
caustics.'' It is a little more difficult to see the existence of
``inner caustics'' in addition to the outer caustics; they can not
be seen in Fig. \ref{fig:phase}. Inner caustic rings occur if the
CDM particles carry angular momentum with respect to the center.
In the limit of zero angular momentum a caustic ring becomes a
caustic point at the center, whereas the outer caustics are
unaffected.

Let us first discuss the case where the overdensity is spherically
symmetric and all the CDM particles have zero angular momentum.
Then, the particles move in and out of the galaxy in radial
orbits. As a result, the galactic center is a caustic point and
the density associated with each flow diverges as $\frac{1}{r^2}$,
where $r$ is the radial coordinate. Caustic point is a degenerate
case since we assumed spherical symmetry and radial orbits.

To see the inner caustics, one has to consider the flow of the
particles which carry angular momentum with respect to the center.
In this case, as will be seen in Section \ref{ExistenceOCDP},
outer caustics which surround the galaxy also exist.

\begin{figure}[ht] \centering
\includegraphics[height=9cm,width=13cm]{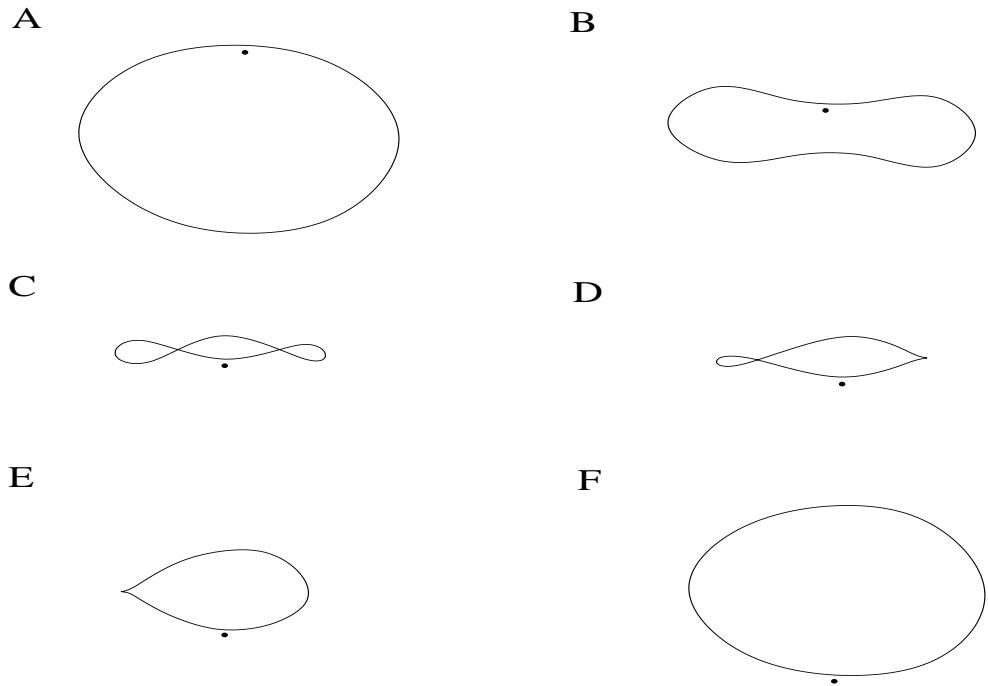}
\caption{Infall of a turnaround sphere.  The closed lines are at
the intersections of the sphere with the plane of the figure at
six successive times.  The sphere has net angular momentum about
the vertical axis.  It crosses itself between frames B and C. The
cusps in frames D and E are at the intersection of a caustic ring
with the plane of the figure. The dot, which represents a particle
in the flow, is introduced to keep track of the sequences in the
flow. The whole infall sphere completes the process of turning
inside out just after frame E. } \label{fig:fig3}
\end{figure}

Let us now consider the case where the collisionless CDM particles
move in and out of the galaxy only under the effect of gravity and
rotate due to the angular momentum they carry. Figure
\ref{fig:fig3} depicts the successive time frames of a set of CDM
particles falling through a galaxy. This sequence is inevitable,
for a flow in and out of the potential well of a galactic halo. In
Fig. \ref{fig:fig3}.A, the particles are about to fall onto the
galaxy for the first time in their history. They make up a closed
2D surface, a topological ``sphere,'' surrounding the galaxy in 3D
physical space. We call this surface the first ``turnaround
sphere'' (in fact, Fig. \ref{fig:fig3} depicts the intersection of
this turnaround sphere with the plane of the figure, as it evolves
in time). To be more definite, we assume that the particles carry
net angular momentum about the vertical axis in this plane (i.e.,
the turnaround sphere is spinning about the vertical). The
particles near the top (bottom) of the sphere in frame A, carry
little angular momentum and end up near the bottom (top) of the
sphere in frame F, after falling through the galaxy. The particles
near the equator carry the most angular momentum. When the
turnaround sphere crosses itself near frame C, they form a ring in
the spatial section perpendicular to the plane of the figure. The
radius of the ring decreases in time down to some minimum value,
reached near frame D, and then increases again during the out
flow. Outside this ring, the region inside the sphere which has not
turned itself inside out makes a doughnut. When this doughnut is
pinched out in spacetime (in frames C and D), as will be shown
later in this section, a caustic ring occurs in space. We will
designate this ring as the location of the ``inner caustic.'' The
radius of the sphere at the second turnaround in frame F is
smaller than the first turnaround radius because of the deepening
of the potential well of the galaxy due to the infall occurring in
the meantime. The second turnaround sphere falls back in,
repeating the same sequence of Fig. \ref{fig:fig3}, although it
gets fuzzier due to the gravitational scattering of
inhomogeneities in the galaxy, and so on. There is another generic
surface caustic associated with the $m$th turnaround, where
$m=2,3,4,.\; .\; .\; .$ We will call these ``outer caustics.'' At
the location of the outer caustics, the number of flow changes by
two (this is also true for the inner caustic rings as well), and
particles pile up since the phase space sheet in this location is
tangent to velocity space (Fig. \ref{fig:phase}). There is no
change in the number of flow near the first turnaround $(m=1)$,
hence there is no caustic associated with it. Outer caustics, in
fact, are located near where the particles of a given outflow
reach their maximum radius before falling back in. To illustrate
this, let us parametrize the flow at a given time $t$ by $\vec{x}
(\alpha,\beta, t_0;t)$, where $t_0$ is the time the particle was
at last turnaround, $\alpha$ and $\beta$ label the particles on
the turnaround sphere at that time. For example, polar and
azimuthal angles, which tell us where the particle was on the
turnaround sphere, can be used to label the particles. The vector,
$\vec{x}$, is therefore the location of the particle
$\vec{\alpha}\equiv(\alpha,\beta, t_0)$ at time $t$ in physical
space. Let us define: $\vec{x}_{t_0} \equiv {\partial \vec{x}\over
\partial t_0}~,~\vec{x}_\alpha \equiv {\partial \vec{x}\over
\partial\alpha}$, $\vec{x}_\beta \equiv
{\partial\vec{x}\over\partial\beta}~,$ and $\dot{\vec{x}} \equiv
{\partial\vec{x}\over\partial t}$. Therefore, the determinant
$D(t_0, \alpha, \beta;
t)=\det\left(\frac{\partial\vec{x}}{\partial\vec{\alpha}}\right)$
can be written as
\begin{equation}
D =~\vec{x}_{t_0}~\cdot~(\vec{x}_\alpha\times \vec{x}_\beta)\ .
\label{3.2n}
\end{equation}
The remaining discussion will focus on the flow at a fixed time
$t$ (we will discuss on the snapshot of the flow at time $t$). As
a result, we do not show the $t$ dependence any further. The
existence of outer and inner caustics in the presence of angular
momentum will be proven in Sections \ref{ExistenceOCDP} and
\ref{sect:ExistenceCRDP}, respectively.

\subsection{Outer Caustic Spheres}
\subsubsection{The Existence of Outer Caustic Spheres}
\label{ExistenceOCDP}

Let us reconsider now the evolution of the spheres in the frames
of Fig. \ref{fig:fig3}.A-F and, after frame F. In the beginning,
all the later infall spheres, which have greater $t_0$, are
outside the sphere $\{\vec{x} (\alpha,\beta,
t_0)~:~\forall\alpha,\beta\}$ of frame B, and $t_0$ increases in
the outward direction. Because the particles on the sphere are
moving inward (hence $\dot{\vec{x}}(\alpha,\beta,t_0)$ is pointing
inward for all $(\alpha, \beta)$), the vectors
$\{{\vec{x}}_{t_0}(\alpha, \beta, t_0)~:~\forall\alpha,\beta\}$
point outward. In frame E the initial infall sphere is just about
to complete the process of turning itself inside out. Except for
the particle at the cusp, all the particles on the sphere have
already started their out flow. After frame E the process is
completed. Hence, inside the sphere there are later infall spheres
and $t_0$ increases in the inward direction. Thus, even though the
particles on the sphere in frame F flowing outward (hence
$\dot{\vec{x}}(\alpha,\beta, t_0)$ is pointing outward for all
$(\alpha, \beta)$), the vectors $\{{\vec{x}}_{t_0}(\alpha, \beta,
t_0)~:~\forall\alpha,\beta\}$ point inward. Note that
$\dot{\vec{x}}$ is always in the direction of motion and
${\vec{x}_{t_0}}$ points opposite to it, since $t_0$ decreases in
the direction of motion (particles in the direction of the motion
started to fall earlier, hence they have smaller $t_0$). After
frame F, $\vec{x}_{t_0}(\alpha, \beta, t_0)$ will remain inward
until the particle $\vec{x}(\alpha, \beta, t_0 )$ reaches the next
turnaround where it starts to fall back again. After the
turnaround the motion is inward and $\vec{x}_{t_0}(\alpha, \beta,
t_0)$ is outward. Therefore, at some point in the evolution
$\vec{x}_{t_0}(\alpha, \beta, t_0)$ either vanishes or becomes
parallel to the sphere $\{\vec{x} (\alpha,\beta,
t_0)~:~\forall\alpha,\beta\}$. At that point $D=0$, since
$\vec{x}_\alpha\times \vec{x}_\beta$ is perpendicular to the
sphere. At a given fixed time the collection of all these points
makes up a closed continuous surface (topological sphere) in
space. Or, we can go back in time and choose the initial sphere
$\{\vec{x} (\alpha',\beta', t_0(\alpha',
\beta'))~:~\forall\alpha',\beta'\}$ such that at the future given
fixed time $\{\vec{x}_{t_0} (\alpha',\beta', t_0(\alpha',
\beta'))~:~\forall\alpha',\beta'\}$ is parallel to $\{\vec{x}
(\alpha',\beta', t_0(\alpha', \beta'))~:~\forall\alpha',\beta'\}$,
where $D=0$. Therefore, the initial sphere $\{\vec{x}
(\alpha',\beta',t_0 (\alpha',\beta'))~:~\forall\alpha',\beta'\}$
will have  simultaneously vanishing $D$ at the given fixed time,
making up a caustic which we call an $outer$ caustic.

To visualize that we really have a closed continuous caustic
surface in the flow of CDM particles in three dimensional space,
recall that the particles that fell in can not reach their last
turn around radius again. Therefore, at any radial direction
$\hat{r}$ from the center of the galactic halo, there are
locations where the number of flows jumps from $n+2$ to $n$, where
$n$ is an odd number. As $\vec{r}$ traces out all the directions
in the sky, all these locations make up a closed surface (for a
fixed $n$) due to the continuity of the flow. In phase space, the
3D phase space sheet folds back at the locations where the
surfaces are. Hence, at the folds the velocity space is tangent to
the phase space sheet and in physical space particles pile up at
the locations of the corresponding folds. The number density
diverges (in the limit $\delta v=0$) at the surfaces because it is
the integral of the phase space density over velocity space.

\subsubsection{Density Profiles of Outer Caustic Spheres}
\label{sect:denprofofoutercaustics} Outer caustics are closed
surfaces (topological spheres) near the $(n+1)$th turn-around
radii with $n=1, 2, 3$. . . . Indeed the number of flows changes
by two at the turnarounds because of the fall back of the
particles. There is no caustic associated with the first
turn-around (Fig. \ref{fig:phase}).

The outer caustics are described by simple ``fold'' ($A_2$)
catastrophes. Their density profile is \beeq d_n
(\sigma)=\frac{A_n}{\sqrt{\sigma}}\,\Theta (\sigma)
\label{densityprofile} \eneq for small $\sigma$, where $\sigma$ is
the distance to the caustic, $\Theta (\sigma)$ is the Heaviside
function, and $A_n$ is a constant which we call the ``fold
coefficient.'' We choose $\sigma >0$ on the side with two extra
flows (i.e., towards the galactic center). Therefore, when an
outer caustic is approached from the inside, the density diverges
as $\sim\frac{1}{\sqrt{\sigma}}$, abruptly falling to zero on the
outside. The observation \cite{malin} of arc-like shells
surrounding giant elliptical galaxies can be interpreted
\cite{quinn} as outer caustics in the distribution of baryonic
matter falling onto those galaxies.

To estimate $A_n$ in Eq. \ref{densityprofile}, consider the time
evolution of CDM particles which are falling out of a galactic
halo for the $n$th time and then falling back in.  Let $R_n$ be
the turn-around radius. We assume that the rotation curve of the
galaxy is flat near $r=R_n$ with time-independent rotation
velocity $v_{\rm rot}$.  The gravitational potential is then \beeq
V(r)=v^2_{\rm{rot}}\ln\left(\frac{r}{R_n}\right) ~~~\ . \eneq The
particles have a trajectory $r(t)$ such that \be
\left|\frac{dr}{dt}\right|=\sqrt{2\left(E-V(r)\right)}=
v_{\rm{rot}}\sqrt{2\ln\left(\frac{R_n}{r}\right)} ~~~\ ,
\label{drdt} \ee where $E$ is the energy per unit mass.  Equation
\ref{drdt} neglects the angular momentum of the particles.  This
is a good approximation at turnaround since the particles are far
from their distance of closest approach to the galactic center.
Finally, we assume that the flow is stationary.  In that case the
number of particles flowing per unit solid angle and per unit
time, $\frac{dN}{d\Omega dt}$, is independent of $t$ and $r$, and
the caustic is located exactly at the $(n+1)$th turnaround radius
$R_n$.

Let us emphasize that none of the assumptions spherical symmetry,
flat rotation curve, time independence of the gravitational
potential, radial orbits, and stationarity of the flow affect the
existence of the outer caustics or the fact that they have the
density profile given by Eq. \ref{densityprofile}. The assumptions
are made only to obtain estimates of the coefficients $A_n$.

The mass density of particles, $d_n(r)$, follows from the
equality: $2\frac{dN}{d\Omega dt}dt= {d_n(r) \over m} r^2 dr$,
where $m$ is the mass of each particle.  The factor of $2$ appears
because there are two distinct flows, out and in. Using Eq.
\ref{drdt}, we obtain the density distribution near the $n$th
outer caustic: \beeq d_n(r)=2\frac{dM}{d\Omega
dt}\Bigg|_{n}\,\,\frac{1}{r^2
v_{\rm{rot}}\sqrt{2\ln{\left(\frac{R_n}{r}\right)}}} ~~~\ ,
\label{dnr} \eneq where $\frac{dM}{d\Omega dt}\equiv
m\frac{dN}{d\Omega dt}$. Near the caustic,
$\ln{\left(\frac{R_n}{r}\right)}=\frac{\sigma}{R_n}$ where $\sigma
= R_n - r$.  Inserting this into Eq. \ref{dnr} and comparing it
with Eq. \ref{densityprofile}, we find \beeq
A_n=\frac{\sqrt{2}}{v_{\rm{rot}}}\frac{dM}{d\Omega dt}\Bigg|_n
R^{-3/2}_n ~~~\ . \label{AN} \eneq Estimates of $R_n$ and
$\frac{dM}{d\Omega dt}\Big|_n$ can be extracted from ref.
\cite{STW1,STW2} for the case of self-similar infall \cite{FG,B}.
The infall is called {\it self-similar} if it is time-independent
after all distances are rescaled by the turn-around radius $R(t)$
at time $t$ and all masses are rescaled by the mass $M(t)$
interior to $R(t)$. In the case of zero angular momentum and
spherical symmetry, the infall is self-similar if the initial
overdensity profile has the form $\frac{\delta
M_i}{M_i}=\left(\frac{M_0}{M_i}\right)^\epsilon$ where $M_0$ and
$\epsilon$ are parameters \cite{FG}. In CDM theories of large
scale structure formation, $\epsilon$ is expected to be in the
range $0.2$ to $0.35$ \cite{STW1,STW2}. In that range, the
galactic rotation curves predicted by the self-similar infall
model are flat \cite{FG}.

The $R_n$ and $\frac{dM}{d\Omega dt}\Big|_n$ do not depend sharply
upon $\epsilon$.  Our estimates are for $\epsilon = 0.2$ because
they can be most readily obtained from ref. \cite{STW1,STW2} in
that case. For $\epsilon=0.2$, the radii of the $(n+1)$th
turnaround spheres are approximately \be \{R_n: n=1,2,.\; .\; .\}
\simeq (240,~120,~90,~70,~60,.\; .\; .)~{\rm kpc}
\cdot\left(\frac{v_{\rm{ rot}}}{220\,{{\rm km/s}}}\right)
\left(\frac{0.7}{h}\right)~~~\ . \label{Rn} \ee Moreover, one has
\be \left.\frac{dM}{d\Omega dt} \right|_n\frac{1}{v_{\rm{ rot}}}
=F_n\,\sqrt{2}\,\frac{v^2_{\rm { rot}}}{4\pi G}~~~\ , \label{Fn}
\ee with \be \{F_n : n=1,2,.\; .\; .\} \simeq (20,~8,~5,~4,~3,.\;
.\; .)~10^{-2}~~~\ . \label{Fne} \ee Combining Eqs. \ref{AN} -
\ref{Fne}, we find \be \{A_n :n=1,2,.\; .\; .\} \sim
(2,~2,~2,~3,~3,.\; .\; .)\cdot
{\rm \frac{10^{-5} \,gr}{cm^2\,kpc^{1/2}}}\nonumber\\
\cdot\left(\frac{v_{\rm{ rot}}}{220\,{{\rm km/s}}}\right)^{1/2}
\left(\frac{h}{0.7}\right)^{3/2}\; . \label{An} \ee Generally the
outer caustics are ``concave'' (i.e., they are curved toward the
side which has two extra flows).  Their radius of curvature is of
order $R_n$.  The lensing by concave caustic surfaces is discussed
in Section \ref{subsection:scvf}.

Next, we will discuss {\it inner} caustics which, as we will soon
see, have the shape of rings (closed tubes) and are located near
where the particles with the most angular momentum in a given
inflow reach their distance of closest approach to the galactic
center before moving back out of the galaxy.

\subsection{Inner Caustic Rings}
\subsubsection{The Existence of Inner Caustic Rings}
\label{sect:ExistenceCRDP}

During each infall-outfall sequence, the sphere of Fig.
\ref{fig:fig3} turns itself inside out. For example, a particle
(represented by a dot throughout the frames of Fig.
\ref{fig:fig3}) which is part of the flow and is just inside the
sphere in frame A, appears outside the sphere in frame F. There
is, therefore, a ring in space-time of points which are inside the
sphere last. The intersection of this space-time ring with the
plane of the figure is at two space-time points, one located at
the cusp in frame D, the other at the cusp in frame E. Although
the cusp in frame D is smoothed out in frame E, and the cusp in
frame E does not exist in frame D, since there is a continuous
flow of spheres falling in and out that are continuous
deformations of one another, the cusp-ring just defined is a
persistent feature in space. Here, we should note that it is
possible mathematically to turn a sphere inside out by means of
continuous deformations which allow the surface to pass through
itself without puncturing, ripping, creasing, pinching, or
introducing any sharp points in to the intermediate stages. This
process is called the ``sphere eversion''
\cite{Smale,Phillips,FrancisMorin} in topology. In our problem,
however, the particles of the sphere are constrained to move
purely under the effect of gravity and the angular momentum they
carry. When a sphere is pushed bottom up and top down through
itself as in Fig. \ref{fig:fig3}.B, a ring shaped fold occurs
where the sphere intersects itself as in Fig. \ref{fig:fig3}.C.
This ring-fold can not be eliminated without creating a sharp
crease (the cusps of Fig. \ref{fig:fig3}.D and Fig.
\ref{fig:fig3}.E) in the physical process we are considering. In
the mathematical-sphere-eversion process, the ring is pulled down
on the sphere becoming smaller and then the ring is squeezed so
that its two sides touch and merge.

\begin{figure}[ht] \centering
\includegraphics[height=9cm,width=9cm]{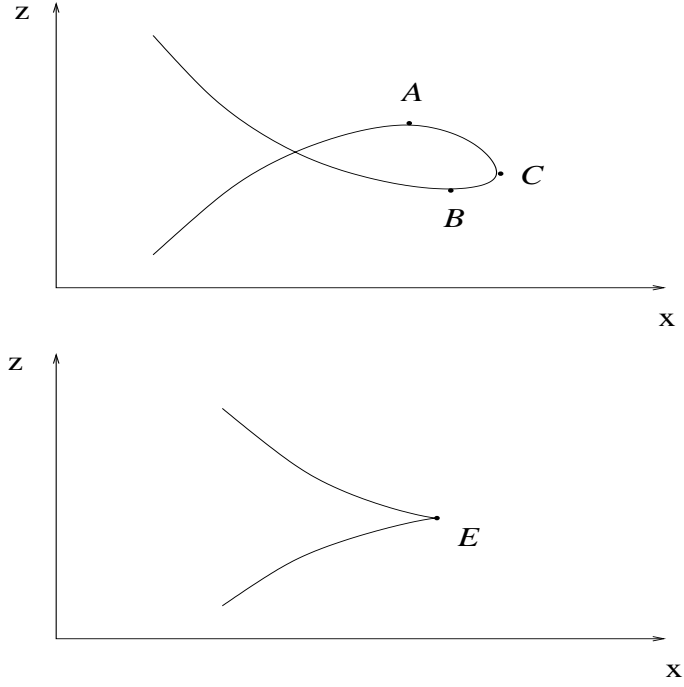}
\caption{An infall sphere near the location where and when it
completes the process of turning itself inside out.  The curves
are the intersections of the sphere with the plane of the figure
at two successive times, corresponding to frames C and D in Fig.
\ref{fig:fig3}. Parameter $\alpha$ labels points along the line.
${\partial z\over\partial\alpha} = 0$ at points $A$ and $B$.
${\partial x\over\partial\alpha} = 0$ at point $C$.  Points $A$,
$B$ and $C$ move to point $E$, which thus is the location of a
caustic since $D=0$ there.} \label{fig:fig4}
\end{figure}

We, now want to show that the cusp-ring, as defined early in the
previous paragraph, is a caustic location (cusps in Fig.
\ref{fig:fig3}.D-E are at the intersections of a cusp-ring with
the plane of the figure). Consider Fig. \ref{fig:fig4} that zooms
in to the neighborhood of space and time where the infall sphere
is completing the process of turning itself inside
out. Coordinates $x$ and $z$ are chosen as shown in the
figure. The $\hat{y}$ direction is into the plane of the figure
and parallel to the ring at point $E$. We parametrize the flow in
a small neighborhood of the ring by $\vec{\alpha} = (\alpha,
\beta, t_0)$ such that ${\partial x\over
\partial\beta} = {\partial z\over \partial\beta} = 0$. As before,
$t_0$ labels successive infall spheres and may be taken to be the
time of their last turnaround. Thus:
\begin{equation}
D = {\partial y\over \partial\beta} \left({\partial
x\over\partial\alpha}~{\partial z\over \partial t_0} - {\partial
x\over
\partial t_0}~{\partial z\over \partial\alpha}\right)\ .
\label{3.1}
\end{equation}
In the top figure of Fig. \ref{fig:fig4}, the infall sphere is
about to reach to the ring. Parameter $\alpha$ labels points
(particles) along the curve. At points $A$ and $B$, ${\partial
z\over\partial\alpha} = 0$ because the tangent vectors to the
curve at these points are parallel to the $\hat{x}$-direction. At
point $C$, on the other hand, ${\partial x\over\partial\alpha} =
0$ because the tangent at $C$ is parallel to the
$\hat{z}$-direction. In the bottom figure of Fig. \ref{fig:fig4},
the infall sphere reaches the ring. Points $A$, $B$ and $C$ of the
top figure have moved to point $E$ in the bottom figure. Hence
${\partial z\over
\partial\alpha} = {\partial x\over\partial\alpha} = 0$ and,
therefore, from Eq. \ref{3.1}, we have $D=0$ at point $E$. Thus,
$E$ is a caustic location. Since, in general, $D$ only has a
simple zero at point $E$, $E$ is in fact a location of a {\it
surface} caustic. In the discussion that follows, we will show
that the complete inner caustic surface is a closed tube and $E$
is a point on this tube.

To see the tube topology, consider Fig. \ref{fig:fig3}.B. The
vector $\vec{x}_{t_0}=\frac{\partial \vec{x}}{\partial t_0}$ (not
shown explicitly) points outward everywhere on the infall sphere
since the later (earlier) infall spheres, which have greater
(smaller) $t_0$, are outside (inside) of this one. In other words,
$t_0$ increases in the outward direction. In Fig.
\ref{fig:fig3}.F, however, where the sphere completes the process
of turning itself inside out, all the later (earlier) infall
spheres are inside (outside) of this one. Therefore,
$\vec{x}_{t_0}$ points inside everywhere on this sphere. This
implies that during the infall-outfall, $\vec {x}_{t_0}$ either
vanishes or becomes parallel to the sphere at some space-time
points. $D=0$ at such points, hence they are the location of
caustics. Since $D$ is a continuous function of
$\vec{\alpha}=(\alpha, \beta, t_0)$, the caustic must lie on a
closed surface (vanishing of $D$ constrains the three parameters
leaving only two independent ones, hence we have a surface). Now,
if we follow the motion of points which are near the North (South)
pole of the sphere in Fig. \ref{fig:fig3}.B and, end up near the
South (North) pole in Fig. \ref{fig:fig3}.E, we see that
$\vec{x}_{t_0}$ always points up (down). It does not reverse its
direction. Hence, $\vec{x}_{t_0}$ does not vanish and is not
parallel to the sphere at any time between these two frames for
these points. Therefore, within a cylindrical region extending
from North to South in the spatial volume under consideration $D$
does not vanish. This implies that inner caustics do not exist in
this cylindrical region. On the other hand, for the points near
the equator in Fig. \ref{fig:fig3}, $\vec{x}_{t_0}$  points
outward during infall and points inward during outfall. For
example, at the East (West) pole in Fig. \ref{fig:fig3},
$\vec{x}_{t_0}$ points outward from frame A to the the end of
frame D (frame E) and points inward in frames E and F (in frame
E). Thus if we track a point near the equator, at some time
$\vec{x}_{t_0}$ either vanishes or is parallel to the sphere,
hence $D$ vanishes. The points where this happens lie on a closed
surface which wraps outside of the previously defined cylinder.
Therefore, the surface is a tube located near the equator, where
the particles with the most angular momentum are at their distance
of closest approach.

According to the Brouwer's Hairy Ball (or Fixed Point) Theorem, an
everywhere nonzero tangent vector field on the 2-sphere does not
exist. Any continuous tangent vector field on the sphere must have
at least one point where the vector is zero. Finding examples is
easy. Since the sphere with one point removed is homeomorphic to
the plane, using the fact that a constant (and thus nonzero)
vector field on the plane exists, we can obtain a vector field on
the once-punctured sphere. Or, consider a line L tangent to the
unit sphere S in $R^3$ at the point $N$. For every plane P
containing L, consider the intersection of P with S. These create
a family of circles C on S all separated from each other, except
that they all contain the point $N$. Now define a vector field on
S at each point x, using the tangent direction of the unique
circle C through x, with a magnitude equal to the distance from
the point $N$. Hence, the zero vector is at $N$. Another way of
stating the theorem is that given a ball with hairs all over it,
it is impossible to comb the hairs continuously and have all the
hairs lay flat so that they change direction smoothly over the
whole surface. At least one hair must be sticking straight up.
This implies, as a corollary, that somewhere on the surface of the
Earth, at any moment there is at least one point with zero
horizontal wind velocity.

Consider now the angular momentum distribution on the turnaround
sphere. This is a special continuous two-dimensional vector field
on the sphere which has two zeros at the poles where the net
angular momentum axis crosses the sphere. When we tracked the
particles in the neighborhood of these angular momentum zeros, we
found that their ${\vec{x}}_{t_0}$ does not vanish and is not
parallel to the sphere at any time during the infall-outfall
process. Therefore the inner caustics appear only in the flow of
particles which are some distance away from both zeros. Since that
set of particles has the topology of a closed ribbon ($I\times
S^1$ where $I$ is a closed interval in $R$), and the previously
defined caustic ring (the set of points $E$ which are in the
turnaround sphere last) goes around this closed ribbon once, the
inner caustic must be a closed surface with one handle (i.e., a
closed tube).

\begin{figure}[ht]
\centering
\includegraphics[height=10cm,width=10cm]{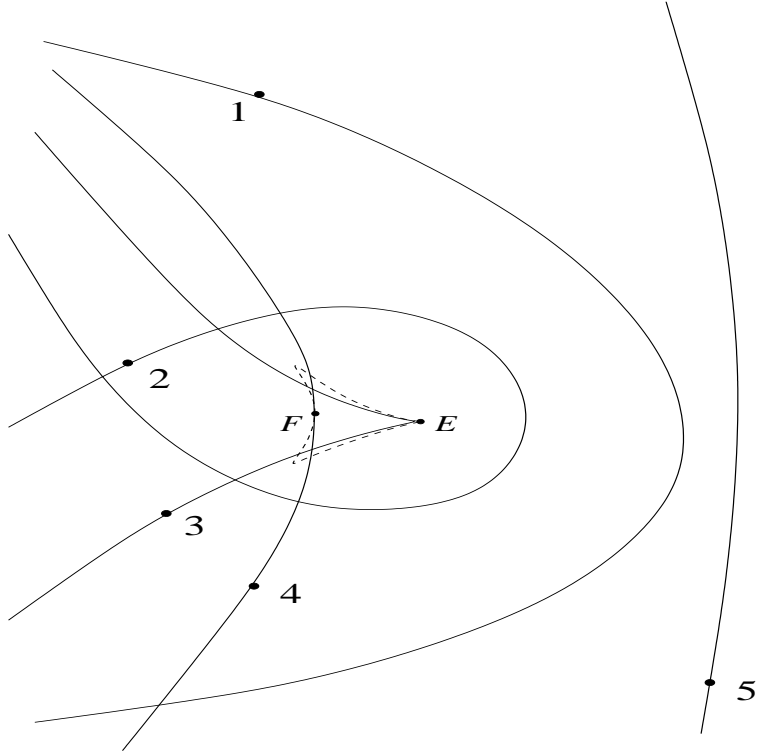}
\caption{Qualitative description of the flow near a caustic ring.
The solid lines are at the intersection of five simultaneous
infall spheres corresponding to different initial times
$t_{05}<t_{04}<.\; .\; .\; <t_{01}$. The five numbered points are
at $\vec{x}(\alpha, t_{0k}), k = 1,.\; .\; .\; 5$, for some value
of $\alpha$. Points $E$ and $F$ are defined in the text.  The
closed dashed line is at the intersection of the caustic tube with
the plane of the figure.  There are four flows inside the dashed
line whereas outside there are two.  The galactic center is to the
left of the figure.} \label{fig:fig5}
\end{figure}

Figure \ref{fig:fig5} depicts the snapshot of the flow near the
caustic. The curves are where the simultaneous (same $t$) infall
spheres corresponding to five different initial times:
$t_{01}>t_{02}>.\; .\; .
>t_{05}$ intersect the plane of the figure near the caustic.
The five numbered dots show the positions $\vec{x}(\alpha,t_{0k}),
k = 1,.\; .\; .\; 5$, for a fixed $\alpha$ and five different
initial times and following the dots gives one a qualitative
picture of the flow in time. The $t_{01}$ sphere is falling in but
has not yet crossed itself, as in frame Fig. \ref{fig:fig3}.B. The
$t_{02}$ sphere has crossed itself but has not yet completed the
process of turning itself inside out, as in frame Fig.
\ref{fig:fig3}.C. The $t_{03}$ sphere is just completing the
process of turning itself inside out, as in frame Fig.
\ref{fig:fig3}.D. The cusp at point $E$ is the location of a
caustic for the reason given earlier. Let $\alpha_{E}$ be the
value of $\alpha$ at $E$ and hence, $\vec{x}(\alpha_{E}, t_{03})$
be the position of $E$ in space. The particles at the cusp are
moving to the left. Thus at $(\alpha_{E}, t_{03})$,
$\dot{\vec{x}}$ is pointing to the left, whereas $\vec{x}_{t_0}$
is pointing to the right of the $t_{03}$ infall sphere. Although
time $t$ increases in the direction of motion, initial time $t_0$
decreases (earlier infall spheres are older, hence $t$ increases
in the direction of motion, or equivalently they have smaller
birthday $t_0$). Therefore $\frac{\partial}{\partial t_0}\sim
-\frac{\partial}{\partial t}$.

For smaller initial times, such as $t_{05}$, the spheres move to
the right; hence $\dot{\vec{x}}$ points to the right of the infall
sphere, whereas $\vec{x}_{t_0}$ points to the left. Let $t_{04}$
be the initial time and $F$ be the point where
$\vec{x}_{t_0}(\alpha_{E},t_0)$ crosses the $t_{04}$ sphere. Then,
according to Eq. \ref{3.2n},
$D=~\vec{x}_{t_0}~\cdot~(\vec{x}_\alpha\times \vec{x}_\beta)$
vanishes at $F$ because $\vec{x}_{t_0}$ is on the sphere and
$(\vec{x}_\alpha\times \vec{x}_\beta)$ is perpendicular to the
sphere at $F$. Thus $F$ is the location of a caustic as well. Now,
let us consider any other point which is far from both $E$ and
$F$. At such a point there are two flows because the sphere passes
such a point twice, once on the way in and once on the way out.
Consider also a point between $E$ and $F$. At such a point there
are four flows because the sphere passes by four times. We can
count these flows on Fig. \ref{fig:fig5} at fixed $t$ as follows:
twice for initial time $t_0$ between $t_{02}$ and $t_{03}$ (down
and up flows), once between $t_{03}$ and $t_{04}$ (in flow), and
once between $t_{04}$ and $t_{05}$ (outflow). Therefore, there is
a finite compact region whose boundary in the plane of Fig.
\ref{fig:fig5} is shown by the dashed lines. Inside the region
there are four flows (down, up, in and out), while outside there
are two (in the left of the region, down and up; in the right, in
and out). The boundary of this region is the location of the tube
caustic. If the successive spheres all fall in exactly the same
way, the caustic ring is stationary. In general, however, the
trajectories of successive spheres change slowly in time $t$, due
to the growth of the halo during the infall and inhomogeneities.
As a result the caustic ring moves about.

\subsubsection{Axially Symmetric Caustic Ring}
\label{sect:axiallysymmetricrings} In this section we assume that
the dark matter flow in and out of the galaxy is axially
(rotationally) symmetric about $\hat{z}$ and reflection symmetric
under $z\rightarrow -z$. We use the following parametrization of
the flow. Let $R(t_0)$ be the turnaround radius in the $z=0$ plane
at time $t_0$. Then let $\vec{x} (\theta_0, \varphi_0, t_0; t)$ be
the position at time $t$ of the particle that was  at the location
of polar and azimuthal angles $(\theta_0, \varphi_0)$ on the
sphere of radius $R(t_0)$ at time $t_0$. With this choice, we keep
the size of the sphere that we use to label the particles
comparable to the size of the growing galaxy. The density is given
by Eqs. \ref{2.2} and \ref{2.3} with $\vec{\alpha} = (\theta_0,
\varphi_0, t_0)$. There are $n$ flows outside the closed tube,
whereas there are $n+2$ flows inside, as described in a general
fashion in the previous section.

Although the flow is in 3D space, in general, the dimension along
the tube is irrelevant as far as the caustic properties are
concerned. Particles are actually moving in this direction,
however the motion is a simple rotation along the azimuthal
direction. To discover the essential properties of the caustic
tubes, it is sufficient to consider the cross-section
perpendicular to the trivial direction. Thus, the flow is in
effect two dimensional, even in the absence of axial symmetry. The
essential properties of caustics are invariant under continuous
deformations. To proceed further, we will assume axial symmetry of
the flow in this section. Therefore, let's throw away the
irrelevant $\varphi$ coordinate and choose $\rho(\alpha, t_0; t)$
and $z(\alpha, t_0; t)$ as the cylindrical coordinates at time $t$
of the ring of particles which start to fall-in with the polar
angle $\theta_0 = {\pi\over 2} - \alpha$ at initial time $t_0$. We
then can write the number of particles\beeq N\!\!=\!\!\int
\!\!\rho d\rho dz 2\pi d(\rho, z, t)\!=\!\int\!\! d\alpha dt_0
d\phi \frac{d^3N}{d\alpha dt_0 d\phi}(\alpha, t_0,
\phi)\!=\!\int\!\! d\alpha dt_0 \frac{d^2N}{d\alpha dt_0}(\alpha,
t_0)\;\;\; .\label{2.8}\eneq In the last equation, changing
variables $(\alpha, t_0)\rightarrow(\rho, z)$, we obtain\beeq
N=\int d\rho\; dz\; \sum_{j=1}^{n(\rho,\; z,\;
t)}\frac{d^2N}{d\alpha dt_0}(\alpha,
 t_0)\frac{1}{|D_2(\alpha, t_0)|}\Bigg|_{(\alpha,\;
 t_0)=(\alpha,\;
t_0)_j(\rho,\; z,\; t)}\; ,\eneq where the determinant of the
Jacobian \beeq D_2(\alpha, t_0)=\det\left(\frac{\partial(\rho,
z)}{\partial(\alpha, t_0)}\right)\label{2.9}\; ,\eneq and
$(\alpha, t_0)_j$, with $j = 1.\; .\; .n$, are the solutions of
$\rho = \rho (\alpha, t_0; t)~,~z= z(\alpha, t_0; t)$, and $n$ is
a function of $\rho$, $z$, and $t$. Comparing Eqs. \ref{2.8} and
\ref{2.9}, we find
\begin{equation}
d(\rho,z,t) = {1\over 2\pi\rho}~\sum_{j=1}^{n(\rho,\; z,\;
t)}~{d^2N\over d\alpha dt_0}~(\alpha, t_0)~{1\over\mid D_2(\alpha,
t_0)\mid}\Biggl|_{(\alpha, t_0) = (\alpha, t_0)_j(\rho,\; z,\;
t)}\; . \label{4.1}
\end{equation}
Under a reparametrization of the flow $(\alpha, t_0) \rightarrow
[\alpha^\prime (\alpha, t_0),~t_0^\prime (\alpha, t_0)]$, the
determinant transforms according to
\begin{equation}
D_2 (\alpha, t_0) = D_2^\prime (\alpha^\prime, t_0^\prime)~
\det\left({\partial (\alpha^\prime, t_0^\prime)\over
\partial (\alpha, t_0)}\right)\ .
\label{4.3}
\end{equation}
In particular, for an $\alpha$-dependent time shift: \be
\alpha^\prime = \alpha,~t_0^\prime = t_0 + \Delta t_0 (\alpha)\; ,
\label{alphadepshift} \ee the Jacobian of the transformation
$\det\left({\partial (\alpha^\prime, t_0^\prime)\over
\partial (\alpha, t_0)}\right)=1$, and
\begin{equation}
D_2 (\alpha, t_0) = D_2^\prime (\alpha^\prime, t_0^\prime)\; .
\label{4.4}
\end{equation}
Therefore, reparametrizations of type Eq. \ref{alphadepshift}
leaves the determinant $D_2$ unchanged. \subsubsection{Flow near
an Axially Symmetric Caustic Ring}
\label{subsect:flowatthecaustic} In section \ref{CR} we
qualitatively described the CDM flow near a ring caustic, as
summarized in Fig. \ref{fig:fig5}. In this section, we give a
quantitative description of ring caustics in the case of axial and
reflection symmetry. Reflection symmetry is automatic if the
caustic is tight, namely if the transverse dimensions $p$ and $q$
of the caustic are small compared to the tube radius. We will
assume that the caustics are tight.

Recall that the flow is reparametrization invariant. In Fig.
\ref{fig:fig5} we describe the flow at given time $t$. The time of
observation is fixed. However, we use initial time $t_0$ to label
the particles. Now, let us introduce a Cartesian coordinate system
on Fig. \ref{fig:fig5} by choosing the point $E$ as the origin,
the horizontal axis as the $x$-axis, and the vertical axis as the
$z$-axis. In the reflection symmetric case, points $E$ and $F$
have $z=0$. Let us label the particles on the $z=0$ axis by
$(\alpha,\tau=0)$ and relabel the flow by shifting the initial
time $t_0$ in an $\alpha$ dependent way: $\alpha\rightarrow
\alpha, ~t_0 \rightarrow \tau = t_0 + \Delta t_0 (\alpha)$ such
that $z (\alpha, \tau = 0) = 0$ for all
$\alpha$.\begin{figure}[ht] \centering
\includegraphics[height=7cm,width=9cm]{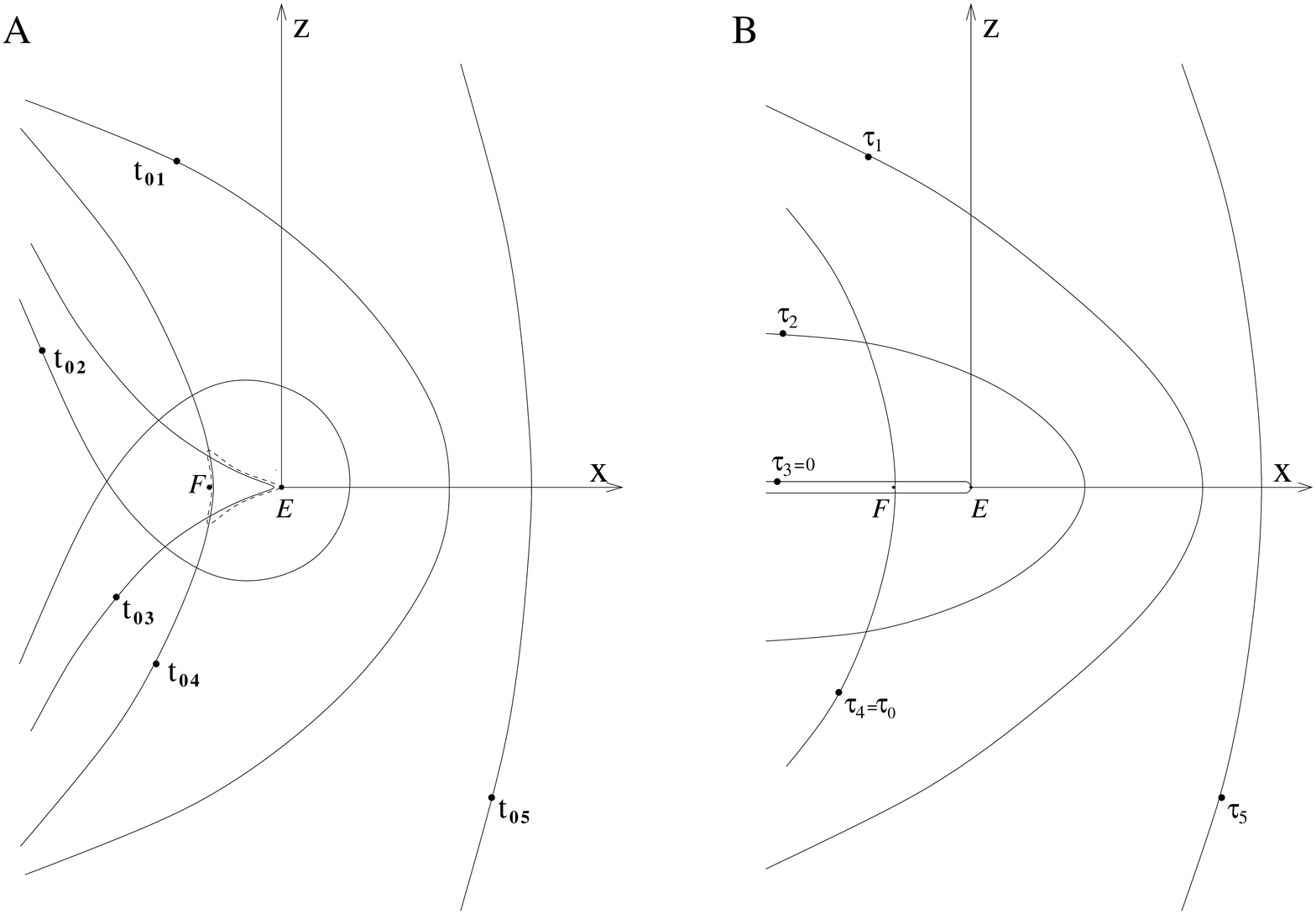}
\caption{Qualitative descriptions of the reflection symmetric flow
near a caustic ring. A) Same as Fig. \ref{fig:fig5} except that
now the flow has reflection symmetry and we introduce a coordinate
system whose origin is at the cusp $E$. $\alpha=0$ at $E$. The
$x$-coordinates of the particles labelled by $\pm\alpha$ are the
same whereas their $z$-coordinates have the opposite sign. B) We
reparametrize the flow by shifting $t_0$ in an $\alpha$ dependent
way: $t_0 \rightarrow \tau = t_0 + \Delta t_0 (\alpha)$ such that
$z (\alpha, \tau = 0) = 0$ for all $\alpha$. We label the particle
at the cusp by $(\alpha, \tau)=(0, 0)$ and all the other particles
on the negative $x$-axis by $(\alpha, \tau=0)$. Due to the
reflection symmetry, we have $(z, x)\rightarrow (-z, x)$ under
$(\alpha, \tau)\rightarrow (-\alpha, \tau)$. This fixes the
labelling for the rest of the particles in the flow.}
\label{fig:sphereatcaustic}
\end{figure}Physically this means that we relabel the particle going through
the cusp at $E$ by $(\alpha, \tau)=(0, 0)$ and all the other
particles on the negative $x$-axis $(z=0)$, by $(\alpha, \tau=0)$.
Due to the reflection symmetry, we must have $(z, x)\rightarrow
(-z, x)$ under $(\alpha, \tau)\rightarrow (-\alpha, \tau)$. We are
interested in the caustics in the flow. The properties of a
caustic depend on the flow near the caustic only. It does not
matter what happens in the flow away from the caustic. Therefore,
we want to expand the flow ($x$ and $z$) near the caustic in
powers of the parameters $\alpha$ and $\tau$ keeping terms up to
second order only. Let's start with the shell of particles that is
going through the cusp. This particular shell is in fact a line
segment on the $x$-$z$ plane where $z=0$ and $x<0$. Recall that
the parameter $\tau$ is chosen as zero for this line. Hence, the
parametrization satisfying the above symmetry requirements is
\beeq z=0\hskip 1cm x=-\frac{1}{2}s\;\alpha^2\ ,\eneq for this
line of particles of the flow, where $s$ is an arbitrary constant.
Now, lets expand the flow around the cusp where $\alpha=\tau=0$ by
writing the most general second order polynomials: \be z&=&c_1\;
\alpha + c_2\;\tau + c_3\;\alpha^2 + c_4\;\alpha\tau + c_5\;\tau^2\nonumber\ ,\\
x&=&d_1\;\alpha + d_2\;\tau + d_3\;\alpha^2 + d_4\;\alpha\tau +
d_5\;\tau^2 \label{secondordexpxz}\ . \ee Due to  reflection
symmetry, $z$ must be odd whereas $x$ must be even under $(\alpha,
\tau)\rightarrow (-\alpha, \tau)$. We then have $c_2=c_3=c_5=0$ in
$z$ and $d_1=d_4=0$ in $x$. Applying the condition that at
$\tau=0$, $z=0$ and $x=-\frac{1}{2}s\alpha^2$, we identify $c_1=0$
and $d_3=-\frac{1}{2}s$. Renaming the constants as $c_4\equiv b$,
$d_2\equiv-u\tau_0$, $d_5\equiv\frac{1}{2}u$ where $u$ and
$\tau_0$ are constants, we have \be z&=&b\alpha\tau\
,\label{4.5.a}\\
x&=&\frac{1}{2}u\left[(\tau-\tau_0)^2-\tau_0^2\right]-\frac{1}{2}s\alpha^2\
.\label{flowx-z}\ee If we call $\rho_0$ the $\rho$-coordinate of
point E where $\alpha=\tau=0$, we have $x=\rho-\rho_0$. Equation
\ref{flowx-z} can then be written as \beeq
\rho=a+\frac{1}{2}u(\tau-\tau_0)^2-\frac{1}{2}s\alpha^2\,
,\label{4.6}\eneq where we introduce a new constant \beeq a\equiv
\rho_0-\frac{1}{2}u\tau_0^2\equiv \rho_0-p\ , \eneq which will
turn out to be the inner tube radius ($\rho$-coordinate of point F
where $\alpha=0$ and $\tau=\tau_0$) and $p=\frac{1}{2}u\tau_0^2$,
the longitudinal dimension of the tube. The parameters $b,~u$ and
$s$ are positive.

The parameter $b$ is positive because the flow is from top to
bottom, $\frac{dz}{dt}<0$, (bottom to top, $\frac{dz}{dt}>0$) for
particles with $\alpha > 0~(\alpha < 0)$. Remember that particles
closer to the $z=0$ plane reached the turn around sphere earlier;
therefore, they have smaller $t_0$, and hence smaller $\tau$.
Although time $t$ increases in the direction of motion, $\tau$
decreases. Hence, for particles with $\alpha > 0~(\alpha < 0)$,
$\frac{d z}{d\tau}\sim-\frac{d z}{dt}>0$ $(\frac{d z}{d\tau}\sim
-\frac{d z}{dt}<0)$. Since $\frac{dz}{d\tau}=b\alpha$, the
constant $b>0$. Thus, $b\alpha$ is the negative of the velocity in
the $z$ direction of the particle coming in at an angle $\alpha$.
For particles with $\alpha=0$ the velocity in the $z$ direction is
zero, because they move in the $z=0$ plane.

The parameter $u$ is positive because the particles with $\alpha =
0$ are accelerated outward by the angular momentum barrier, hence
$\rho=a+\frac{1}{2}u(\tau-\tau_0)^2 \geq a$. Thus, $u$ is the
acceleration of the particles away from the distance of closest
approach. The parameter $s$ is positive because at $\tau = 0$, the
particles with $\alpha \neq 0$ are at
$\rho=\rho_0-\frac{1}{2}s\alpha^2 < \rho_0$. The parameter
$\tau_0$, which is the time it takes a particle to go from the
cusp to the radius of closest approach, can either be positive or
negative. Since $p=\frac{1}{2}u\tau^2_0>0$, $a=\rho_0-p$ is always
smaller than $\rho_0$ (i.e., point $F$ is always closer to the
galactic center than point $E$, as noted earlier).

The determinant of the Jacobian $D_2(\alpha,\tau)\equiv\det
\cal{D}(\alpha,\tau)$ where \be
\cal{D}(\alpha,\tau)\equiv\left(\begin{array}{cc}
{\dd\rho\over\dd\alpha}&~~~{\dd z\over\dd\alpha}\\
\\
{\dd\rho\over\dd\tau}&~~~ {\dd z\over\dd\tau}\\
\end{array}
\right)=\left(\begin{array}{cc}
{-s\alpha}&~~~{b\tau}\\
\\
{u(\tau-\tau_0)}&~~~ {b\alpha}\\
\end{array}
\right) \; ,\label{Jaco} \ee is
\begin{equation}
D_2 (\alpha,\tau) = - b[u\tau (\tau -\tau_0) + s\alpha^2]\; ,
\label{4.7}
\end{equation}
and vanishes for
\begin{equation}
\alpha = \pm \sqrt{{u\over s} \tau (\tau_0 - \tau)} \;
,\label{4.8}
\end{equation}
with $0 < \tau < \tau_0$ if $\tau_0 > 0$, and $\tau_0 < \tau < 0$
if $\tau_0 < 0$. Since the CDM density diverges where $D_2$
vanishes, by substituting Eq. \ref{4.8} into Eqs. \ref{4.5.a} and
\ref{4.6}, we find parametric ($\tau$ = parameter) equations
describing the cross-section of the caustic surface in the $(\rho
, z)$-plane: \beeq \rho = a + {1\over 2} u (\tau - \tau_0) (2\tau
- \tau_0)\; ,\nonumber\eneq \beeq z = \pm b\sqrt{{u\over s}
\tau^3(\tau_0 - \tau)}~~~~~ \ . \label{4.9} \eneq

Figure \ref{fig:fig6} shows a plot of the cross-section.
Both $\rho(\tau)$ and $z(\tau)$ are extremized for
$\tau=\frac{3}{4}\tau_0$, where $\rho=a-\frac{1}{8}p$ and $z=\pm
\frac{\sqrt{27}}{8}\frac{b}{\sqrt{us}}p$. At $\tau=0$, $\rho=a+p$
and $z=0$. At $\tau=\tau_0$, $\rho=0$ and $z=0$. Thus, the
dimensions of the cross section in the $\hat{\rho}$ and $\hat{z}$
directions are \be p = {1\over 2} u\tau_0^2\hskip 0.5cm {\rm
and}\hskip 0.5cm q = {\sqrt{27}\over 4}~{b\over\sqrt{us}}~p~~\;
,\ee respectively. The caustic has three cusps: one at $\tau=0$
where $(\rho , z)=(a+p , 0)$; and two at $\tau=\frac{3}{4}\tau_0$
where $(\rho , z)=(a-\frac{1}{8}p , \pm \frac{1}{2}q)$.

The three cusps are related by $Z_3$ transformations. To show
this, let us rescale the parameters:
\begin{equation}
T = {\tau\over\tau_0}\ ,~~~A = \sqrt{{s\over u}}~{1\over
\tau_0}~\alpha\ , \label{4.10}
\end{equation}
and rescale and shift (in the $\hat{\rho}$ direction) the
coordinates:
\begin{equation}
Z = {2 \over b}~\sqrt{{s \over u}}~{1\over \tau_0^2}~z~,~~~ X =
{2\over u\tau_0^2}~(\rho -a) - {1\over 4}\ , . \label{4.11}
\end{equation}
Eqs. \ref{4.5.a} and \ref{4.6} then become
\begin{equation}
Z = 2AT~,~~~X = (T -1)^2 - A^2 - {1\over 4}\ . \label{4.12}
\end{equation}
These relations are invariant under the discrete transformation:
\begin{eqnarray}
Z^\prime &=&-X\, \sin{(-2\pi/3)}+Z\cos{(-2\pi/3)}= - {1\over 2} Z
+
{\sqrt{3}\over 2} X~~,\nonumber\\
X^\prime &=& X\cos{(-2\pi/3)}+Z\cos{(-2\pi/3)}=- {\sqrt{3}\over 2} Z - {1\over 2} X~~,\nonumber\\
T^\prime &=& - {1\over 2} T + {\sqrt{3}\over 2} A
+\frac{3}{4}~~,~~~ A^\prime = - {\sqrt{3}\over 2} T - {1\over 2} A
+\frac{\sqrt{3}}{4}\ .
\end{eqnarray}
Notice that the cube of this transformation is the identity:
$Z^{\prime\prime\prime}=Z ,\; X^{\prime\prime\prime}=X ,\;
A^{\prime\prime\prime}=A$, and $T^{\prime\prime\prime}=T$. In the
X-Z plane, the transformation is a rotation by $120^o$.  It
transforms the three cusps of the caustic into one another. Thus,
after the rescaling, the caustic has a $Z_3$ symmetry. Hereafter,
we will call the shape of Fig. \ref{fig:fig6} ``tricusp.'' In the
language of Catastrophe Theory, the tricusp is a $D_{-4}$
catastrophe. The apparent reason for the two cusps which are not
on the $\hat{\rho}$-axis in Fig. \ref{fig:fig6} is that both \beeq
{d\rho\over d\tau}=\frac{u}{2}(4\tau-{3}\tau_0)\; ,\eneq and \beeq
{dz \over d\tau}=\pm
\frac{b}{2}\sqrt{\frac{u}{s}}\;\frac{(3\tau_0-4\tau)}{\sqrt{\frac{\tau_0}{\tau}-1}}\,
,\eneq vanish for $\tau = {3 \over 4} \tau_0$. The simultaneous
vanishing of ${d\rho \over d\tau}$ and ${dz \over d\tau}$ is,
however, not an accident peculiar to our assumptions of symmetry
and/or our limiting the expansion of $\rho$ and $z$ to terms of
second order in $\alpha$ and $\tau$. This claim is easy to prove.
Notice that the equation for the location of a generic caustic in
2 dimensions:
\begin{equation}
D_2(\alpha,\tau) = {\partial \rho \over \partial \alpha} {\partial
z \over \partial \tau} - {\partial \rho \over \partial \tau}
{\partial z \over \partial \alpha} = 0~~~\ , \label{4.21n}
\end{equation}
defines $\alpha(\tau)$ such that $[\rho(\tau) =
\rho(\alpha(\tau),\tau), z(\tau) = z(\alpha(\tau),\tau)]$ is a
parametric representation of the caustic location.  Wherever
$\rho(\tau)$ has an extremum, $z(\tau)$ is expected to have an
extremum as well, since
\begin{equation}
{d \rho \over d \tau} = {\partial \rho \over \partial \alpha} {d
\alpha \over d \tau} + {\partial \rho \over \partial \tau} = 0\
,\eneq implies\beeq  {\partial \rho \over \partial \tau}
=-{\partial \rho \over \partial \alpha} {d \alpha \over d \tau} \
.\label{4.22n}
\end{equation}
And Eq.~(\ref{4.21n}) yields
\begin{equation}
{\partial \rho \over \partial \alpha} \left[{\partial z \over
\partial \tau} + {\partial z \over \partial \alpha} {d \alpha
\over d\tau} \right]= 0 \ ,\eneq which implies\beeq {dz \over
d\tau} = {\partial z \over
\partial \alpha} {d \alpha \over d \tau} + {\partial z \over
\partial \tau} = 0 \, ,\label{4.23n}
\end{equation}
if ${\partial \rho \over \partial \alpha}$ and ${d \alpha \over d
\tau}$ are finite. In the same way, it can be shown that the
vanishing of $\frac{dz}{d\tau}$ implies that
$\frac{d\rho}{d\tau}=0$. The cusp at point $E$ may appear to have
a different origin.  Its apparent reason is that $z \sim \pm
\tau^{3 \over 2}$ near $\tau = 0$ which is at the boundary of the
range of $\tau$.  This, however, is an artifact of the
parametrization we have used. Indeed, we saw that the three cusps
can be transformed to each other due to a $Z_3$ symmetry, hence
the cusp on the $\hat{\rho}$-axis can be given the same
parametrization as the other two. In conclusion, the appearance of
three cusps in the cross-section of a ring caustic is neither an
accidental consequence of the assumed axial and reflection
symmetries nor is it due to the expansion including only second
order terms in Eq. \ref{secondordexpxz}.

Inside (outside) the tricusp, there are four (two) flows.  To
count the distinct flows, let us restrict ourselves to the $z=0$
plane, where either $\alpha = 0$ or $\tau = 0$.  If $\alpha = 0$,
then $\rho = a + {1\over 2} u (\tau - \tau_0)^2 > a$.  If $\tau =
0$, then $\rho = a + {1\over 2} u \tau_0^2 - {1\over 2} s\alpha^2
< \rho_0$. Thus, for $z=0$ and $\rho > a$, with $\alpha=0$, there
are two flows parametrized by
\begin{equation}
(\alpha,\tau_+) = \left(0,\tau_0 + \sqrt{{2\over u} (\rho -
a)}\,\right) , \; (\alpha,\tau_-) = \left(0, \tau_0 -
\sqrt{{2\over u} (\rho - a)}\,\right) \ .\label{4.14a}
\end{equation}
Both of the flows have \beeq{\partial z\over \partial
t}(\alpha=0)=- {\partial z\over \partial \tau}(\alpha=0) = 0 \
,\eneq whereas \beeq{\partial \rho\over
\partial t}(\alpha=0 , \tau_+) =
-{\partial \rho\over
\partial \tau}(\alpha=0 , \tau_+) =- u(\tau_+-\tau_0)=- \sqrt{2u(\rho - a)}\
,\nonumber\eneq\beeq {\partial \rho\over
\partial t}(\alpha=0 , \tau_-) = -{\partial \rho\over
\partial \tau}(\alpha=0 , \tau_-) = - u(\tau_--\tau_0)= +\sqrt{2u(\rho - a)}\ . \label{4.14b}
\eneq Thus, we call the flow $(\alpha, \tau_+)$, for which
${\partial \rho\over
\partial t} <0$, the ``in'' flow and the flow
$(\alpha, \tau_-)$, for which ${\partial \rho\over
\partial t} >0$, the ``out''
flow. For $z=0$, $(\alpha\not= 0, \tau=0)$ and hence
$\rho=\rho_0-\frac{1}{2}s\alpha^2 <\rho_0$, there are two other
flows parametrized by
\begin{equation}
(\alpha_+, \tau)=\left(\sqrt{{2\over s} (\rho_0 - \rho)}\; ,\;
0\right)\; ,\; (\alpha_-, \tau)=\left( -\sqrt{{2\over s} (\rho_0 -
\rho)}\; ,\; 0\right)\ , \label{4.15a}
\end{equation}
for both of which \beeq{\partial \rho \over \partial t}(\tau=0) =
- {\partial \rho \over \partial \tau}(\tau=0) = u\tau_0\ ,\eneq
whereas \be{\partial z\over \partial t}(\alpha_+ , \tau=0)&=&-
{\partial z\over \partial \tau}(\alpha_+ , \tau=0) = - b\alpha_+=-
b \sqrt{{2\over s} (\rho_0 -\rho)}\ ,\nonumber\\
{\partial z\over\partial t}(\alpha_- , \tau=0)&=&-{\partial z\over
\partial \tau}(\alpha_- , \tau=0) =- b\alpha_-=+b \sqrt{{2\over s}
(\rho_0 -\rho)} \ . \label{4.15b} \ee Thus, we call the flow
$(\alpha_+, \tau=0)$, for which ${\partial z\over
\partial t} <0$  as the ``down'' flow and the flow
$(\alpha_-, \tau=0)$, for which ${\partial z\over
\partial t} >0$  as the ``up''
flow. Therefore, in the $z=0$ plane, there are four flows (in,
out, down and up) for $a < \rho < \rho_0$, whereas there are two
flows (down and up) for $\rho < a$, and two flows (in and out) for
$\rho > \rho_0$.

Away from the cusps, the caustic is a generic surface caustic, as
described in Section \ref{sec:caustic surfaces}.  Thus, if one
approaches the boundary of the tricusp from the inside and away
from any of the cusps, as will be shown for a sample point below,
the density diverges as $d \sim {1\over\sqrt{\sigma}}$ where
$\sigma$ is the distance to the boundary.  If the boundary is
approached from the outside, away from any of the cusps, then the
density remains finite until the boundary is reached.  To
illustrate this, let us again restrict ourselves to the $z=0$
plane and calculate $D_2$ as $\rho - a \rightarrow 0_+$ and $\rho
- a \rightarrow 0_-$. In the former case there are four flows (up,
down, in and out), while in the latter there are only two (up and
down). The caustic is caused by the appearance of the extra in and
out flows in the region $\rho>a$. For $z=0$, and $\rho - a
\rightarrow 0_+$, the in and out flows have $(\alpha, \tau)=(0,
\tau_0\pm\sqrt{\frac{2}{u}(\rho- a)})$, respectively. Therefore
the determinant (\ref{4.7}) is \beeq D_2=\mp
b\tau\sqrt{2u(\rho-a)}\; .\eneq Since $\tau\rightarrow\tau_0$ as
$\rho - a \rightarrow 0_+$ when $\alpha=0$, $D_2 \simeq \mp
b\tau_0 \sqrt{2u(\rho - a)}$ for the in and out flows, the density
associated with these flows diverges as ${1\over\sqrt{\rho - a}}$.
The two other flows (down and up) yield finite contributions to
the density. At $z=0$, for the down and up flows $(\alpha_\pm ,
\tau)=(\pm\sqrt{\frac{2}{s}(\rho_0-\rho)}\;,0)$. Hence, the
determinant\beeq D_2=-bs\alpha^2_\pm=-2b(\rho_0 -\rho)=-2b(a+p
-\rho)\; .\eneq Therefore, in the limits $\rho - a \rightarrow
0_+$, $(\rho>a)$, and $\rho - a \rightarrow 0_-$, $(\rho<a)$,
$|D_2|\rightarrow 2bp$ for the down and up flows. The contribution
of each flow to the density, in both regions, is $\frac{1}{2bp}$.

Near the cusps, the behavior depends upon the direction of
approach. For $z=0$ and $\rho - \rho_0 \rightarrow 0_-$ (inside
the caustic tube), there are four flows: in, out, up and down. For
$\rho - \rho_0 \rightarrow 0_+$ (outside the caustic tube), there
are two flows: in and out. In the region $\rho<\rho_0$, in
addition to the extra down and up flows, which both diverge in the
limit $\rho-\rho_0\rightarrow 0_-$, one of the common flows (out
flow if $\tau_0>0$, in flow if $\tau_0<0$) of the regions also
diverges in both limits $\rho-\rho_0\rightarrow 0_\pm$, whereas
the other (in flow if $\tau_0>0$, out flow if $\tau_0<0$) remains
finite. Notice that near the cusp (unlike the case $\rho-a<0$
where both of the down and up flows remain finite as one
approaches the caustic from the outside), there is a flow which
diverges as one approaches the caustic from the outside of the
tube. Hence, the cusps are more divergent then the folds.

Let us first calculate the density due to the down and up flows
which exist only in the region $\rho<\rho_0$, for $z=0$ and $\rho
- \rho_0 \rightarrow 0_-$. Recall (Eq. \ref{4.15a}) that in the
$z=0$ plane, for the up and down flows, $\tau=0$ and
$\alpha^2=\frac{2}{s}(\rho-\rho_0)$. Then,
$D_2=-bs\alpha^2=2b(\rho_0 -\rho)$, and hence $d \sim {1\over
\rho_0 - \rho}$. Let us now calculate the density due to the in
and out flows which exist for both $\rho<\rho_0$ and
$\rho>\rho_0$. For $\rho - \rho_0\rightarrow 0_-$, let us write
$\rho\equiv \rho_0-\sigma$ where $\sigma\rightarrow 0_+$. Then,
Eq. \ref{4.14a} gives for the in and out flows\beeq (\alpha ,
\tau_{\pm})=(0 , \tau_0\pm\sqrt{\tau_0^2-\frac{2}{u}\sigma}\;)\; ,
\eneq respectively. Thus, using Eq. \ref{4.7}, we find\beeq
D_2=-bu\left[\pm
\tau_0|\tau_0|\sqrt{1-\frac{\sigma}{p}}+\tau_0^2-\frac{2}{u}\sigma\right]\
.\label{|D2|}\eneq If $\tau_0>0$ the in flow gives, to first order
in $\sigma$, \beeq D_2\simeq -bu\left[
2\tau_0^2-\frac{3}{u}\sigma\right]\ .\eneq As $\sigma\rightarrow
0$, $D_2\rightarrow -2bu\tau_0^2=-4bp$. Hence the density of the
in flow $d\sim \frac{1}{4bp}$ is finite for $\tau_0>0$. The out
flow, however, has a divergent density for $\tau_0>0$. Since
$D_2\simeq b\sigma=b(\rho_0-\rho)$, the density is
$d\sim\frac{1}{\rho_0-\rho}$. If $\tau_0<0$, the flows are time
reversed: the out flow becomes the in flow and vice versa. Indeed,
Eq. \ref{|D2|} gives $D_2\simeq b\sigma$, hence $d\sim
\frac{1}{\rho_0-\rho}$ for the in flow, and $D_2\simeq
-bu[2\tau_0^2 -\frac{3}{u}\sigma]$, which yields $d\sim
\frac{1}{4bp}$ for the out flow. For the region $\rho>\rho_0$, in
the limit $\rho-\rho_0\rightarrow 0_+$, $\rho=\rho_0+\sigma$ where
$\sigma\rightarrow 0_+$. Then, for the in and out flows $(\alpha ,
\tau_\pm)=(0 ,\tau_0\pm\sqrt{\tau_0^2+\frac{2}{u}\sigma} \;)$ and
$D_2=-bu[\pm
\tau_0|\tau_0|\sqrt{1+\frac{\sigma}{p}}+\tau_0^2+\frac{2}{u}\sigma]$.
Therefore, the in and out flows have the same density as for the
in and out flows where $\rho -\rho_0\rightarrow 0_-$. If
$\tau_0>0$, the in flow is finite: $d\sim\frac{1}{4bp}$ and the
out flow is divergent: $d\sim\frac{1}{\rho - \rho_0}$. If
$\tau_0<0$, flows are time reversed, hence the in flow has $d\sim
\frac{1}{\rho-\rho_0}$ and the out flow has $d\sim \frac{1}{4bp}$.

Finally, let us calculate the density near the cusp at $\rho =
\rho_0$ and as $z\rightarrow 0_\pm$. Since $\rho= \rho_0
-u\tau_0\tau +\frac{u}{2} \tau^2-{s\over 2} \alpha^2$, at
$\rho=\rho_0$ where $(\tau,\alpha)=(0, 0)$, \beeq
-u\tau_0\tau+\frac{u}{2} \tau^2-{s\over 2} \alpha^2= 0\;
.\label{sifir}\eneq Near the cusp, where $(\rho, z)=(\rho_0,
0_\pm)$, $(\tau, \alpha)\not=(0, 0)$ and $\alpha$ and $\tau$ are
independent since the point of interest is not exactly on the
caustic surface. The determinant
$D_2=-b[u\tau^2-u\tau_0\tau+s\alpha^2]$ can be written at
$\rho=\rho_0$, using Eq. \ref{sifir} as\beeq
D_2=-bu\left[2\tau^2-3\tau_0\tau\right] \, .\eneq Since
$\frac{z}{\alpha}=b\tau$, $D_2$ at $\rho_0$ can also be written as
\beeq D_2=-u\left[\frac{2}{b}\frac{z^2}{\alpha^2}
-3\tau_0\frac{z}{\alpha}\right]\; \label{D2z}.\eneq Since near the
cusp $\frac{z}{\alpha}=b\tau\rightarrow 0$, $(\frac{z}{\alpha})^2$
can be neglected next to $\frac{z}{\alpha}$ in the above equation.
This is equivalent to neglecting the $u\tau^2$ term next to
$u\tau$ in Eq. \ref{sifir}. Therefore, from Eq. \ref{sifir} and
Eq. \ref{D2z} we obtain \beeq \alpha^2\simeq
-2\frac{u}{s}\tau_0\tau=-2\frac{u}{bs}\tau_0\frac{z}{\alpha}\Rightarrow
\alpha\simeq \left[-2\frac{u}{bs}\tau_0 z\right]^{1/3}\; ,\eneq
and \beeq D_2\simeq 3u\tau_0\frac{z}{\alpha}=-3 \left[{u^2 b
\tau_0^2 s |z|^2\over 2}\right]^{1/3}\; ,\eneq for {\it one} of
the flows. Hence $d\sim {1\over\mid z\mid^{2/3}}$ for this flow.
Thus, we find that, the density also diverges in this particular
limit near the cusp.

The tube caustic collapses to a line caustic in the limit
$\tau_0\rightarrow 0$ with ${b\over\sqrt{us}}$ fixed.  In this
limit \beeq p=\frac{1}{2}u\tau_0^2\rightarrow 0\; ,\;\;\;
q=\frac{\sqrt{27}}{4}\frac{b}{\sqrt{us}}p\rightarrow 0\; ,
\nonumber\eneq \beeq D_2 \rightarrow -b(u\tau^2 + s\alpha^2)\;
,\;\;\; \rho\rightarrow a+\frac{1}{2}(u\tau^2-s\alpha^2) \; .\eneq
Therefore, we can rewrite\beeq D_2\!=\!-2\sqrt{b^2 \left[ (\rho
-a)+s\alpha^2
\right]^2}\!=\!-2\sqrt{-b^2\left[(\rho-a)^2+s^2\alpha^4+s\alpha^2(u\tau^2-s\alpha^2)
\right]}\; ,\label{r-a}\eneq where we replaced $\rho -a$ by
$\frac{1}{2}(u\tau^2-s\alpha^2)$ using Eq. \ref{r-a}. Since
$z=b\alpha\tau$, we might write \beeq D_2= -2 \sqrt{b^2(\rho -a)^2
+ usz^2} \label{4.16}~~~\  .\eneq Hence the density
\begin{equation}
d(\rho, z) = {1\over 2\pi\rho}~{d^2N\over d\alpha dt_0}
~{1\over\sqrt{b^2(\rho -a)^2 + usz^2}}~~~\ , \label{4.17}
\end{equation}
in the $\tau_0\rightarrow 0$ limit. When $p$ and $q$ are finite,
but much smaller than $a$, Eq. \ref{4.17} is approximately valid
for $p, q\ll\sigma\ll a$ since the terms of order $\tau_0$ in
Eqs.~\ref{4.5.a}, \ref{4.6} and \ref{4.7} are negligible in this
regime. \subsubsection{Differential Geometry of an Axially
Symmetric Caustic Ring} \label{difgeo}

In this section, the differential geometric properties of the
axially symmetric caustic ring are studied. At the end, we
calculate the principal curvature radii, which are needed for
lensing applications, at an arbitrary point on the ring surface.
Any surface in $R^3$ is uniquely determined by certain local
invariant quantities called first and second fundamental forms.
Following \cite{Lipschutz}, let us derive the fundamental forms
for the ring surfaces.

Let $\vec{X}=\vec{X}(\tau, \phi)$ be a coordinate patch on a
surface. The differential of the map $\vec{X}=\vec{X}(\tau, \phi)$
at $(\tau, \phi)$ is a one to one linear map \beeq
d\vec{X}=\vec{X}_\tau d\tau +\vec{X}_\phi d\phi\ ,\eneq with
$\vec{X}_\tau \equiv \frac{\partial \vec{X}}{\partial\tau}\ ,
\vec{X}_\phi \equiv \frac{\partial \vec{X}}{\partial\phi}$, of the
vectors $(d\tau, d\phi)$, which extend from $(\tau, \phi)$ to
$(\tau+d\tau, \phi+d\phi)$ in the $\tau\phi$-plane, on to the
vectors $\vec{X}_\tau d\tau +\vec{X}_\phi d\phi$ parallel to the
tangent plane of the surface at $\vec{X}(\tau, \phi)$. Notice
that, we are using the symbols $d\tau$ and $d\phi$ both for the
differentials of the coordinate functions in the $\tau\phi$-plane
and for the components of a vector in the $\tau\phi$-plane. In the
same way, we denote $\vec{X}_\tau d\tau +\vec{X}_\phi d\phi$ by
$d\vec{X}$. Recall that, since \beeq\vec{X}(\tau+d\tau ,
\phi+d\phi)=\vec{X}(\tau, \phi)+ d\vec{X}+O(d\tau^2, d\tau d\phi,
d\phi^2)\; ,\eneq the vector $d\vec{X}$ is a first order
approximation to the vector $\vec{X}(\tau+d\tau ,
\phi+d\phi)-\vec{X}(\tau, \phi)$ from the point $\vec{X}(\tau,
\phi)$ on the patch, to the neighboring point $\vec{X}(\tau+d\tau
, \phi+d\phi)$. Consider the quantity:\be I\equiv
ds^2&\equiv&(\vec{X}_\tau d\tau +\vec{X}_\phi d\phi)\cdot
(\vec{X}_\tau d\tau +\vec{X}_\phi d\phi)\nonumber\\
&=&(\vec{X}_\tau\cdot\vec{X}_\tau)d^2\tau+2(\vec{X}_\tau\cdot\vec{X}_\phi)d\tau
d\phi +(\vec{X}_\phi\cdot\vec{X}_\phi)d^2\phi\nonumber\\
&\equiv& {\bf E}\; d^2\tau+2{\bf F}\; d\tau d\phi +{\bf G}\;
d^2\phi\ .\ee The function $I$ is called the first fundamental
form of $\vec{X}=\vec{X}(\tau, \phi)$. The first fundamental
coefficients ${\bf E}$, ${\bf F}$ and ${\bf G}$ are functions of
$\tau$ and $\phi$. Let us find the first fundamental form (i.e.,
the metric), on the caustic surface. The three vector \beeq
\vec{X}(\tau,
\phi)=\rho(\tau)\cos{(\phi)}\hat{x}+\rho(\tau)\sin{(\phi)}\hat{y}+z(\tau)\hat{z}
\label{Xtf}\eneq completely describes the surface. The functions
$\rho(\tau)$ and $z(\tau)$ are given in Eq. \ref{4.9}, which we
copy here \beeq \rho = a + {1\over 2} u (\tau - \tau_0) (2\tau -
\tau_0)\; ,\hskip 1cm z = \pm b\sqrt{{u\over s} \tau^3(\tau_0 -
\tau)}~~~~~ \ .\nonumber \eneq Since \beeq
\vec{X}_\tau=\rho'\cos(\phi)\hat{x}+\rho'\sin(\phi)\hat{y}+z'\hat{z}\
, \vec{X}_\phi=-\rho\sin(\phi)\hat{x}+\rho\cos(\phi)\hat{y}\
,\eneq where prime denotes $\frac{\partial}{\partial\tau}$, the
first fundamental form coefficients for the caustic ring are\beeq
{\bf E}=\vec{X}_\tau\cdot\vec{X}_\tau=\rho'\;^2+z'\;^2\; ,\hskip
0.5cm {\bf F}=\vec{X}_\tau\cdot\vec{X}_\phi=0\; ,\hskip 0.5cm {\bf
G}=\vec{X}_\phi\cdot\vec{X}_\phi=\rho^2\ .\eneq Therefore, \beeq
ds^2=(\rho'\;^2+z'\;^2)d\tau^2+\rho^2d\phi^2\ .\eneq Using Eq.
\ref{4.9} the explicit metric of the surface is found \beeq
ds^2=\frac{1}{4}\left[u{(3\tau_0-4\tau)^2[
u+\frac{b^2\tau}{s(\tau_0-\tau)}]}d\tau^2+[2a +u(\tau_0
-\tau)(\tau_0-2\tau)]^2d\phi^2\right]\ .\eneq The area
${{\mathcal{A}}}$ of the caustic surface can be obtained readily
since\beeq
{{\mathcal{A}}}=2\int_0^{2\pi}d\phi\int_0^{\tau_0}d\tau\sqrt{\det{g}}\eneq
where the determinant of the metric is ${\bf E}{\bf G}-{\bf F}^2
>0$. Because $I$ is positive definite, its coefficients must satisfy
${\bf E}>0$, ${\bf G}>0$, and ${\bf E}{\bf G}-{\bf F}^2
>0$. This can easily be verified. Since $\vec{X}_\tau$ and
$\vec{X}_\phi$ are independent, $\vec{X}_\tau\not= 0,
\vec{X}_\phi\not= 0$, and hence ${\bf
E}=\vec{X}_\tau\cdot\vec{X}_\tau=|\vec{X}_\tau|^2>0$, ${\bf
G}=\vec{X}_\phi\cdot\vec{X}_\phi=|\vec{X}_\phi|^2>0$. The
determinant\be \det(g)={\bf E}{\bf G}-{\bf
F}^2&=&(\vec{X}_\tau\cdot\vec{X}_\tau)(\vec{X}_\phi\cdot\vec{X}_\phi)
-(\vec{X}_\tau\cdot\vec{X}_\phi)(\vec{X}_\tau\cdot\vec{X}_\phi)\nonumber\\
&=&(\vec{X}_\tau\times\vec{X}_\phi)\cdot(\vec{X}_\tau\times\vec{X}_\phi)
=|\vec{X}_\tau\times\vec{X}_\phi|^2>0\ .\label{detEF-G}\ee On the
caustic surface we have \beeq \det(g)=(\rho'\;^2+z'\;^2)\rho^2
=\frac{u}{16}{(3\tau_0-4\tau)^2}{(u+\frac{b^2\tau}{s(\tau_0-\tau)})}\left[2a
+u(\tau_0^2 -3\tau_0\tau+2\tau^2)\right]^2\; .\nonumber\eneq For
simplicity, if we choose $\frac{us}{b^2}=1$, we find the area
as\beeq {{\mathcal{A}}}=\frac{8\pi}{105}\;p\;(35 a +29p)\; .\eneq
The non-vanishing, independent components of the  Riemann and
Ricci tensors of the caustic ring surface are\beeq
R_{\tau\phi\tau\phi}=\frac{\rho\; z'(\rho'
z''-z'\rho'')}{(\rho'\;^2+z'\;^2)}=-\frac{ub^2}{8}
\frac{\tau_0(3\tau_0-4\tau)(2a+u(\tau_0^2-3\tau_0\tau+2\tau^2))}
{(\tau_0-\tau) (b^2\tau+us(\tau_0-\tau))}\, ,\eneq \beeq
R_{\tau\tau}=\frac{z'(\rho'
z''-z'\rho'')}{\rho(\rho'\;^2+z'\;^2)}=-\frac{ub^2}{2}
\frac{\tau_0(3\tau_0-4\tau)}{(\tau_0-\tau)
(b^2\tau+us(\tau_0-\tau))(2a+u(\tau_0^2-3\tau_0\tau+2\tau^2))} \,
,\nonumber\eneq \beeq R_{\phi\phi}=\frac{\rho z'(\rho'
z''-z'\rho'')}{(\rho'\;^2+z'\;^2)^2}=-\frac{sb^2}{2}
\frac{\tau_0(2a+u(\tau_0^2-3\tau_0\tau+2\tau^2))} {(3\tau_0-4\tau)
(b^2\tau+us(\tau_0-\tau))^2}\, , \eneq respectively. Components of
the Ricci tensor can also be written as \beeq
R_{ij}=g_{ij}\frac{z'(\rho'
z''-z'\rho'')}{\rho(\rho'\;^2+z'\;^2)^2}\, .\eneq Therefore, the
Ricci scalar is\beeq R=2\frac{z'(\rho'
z''-z'\rho'')}{\rho(\rho'\;^2+z'\;^2)^2}=\frac{-4sb^2\tau_0}
{(3\tau_0-4\tau)(b^2\tau+us(\tau_0-\tau))^2
(2a+u(\tau_0^2-3\tau_0\tau+2\tau^2))}\; .\nonumber\eneq The
Gaussian curvature of the surface is calculated as\beeq
{{\mathcal{K}}}=\frac{R_{\tau\phi\tau\phi}}{\det(g)}=\frac{z'(\rho'
z''-z'\rho'')}{\rho(\rho'\;^2+z'\;^2)^2}=\frac{1}{2}R\
,\label{GaussKfirst}\eneq and will also be obtained as the product
of the principal curvatures later in this section.

In order to calculate the principal curvature radii and the Gauss
curvature of the surface, we need to calculate the second
fundamental form $II$. To find $II$, we need the unit normal
vector $\hat{N}=\frac{\vec{X}_\tau \times
\vec{X}_\phi}{|\vec{X}_\tau \times \vec{X}_\phi|}$, which is a
function of $\tau$ and $\phi$, with differential
$d\hat{N}=\hat{N}_\tau d\tau +\hat{N}_\phi d\phi$. Observe that
$0=d(1)=d(\vec{N}\cdot\vec{N})=2\, d\vec{N}\cdot\vec{N}$. Thus
$d\vec{N}$ is parallel to the tangent plane at $\vec{X}$. The
quantity \be II&\equiv& -d\vec{X}\cdot d\hat{N}=-(\vec{X}_\tau
d\tau+\vec{X}_\phi d\phi)\cdot (\hat{N}_\tau d\tau +\hat{N}_\phi
d\phi)\nonumber\\
&=&-\vec{X}_\tau\cdot\hat{N}_\tau
d^2\tau-(\vec{X}_\tau\cdot\hat{N}_\phi
+\vec{X}_\phi\cdot\hat{N}_\tau)d\tau d\phi
-\vec{X}_\phi\cdot\hat{N}_\phi d^2\phi\nonumber\\
&\equiv&{\bf L}\; d^2\tau+2{\bf M}\; d\tau d\phi+{\bf N}\;
d^2\phi\; \label{formII}\ ,\ee is called the second fundamental
form of $\vec{X}=\vec{X}(\tau, \phi)$. {\bf L}, {\bf M} and {\bf
N} are called the second fundamental coefficients, and are
continuous functions of $\tau$ and $\phi$. We can express the
second fundamental coefficients  in terms of $\hat{N}$ and the
second derivatives of $\vec{X}$, using the fact that
$\vec{X}_\tau$ and $\vec{X}_\phi$ are perpendicular to $\hat{N}$
for all $(\tau , \phi)$:\be 0&=&(\vec{X}_\tau\cdot\hat{N})_\tau
=\vec{X}_{\tau\tau}\cdot\hat{N} +\vec{X}_{\tau}\cdot\hat{N}_\tau
\Rightarrow {\bf
L}=-\vec{X}_{\tau}\cdot\hat{N}_\tau=\vec{X}_{\tau\tau}\cdot\hat{N}\ \label{bfL}\\
0&=&(\vec{X}_\phi\cdot\hat{N})_\phi
=\vec{X}_{\phi\phi}\cdot\hat{N} +\vec{X}_{\phi}\cdot\hat{N}_\phi
\Rightarrow {\bf
N}=-\vec{X}_{\phi}\cdot\hat{N}_\phi=\vec{X}_{\phi\phi}\cdot\hat{N}
\ .\label{bfN}\ee Using the remaining two identities\be
0&=&(\vec{X}_\tau\cdot\hat{N})_\phi=\vec{X}_{\tau\phi}\cdot\hat{N}
+\vec{X}_{\tau}\cdot\hat{N}_\phi \Rightarrow
-\vec{X}_{\tau}\cdot\hat{N}_\phi=\vec{X}_{\tau\phi}\cdot\hat{N} \\
0&=&(\vec{X}_\phi\cdot\hat{N})_\tau=\vec{X}_{\phi\tau}\cdot\hat{N}
+\vec{X}_{\phi}\cdot\hat{N}_\tau \Rightarrow
-\vec{X}_{\phi}\cdot\hat{N}_\tau=\vec{X}_{\phi\tau}\cdot\hat{N}\ .
\ee we find \beeq {\bf
M}=-\frac{1}{2}(\vec{X}_\tau\cdot\hat{N}_\phi
+\vec{X}_\phi\cdot\hat{N}_\tau)=\vec{X}_{\tau\phi}\cdot\hat{N}=\vec{X}_{\phi\tau}\cdot\hat{N}\
. \label{bfM}\eneq Therefore, the second fundamental form \be
II&=&{\bf L} d^2\tau+2{\bf M}d\tau d\phi +{\bf
N}d^2\phi\nonumber\\
&=&\vec{X}_{\tau\tau}\cdot \hat{N}
d^2\tau+2\vec{X}_{\tau\phi}\cdot \hat{N}d\tau d\phi
+\vec{X}_{\phi\phi}\cdot \hat{N} d^2\phi=d^2\vec{X}\cdot\hat{N}\
,\ee where we define $d^2\vec{X}\equiv\vec{X}_{\tau\tau}
d^2\tau+2\vec{X}_{\tau\phi}d\tau d\phi
+\vec{X}_{\phi\phi}d^2\phi$. For the caustic ring, the unit normal
\beeq \hat{N}=\frac{\vec{X}_\tau \times
\vec{X}_\phi}{|\vec{X}_\tau \times
\vec{X}_\phi|}=\frac{\rho'\hat{z}-z'(\cos\phi\;\hat{x}+\sin\phi\;\hat{y})}
{\sqrt{\rho'\;^2+z'\;^2}}\; ,\eneq and the second fundamental
coefficients are \begin{equation}{\bf
L}=\vec{X}_{\tau\tau}\cdot\hat{N}=\frac{\rho'
z''-z'\rho''}{\sqrt{\rho'\; ^2+z'\; ^2}}\; ,\;\;\; {\bf
N}=\vec{X}_{\phi\phi}\cdot\hat{N}=\frac{z'\rho}{\sqrt{\rho'\;
^2+z'\; ^2}}\; , \nonumber\end{equation}
\begin{equation}
{\bf M}=\vec{X}_{\tau\phi}\cdot\hat{N}=0\; .\end{equation}

Now, let $P$ be a point on the surface, $\vec{X}=\vec{X}(\tau,
\phi)$ a patch containing $P$, and $\vec{X}=\vec{X}(\tau(t),
\phi(t))$ a curve $C$ through $P$ where $t$ is an arbitrary real
parameter in an interval $I$. For instance, \beeq
s=s(t)\equiv\int_{t_0}^t \Big|\frac{d\vec{X}}{dt}\Big| \; dt \
,\label{ds/dt}\eneq where the constant $t_0\in I$ and $t_0<t$, can
be used to parametrize the curve. Notice that, $s(t)$ is the
length of the arc segment of the curve between $\vec{X}(t_0)$ and
$\vec{X}(t)$, hence $X(s)$ is called ``$\it{a}$ natural
representation'' of a curve. Note also that, a representation in
terms of arc length is not unique, because, it depends on the
chosen initial point $t_0$ (where $s=0$). The tangent vector
$\vec{\bf t}$ to $C$ at $\vec{X}(s)$ is defined as\beeq \vec{\bf
t}={\vec{\bf t}}(s)\equiv\frac{d \vec{X}}{d s}(s)\ ,\eneq where ${
\vec{\bf t}}$ is a unit vector since \beeq {\vec{\bf t}}=\frac{d
\vec{X}}{d s}=\frac{d \vec{X}}{d t}\frac{d t}{d s}={\frac{d
\vec{X}}{d t}}\Big{/}{\frac{d s}{d t}} \ .\label{bft}\eneq From
Eq. \ref{ds/dt}, however, we obtain \beeq
\frac{ds}{dt}=\Big|\frac{d\vec{X}}{dt}\Big|\ ,\label{ds/dt |dx/dt|
}\eneq hence Eq. \ref{bft} becomes \beeq {\vec{\bf t}}=\frac{d
\vec{X}}{d s}={\frac{d \vec{X}}{d t}}\Big{/}\Big|{\frac{d
\vec{X}}{d t}}\Big|\ ,\label{tdxdt}\eneq and, $|\vec{\bf t}|=1$,
as claimed. We will define all the geometric quantities along the
curve in terms of a natural representation, as in the case of the
unit tangent vector $\vec{\bf t}$, however, all of these
quantities can be re-expressed in terms of an arbitrary parameter
$t$, by using the chain rule and Eq. \ref{ds/dt |dx/dt| }.

The first derivative of the tangent vector $\vec{\bf t}(s)$,
denoted by $\vec{\bf k}(s)$, is called the curvature vector on $C$
at the point $\vec{X}(s)$, hence\beeq\vec{\bf k}=\vec{\bf k}(s)=
\frac{d \vec{\bf t}}{ds}(s)=\frac{d^2 \vec{X}}{d s^2}(s)\ ,\eneq
or, using Eq. \ref{ds/dt |dx/dt| } \beeq \vec{\bf
k}=\frac{d\vec{\bf t}}{ds}=\frac{d\vec{\bf
t}}{dt}\Big/\Big|\frac{d \vec{X}}{d t}\Big| \
.\label{kdtdsdt}\eneq Since $\vec{\bf t}$ is a unit vector,
$\vec{\bf k}=\frac{d \vec{\bf t}}{ds}$ is orthogonal to $\vec{\bf
t}$: \beeq 0=d(1)=\frac{d}{ds}(\vec{\bf t}\cdot\vec{\bf
t}\;)=2\frac{d\vec{\bf t}}{ds}\cdot\vec{\bf t} \ .\nonumber\eneq
Therefore $\vec{\bf k}(s)$ is parallel to the normal plane. The
direction of $\vec{\bf k}$ is the direction in which the curve is
turning. The magnitude of the curvature vector $\kappa=|\vec{\bf
k}(s)|$ is called the curvature of $C$ at $\vec{X}(s)$. In fact,
geometrically, $\kappa$ is equal to the rate of change of the
direction of the tangent vector with respect to the arc length. To
prove the claim let us call the angle between the unit tangent
$\vec{\bf t}(s)$ at $\vec{X}(s)$ and $\vec{\bf t}(s+\Delta s)$ at
a neighboring point $\vec{X}(s+\Delta s)$ as $\Delta\theta$. Since
$\vec{\bf t}$ is a unit vector, $|\vec{\bf t}(s+\Delta s)-\vec{\bf
t}(s)|$ is the base of an isosceles triangle with sides of unit
length. Hence $|\vec{\bf t}(s+\Delta s)-\vec{\bf t}(s)|=
2\sin{(\frac{\Delta\theta}{2})}=\Delta\theta+O\left((\Delta\theta)^2\right)$.
Then, \beeq \kappa=\Big|\frac{d\vec{\bf
t}}{ds}\Big|=\Big|\lim_{\Delta s\rightarrow 0}\frac{\vec{\bf
t}(s+\Delta s)-\vec{\bf t}(s)}{\Delta s}\Big|=\lim_{\Delta
s\rightarrow
0}\frac{\Delta\theta+O\left((\Delta\theta)^2\right)}{\Delta s} \ ,
\eneq or, we may write \beeq \kappa=\lim_{\Delta s\rightarrow
0}\left[\frac{\Delta\theta}{\Delta s}\left(
1+\frac{O\left((\Delta\theta)^2\right)}{\Delta\theta}\right)\right]\
.\nonumber\eneq Since \beeq\lim_{\Delta s\rightarrow 0}
\Delta\theta=0 \Rightarrow\lim_{\Delta s\rightarrow 0}
\frac{O\left((\Delta\theta)^2\right)}{\Delta\theta}=0\ ,\eneq the
curvature \beeq\kappa=\lim_{\Delta s\rightarrow
0}\frac{\Delta\theta}{\Delta s}=\frac{d\theta}{ds}\ .\eneq Thus,
along a curve that has a rapidly changing tangent direction with
respect to arc length, such as a with a small radius, the
curvature is large. The reciprocal of the curvature is denoted by
\beeq R=\frac{1}{\kappa}=\frac{1}{|\vec{\bf k}(s)|}\ ,\eneq and is
called the radius of curvature at $\vec{X}(s)$. A point on $C$
where the curvature vector $\vec{\bf k}=0$, is called a point of
inflection. Thus, at a point of inflection the curvature $\kappa$
is zero and $R$ is infinite. If $\kappa$ is identically zero along
a curve $C$, then $\frac{d\vec{\bf t}}{ds}=0$, and by integrating
we find, $\vec{\bf t}=\frac{d\vec{X}}{ds}=\vec{a}$, where
$\vec{a}=const.\not=0$. Thus $\vec{X}=\vec{a}s+\vec{b}$, where
$\vec{a}=const.$ which implies that $C$ is a straight line through
$\vec{b}$, and parallel to $\vec{a}$. Furthermore, on a surface,
we can define a normal curvature vector. The normal curvature
vector to $C$ at $P$, denoted by $\vec{\bf{k}}_n$, is the vector
projection of the curvature vector $\vec{\bf{k}}$ of $C$ at $P$ on
to the unit normal vector $\hat N$ of the surface at $P$:\beeq
\vec{\bf{k}}_n=(\vec{\bf{k}}\cdot \hat{N})\hat{N}\ .\eneq The
component of $\vec{\bf{k}}_n$ in the direction of $\hat{N}$,
denoted by $\kappa_n\equiv \vec{\bf{k}}\cdot \hat{N}$, is called
the normal curvature of $C$ at $P$. The normal curvature can be
re-expressed using Eq. \ref{kdtdsdt} \beeq
\kappa_n=\vec{\bf{k}}\cdot \hat{N}=\frac{d\vec{\bf
t}}{dt}\cdot\hat{N}\Big/\Big|\frac{d\vec{X}}{dt}\Big|\ .\eneq Now,
using the fact that $\vec{\bf t}$ is perpendicular to $\hat N$
along $C$: \beeq \frac{d}{dt}(\vec{\bf t}\cdot \hat
N)=0\Rightarrow \frac{d\vec{\bf t}}{dt}\cdot \hat N = -\vec{\bf
t}\cdot \frac{d\hat N}{dt}\ ,\eneq we find\beeq \kappa_n
=-\vec{\bf t}\cdot \frac{d \hat
N}{dt}\Big/\Big|\frac{d\vec{X}}{dt}\Big|=-\frac{d\vec{X}}{dt}\cdot
\frac{d \hat N}{dt}\Big/\Big|\frac{d\vec{X}}{dt}\Big|^2\ ,\eneq
where we replaced $\vec{\bf t}$ using Eq. \ref{tdxdt}. The above
equation can be written more explicitly as\beeq \kappa_n = -\left[
\vec{X}_\tau\frac{d\tau}{dt}+\vec{X}_\phi\frac{d\phi}{dt}\right]\cdot\left[
\hat{N}_\tau\frac{d\tau}{dt}+\hat{N}_\phi\frac{d\phi}{dt}\right]\Big/\left[
\vec{X}_\tau\frac{d\tau}{dt}+\vec{X}_\phi\frac{d\phi}{dt}\right]\cdot\left[
\vec{X}_\tau\frac{d\tau}{dt}+\vec{X}_\phi\frac{d\phi}{dt}\right]\
,\nonumber\eneq and hence\beeq \kappa_n =\frac{{\bf
L}(d\tau/dt)^2+2{\bf M}(d\tau/dt)(d\phi/dt)+{\bf
N}(d\phi/dt)^2}{{\bf E}(d\tau/dt)^2+2{\bf
F}(d\tau/dt)(d\phi/dt)+{\bf G}(d\phi/dt)^2} \ .\eneq Observe that
$\kappa_n$, as a function of ${d\tau}/{dt}$ and ${d\phi}/{dt}$,
depends only upon the ratio $({d\tau}/{dt})/({d\phi}/{dt})$,
namely the direction of $d\tau/d\phi$, hence \beeq
\kappa_n=\frac{{\bf L}(d\tau)^2+2{\bf M}(d\tau)(d\phi)+{\bf
N}(d\phi)^2}{{\bf E}(d\tau)^2+2{\bf F}(d\tau)(d\phi)+{\bf
G}(d\phi)^2}=\frac{II}{I} \ ,\eneq where $(d\tau)^2+(d\phi)^2\not=
0$. The two perpendicular directions on the surface, for which the
$\kappa_n$ take maximum and minimum values, are called ``the
principal directions,'' and the corresponding normal curvatures,
$\kappa_1$ and $\kappa_2$, are called ``the principal
curvatures.'' If $\kappa_n$ has a maximum or minimum, $\kappa^*$,
for $(d\tau^* , d\phi^*)$, then\be
\frac{\partial\kappa_n}{\partial d\tau}\Big|_{(d\tau^*, \;
d\phi^*)}&=&\frac{I\; II_{d\tau}-II\;
I_{d\tau}}{I^2}\Big|_{(d\tau^*, \; d\phi^*)}=0 \; ,\nonumber\\
\frac{\partial\kappa_n}{\partial d\phi}\Big|_{(d\tau^*, \;
d\phi^*)}&=&\frac{I\; II_{d\phi}-II\;
I_{d\phi}}{I^2}\Big|_{(d\tau^*, \; d\phi^*)}=0 \ .\ee Multiplying
both equations by $I$, and using the identity\beeq
\frac{II}{I}\Big|_{(d\tau^*, \; d\phi^*)}=\kappa_n|_{(d\tau^*, \;
d\phi^*)}=\kappa^*\ ,\eneq we obtain \beeq (II_{d\tau}-\kappa^*
I_{d\tau})|_{(d\tau^*, \; d\phi^*)}=0\; ,\;\;\;
(II_{d\phi}-\kappa^* I_{d\phi})|_{(d\tau^*, \; d\phi^*)}=0 \
.\label{II-I}\eneq Since $II_{d\tau}=2({\bf L}\; d\tau +{\bf M}\;
d\phi)$, $I_{d\tau}=2({\bf E}\; d\tau +{\bf F}\; d\phi)$,
$II_{d\phi}=2({\bf M}\; d\tau +{\bf N}\; d\phi)$ and
$I_{d\phi}=2({\bf F}\; d\tau +{\bf G}\; d\phi)$, Eq. \ref{II-I}
yields\be ({\bf L}\; d\tau^* +{\bf M}\; d\phi^*)-\kappa^*({\bf
E}\; d\tau^* +{\bf F}\; d\phi^*)=0\nonumber\ ,\\ ({\bf M}\;
d\tau^* +{\bf N}\; d\phi^*)-\kappa^*({\bf F}\; d\tau^* +{\bf G}\;
d\phi^*)=0\ .\ee The above is a homogeneous system of equations
and will have a nontrivial solution $(d\tau^*, d\phi^*)$ (or
nontrivial principal direction $d\tau^*/d \phi^*)$ if \be
\det\left(\begin{array}{cc}
{{\bf L}-\kappa^* {\bf E}}&~~{\bf M}-\kappa^* {\bf F}\\
{\bf M}-\kappa^* {\bf F}&~~{\bf N}-\kappa^* {\bf G}\\
\end{array}
\right)=0 \ .\ee Therefore, by expanding, we find that a number
$\kappa$ is a principal curvature, if it is a solution of the
equation \beeq ({\bf E}{\bf G}-{\bf F}^2)\kappa^2 -({\bf E}{\bf
N}+{\bf G}{\bf L}-2{\bf F}{\bf M}))\kappa+({\bf L}{\bf N}-{\bf
M}^2)=0\ . \label{secordkappa}\eneq The discriminant $\delta$ of
Eq. \ref{secordkappa} is\be \delta\!\!\!\!&=&\!\!\!\!({\bf E}{\bf
N}+{\bf G}{\bf L}-2{\bf F}{\bf M})^2-4({\bf
E}{\bf G}-{\bf F}^2)({\bf L}{\bf N}-{\bf M}^2)\ ,\nonumber\\
\!\!\!\!&=&\!\!\!\!({\bf E}{\bf N}-{\bf G}{\bf L})^2+4({\bf E}{\bf
M}-{\bf F}{\bf L})({\bf G}{\bf M}-{\bf F}{\bf N})\\
\!\!\!\!&=&\!\!\!\!4\left(\frac{{\bf E}{\bf G}-{\bf F}^2}{{\bf
E}^2}\right)({\bf E}{\bf M}-{\bf F}{\bf L})^2+\left({\bf E}{\bf
N}-{\bf G}{\bf L}-\frac{2{\bf F}}{{\bf E}}({\bf E}{\bf M}-{\bf
F}{\bf L})\right)^2 \ .\ee The discriminant $\delta\geq 0$ since
${\bf E}{\bf G}-{\bf F}^2>0$ (Eq. \ref{detEF-G}). We have
$\delta=0$ if and only if ${\bf E}{\bf M}-{\bf F}{\bf L}=0$ and
${\bf E}{\bf N}-{\bf G}{\bf L}-\frac{2{\bf F}}{{\bf E}}({\bf
E}{\bf M}-{\bf F}{\bf L})=0$, which implies ${\bf E}{\bf M}-{\bf
F}{\bf L}=0$ and ${\bf E}{\bf N}-{\bf G}{\bf L}=0$, or
equivalently, $\frac{\bf E}{\bf L}=\frac{\bf F}{\bf M}=\frac{\bf
G}{\bf N}\equiv{\bf\lambda}$. Thus, Eq. \ref{secordkappa} has
either a single real root, $\kappa=\frac{1}{{\bf{\lambda}}}$, if
$\delta=0$, or two distinct real roots, $\kappa_1$ and $\kappa_2$.
Dividing Eq. \ref{secordkappa} by ${\bf E}{\bf G}-{\bf F}^2>0$, we
obtain\beeq \kappa^2-2{{\mathcal{H}}}\kappa
+{{\mathcal{K}}}=0\label{kappa2-2H+K}\ ,\eneq where \beeq
{{\mathcal{H}}}\equiv \frac{1}{2}(\kappa_1 +\kappa_2)=\frac{{\bf
E}{\bf N}+{\bf G}{\bf L}-2{\bf F}{\bf M}}{2({\bf E}{\bf G}-{\bf
F}^2)}\ ,\label{meancurvH}\eneq is the average of the principal
curvatures and is called the mean curvature at $P$, and \beeq
{{\mathcal{K}}}\equiv\kappa_1\kappa_2=\frac{{\bf L}{\bf N}-{\bf
M}^2}{{\bf E}{\bf G}-{\bf F}^2}\ ,\label{GausscurvK}\eneq is the
product of the principal curvatures and is called the Gaussian
curvature at $P$. Hence the roots of Eq. \ref{kappa2-2H+K}
are\beeq \kappa_{1,\; 2}={{\mathcal{H}}}\pm
\sqrt{{{\mathcal{H}}}^2-{{\mathcal{K}}}}\; ,\nonumber\eneq\beeq
\kappa_{1,\; 2}=\frac{{{\bf E}{\bf N}+{\bf G}{\bf L}-2{\bf F}{\bf
M}}\pm \sqrt{({\bf G}{\bf L}-{\bf E}{\bf N})^2+4({\bf E}{\bf
M}-{\bf F}{\bf L})({\bf G}{\bf M}-{\bf F}{\bf N})}}{2({\bf E}{\bf
G}-{\bf F}^2)}\nonumber\ .\eneq For the caustic surface, we have
${\bf F}={\bf M}=0$; therefore\beeq \kappa_{1,\;
2}=\frac{1}{2}\left[\frac{\bf L}{\bf E}+\frac{\bf N}{\bf
G}\pm\frac{|{\bf G}{\bf L}-{\bf E}{\bf N}|}{{\bf E}{\bf
G}}\right]\ , \eneq and hence, we choose $\kappa_1=\frac{\bf
L}{\bf E}$ and $\kappa_2=\frac{\bf N}{\bf G}$. When explicitly
calculated in terms of the caustic parameters, they yield \beeq
\kappa_1=\frac{\bf L}{\bf
E}=\frac{\rho'z''-z'\rho''}{(\rho'\;^2+z'\;^2)^{{3}/{2}}}
=\mp\frac{1}{2}\frac{b}{\sqrt{us}}\frac{1}{p}\sqrt{\frac{\tau_0}{\tau}}
\left[3-4\frac{\tau}{\tau_0}\right]^{-1}\left[1-\left(1-\frac{b^2}{us}\right)
\frac{\tau}{\tau_0}\right]^{-3/2}\
,\label{prinkappa1}\nonumber\eneq \beeq \kappa_2=\frac{\bf N}{\bf
G}=\frac{z'}{\rho\sqrt{\rho'\;^2
+z'\;^2}}=\pm\frac{1}{\rho}\left[1+\frac{us}{b^2}\left(\frac{\tau_0}{\tau}-1\right)\right]^{-1/2}
\; ,\label{prinkappa2}\eneq where $\rho$ is given explicitly in
terms of the parameters in Eq. \ref{4.9}. Then, we can verify the
Gaussian curvature found in Eqs. \ref{GaussKfirst}: \beeq
{{\mathcal{K}}}=\kappa_1 \kappa_2=\frac{z'(\rho' z''-z'
\rho'')}{\rho(\rho'\; ^2+z'\; ^2)^2}=-\frac{b^2\;
s}{\rho\left(3-4\frac{\tau}{\tau_0}
\right)\left(us(\tau_0-\tau)+b^2\tau\right)^2}\; .\eneq

The principal curvatures, $\kappa_1$ and $\kappa_2$, are the
inverse of the principal curvature radii $R_1$ and $R_2$. The
lensing by a caustic surface depends on the curvature radius $R$
of the surface along the line of sight. Therefore, we will need
the curvature radii of the caustic ring surface at an arbitrary
point $(\alpha_1 (\tau_1), \tau_1)$ on the surface. $R$ is given
by the Euler's theorem: \beeq {1 \over R} = {(\cos{\omega})^2
\over R_1} + {(\sin{\omega})^2 \over R_2} \, ,\label{Rad} \eneq
where \beeq R_1 (\tau_1) =\frac{1}{\kappa_1 (\tau_1)}= -2
{\sqrt{su} \over b} p \sqrt{{\tau_1 \over \tau_0}}\, |3 - 4
{\tau_1 \over \tau_0}|\, \left(1 - (1 - {b^2 \over su}){\tau_1
\over \tau_0}\right)^{3 \over 2} \label{R1} \eneq in the direction
of the cross-sectional plane of the caustic ring, \beeq R_2
(\tau_1) =\frac{1}{\kappa_2 (\tau_1)}= \pm {\sqrt{su} \over b}
\sqrt{{\tau_0 \over \tau_1} - 1 + {b^2 \over su}} \left(a + {u
\over 2}(\tau_0 - \tau_1)(\tau_0 - 2 \tau_1)\right)~~, \label{R2}
\eneq in the direction perpendicular to the cross-sectional plane,
and $\omega$ is the angle between the line of sight and the
direction associated with $R_1$.  In this dissertation, we adopt
the convention that $R$ is positive (negative) if, along the line
of sight, the surface curves toward (away from) the side with two
extra flows. If $R$ is positive, the surface is called ``concave''
(Fig. \ref{fig:concave}). If $R$ is negative, the surface is
called ``convex'' (Fig. \ref{fig:convex}). In Eq. \ref{R2}, the
$+$ sign pertains if $~0 \leq {\tau_1 \over \tau_0} \leq {3 \over
4}$; the $-$ sign pertains if $~{3 \over 4} \leq {\tau_1 \over
\tau_0} \leq 1$. $R_1$ is always negative except at the three
cusps, where it vanishes. $R_2$ diverges at the cusp in the $z=0$
plane.

For  $0 \leq {\tau_1 \over \tau_0} \leq {3 \over 4}$, there is a
pair of lines of sight for which the curvature vanishes.  They are
at angles: \beeq \omega = \pm \arctan{\sqrt{- {R_2 \over R_1}}}
\eneq relative to the cross-sectional plane.  Gravitational
lensing by a fold of zero curvature is discussed in Section
\ref{lensingzerocurvature}.

\subsubsection{Density Profiles of Axially Symmetric Caustic Rings}

As we have seen the inner caustics \cite{cr} are closed tubes
whose cross-section, shown in Fig. \ref{fig:fig6}, is a $D_{-4}$
catastrophe \cite{sing}. They are located near where the particles
with the most angular momentum in a given inflow are at their
distance of closest approach to the galactic center. For
simplicity, we study caustic rings which are axially symmetric
about the $z$-direction as well as reflection symmetric with
respect to the $z=0$ plane. In galactocentric cylindrical
coordinates, the flow at such a caustic ring is described by Eqs.
\ref{4.5.a} and \ref{4.6}.

Recall that the physical space density is given by Eq. \ref{4.1}:
\beeq d(\rho,z)=\frac{1}{\rho}\sum_{j=1}^{n}\frac{dM}{d\Omega
d\tau}(\alpha\, , \tau){\left.\frac{\cos{(\alpha)}}{\left|D_2
(\alpha\, , \tau)\right|}\right|_{\left(\alpha_j(\rho,z)\, , \,
\tau_j(\rho, z)\right)}}~~~\ , \label{km} \eneq where we define
$\frac{dM}{d\Omega d\tau} = \frac{dM}{2 \pi \cos{(\alpha)} d\alpha
d\tau}$ as the mass falling in per unit time and unit solid angle.
The parameters $\alpha_j(\rho,z)$ and $\tau_j(\rho, z)$, with $j =
1,.\; .\; ., n$, are the solutions of $\rho(\alpha, \tau) = \rho$
and $z(\alpha, \tau) = z$, where $n(\rho, z)$ is the number of
flows at $(\rho,z)$. Outside the caustic tube $n=2$, whereas
inside $n=4$. The determinant $D_2(\alpha,\tau)$ is defined in Eq.
\ref{Jaco}. In the limit of zero velocity dispersion, the density
of dark matter particles is infinite at the location of caustic
surfaces. Thus, the location of the caustic ring surface is
obtained by demanding that $D_2(\alpha,\tau)= 0$ in Eq. \ref{4.7},
which implied $\alpha(\tau) = \pm \sqrt{{u \over s} \tau (\tau_0 -
\tau)}$ with $0 \leq {\tau \over \tau_0} \leq 1$ (Eq. \ref{4.8}).
Hence, we obtained $\rho(\tau)=a+{u\over
2}(\tau-\tau_0)(2\tau-\tau_0)$ and $z(\tau)= \pm b\sqrt{{u\over
s}\tau^3(\tau_0-\tau)}$ in Eqs. \ref{4.9}, for the location of the
tricusp outline.

\begin{figure}[ht] \centering
\includegraphics[height=5.65cm,width=12cm]{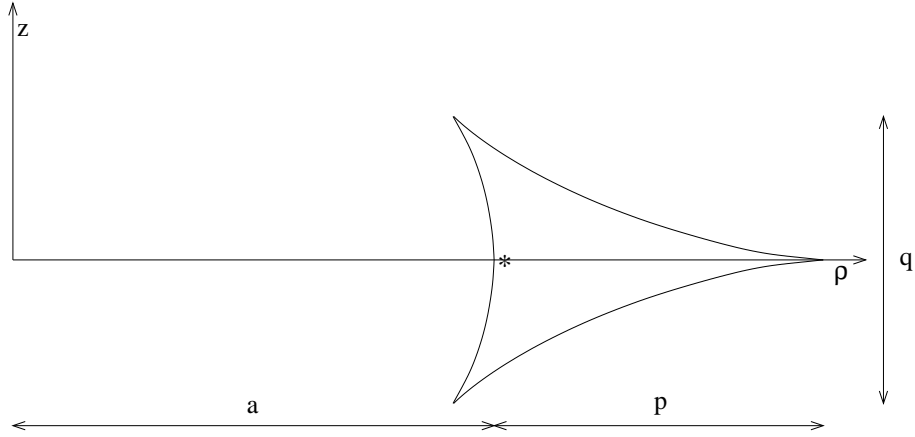}
\caption{Cross-section of a caustic ring in the case of axial and
reflection symmetry. The transverse dimensions in the $\hat{\rm
\rho}$ and $\hat{\rm z}$ directions are p and q, respectively. The
ring radius is a. The star indicates the sample location
($\rho\simeq a , z=0$). For clarity, we took $p,q \sim a$. For
actual caustic rings, $p,q \ll a$.} \label{fig:tricusp}
\end{figure}

Near the surface of a caustic ring, but away from the cusps, the
density profile is that of a simple fold:
$d(\sigma)=\frac{A}{\sqrt{\sigma}}\Theta(\sigma)$, with $\sigma>0$
inside the tricusp.  Next, we calculate the fold coefficient $A$
at arbitrary points other than the cusps, on the surface of a
caustic ring. As a warm-up, we start with a special point, marked
by a star in Fig. \ref{fig:tricusp}.  We then obtain the density
profile near the cusps.

{\bf A sample point} \label{subsection:aspecialpoint}

As an example, we determine the fold coefficient $A$ at $(\rho ,
z)=(a, 0)$ (Fig. \ref{fig:tricusp}).

Setting $\alpha=z=0$, $\sigma=\rho-a=\frac{1}{2}u(\tau-\tau_0)^2$
and $\tau\simeq\tau_0$, we have \beeq |D_2(\tau)|\simeq 2
b\sqrt{p\sigma} ~~~\ . \eneq Including the factor 2 for in and out
flows, Eq. \ref{km} yields \beeq d(\sigma)=\frac{dM}{d\Omega
dt}\frac{1}{a b}\frac{1}{\sqrt{p\sigma}} ~~~\ . \eneq Therefore,
\beeq A_0=\frac{d^2M}{d\Omega dt}\frac{1}{a
b}\frac{1}{\sqrt{p}}~~~\ , \label{A_nc} \eneq where the subscript
${0}$ is used to indicate that $A$ is being evaluated at the
sample point.

To obtain a numerical estimate for $A_0$, we again use the
self-similar infall model \cite{FG,B,STW1,STW2} with $\epsilon =
0.2$.  For the $n$th ring, we have \cite{cr}, \beeq \{a_n: n=1,
2,.\; .\; .\}\simeq(39,~19.5,~13,~10,~8,.\; .\; .){\rm kpc}\cdot\!
\left(\frac{j_{\rm{max}}}{0.27}\right)\!\left(\frac{0.7}{h}\right)\!
\left(\frac{v_{\rm{rot}}}{220\,{\rm km/s}}\right)\; , \label{a_n}
\eneq where $j_{\rm {max}}$ is a parameter, with a specific value
for each halo, which is proportional to the amount of angular
momentum that the dark matter particles have \cite{STW1,STW2}.
Also, \beeq \left.\frac{d^2 M}{d\Omega dt}\right|_n =f_n
v_n\frac{v^2_{\mbox{rot}}} {4\pi G}~~~\ , \label{f} \eneq where
$v_n$ is the velocity of the particles in the $n$th caustic ring,
and the dimensionless coefficients $f_n$ characterize the density
of the $n$th in and out flow.  In the self-similar model
\cite{cr}, \beeq \{f_n: n=1, 2,.\; .\; .\}\simeq
(13,~5.5,~3.5,~2.5,~2,.\; .\; .)\cdot 10^{-2} \eneq for
$\epsilon=0.2$. The $f_n$ are like the $F_n$ in Eq. \ref{Fn}, but
they describe the $n$th in and out flow near the caustic ring,
whereas the $F_n$ describe that flow near turnaround.

Combining Eqs. \ref{A_nc} and \ref{f}, we have \beeq A_{0,n}
=\frac{v^2_{ \mbox{rot}}}{4\pi
G}\,\frac{f_n}{a_n}\,\frac{v_n}{b_n}\,\frac{1}{\sqrt{p_n}}~~~\ .
\label{Anv2} \eneq It was shown in reference \cite{sing} that
$b_n$ and $v_n$ are of the same order of magnitude.  Moreover,
Sikivie \cite{milk} interpreted the ten rises in the rotation
curve of the Milky Way as the effect of caustic rings. In that
case, the widths $p_n$ of caustic rings are determined from the
observed widths of the rises.  Typically one finds $p_n\sim 0.1\,
a_n$. Using this and $v_n \sim b_n$, Eq. \ref{Anv2} yields the
estimates \be \{A_{0,n}: n=1, 2,.\; .\; .\}\sim (3,~4,~4,~5,~5,.\;
.\; .)\cdot
\frac{10^{-4}\,{\rm gr}}{\rm cm^2\,kpc^{1/2}}\nonumber\\
\cdot\left(\frac{0.27}{j_{\rm{max}}}\right)^{3/2}
\left(\frac{h}{0.7}\right)^{3/2} \left(\frac{v_{\mbox{rot}}}{220\,
{\rm km/s}}\right)^{1/2}~~~\ . \label{Anring} \ee At the point
under consideration, $(\rho , z)=(a, 0)$, the surface of the
caustic ring is convex for all lines  of sight (i.e., all tangents
at that point are on the side with two extra flows).  If the line
of sight is in the $z=0$ plane, the curvature radius is $a$.  If
the line of sight is perpendicular to the $z=0$ plane, the
curvature radius is $2 {b^2 \over su} p$. Lensing by a convex
caustic surface is discussed in Section \ref{subsection:scvf}.

To obtain the lensing properties of the caustic ring surface at an
arbitrary point, we need the curvature radii $R$ given in Eq.
\ref{Rad} and the coefficient $A$ at all locations. We derive $A$
everywhere on the surface in the next two sections.

{\bf The fold coefficient everywhere}

We choose an arbitrary point on the tricusp (i.e., on the surface
of the caustic ring).  Its parameters are $(\alpha_1, \tau_1)$
with $\alpha_1$ given in terms of $\tau_1$ by Eq. \ref{4.8}.  The
physical coordinates $(\rho_1, z_1)$ are given in terms of
$\tau_1$ by Eqs. \ref{4.9}. We assume that the point is not at one
of the three cusps.  The latter are located at $\tau_1=0$, at
$\tau_1=\frac{3}{4}\tau_0$ with $\alpha_1>0$, and at
$\tau_1=\frac{3}{4}\tau_0$ with $\alpha_1<0$.

The vanishing of $D_2(\tau_1)=\det {\cal D}(\tau_1)$ implies the
existence of a zero eigenvector of the matrix \be {\cal
D}(\tau_1)=\left(\begin{array}{cc}
{-s\alpha_1}&{b\tau_1}\\
{u(\tau_1-\tau_0)}& {b\alpha_1}\\
\end{array}
\right)~~~\ . \ee Let us define $\theta_1(\tau_1)$ such that \be
{\cal D}(\tau_1)\left(\begin{array}{c}
{\sin{(\theta_1)}}\\
{\cos{(\theta_1)}}\\
\end{array}
\right)=0~~~\ . \label{dt1} \ee We have \be
\sin{(\theta_1)}=\frac{b\tau_1}{\sqrt{(b\tau_1)^2+(s\alpha_1)^2}}
\hskip 0.5 cm \mbox{and} \hskip 0.5 cm
\cos{(\theta_1)}=\frac{s\alpha_1}{\sqrt{(b\tau_1)^2+(s\alpha_1)^2}}~~~\
. \ee We define new Cartesian coordinates $(\sigma , \eta)$
related to $(\rho - \rho_1, z - z_1)$ by a rotation of angle
$\theta_1+{\pi\over 2}$: \be \left(\begin{array}{c}
{\sigma}\\
{\eta}\\
\end{array}
\right)=  \left(\begin{array}{cc}
{-\sin{(\theta_1)}}&{-\cos{(\theta_1)}}\\
{\cos{(\theta_1)}}&{-\sin{(\theta_1)}}\\
\end{array}
\right)\left(\begin{array}{c}
{\rho-\rho_1}\\
{z -z_1}\\
\end{array}
\right)\; . \label{mateq} \ee We now show that $\sigma$ is the
coordinate in the direction orthogonal to the caustic surface at
$(\rho_1, z_1)$.

\begin{figure}[ht] \centering
\includegraphics[height=6.5cm,width=13cm]{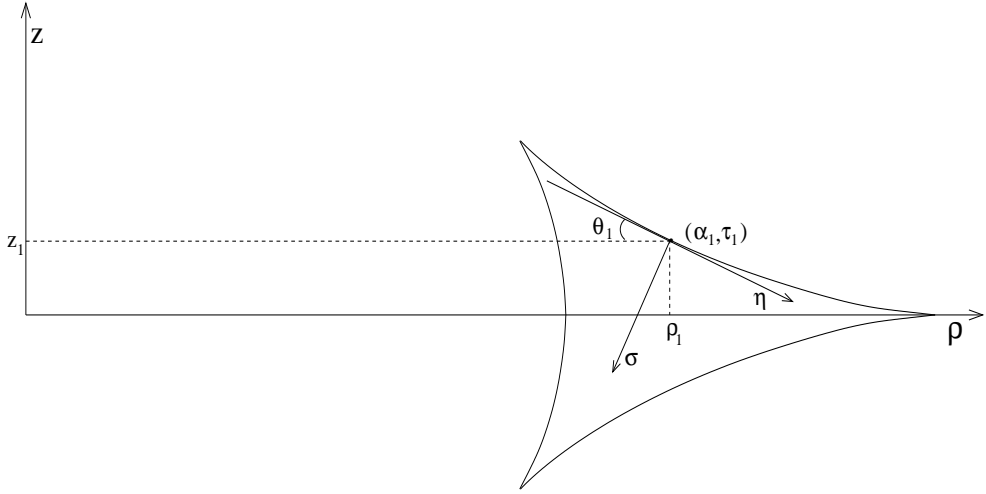}
\caption{An arbitrary point on the tricusp is labeled by $\tau_1$
with $\alpha_1 = \alpha_1 (\tau_1)$.  Its physical coordinates are
$(\rho_1, z_1)$. A new Cartesian coordinate system ($\sigma ,
\eta$) is defined there such that $\hat{\sigma}$ is perpendicular
to the caustic surface.  It is rotated relative to the $(\rho, z)$
cooordinates by an angle $\theta(\tau_1)+{{\pi}\over 2}$.}
\label{fig:tricusp1}
\end{figure}

Consider small deviations about $(\alpha_1 , \tau_1)$ in parameter
space: $(\alpha,\tau)=(\alpha_1+\Delta \alpha, \tau_1+\Delta
\tau)$. Equation \ref{Jaco} implies \be \left(\begin{array}{c}
{{\Delta\rho}}\\
{{\Delta z}}\\
\end{array}
\right)= {\cal D}^T(\tau_1)\left(\begin{array}{c}
{{\Delta\alpha}}\\
{{\Delta\tau}}\\
\end{array}
\right)+O(\Delta\alpha^2 , \Delta\tau^2 ,
\Delta\alpha\Delta\tau)~~~\ , \label{DrDz} \ee where $T$ indicates
transposition. The expansion of $\sigma$ in powers of $\Delta
\alpha$ and $\Delta \tau$ yields \beeq \sigma=O(\Delta\alpha^2 ,
\Delta\tau^2 , \Delta\alpha\Delta\tau)\; , \eneq because the first
order terms vanish: \be \left.\frac{\dd \sigma}{\dd
\alpha}\right|_{(\alpha_1,\tau_1)}\Delta \alpha+\left.\frac{\dd
\sigma} {\dd \tau}\right|_{(\alpha_1,\tau_1)}\Delta \tau =
-(\Delta\alpha\,\,\,\, \Delta\tau)~{\cal
D}(\tau_1)\left(\begin{array}{c}
{\sin{(\theta_1)}}\\
{\cos{(\theta_1)}}\\
\end{array}
\right)=0~~~\ . \ee The fact that $\sigma$ is second order in
$\Delta\alpha$ and $\Delta\tau$ shows that $\sigma$ is the
coordinate in the direction perpendicular to the caustic surface,
and \beeq \hat{\sigma}=-\sin{\left(\theta_1\right)}\hat{\rho}
-\cos{\left(\theta_1\right)}\hat{z} \label{s} \eneq is the unit
normal to the surface, pointing inward. $\theta(\tau_1)$ is the
angle between the $\rho$ axis and the tangent line at $(\rho_1,
z_1)$ (Fig. \ref{fig:tricusp1}). To obtain the density profile of the caustic near the point under
consideration, we need $D_2(\eta, \sigma)$ to order
$\sqrt{\sigma}$. So we calculate $\sigma$ to second order in
powers of $\Delta \alpha$ and $\Delta \tau$, and $D_2$ and $\eta$
to first order.  We find \be
\sigma=\frac{1}{2}(\Delta\alpha\,\,\,\,\Delta\tau)
\left(\begin{array}{cc}
{s\sin{(\theta_1)}}&{-b\cos{(\theta_1)}}\\
{-b\cos{(\theta_1)}}&{-u\sin{(\theta_1)}}\\
\end{array}\right)\left(\begin{array}{c}
{{\Delta\alpha}}\\
{{\Delta\tau}}\\
\end{array} \right)\; ,
\label{y'} \ee and
\begin{eqnarray}
D_2 &=&-b\left[2s\alpha_1\Delta\alpha +
u(2\tau_1-\tau_0)\Delta\tau\right]\nonumber\\
\eta &=& [u(\tau_1 - \tau_0) \cos{\theta_1} - b \alpha_1
\sin{\theta_1}] \Delta \tau - [s \alpha_1 \cos{\theta_1} + b
\tau_1 \sin{\theta_1}] \Delta \alpha \; . \label{D2e}
\end{eqnarray}
Eqs. \ref{D2e} can be inverted to obtain $\Delta\alpha$ and
$\Delta\tau$ as functions of $D_2$ and $\eta$. When the result is
inserted into Eq. \ref{y'}, we obtain \beeq \label{density}
\sigma(D_2,\eta)={b \over 2\sqrt{(b\tau_1)^2+(s\alpha_1)^2}}\,
{1\over u\tau_1|3\tau_0-4\tau_1|}\left(\tau_1({D_2\over b})^2 -
{us\tau_0\tau_1^2 \eta^2 \over (b\tau_1)^2+(s\alpha_1)^2}
\right)\, . \eneq This implies \beeq D_2(\eta, \sigma)=
\sqrt{\frac{b^2}{1+(\frac{b^2}{us}-1)\frac{\tau_1}{\tau_0}}\eta^2
+
2b^2u|3\tau_0-4\tau_1|\sqrt{\frac{us}{b^2}\,(\tau_0-\tau_1)\tau_1+\tau_1^2}
\,\,\,\sigma} \; . \eneq Along the $\hat{\sigma}$ direction
($\eta=0$), we have \be D_2(\sigma)= 2 C(\tau_1)\, b\sqrt{\sigma
p}~~~\ , \label{D2sC} \ee with \beeq C(\tau_1)=\sqrt{ |3 -4
{\tau_1 \over \tau_0}| \sqrt{{{us}\over{b^2}}(1- {\tau_1 \over
\tau_0}){\tau_1 \over\tau_0} +({\tau_1 \over \tau_0})^2}}~~~~\ .
\eneq Combining Eqs. \ref{km} and \ref{D2sC}, and minding the
factor of two because two flows contribute, we have \beeq
d(\tau_1, \sigma) = {A(\tau_1) \over \sqrt{\sigma}} \Theta(\sigma)
\eneq with \beeq A(\tau_1)= \frac{d^2M}{d\Omega dt}
\frac{1}{bC(\tau_1)\sqrt{p}}\frac{\cos{(\alpha_1)}}{\rho(\tau_1)}~~~\
. \eneq In terms of $A_{0,n} = A_n(\tau_1 = \tau_0)$, for which
estimates are provided in Eq. \ref{Anring}, we have \beeq A_n
(\tau_1) = A_{0,n} {a_n \over \rho_n(\tau_1)}
{\cos{\alpha_1(\tau_1)} \over C_n(\tau_1)}~~~~~\ . \label{At1}
\eneq Note that $A(\tau_1)$ diverges at each of the three cusps
because $C$ vanishes there.  The caustic ring parameters
($a,~\tau,~b,~u,~s$) are related to the velocity distribution of
the flow at last turnaround \cite{sing}.

{\bf Density near a cusp} \label{subsection:densitynearacusp}

In this section, we derive the dark matter density profile near a
cusp. For the sake of convenience, we choose the cusp in the $z=0$
plane at $\rho = a + p \equiv \rho_0$, where $\alpha = \tau = 0$.
Very close to the cusp, we may neglect the term of order $\tau^2$
in Eq. \ref{4.6}. Hence, \beeq z=b\,\alpha\tau\,\, ,\hskip 0.5cm
\rho=\rho_0 -u\tau_0 \tau -\frac{s}{2}\alpha^2\; . \label{kem2}
\eneq The term of order $\alpha^2$ cannot be neglected. We define
new dimensionless quantities \cite{sing}: \beeq
A\equiv\frac{\alpha}{\tau_0}\sqrt{\frac{s}{u}}\,\, ,\hskip 0.5 cm
T\equiv\frac{\tau}{\tau_0}\,\, ,\hskip 0.5 cm
X\equiv{{\rho-\rho_0}\over p}\,\, ,\hskip 0.5 cm
Z\equiv\frac{z\sqrt{\zeta}}{p}\; , \label{capvar} \eneq where
$\zeta = {su \over b^2}$.  In terms of these, Eq. \ref{kem2}
becomes \be Z&=&2AT\;,\label{kemn1}\\ X&=&-2T-A^2\;.\label{kemn2}
\ee The determinant of the Jacobian in Eq. \ref{4.7} becomes \beeq
D_2(A,T)= 2bp\left(T-A^2\right)\; . \label{dat} \eneq Substitution
of Eq. \ref{kemn1} into Eq. \ref{kemn2} yields the third order
polynomial equation: \beeq A^3+X A+ Z=0\; . \label{pol} \eneq The
discriminant is: \beeq
\delta=\left(\frac{Z}{2}\right)^2+\left(\frac{X}{3}\right)^3\; .
\label{discr} \eneq If $\delta>0$, the cubic equation has one real
root, and two complex roots which are complex conjugates of each
other.  If $\delta < 0$, all the roots are real and unequal.  For
$\delta=0$, all the roots are real and at least two are equal. The
number of real roots is the number of flows at a given location.
The tricusp has two flows outside and four inside.  In the
neighborhood of a cusp, however, one of the flows of the tricusp
is nonsingular and does not participate in the cusp caustic.  To
include the root corresponding to the nonsingular flow near $(z,
\rho) = (0, \rho_0)$, one must keep the term of order $\tau^2$ in
Eq. \ref{kem2}.

The equation for the caustic surface in physical space is
$\delta=0$. Indeed, Eq. \ref{dat} implies that at the caustic
$T=A^2$. Therefore, $Z=2T^{3/2}$ and $X=-3T$. A straightforward
calculation shows that \beeq
\delta=\frac{D_2}{54\,bp}(A^4+7A^2T-8T^2)\; . \eneq Thus $D_2=0$
implies $\delta=0$, however, the converse is not true: $\delta=0$
does not imply $D_2=0$ because not all flows at the location of
the caustic surface are singular.

Eq. \ref{km} for the density becomes \beeq d = {1 \over 2 \rho_0 b
p} {{d^2M}\over{d\Omega dt}} \sum_{j=1}^n {1 \over |T - A^2|_j}\;
, \label{dTA} \eneq where the sum is over the flows (i.e., the
real roots of the cubic polynomial Eq. \ref{pol}).  If $\delta>0$,
the one real root is \beeq A=(-{Z\over
2}+\sqrt{\delta})^{1/3}+(-{Z\over 2}-\sqrt{\delta})^{1/3}.
\label{A1} \eneq This describes the one flow outside the cusp.
Using Eqs. \ref{kemn2} and \ref{A1} in Eq. \ref{dTA}, we obtain
\beeq d=\frac{1}{\rho_0\, b\, p}\frac{d^2M}{d\Omega dt}
\frac{1}{|X-3(-\frac{Z}{2}+\sqrt{\delta})^{2/3}
-3(\frac{Z}{2}+\sqrt{\delta})^{2/3}|} ~~~\ . \eneq Just above or
below the cusp, where $X=0$ and $|Z|\ll 1$, we have \beeq
d=\frac{1}{3\, b\, p\, \rho_0}\frac{d^2M}{d\Omega
dt}\frac{1}{|Z|^{2/3}}~~~\ . \eneq On the other hand, if we
approach the cusp in the plane of the ring from the outside, where
$Z=0$ and $0<X\ll 1$, we find \beeq d=\frac{1}{b\,
\rho_0}\frac{d^2M}{d\Omega dt}\frac{1}{(\rho-\rho_0)}~~~\ .
\label{fir} \eneq Next, we calculate the density inside the cusp,
where $\delta <0$. The three real roots of the polynomial Eq.
\ref{pol} are \be
A_1&=&2\sqrt{\frac{-X}{3}}\cos{\theta}\\
A_2&=&2\sqrt{\frac{-X}{3}}\cos{(\theta +\frac{2\pi}{3})}\\
A_3&=&2\sqrt{\frac{-X}{3}}\cos{(\theta +\frac{4\pi}{3})}\; , \ee
where $\cos{3\theta}\equiv
-\frac{Z}{2}\left(-\frac{3}{X}\right)^{3/2}$ and
$0\leq\theta\leq{\pi\over 3}$.  Inserting them into Eq. \ref{dTA}
and using Eq. \ref{kemn2}, we obtain \beeq d=\frac{1}{b\, p\,
\rho_0}\frac{d^2M}{d\Omega
dt}\left(\frac{1}{-X}\right)\left(\frac{1}{4\cos^2(\theta)-1}
+\frac{1}{4\cos^2(\theta+\frac{2\pi}{3})-1}+
\frac{1}{1-4\cos^2(\theta+\frac{4\pi}{3})}\right)~~~\ , \eneq
where each term in the parentheses corresponds to one of the three
flows. Adding the individual flow densities yields \beeq
d=\frac{2}{b\, p\, \rho_0}\frac{d^2M}{d\Omega
dt}\frac{1}{|X|}\frac{1}{(\sqrt{3}-\tan{\theta})
\sin{2\theta}}~~~\ . \eneq If we approach the cusp in the galactic
plane from the inside, where $Z=0$ and $0< -X \ll 1$, we have
\beeq d=\frac{2}{b\, \rho_0}\frac{d^2M}{d\Omega
dt}\frac{1}{(\rho_0 -\rho)}~~~~, \eneq which is twice the result
in Eq. \ref{fir}. The gravitational lensing properties of a cusp
are derived, for a special line of sight, in Section
\ref{subsection:lensingbycusp}.

\newpage

\section{Gravitational Lensing by Dark Matter Caustics}
\label{chap:lensing}


Gravitational lensing techniques have proven useful in studying
the distribution of dark matter in the universe.  They have been
used to reveal the existence of massive compact halo objects
(MACHOs) in galaxies \cite{m1,m2,m3,m4}, and to constrain the mass
distribution in galaxy clusters \cite{clu1,clu2,clu3}.  In this
section, we calculate the gravitational lensing properties
\cite{lensingbycaustics} of dark matter caustics
\cite{cr,sing,Tre}. Gravitational lensing by dark matter caustics
has been discussed by Hogan \cite{Hogan}. We confirm Hogan's
results for the case he considered, called the ``concave fold'' in
our nomenclature.  We study additional cases and introduce a
formalism to facilitate the calculations.

\begin{figure}[ht] \centering
\includegraphics[height=4.7cm,width=14cm]{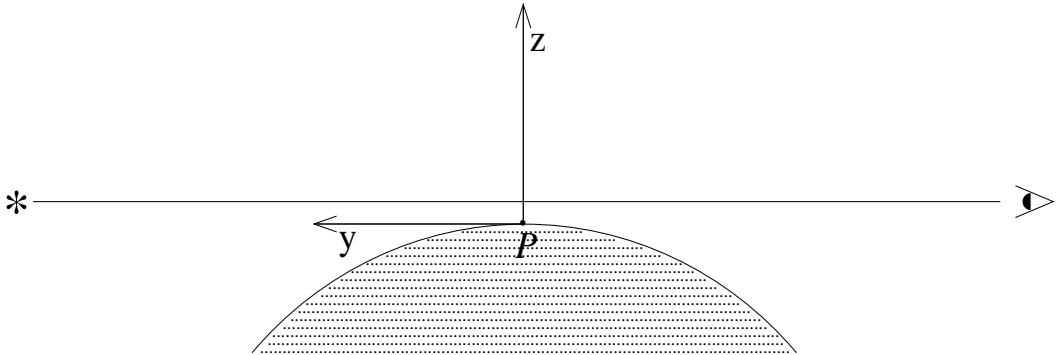}
\caption{Lensing by a concave fold. The arc is the intersection of
the caustic surface with the plane containing the normal
($\hat{z}$) to the surface and the line of sight ($\hat{y}$).  The
shaded area indicates the side with the two extra flows.}
\label{fig:concave}
\end{figure}

\begin{figure}[ht] \centering
\includegraphics[height=4.7cm,width=13cm]{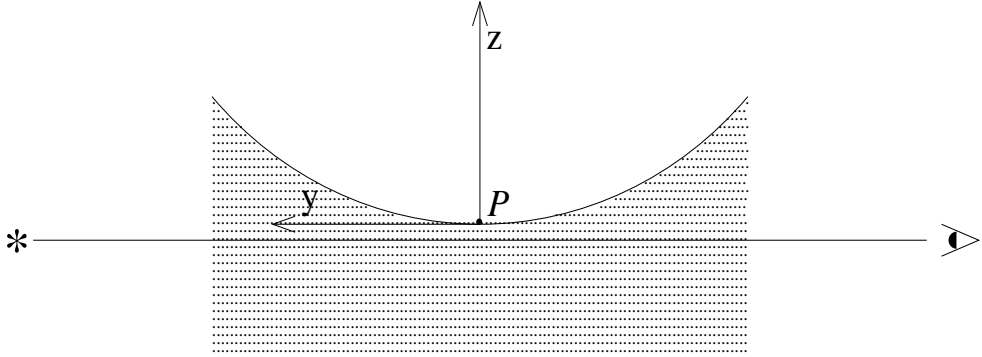}
\caption{Lensing by a convex fold.  Same as Fig. \ref{fig:concave}
except that now the caustic surface curves away from the side with
two extra flows.} \label{fig:convex}
\end{figure}

\begin{figure}[ht]
\centering
\includegraphics[height=11cm,width=9cm]{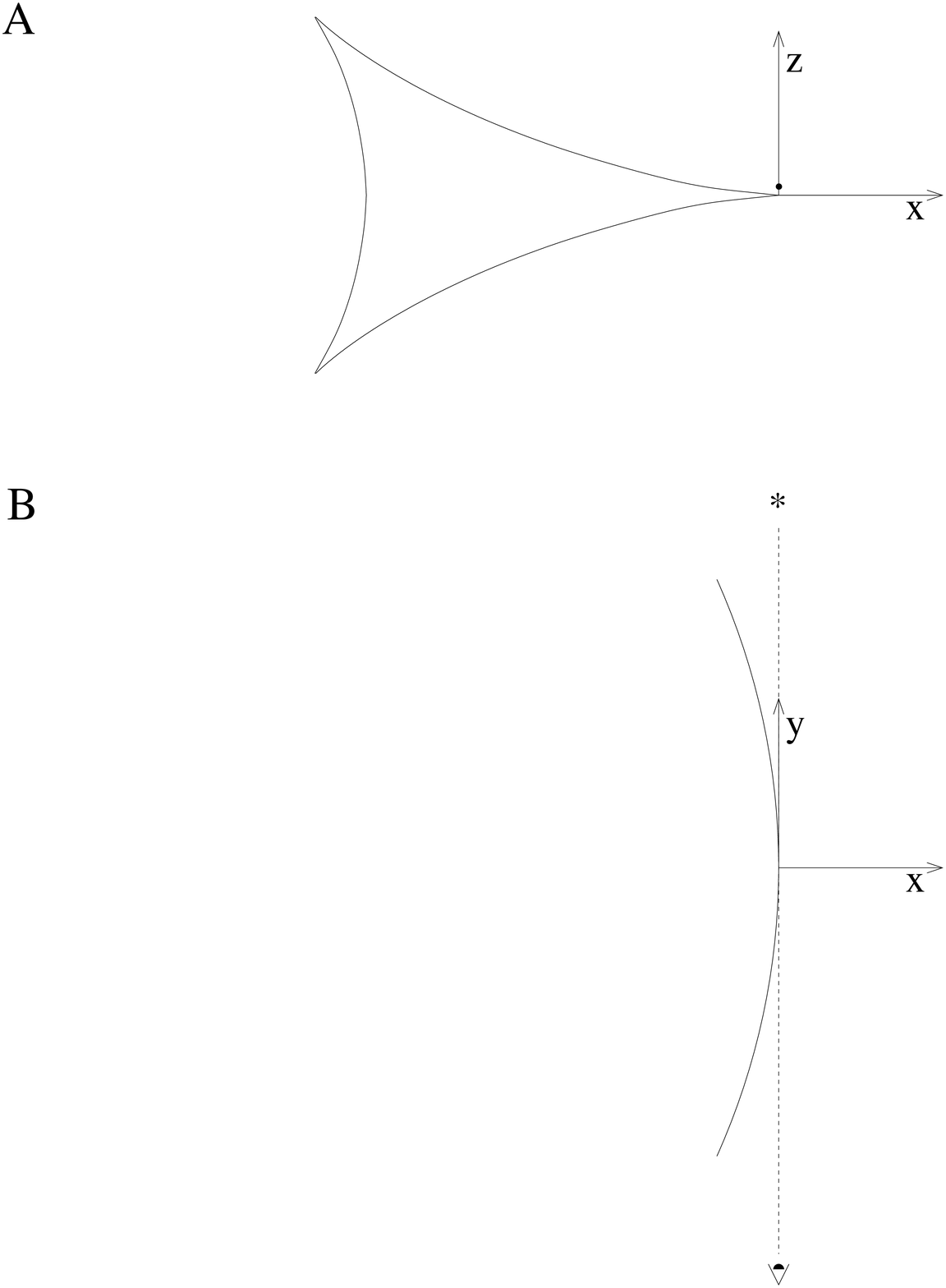}
\caption{Lensing by the caustic ring cusp at ($z, \rho) = (0,
\rho_0$), with line of sight parallel to the $z=0$ plane.  We
define $x \equiv \rho - \rho_0$. A) Side view in the direction of
the line of sight.  The latter is represented by the dot near
$x=z=0$. B) Top view.  The curve is the location of the cusp in
the $z=0$ plane.} \label{fig:tricusp2}
\end{figure}

The gravitational lensing effects of a caustic surface are largest
when the line of sight is near tangent to the surface because the
contrast in column density is largest there.  The effects depend
on the curvature of the caustic surface at the tangent point in
the direction of the line of sight: the smaller the curvature, the
larger the effects.  A caustic is an oriented surface because one
side has two more flows than the other.  We will consider three
cases of gravitational lensing by a smooth caustic surface. In the
first case, the line of sight is near tangent to a caustic surface
which curves toward the side with two extra flows, as in Fig.
\ref{fig:concave}. We call such a surface
``concave.''In the second case, the surface is ``convex'' (i.e., it curves away
from the side with two extra flows as in Fig.
\ref{fig:convex}).In the third case, the caustic surface has zero curvature at the
tangent point (the radius of curvature is infinite), but the
tangent line is entirely outside the side with two extra flows.
Caustic surfaces may have cusps.  The outer dark matter caustics
of galactic halos are topological spheres which have no cusps, but
the inner dark matter caustics of galactic halos are closed tubes
whose cross-section is a $D_{-4}$ catastrophe (tricusp; Fig.
\ref{fig:tricusp}). The fourth case of gravitational lensing which
we consider has a line of sight near a cusp, and parallel to the
plane of the cusp (Fig. \ref{fig:tricusp2}).

Gravitational lensing produces a map of an object surface onto an
image surface.  The magnification is the inverse of the Jacobian
of this map.  Because dark matter caustics have well defined
density profiles, it is a neat mathematical exercise to calculate
their gravitational lensing characteristics. The images of
extended sources may show distortions that can be unambiguously
attributed to lensing by dark matter caustics in the limit of
perfect observations.  We see that in three of the cases
considered, a point source can have multiple images.  In those
cases, when two images merge, the Jacobian of the map vanishes and
the magnification diverges.  So, at least in theory, it seems that
gravitational lensing is a very attractive tool for investigating
dark matter caustics.  Observation of the calculated lensing
signatures would give direct evidence for caustics and CDM.

We have been particularly motivated by the possibility
\cite{Hogan} that the observer might be able to distinguish
between axions and WIMPs by determining the distance over which
the caustics are smeared.  The nearby caustic, whose position is
revealed according to ref. \cite{milk} by a triangular feature in
the IRAS map of the Milky Way plane, is only 1 kpc away from us in
the direction of observation.  By observing the gravitational
lensing due to that caustic, one may be able to measure $\delta x$
as small as $10^{13}$ cm, assuming an angular resolution of $3
\cdot 10^{-9}$ radians.  If $\delta x$ turned out to be that
small, the WIMP dark matter hypothesis would be severely
challenged (Eq. \ref{dxW}).  Unfortunately, as will be shown
below, the gravitational lensing due to a caustic only a kpc away
from us is too weak to be observed with present instruments.  It
is well known that gravitational lensing effects are proportional
to ${D_L D_{LS} \over D_S}$ where $D_S, D_L$ and $D_{LS}$ are
respectively the distances from the observer to the source, from
the observer to the lens, and from the lens to the source. We see
below that, for the gravitational lensing effects of dark matter
caustics to be observable with present technology, the lenses and
sources must be as far as possible, at the cosmological distances
of order Gpc.  Even then, the observation of such effects will be
difficult. Unfortunately, at Gpc distances it is not possible to
measure $\delta x$ as small as Eqs. \ref{dxa} and \ref{dxW} with
foreseeable technology.  So it seems unlikely that one will be
able to distinguish between dark matter candidates on the basis of
the gravitational lensing characteristics of the caustics they
form.  Henceforth, unless otherwise stated, the velocity
dispersion is set equal to zero.

\subsection{General Formalism} \label{sec:gf}

The first part of this section gives a brief account of the
gravitational lensing formalism \cite{lens}. In the second part we
show how this formalism can be streamlined in the special case of
lensing by dark matter caustics.

In linear approximation, the deflection angle $\vec\theta$ of a
light ray due to a gravitational field is given by \beeq
\vec\theta =\vec{\nabla} {2\over{c^2}}\int \Phi~dy \label{v} \;
,\eneq where $\Phi$ is the Newtonian potential. We choose the
$y$-axis in the direction of propagation of light. The deflection
angle $\vec{\theta}$ is related to the angular shift
$~\vec{\xi_I}-\vec{\xi_S}~$ on the sky of the apparent direction
of a source: \beeq \vec\theta (\vec{\xi_{
I}})={{D_S}\over{D_{LS}}}(\vec{\xi_I}-\vec{\xi_S}) \label{fund}
\eneq where $D_S$ and $D_{LS}$ are the distances of the source to
the observer and to the lens respectively. ${\vec{\xi}}_{ S}$ is
the angular position of the source in the absence of the lens
while ${\vec{\xi}}_{ I}$ is the angular position of the image with
the lens present. The angles carry components in the $x$ and
$z$-directions: $\vec{\theta}= (\theta_x , \theta_z)$, $\vec{\xi}
= (\xi_x,\xi_z)$, etc.  Unless otherwise stated, we mean by a
vector a quantity with components in the $x$- and $z$-directions.
We have $\vec{x}=(x, z)=D_L{\vec{\xi}}_I$.

It is convenient to introduce a 2D potential \beeq
\psi({\vec{\xi}}_{ I})=\frac{2D_{LS}}{c^2\, D_L\, D_S}\int
dy~\Phi~~~\ , \eneq so that \beeq \vec\theta={D_{ S} \over D_{
LS}} \vec{\nabla}_{\xi_{ I}}\psi({\vec{\xi}}_{ I})~~~\ ,
\label{th} \eneq where $\vec{\nabla}_{{\xi}_{ I}} =
D_L\vec{\nabla}$, and $D_L$ is the distance of the observer to the
lens. Then, Eq. \ref{fund} becomes \beeq \vec{\xi}_{
I}=\vec{\xi}_{ S}+\vec{\nabla}_{\xi_{ I}}\psi(\vec{\xi}_{ I})~~~\
. \label{lens} \eneq It gives the map $\vec{\xi}_{ S}(\vec{\xi_{
I}})$ from the image plane to the source plane. The inverse map
may be one to one, or one to many. In the latter case, there are
multiple images and infinite magnification when a pair of images
merge.

Our problem is to find the image map of a point source when the
matter distribution is given. The 2D gravitational potential
$\psi$ obeys the Poisson equation: \beeq \nabla^2_{\xi_{
I}}\psi={{8\pi G}\over{c^2}}{{D_L D_{LS}}\over{D_{S}}}\Sigma= 2
{{\Sigma}\over{\Sigma_c}}~~~~\ , \label{Poisson} \eneq where
$\Sigma(\xi_{ Ix}, \xi_{ Iz})$ is the column density (i.e., the
integral of the volume density along the line of sight): \beeq
\Sigma(\vec{\xi}_{ I})=\int dy\, d(D_L\xi_{ Ix},y,D_L\xi_{
Iz})~~~\ , \label{S} \eneq and $\Sigma_c$ is the critical surface
density \beeq \Sigma_c={{c^2D_{S}} \over{4\pi G D_L D_{LS}}}=0.347
\, {\rm{g/cm^2}} \left({{D_{S}}\over{D_L D_{LS}}}\,{\rm{Gpc}}
\right)\, . \label{scri} \eneq A uniform sheet of density
$\Sigma_c$ focuses radiation from the source to the observer.

Equation \ref{Poisson} is solved by \beeq \psi(\vec{\xi_{ I}})
=\frac{1}{\pi\Sigma_c}\int~d^2\xi'_{ I}~ \Sigma(\vec{\xi}'_{ I})
\ln{|\vec{\xi}_{ I}-\vec{\xi}'_{ I}|}~~~~~~\ , \label{psi} \eneq
and hence, \beeq \Delta\vec{\xi}\equiv \vec{\xi}_{ I}-\vec{\xi}_{
S} =\vec{\nabla}_{\xi_I} \psi (\vec{\xi}_{ I})
={1\over{\pi\Sigma_c}}\int~d^2\xi'_{ I}\, \Sigma(\vec{\xi}'_{
I}){{\vec{\xi}_{ I}-\vec{\xi}'_{ I}} \over{(\vec{\xi}_{
I}-\vec{\xi}'_{ I})^2}} ~~~\ . \label{shift} \eneq The image
structure, distortion, and magnification are given by the Jacobian
matrix of the map $\vec\xi_S(\vec\xi_I)$ from image to source:
\beeq K_{ij}\equiv {{\partial \xi_{ Si}}\over{\partial \xi_{
Ij}}}=\delta_{ij}-\psi_{ij} \label{K} ~~~\ , \eneq where
$\psi_{ij}\equiv\frac{\partial^2\psi} {\partial \xi_{
Ii}\partial\xi_{ Ij}}$. Because gravitational lensing does not
change surface brightness, the magnification ${\mathcal{M}}$ is
the ratio of image area to source area. Therefore, \beeq
{\mathcal{M}}={1\over{|\det{(K_{ij})}|}}~~~\ . \eneq To first
order, for $\psi_{ij} \ll 1$, \beeq
{\mathcal{M}}=1+\nabla_{\xi_I}^2\psi = 1+2{\Sigma\over\Sigma_c}
~~~\ . \label{mag} \eneq To obtain the largest lensing effects, we
wish to minimize $\Sigma_c$, given in Eq. \ref{scri}.  For fixed
$D_S$, the minimum occurs when the lens is situated half-way
between the source and the observer. Also, $D_S$ should be as
large as possible.  For our estimates, we will assume that the
source is at cosmological distances (e.g., $2D_L=2D_{LS}=D_S = 1
{\rm Gpc}$, in which case $\Sigma_c = 1.39$ g/cm$^2$).

For a general mass distribution, the gravitational lensing effects
are obtained by first calculating the column density, Eq. \ref{S},
and then the image shift, Eq. \ref{shift}.  This procedure can be
simplified when the lens is a dark matter caustic.  We are
interested in lines of sight which are tangent to a caustic
surface, because the column density $\Sigma$ has the highest
contrast there.  We assume that, in the neighborhood of the
tangent point, the flow is independent of $y$ except for a shift
$\vec{x}(y)$ of the caustic surface with $y$. The density can then
be written as \beeq d(x,y,z)=d(x-x(y),z-z(y))~~~\ , \label{den}
\eneq where $d(x,z)$ is the density of the 2D flow in the plane
orthogonal to the line of sight. The flow in the $y$-direction is
irrelevant because lensing depends only on the column density
$\Sigma$.

In the limit of zero velocity dispersion, a 2D flow is specified
by giving the spatial coordinates $\vec{x}(\alpha,\beta,t)$ of the
particle labeled $(\alpha,\beta)$ at time $t$, for all
$(\alpha,\beta,t)$. The labels $\alpha$ and $\beta$ are chosen
arbitrarily. At a given time $t$,
$\vec{x}(\alpha,\beta,t)=\vec{x}$ has a discrete number of
solutions $(\alpha, \beta)_j(\vec{x},t)$ labeled by $j=1,.\; .\;
., n(\vec{x}, t)$, where $n$ is the number of distinct flows at
$(\vec{x}, t)$. Here, $t$ is the time at which the light ray
passes by the caustic. Henceforth we will not show the time
dependence explicitly. The particle density in physical space is
\beeq d(x, z)=\sum_{j=1}^{n(\vec{x})}~{{d^2 \Lambda}\over{d\alpha
d\beta}}~ {1\over{{|{{\dd (x, z)}\over{\dd (\alpha,\beta)}}|}}}
\Bigg|_{(\alpha,\beta)={(\alpha,\beta)}_j(\vec{x})} ~~~\  ,
\label{dxi} \eneq where $\Lambda$ is the mass per unit length in
the direction of the line of sight and  ${{d^2
\Lambda}\over{d\alpha d\beta}}$ is the $\Lambda$ density in
parameter space.  Inserting Eqs. \ref{S} and \ref{den} into Eq.
\ref{shift} we obtain\beeq \vec{\nabla}_{\xi_I}\psi(\vec{\xi_{
I}}=\vec{x}/D_L)= \frac{1}{\pi\Sigma_c}\int dy\int d^2\xi'_{ I}~
{{\vec\xi_{ I}-\vec\xi'_{ I}}\over {|\vec\xi_{ I} -\vec\xi'_{
I}|^2}}~ d(D_L\xi'_{ Ix} -x(y), D_L\xi'_{ Iz} -z(y)) \hskip 0.2 cm
. \label{np} \eneq When $d$ is replaced by the caustic density Eq.
\ref{dxi}, Eq.\ref{np} becomes
\begin{eqnarray}
\vec{\nabla}_{\xi_I}\psi (\vec{\xi_{ I}}=\vec{x}/D_L)\!\!\!&=&\!\!\!\nonumber\\
\frac{1}{\pi\Sigma_c D_L}\int dy \int
dx'\,dz'\!\!\!\!\!\!&&\!\!\!\!\!\!
{{\vec{x}-\vec{x}'}\over{(\vec{x}-\vec{x}')^2}}~ \sum_{j=1}^n
{1\over{{|{{\dd (x', z')}\over{\dd(\alpha,\beta)}}|}}} \frac{d^2
\Lambda}{d\alpha d\beta}
\Bigg|_{(\alpha,\beta)={(\alpha,\beta)}_j(\vec{x}'
-\vec{x}(y))}~~~\ . \label{ara}
\end{eqnarray}
Changing variables from $(x',z')$ to $(\alpha,\beta)$ and assuming
that the density in parameter space varies only slowly over the
region of integration, we rewrite Eq. \ref{ara} as \beeq
\vec{\nabla}_{\xi_I}\psi (\vec{\xi_{ I}}=\vec{x}/D_L)=
\frac{1}{\pi\Sigma_c D_L} \frac{d^2 \Lambda}{d\alpha d\beta}\int
dy\int d\alpha\, d\beta\;
\frac{\vec{x}-\vec{x}(\alpha,\beta)-\vec{x}(y)}
{|\vec{x}-\vec{x}(\alpha,\beta)-\vec{x}(y)|^2} ~~~\ .
\label{gradpot} \eneq Further simplification is achieved by
defining the complex integral: \beeq I(x, z)\equiv\int d\alpha
d\beta~\frac{1}{x-x (\alpha ,\beta) +i(z-z(\alpha , \beta))}~~~~~\
, \label{I} \eneq in terms of which the shifts are given by \beeq
\Delta \xi_{ x}=\frac{\partial\psi}{\partial\xi_{ Ix}}=
\frac{1}{\pi\Sigma_c D_L}\frac{d^2 \Lambda}{d\alpha d\beta}
~\mbox{Re}\,\int dy\,\, I(\vec{x}-\vec{x}(y))~~~\
,\nonumber\eneq\beeq \Delta \xi_{ z}=
\frac{\partial\psi}{\partial\xi_{ Iz}} = -\frac{1}{\pi\Sigma_c
D_L}\frac{d^2 \Lambda}{d\alpha d\beta} ~\mbox{Im}\,\int dy\,\,
I(\vec{x}-\vec{x}(y))~~~\ . \label{LCI1} \eneq Eqs. \ref{I} and
\ref{LCI1} are useful when the caustic has contrast in the two
dimensions transverse to the line of sight (e.g., near a cusp).

In many applications, however, the caustic has contrast in only
one of the dimensions transverse to the line of sight.  Choosing
$\hat x$ to be the trivial direction, Eq. \ref{Poisson} is reduced
to \beeq {{d^2\psi}\over{d\xi^2_{ I}}}(\xi_{ I})= {2 \over
\Sigma_c}~\Sigma(z=D_L\xi_{ I})~~~\ , \label{Poi2} \eneq and the
column density is given by \beeq \Sigma(\xi_{
I})=\int~dy~d(z-z(y))~~~\ . \label{col1} \eneq The flow is now
effectively one dimensional.  Its physical space density is given
by \beeq d(z)=\sum_{j=1}^{n(z)}~{{d \Lambda}\over{d\alpha}}~
{1\over{{|{{dz}\over{d\alpha}}|}}}\Bigg|_{\alpha={\alpha}_j(z)}~~~\
, \label{d1D} \eneq where $\Lambda$ is the mass surface density in
the two trivial directions ($x$ and $y$) and ${d\Lambda \over
d\alpha}$ is the $\Lambda$ density in parameter space.  The 1D
Green's function is $G={1 \over 2}(|\xi|+a\xi)+b$, where $a$ and
$b$ are arbitrary constants. The shift is \beeq
{{d\psi}\over{d\xi_{ I}}}(\xi_{ I})= {1\over{\Sigma_c}}\int
d\xi'_{ I}~\Sigma(\xi'_{ I})~ ({\rm Sign}(\xi_{ I}-\xi'_{
I})+a)~~~\ . \label{sign} \eneq The constant $a$ causes an overall
shift of the image, which does not concern us. We choose $a=-1$.
Repeating the steps done earlier for the 2D case, Eq. \ref{sign}
is re-expressed as \beeq \Delta\xi = -{{2}\over{\Sigma_c D_L}}{{d
\Lambda}\over{d\alpha}} \int dy \int d\alpha~\Theta
(-z+z(\alpha)+z(y))~~~\ . \label{theeq} \eneq In the next section
Eq. \ref{theeq} is used to calculate the shifts due to simple
folds of dark matter flows, and Eq. \ref{LCI1} is used to
calculate the shifts due to a cusp.

\subsection{Applications} \label{sec:fc}

\subsubsection{Lensing by a Concave Fold} \label{subsection:sccf}

We consider a simple fold caustic which has a curvature radius $R$
along the line of sight.  We assume that the surface is concave
(i.e., it is curved in the direction of the two extra flows; Fig.
\ref{fig:concave}). The outer caustics of dark matter halos are
examples of concave caustic surfaces.  The convex case is
discussed in the next subsection.

\begin{figure}[ht] \centering
\includegraphics[height=9cm,width=9cm]{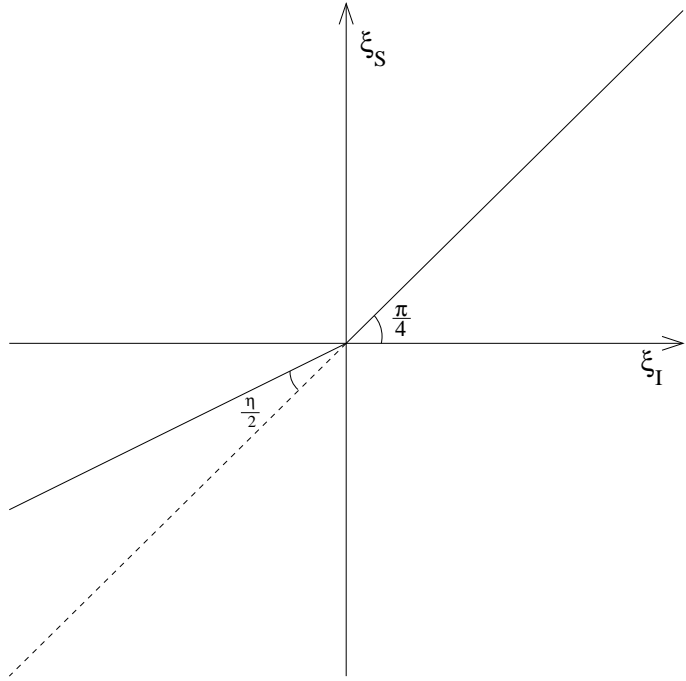}
\caption{Source position $\xi_S$ as a function of image position
$\xi_I$ for lensing by a concave fold. $\eta$ is given by Eq.
\ref{ep} in terms of the fold coefficient and the curvature radius
of the caustic surface.  Estimates of $\eta$ for the outer
caustics of galactic halos are given in Eq. \ref{epe}.}
\label{fig:shiftconcave}
\end{figure}

In the neighborhood of the point $P$ of closest approach of the
line of sight with the caustic surface, we choose coordinates such
that $\hat{z}$ is perpendicular to the surface while $\hat{x}$ and
the direction $\hat{y}$ of the line of sight are parallel. $P$ has
coordinates $x=y=z=0$. In the $y=0$ plane, the flow is given by
$z(\alpha)=-{1\over 2}h\alpha^2$, where $h$ is a positive
constant. Using Eq. \ref{d1D} we find the density in the $y=0$
plane: \beeq d(\sigma) = A~{\Theta(\sigma) \over
\sqrt{\sigma}}~~~\ , \label{Ah} \eneq where $\sigma = -z$, and
\beeq A=\sqrt{{2\over h}}~{{d \Lambda}\over{d\alpha}}~~~\ .
\label{Aa} \eneq For $y \neq 0$, the density is still given by Eq.
\ref{Ah}, but with $\sigma=z(y)-z$ and $z(y)=-\frac{y^2}{2R}$. We
calculate the shift using Eq. \ref{theeq}:
\begin{eqnarray}
\Delta\xi&=&-\frac{2}{\Sigma_c D_L} \frac{d
\Lambda}{d\alpha}\int_{-\sqrt{-2Rz}}^{\sqrt{-2Rz}} dy\int d\alpha~
\Theta \left(-z - {h \over 2} \alpha^2 - {y^2 \over 2R} \right)\nonumber\\
&=& {{4\pi}\over{\Sigma_c}}{{d\Lambda}\over{d\alpha}}\sqrt{{R\over
h}}\, \Theta (-\xi_{ I})\,\xi_{ I}~~~\ . \label{grt2}
\end{eqnarray}
Hence \beeq \Delta\xi =\xi_{ I}-\xi_{ S} = \eta\,\xi_{
I}\,\Theta{(-\xi_{ I})}~~~\ , \label{Dks} \eneq with \beeq \eta =
{{2\pi A\sqrt{2R}}\over{\Sigma_c}}~~~\ . \label{ep} \eneq One can
also obtain this result by calculating the column density: \beeq
\Sigma(\xi_{ I})=\int dy~d(y,z=\xi_{ I}) =\int
dy~\frac{A~\Theta(-\xi_{ I}D_L-{{y^2}\over{2R}})} {\sqrt{-\xi_{
I}D_L -{{y^2}\over{2R}}}}= \pi A\sqrt{2R}\,\Theta (-\xi_{ I}) ~~~\
, \label{cofcod} \eneq and solving Eq. \ref{Poisson}: \beeq \xi_{
I}-\xi_{ S}={{d\psi}\over{d\xi_{ I}}} ={2\over{\Sigma_c}}\int
d\xi_{ I}\,\Sigma(\xi_{ I}) =\eta\,\xi_{ I}\,\Theta(-\xi_{ I})
~~~\ . \label{-} \eneq Eqs. \ref{cofcod}-\ref{-} were first
obtained by Hogan \cite{Hogan}. The agreement with Eq. \ref{Dks}
validates the formalism derived in Section \ref{sec:gf}.  Figure
\ref{fig:shiftconcave} plots the source position $\xi_S$ versus
the image position $\xi_{ I}$.

The magnification is: \beeq {\mathcal{M}}={{d\xi_{
I}}\over{d\xi_{S}}} = 1 + \eta\;\Theta{(-\xi_I)} + 0(\eta^2)~~~\ .
\label{Mm} \eneq When the line of sight of a moving source crosses
the surface of a simple concave fold, the component of its
apparent velocity perpendicular to the fold changes abruptly.
Also, a discontinuity occurs in the magnification of the image.
Both effects are of order $\eta$.

We estimate $\eta$ for the outer caustics of galactic halos using
Eqs. \ref{AN}-\ref{Fne}: \be \{\eta_n=\frac{\sqrt{2}\, v_{{\rm
rot}}^2}{G\,\Sigma_c}\, \frac{F_n}{R_n} \, : n=1, 2,.\; .\;
.\}\sim
(7,~6,~6,~6,~6,.\; .\; .)\cdot 10^{-3}\nonumber\\
\cdot \left(\frac{D_L D_{LS}}{D_S~{\rm Gpc}}\right)
\left(\frac{v_{\rm {rot}}}{220 {\rm km/s}}\right)
\left(\frac{h}{0.7}\right)~~~\ . \label{epe} \ee A magification of
order $10^{-2}$ seems hard to observe, however, the images of
extended sources may be modified in recognizable ways. In
particular, straight jets would be seen with an abrupt bend where
their line of sight crosses a fold. Indeed, the image is stretched
by the factor $1+\eta$ in the direction perpendicular to the
caustic, on the side with the two extra flows.  If the jet makes
angle $\alpha$ with the normal, it appears bent by an angle
$\delta \equiv \frac{1}{2}\eta\sin{\left({2\alpha}\right)}$.
Searching the sky for bends in extended sources may be a realistic
approach to detecting caustic structure in galactic halos
\cite{Hogan}.

\subsubsection{Lensing by a Convex Fold} \label{subsection:scvf}

\begin{figure}[ht]
\centering
\includegraphics[height=9cm,width=9cm]{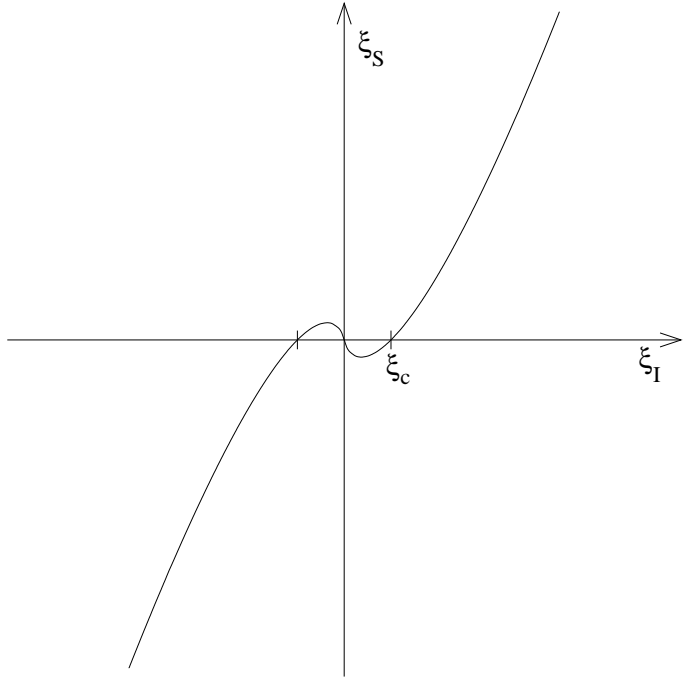}
\caption{Source position $\xi_S$ as a function of image position
$\xi_I$ for lensing by a convex fold.  $\xi_c$ is given in Eq.
\ref{xic}.} \label{fig:shiftconvex}
\end{figure}

By definition, a convex fold is curved in the direction opposite
to the side with two extra flows (Fig. \ref{fig:convex}). Using
the conventions of the previous subsection, we write the equation
for the displacement of the surface along the line of sight as
$z(y)={{y^2}\over{2R}}~$, and that for the flow as
$~z(\alpha)=-{1\over 2}h\alpha^2~$, for small $z$, with $R, h>0$.
Eqs. \ref{Ah} and \ref{Aa} still apply, with $~\sigma = {y^2/2R} -
z~$.  Eq. \ref{theeq} yields the shift: \be \Delta\xi
=-{{8}\over{{\Sigma_c}D_L}}{{d \Lambda}\over{d\alpha}}
\left\{\Theta(-\xi_{ I})\int_0^\infty dy +\Theta(\xi_{
I})\int_{\sqrt{2RD_L\xi_{ I}}}^\infty dy\right\} \sqrt{{2 \over
h}\left({{y^2}\over{2R}}-D_L\xi_{ I}\right)} ~~~\ . \ee We
introduce a cut-off $L$ for the integral over large $y$.  $L$ can
be thought of as the length scale beyond which our description of
the flow is invalid.  The above equation becomes \beeq \Delta\xi
=-\frac{4}{\Sigma_c D_L} \frac{d
\Lambda}{d\alpha}\frac{1}{\sqrt{hR}} \left\{L\sqrt{L^2-2RD_L\xi_{
I}} -2RD_L\xi_{ I}\,\ln{\left(\frac{L +\sqrt{L^2 -2RD_L\xi_{ I}}}
{\sqrt{2RD_L|\xi_{ I}|}}\right)}\right\}. \label{Dxiz} \eneq By
expanding Eq. \ref{Dxiz} in powers of ${1\over L}$, and using Eq.
\ref{Aa}, we obtain the $\xi_{I}$-dependent shift: \beeq
\Delta\xi=\xi_{ I} -\xi_{ S}
=-\frac{\eta^\prime}{\pi}\left[\,\ln{\left({{RD_L|\xi_{ I}|}
\over{2 L^2}}\right)}-1\right]\xi_{ I} ~~~\ , \label{shifc} \eneq
where $\eta^\prime$ (like $\eta$ in Eq. \ref{ep}) is given by
\beeq \eta^\prime = {2 \pi A \sqrt{2 R} \over \Sigma_c}~~~\ .
\label{epp} \eneq The magnification is\beeq
{\mathcal{M}}=\frac{d\xi_{ I}}{d\xi_{ S}}= \left|1 +
\frac{\eta^\prime}{\pi}~ \ln{\left({{R\,D_L\,|\xi_{ I}|} \over{2
L^2}}\right)}\right|^{-1} ~~~\ . \label{magc} \eneq The cut-off
$L$ has an effect on both the magnification and the elongation of
the image in the direction normal to the caustic surface, but that
effect is $\xi_I$ independent.  $L$ has a global effect on the
image, as opposed to an effect localized near $\xi_I = 0$. Figure
\ref{fig:shiftconvex} plots $\xi_{ S}$ versus $\xi_{I}$. It shows
that a convex fold can cause a triple image of a point source.
In particular, when the source is exactly behind the caustic
($\xi_S = 0$), the images are at $\xi_I = -\xi_c$, 0, and
$+\xi_c$, with \beeq \xi_c = {2 L^2 \over R D_L} \exp{(-{\pi \over
\eta^\prime} + 1)}~~~\ . \label{xic} \eneq A point source has a
single image sufficiently far from the caustic, say at $\xi_{
I1}>0$. When the line of sight approaches the caustic surface
tangent point,  two new images appear on top of each other at
$\xi_{ I2}=\xi_{ I3} = -\xi_c/e$. At that moment, the
magnification at $\xi_{I2}$ is infinite, and $\xi_S = \eta^\prime
\xi_c/e \pi$. As the source crosses the caustic, $\xi_{ I2}$ moves
toward $\xi_{ I1}$ and finally merges with it. When $\xi_{ I1} =
\xi_{ I2} = +\xi_c/e$, the magnification diverges again. After
that, only the image at $\xi_{ I3}$ remains.

Let us apply the above results to the line of sight in the plane
of a caustic ring at the sample point, $(z=0, \rho=a)$, discussed
in Section \ref{subsection:aspecialpoint}. Setting $R = a_n$ and
using Eq. \ref{Anv2} we obtain for the $n$th ring \beeq
\eta^\prime_n =\frac{2\pi\, A_{0,n}\sqrt{2a_n}}{\Sigma_c}=
\frac{v^2_{ {\rm rot}}}{\sqrt{2} G\Sigma_c}
\frac{f_n}{\sqrt{a_n\,p_n}}\frac{v_n}{b_n} ~~~\ . \eneq Using Eq.
\ref{Anring} to estimate $A_{0,n}$, we find \beeq \{\eta^\prime_n:
n=1,2,.\; .\; .\}\sim (5,~4,~4,~4,~4,.\; .\; .)
\cdot 10^{-2}~\frac{D_L\,D_{LS}}{D_S\,{\rm Gpc}}\nonumber\\
\cdot\left(\frac{0.27}{j_{\rm{max}}}\right)
\left(\frac{h}{0.7}\right) \left(\frac{v_{{\rm rot}}}{220\,{\rm
km/s}}\right)~~~\ . \eneq For such small values of $\eta^\prime$
the angular distance between the triple images is exponentially
small and unresolvable with present and foreseeable instruments.

The image of an extended object is stretched in the direction
perpendicular to the caustic by the relative amount \beeq
{\mathcal{M}} - 1 = - {\eta^\prime \over \pi} \ln{\left({R D_L
|\xi_I| \over 2 L^2}\right)}~~~\ . \label{Str} \eneq The image is
stretched for $\xi_I < \xi_d$, and compressed for $\xi_I > \xi_d$,
where \beeq \xi_d = {2 L^2 \over R D_L}~~~\ . \label{xid} \eneq In
the case of the sample point $(z=0, \rho=a)$ with line of sight in
the $z=0$ plane, the cut-off length $L$ (i.e., the distance in the
$y$-direction over which our desciption of the flow is valid), is
of order $\sqrt{2 a p}$.  In that case $\xi_d \sim 4 p/D_L$. Since
$p/D_L$ is the transverse angular size of the caustic ring, our
description certainly fails for $\xi_I > p/D_L$.  So, over the
region where our calculation is valid, the image is magnified. The
effects are generically of order one percent, increasing to
several percent when $\xi_I \sim 10^{-3} \xi_d$, for caustic rings
at cosmological distances.

\subsubsection{Lensing by a Fold with Zero Curvature}
\label{lensingzerocurvature} We saw in Section \ref{difgeo} that
the surface of a caustic ring has tangent lines along which the
curvature vanishes.  One may speculate that the lensing effects of
a caustic surface are strongest when the line of sight is tangent
to the surface in a direction of zero curvature, because the line
of sight stays close to the caustic over greater depths in that
case. If the line of sight of an observer looking at the outside
profile of a caustic ring is at some point on the profile in a
direction of zero curvature, then the equation for the
intersection of the caustic surface with the plane containing the
outward normal to the surface ($\hat{z}$) and the line of sight
($\hat{y}$) is $z(y)=-\frac{y^4}{4U}$ where $U$ is positive and
has dimensions of (length)$^3$.  In such a case, the cubic term in
the Taylor expansion of $z(y)$ is absent because the line of sight
remains everywhere outside the caustic ring tube.  We did not
calculate $U$ for caustic rings, but expect $U \sim$ (kpc)$^3$ in
order of magnitude. The flow is given by $z(\alpha)=-{1\over
2}h\alpha^2$ with $h>0$, as before, and Eq. \ref{Aa} holds.

\begin{figure}[ht]
\centering
\includegraphics[height=9cm,width=9cm]{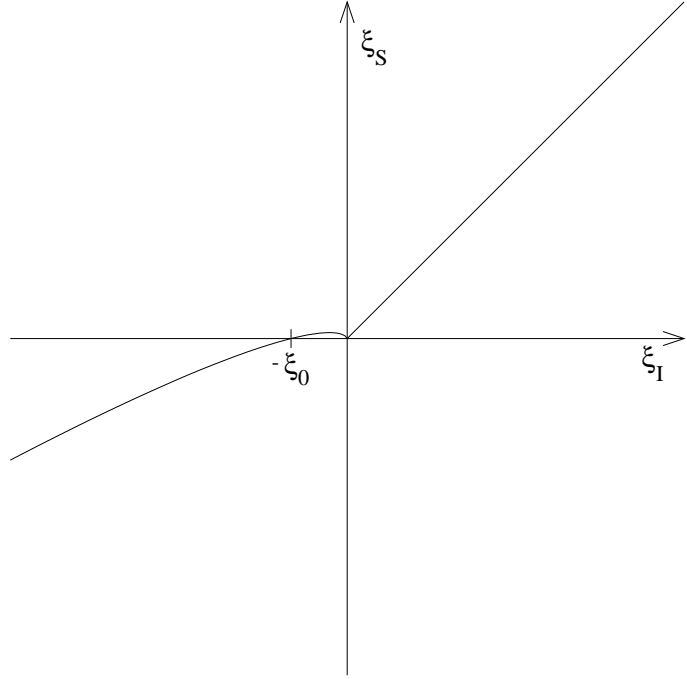}
\caption{Source position $\xi_S$ as a function of image position
$\xi_I$ for lensing by a fold with zero curvature. $\xi_0$ is
given in Eq. \ref{xi0}.} \label{fig:shiftflat}
\end{figure}

Using the above expressions for $z(y)$ and $z(\alpha)$ in Eq.
\ref{theeq}, we find the shift \be \Delta\xi &=&-{2 \over \Sigma_c
D_L}{d \Lambda \over d\alpha} \int dy \int d\alpha~
\Theta\left(-z-{1 \over 2}h\alpha^2-{y^4 \over 4U}\right)\nonumber\\
&=&-{8 \over \Sigma_c D_L} {d \Lambda \over d\alpha}
{\left(-4Uz\right)^{3/4} \over \sqrt{2hU}}
\Theta(-\xi_I)\int_0^1 dt \sqrt{1-t^4}\nonumber\\
&=& -\Theta(-\xi_I) \left(-\xi_0\,\xi_I^3 \right)^{1/4}~~~~\ ,
\label{dxio} \ee where \beeq \xi_0 = (9.89~\frac{A}{\Sigma_c})^4
\frac{U}{D_L}~~~\ . \label{xi0} \eneq The magnification is: \beeq
{\mathcal{M}} =  \left|1 - {3 \over 4}\Theta(-\xi_I)
\left(-\frac{\xi_0}{\xi_I}\right)^{1/4}\right|^{-1} ~~~\ . \eneq
Figure \ref{fig:shiftflat} shows $\xi_{ S}$ versus $\xi_{ I}$.

Triple images occur when $|\xi_I| \leq \xi_0$. Unfortunately, for
the zero curvature tangents of caustic rings envisaged above,
$\xi_0$ is very small.  For $2D_L=2D_{LS}=D_S= {\rm Gpc}$, $A = 3
\cdot 10^{-4} {{\rm gr} \over {\rm cm}^2 {\rm kpc}^{1 \over 2}}~$,
and $U = ({\rm kpc})^3$, one finds $\xi_0 = 4 \cdot 10^{-17}$.
Hence the triple images cannot be resolved.  Also, even at angular
distances as small as $\xi_I \sim 10^{-9}$, the magnification and
image distortion is only of order 1\%.

\subsubsection{Lensing by a Cusp} \label{subsection:lensingbycusp}

In this section, we investigate a line of sight parallel to the
plane ($z=0$) of a caustic ring, and passing near the cusp at
$\rho = \rho_0$ (Fig. \ref{fig:tricusp2}).  We use the 2D lensing
equations derived in Section \ref{sec:gf}. The shifts are given by
Eq. \ref{LCI1} in terms of the complex integral $I$ of Eq.
\ref{I}. Using Eq. \ref{kem2}, we have \beeq I=\int d\alpha\int
d\tau \frac{1}{\rho-a -\frac{1}{2}u\tau_0^2
+u\tau_0\tau+\frac{1}{2}s\alpha^2+i(z-b\alpha\tau)} ~~~\ .
\label{Iint} \eneq Because we are close to the cusp, the term of
order $\tau^2$ is neglected in the denominator of the integrand.
In terms of the parameters defined in Eqs. \ref{capvar}, \beeq
I=\frac{2}{b\sqrt{\zeta}}\int dT\int_{-\infty}^{\infty}
\frac{dA}{A^2- 2 i {AT \over \sqrt{\zeta}} +2T + X +
i\frac{Z}{\sqrt{\zeta}}}~~~~\ . \label{reint} \eneq The
integration over $A$ yields \beeq
I=\frac{2\pi}{b\sqrt{\zeta}}\int_{\frac{|T|}{\sqrt{\zeta}}<C}\,
\frac{dT}{\sqrt{\frac{T^2}{\zeta}+2T+X+i\frac{Z}{\sqrt{\zeta}}}}~~~~,
\label{IASZ} \eneq where\hskip 0.3 cm $C\equiv{\rm
Re}\,\,\sqrt{\frac{T^2}{\zeta}+2T+X+i\frac{Z}{\sqrt{\zeta}}}
>0\,$. Near the cusp ($X, Z \ll 1$) the terms of order $T^2$ can be
neglected since they are unimportant in the denominator of Eq.
\ref{reint}.  Equation \ref{IASZ} becomes then \beeq
I(X,Z)=\frac{2\pi}{b\sqrt{\zeta}}\int\frac{dT}{\sqrt{2T+X
+i\frac{Z}{\sqrt{\zeta}}}}~~~ , \label{2pi} \eneq where the
integration domain is defined by the inequality: \beeq T^4-\zeta
(2T^3+XT^2)-{{\zeta Z^2}\over{4}}<0~~~~\ . \label{relin} \eneq

\begin{figure}[ht] \centering
\includegraphics[height=11cm,width=11cm]{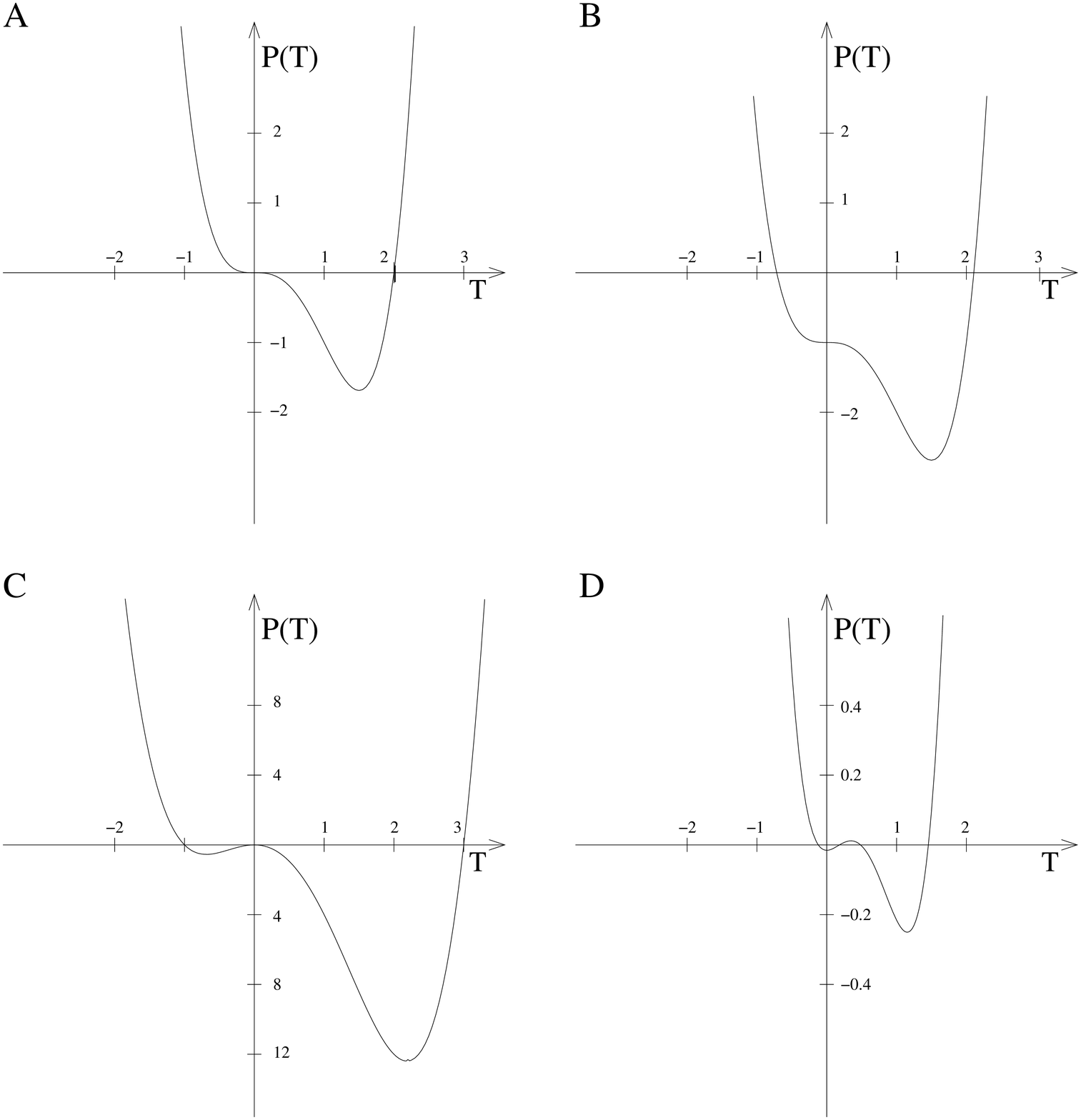}
\caption{Graphs of the quartic polynomial $P(T)\equiv
~T^4-\zeta(2T^3+XT^2)-\frac{\zeta Z^2}{4}~$ for $\zeta =1$ and $(X
, Z)$ = $(0 , 0)$, $(0 , 2)$, $(1/3 , 0)$ and $(-1/3 , 0)$; a)
through d)
 respectively.}
\label{fig:quartic}
\end{figure}

Let us call $T_i\,$ $(i = 1,.\; .\; ., 4)$ the roots of the
polynomial on the left hand side of Eq. \ref{relin}. Near the
cusp, three of the roots $T_i$ are near zero and one of the roots
is close to $2\zeta$. Let us call the latter $T_4$.  In order to
find the roots near $T=0$ we neglect the quartic term and solve
the cubic equation: \beeq T^3+\frac{X}{2}T^2+\frac{Z^2}{8}=0~~~~\
. \label{cubic} \eneq This equation is also obtained by
eliminating $A$ from Eqs. \ref{kemn1} and \ref{kemn2}. Hence the
solutions of Eq. \ref{cubic} are the values of $T$ for the flows
near the cusp. As was discussed in Section
\ref{subsection:densitynearacusp}, the number of flows is
determined by $\delta$, given in Eq. \ref{discr}. Outside the
cusp, where $\delta > 0$, there is a single root $T_1$. Inside the
cusp  where $\delta < 0$, there are three roots $T_1$, $T_2$ and
$T_3$. Figure \ref{fig:quartic} shows the quartic polynomial, Eq.
\ref{relin}, for various values of $X$ and $Z$, and $\zeta = 1$.
For a line of sight just outside the cusp, the quartic polynomial
is negative from $T_1$ to $T_4$ with \beeq T_1=-{1\over
2}\left(\sqrt{\delta} -\frac{|Z|}{2}\,\,\right)^{2/3} -{1\over
2}\left(\sqrt{\delta} +\frac{|Z|}{2}\,\,\right)^{2/3} -{X\over
6}~~~~\ . \label{T1} \eneq Eq. \ref{2pi} becomes in that case:
\beeq
I=\frac{2\pi}{b\sqrt{\zeta}}\Bigg\{\sqrt{2T_4+X+i\frac{Z}{\sqrt{\zeta}}}
-\sqrt{2T_1+X+i\frac{Z}{\sqrt{\zeta}}}\,\Bigg\}~~~~\ .
\label{Ikoc} \eneq

Let us consider the line of sight defined by Fig.
\ref{fig:tricusp2}. Equation \ref{LCI1} gives the shift in the
direction perpendicular to the plane of the cusp as \beeq \Delta
\xi_{z} = - {1 \over \pi \Sigma_c D_L} {d^2 \Lambda \over d\alpha
d\tau}\; {\rm Im} \int dy~I(X - X(y), Z) \label{ano} \eneq where
$X(y) = - {y^2 \over 2 \rho_0 p}$ is the shift of the cusp as a
function of depth.  The integral in Eq. \ref{ano} with the
integrand given by Eqs. \ref{T1} and \ref{Ikoc} can be done
numerically.  Here we only give a rough estimate, to determine the
order of magnitude and qualitative properties of the lensing
effects.

For $X=0$ and $Z\ll 1$, we have \beeq
I=\frac{2\pi}{b}\left[2-\frac{i}{\sqrt{\zeta}} \mbox{Sign}(Z)
|Z|^{1/3}-\frac{1}{2\zeta}|Z|^{\frac{2}{3}}\right]+O(Z)~~~~\ .
\label{Ikocx0} \eneq Because the scaling law $X^3\sim Z^2$ holds
close to the cusp, we expect Eq. \ref{Ikocx0} to be valid as long
as the shift in the $x$-direction $|\Delta X|\leq
|Z|^{\frac{2}{3}}$.  Hence Eq. \ref{Ikocx0} provides a good
estimate of $I(X - X(y), Z)$ over a depth $2 \Delta y$ with
$\Delta y \sim \sqrt{2 \rho_0 p}\, |Z|^{1 \over 3}$.  Therefore
\beeq \Delta \xi_z \sim {2 \over \Sigma_c D_L b} \sqrt{{2 \rho_0 p
\over \zeta}} \frac{d^2\Lambda}{d\alpha d\tau}~|Z|^{2\over 3}~{\rm
Sign}(Z)~~~~\ . \label{dxiz2} \eneq Since we are near $\alpha =
0$, \beeq \frac{d^2\Lambda}{d\alpha d\tau} = {1 \over 2 \pi
\rho_0} {d^2 M \over d\alpha d\tau} = {1 \over \rho_0} {d^2 M
\over d\Omega dt}~~~~\ . \eneq Using this and Eq. \ref{A_nc}, we
obtain \beeq \Delta\xi_{z} \sim \eta''\,|\xi_{Iz}|^{{2}\over
{3}}\,{\rm Sign}(\xi_{Iz})~~~~ , \eneq where \beeq
\eta''\equiv\frac{2\sqrt{2\, a}}{\zeta^{{1}\over {6}}\Sigma_c
D_L^{{1}\over {3}}} \,p^{{1}\over {3}}\,A_{0}~~~~\ . \label{ep''}
\eneq The contribution to the magnification from distortion in the
$z$-direction is therefore: \beeq {\mathcal{M}}_z \sim
\left|1-\eta''\,{{2}\over
{3}}|\xi_{Iz}|^{-{{1}\over{3}}}\right|^{-1}~~~~\ . \eneq Figure
\ref{fig:shiftcusp} plots $\xi_{Sz}$ versus $\xi_{Iz}$:

\begin{figure}[ht] \centering
\includegraphics[height=9cm,width=9cm]{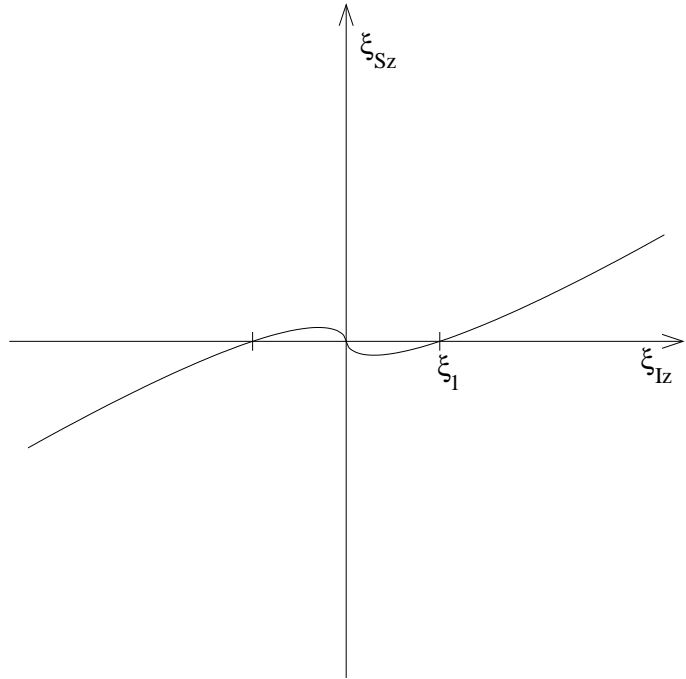}
\caption{Source position $\xi_{Sz}$ as a function image position
$\xi_{Iz}$ for lensing by a cusp along the line of sight described
in Fig. \ref{fig:tricusp2}.  $\xi_1 \sim \eta''^3$ where $\eta''$
is given by Eq. \ref{ep''}.} \label{fig:shiftcusp}
\end{figure}For $\xi_{Iz} \leq \xi_1 \sim \eta''^3$, a point source has
triple images.  For the $n$th caustic ring, $\eta''$ can be
estimated using Eqs. \ref{a_n} and \ref{Anring}, and assuming $p
\simeq 0.1 a,~\zeta \sim 1$: \be \{\eta''_n : n=1, 2,.\; .\;
.\}\sim (1.5,~ 1,~ 0.7,~ 0.6,~ 0.5,.\; .\; .)\cdot
10^{-4}\frac{D_L^{{2}\over {3}}\,D_{LS}}{D_S\,{\rm
Gpc}^{{2}\over {3}}}~\cdot\nonumber\\
\cdot\left(\frac{0.27}{j_{\rm{max}}}\right)^{{2}\over {3}}
\left(\frac{h}{0.7}\right)^{{2}\over {3}} \left(\frac{v_{{\rm
rot}}}{220\,{\rm km/s}}\right)^{4\over 3}~~~\ . \label{ep''est}
\ee At cosmological distances, the typical angular separation
between the images is of order $10^{-12}$.  Unfortunately, here
again the images are too close to be resolved, however, for an
angular distance $\xi_I\sim 10^{-9}$, ${\mathcal{M}}_z$ is of
order 10\%, which may be observable.  The lensing properties of a
cusp for other lines of sight can be calculated numerically using
Eqs. \ref{LCI1} and \ref{Iint}.  The latter equation is replaced
by Eqs. \ref{Ikoc} and \ref{T1}, and $T_4 = 2 \zeta$, if the line
of sight is everywhere outside the region with two extra flows.

\newpage

\section{Conclusions} \label{Conclusions}

We reviewed the leading cold dark matter (CDM) candidates: axions
and weakly interacting massive particles (WIMPs) in Section
\ref{chap:CDMcandidates }. We discussed each candidate's status in
particle physics and cosmology, its production in the early
universe, its energy density and velocity dispersion. We found
that the velocity dispersions of the candidates are negligible on
astronomical scales; hence, they indeed are ``cold.''

In Section \ref{Flow}, we discussed the appearance of caustics in
the flow of CDM particles in and out of a galaxy. Caustics are
surfaces in space where the density of CDM diverges in the limit
of zero velocity dispersion. In this limit the particles lie on a
3D sheet in 6D phase space. The physical space density is the
projection of the phase space sheet onto physical space. Caustics
occur wherever the phase space sheet folds back in phase space. At
the fold the phase space sheet is tangent to velocity space,
particles pile up, and hence the number density, which is the
integral of the phase space density over velocity space, diverges.
Generically caustics are boundary surfaces separating two regions
in physical space. One region has $n$ flows and the other has
$n+2$ flows, where $n$ is an odd number. Because the flow of CDM
particles is continuous, caustic surfaces are topologically
stable. There are two types of CDM caustics in the halos of
galaxies: outer and inner. An outer caustic is a simple fold
($A_2$) catastrophe, located on a topological sphere, surrounding
the galaxy. Outer caustics occur where a given outflow reaches its
furthest distance from the galactic center before falling back in.
We provided estimates of the radii and fold coefficients of the
outer caustics, based upon the self similar infall model. An inner
caustic is a ring (i.e., a closed circular tube whose
cross-section is a $D_{-4}$ catastrophe), with three cusps. Inner
caustics are located near where the particles with the most
angular momentum in a given inflow reach their closest approach to
the galactic center before going back out. Inside the tube there
are $n+4$ flows. Outside the tube there are $n+2$, where $n$ is
again odd. We gave a detailed analysis of caustic rings in the
case where the flow of particles is axially and reflection
symmetric and where the transverse dimensions of the ring is
tight. In that case, the flow, and hence the caustic, is described
by 5 parameters: $a$, $b$, $s$, $\tau_0$ and $u$. The precise
shape of a transverse section of the ring in this limit is shown
in Fig. \ref{fig:fig6}. For the inner caustic rings, the curvature
radii and fold coefficients depend on the position on the surface
of the ring caustic. We derived formulae for these quantities as a
function of this position (parametrized by $\tau_1$) and of the
five caustic ring parameters ($a,~b,~u,~\tau_0,~s$).  A great
fraction of the surface of a caustic ring ($0 < |{\tau_1 \over
\tau_0}| < {3 \over 4}$) is saddle-shaped and therefore has
tangent directions along which the curvature of the surface
vanishes. Finally, we derived the mass density profile near a cusp
caustic. These quantities were estimated using the self similar
infall model.

Gravitational lensing was discussed in Section \ref{chap:lensing}.
In general, the shift in image position is the gradient of a
potential whose 2D Laplacian is the column density.  For an
arbitrary mass distribution, the procedure for calculating the
shift involves two steps.  First the matter density is integrated
along the line of sight to obtain the column density.  Second, the
potential is obtained by convolving the column density with the 2D
Green's function.  In the case of lensing by dark matter caustics,
this procedure can be simplified: the shift is expressed directly
as an integral over the parameter space of the dark matter flow
forming the caustic.  The relevant result is given in Eqs. \ref{I}
and \ref{LCI1} if the caustic has contrast in the two dimensions
transverse to the line of sight, and in Eq. \ref{theeq} if the
caustic has contrast in only one of the dimensions transverse to
the line of sight. We applied our formalism to the gravitational
lensing by dark matter caustics in four specific cases.  In the
first three, the line of sight is tangent to a surface where a
simple fold catastrophe is located. The three cases are
distinguished by the curvature of the caustic surface at the
tangent point in the direction of the line of sight: (1) the
surface curves toward the side with two extra flows; (2) the
surface curves away from the side with two extra flows; and (3)
the surface has zero curvature. In the fourth case (4) studied,
the line of sight is at a specific location close to a cusp
catastrophe and parallel to the plane of the cusp.  We found that
each case has characteristic lensing signatures.  In three of the
cases (2, 3 and 4) there are multiple images and infinite
magnification of point sources when their images merge. In case 1,
there are no multiple images. Unfortunately, the effects are small
even for dark matter caustic lenses at cosmological distances. The
multiple images of a point source cannot be resolved with present
instruments.  Typical magnifications and image distortions are of
order one \% to a few \%. A promising approach may be to observe
the distortions caused by dark matter caustics in the images of
extended sources, such as radio jets. This possiblity was
discussed in case (1).

\newpage
\appendix

\section{Vacua of Non-Abelian Gauge Theory and Instantons}
\label{Inst}

\subsection{Description of Non-Abelian Gauge Theory} Let the
Lagrangian density be ${{\mathcal{L}}}(\Psi(x),
\partial_\mu\Psi(x))$, where $\Psi(x)$ denotes an arbitrary field as a
function of spacetime coordinate $x^\mu = (x^0, x^1, x^2, x^3)$.
We adopt the convention where the spacetime metric tensor has the
components $g_{\mu\nu}={\rm diag}(+1, -1, -1, -1)$. Under a local
(i.e., $x$-dependent) gauge transformation \be
\Psi(x)&&\rightarrow\Psi'(x) = U(x)\Psi(x)\, ,\nonumber\\
\partial_\mu\Psi(x)&&\rightarrow
(\partial_\mu\Psi(x))'=U(x)\partial_\mu\Psi(x)+(\partial_\mu
U(x))\Psi(x)\; , \label{trans2} \ee where $U(x)$ is an element of
a Lie Group $G$. A Lie group is a continuous group generated by
the corresponding Lie algebra which is a vector space whose basis
consists of $N$ generators $T^a (a=1,.\; .\; .,N)$ that are closed
under commutation: $[T^a , T^b ]=if^{abc}T^c$. The structure
constants $f^{abc}$ are real numbers and they completely define
the Lie algebra. We adopt the normalization conditions ${\rm
Tr}(T^a T^b)=\frac{1}{2}\delta^{ab}$. An element $U(x)$ of $G$ is
obtained by exponentiating the algebra: \beeq U(x)=
e^{-i\omega^a(x) T^a} \, . \eneq Because $U$ is parametrized by
$\{\omega^a\}$, the group manifold of $G$ is the space of the
possible values of $\{\omega^a\}$. The correspondence between Lie
group and Lie algebra is many-to-one. In order to have a
one-to-one correspondence between $\{\omega^a\}$ and a group
element, the possible values of $\{\omega^a\}$ must be restricted.

Gauge invariance, at the Lagrangian level, means \beeq
{{\mathcal{L}}}(\Psi(x)', (\partial_\mu\Psi(x))') =
{{\mathcal{L}}}(\Psi(x),
\partial_\mu \Psi(x))\, .
\eneq This requires invariance under global (i.e.,
$x$-independent) transformations. However, since $\partial_\mu
U(x)\not= 0$, in Eq. \ref{trans2}, $\partial_\mu\Psi(x)$ does not
transform in the same way as $\Psi(x)$. This spoils the gauge
invariance of ${{\mathcal{L}}}$, even if ${{\mathcal{L}}}$ is
invariant under global transformations. To preserve the invariance
we look for a covariant derivative $D_\mu\Psi$ with the
transformation property $(D_\mu\Psi(x))'\equiv
D'_\mu\Psi'(x)=U(x)(D_\mu\Psi(x))$. Equivalently, since
$D'_\mu\Psi'(x)=D'_\mu U(x)\Psi$, we want $D'_\mu
U(x)\Psi=U(x)(D_\mu\Psi(x))$ which implies \beeq D'_\mu
U(x)=U(x)D_\mu\; ,\label{Dtransform} \eneq or \beeq D'_\mu
=U(x)D_\mu U(x)^{-1}\label{trD} \, . \eneq To find $D_\mu$, we
introduce a Lie Algebra valued vector field
$A_\mu(x)=A_\mu^a(x)T^a$, and try the ansatz \beeq
D_\mu\Psi(x)=[\partial_\mu -igA_\mu(x)]\Psi(x) \label{an} \, .
\eneq If we demand $A_\mu(x)$ to transform as \beeq
A'_\mu(x)=U(x)A_\mu(x)U^{-1}(x) -\frac{i}{g} (\partial_\mu U(x))
U^{-1}(x)\, , \label{A} \eneq the ansatz Eq. \ref{an} gives \be
(D_\mu\Psi(x))'&=&D'_\mu\Psi'(x)=(\partial_\mu
-igA'_\mu(x))U(x)\Psi(x)\nonumber\\&=&(\partial_\mu U(x))\Psi
+U(x)\partial_\mu\Psi(x)\nonumber\\
&&-ig\left[ U(x)A_\mu(x)U^{-1}(x)-\frac{i}{g}(\partial_\mu
U(x))U^{-1}(x)\right]U(x)\Psi(x)\, ,\ee which yields the required
transformation property Eq. \ref{Dtransform} for
$D_\mu\Psi(x)$:\beeq (D_\mu\Psi(x))' =U(x)[\partial_\mu
-igA_\mu(x)]\Psi(x)=U(x)D_\mu\Psi(x)\, .\eneq Suppose the gauge
field is the vacuum configuration; $A_\mu\equiv 0$. If we perform
a spacetime dependent gauge transformation Eq. \ref{A}, we obtain
\beeq A_\mu\rightarrow A_\mu(x)=-\frac{i}{g}(\partial_\mu
U(x))U^{-1}(x) \label{pure} \, .\eneq The form of $A_\mu(x)$ in
Eq. \ref{pure} is called ``pure gauge.''

For an infinitesimal change \beeq
U_{kl}=\delta_{kl}-iT^a_{kl}\omega^a+O(\omega^2) \,
,\label{Uinfinitesimal} \eneq Eq. \ref{A} becomes \beeq
T^a_{kl}A'^a_\mu=
T^a_{kl}A^a_\mu+i\omega^b\left[T^a,T^b\right]_{kl}A^a_\mu-\frac{1}{g}
T^a_{kl}\partial_\mu\omega^a \; . \eneq Using the commutation
relation $[T^a , T^b]_{kl}=if^{abc} T^c_{kl}$, we find  \beeq
\delta A^a_\mu(x)=f^{abc}\omega^b(x)
A^c_\mu(x)-\frac{1}{g}\partial_\mu\omega^a (x) \label{infgaugA}\,
.\eneq

To make the gauge fields $A_\mu^a$ dynamical variables of the
theory we need to add their free Lagrangian density, which should
satisfy gauge and Lorentz invariance and should be quadratic in
spacetime derivatives of $A_\mu^a$. Thus, we need the
generalization of $F_{\mu\nu}F^{\mu\nu}$ of the Abelian case
(Maxwell's theory) where the field strength tensor is
$F_{\mu\nu}=\partial_\mu A_\nu -
\partial_\nu A_\mu$. We may try to define a field strength tensor
of the same form $\partial_\mu A_\nu^a -
\partial_\nu A_\mu^a$, however, this transforms in a
complicated way under Eq. \ref{infgaugA}. Recall that $F_{\mu\nu}$
is the curvature (commutator of the covariant derivative)
associated with the covariant derivative $D_\mu=\partial_\mu -i e
A_\mu$, hence $F_{\mu\nu}=\frac{i}{e}\left[D_\mu , D_\nu\right]$,
where $e$ is the electric charge. Using this analogy, we find \be
\left[ D_\mu , D_\nu
\right]\Psi(x)\!\!\!\!&=&\!\!\!\!\Big\{\partial_\mu\partial_\nu\Psi(x)-ig\partial_\mu(A_\nu(x)\Psi(x))
-igA_\mu(x)\partial_\nu\Psi(x)-g^2A_\mu(x)A_\nu(x)\Big\}\nonumber\\
\!\!\!\!&-&\!\!\!\!\Big\{ \mu\leftrightarrow\nu\Big\}=-i g \Big\{
\Big(\partial_\mu A_\nu(x) - \partial_\nu A_\mu(x) -ig\left[
A_\mu(x) ,
A_\nu(x)\right]\Big)\Psi(x)\nonumber\\
\!\!\!\!&-&\!\!\!\!\Big( A_\nu(x)\partial_\mu\Psi(x) +
A_\mu(x)\partial_\nu\Psi(x)
-(\mu\leftrightarrow\nu)\Big)\Big\}\nonumber\\
\!\!\!\!&=&\!\!\!\!-ig\Big( \partial_\mu A_\nu(x) - \partial_\nu
A_\mu(x) -ig [A_\mu (x) , A_\nu
(x)]\Big)\Psi(x)\nonumber\\
\!\!\!\!&\equiv&\!\!\!\!-igG_{\mu\nu}(x)\Psi(x)=-ig
T^aG^a_{\mu\nu}(x)\Psi(x) \label{cov} \, ,\ee where \beeq
G^a_{\mu\nu}(x)=\partial_\mu A^a_\nu(x) -\partial_\nu A^a_\mu(x)
+gf^{abc}A^b_\mu(x)A^c_\nu(x) \; .\eneq Let us check that the
above form of the field strength tensor transforms in a simple
way, determined only by $f^{abc}$. Because of the transformation
property Eq. \ref{trD} of the covariant derivative , we have \beeq
\Big( [D_\mu , D_\nu]\Psi(x)\Big)'=U(\omega(x))\Big([D_\mu ,
D_\nu]\Psi(x)\Big) \label{gtranF}\, .\eneq Using Eq. \ref{cov} in
Eq. \ref{gtranF}, we find that
$T^aG'^a_{\mu\nu}U(\omega)\Psi=U(\omega)T^aG^a_{\mu\nu}\Psi$, or
\beeq T^aG'^a_{\mu\nu}(x)=U(\omega(x))\Big(
T^aG^a_{\mu\nu}(x)\Big)U^{-1}(\omega(x))\label{G'}\; . \eneq Then,
the infinitesimal transformation is calculated by inserting Eq.
\ref{Uinfinitesimal} into Eq. \ref{G'}: \beeq \delta
G^a_{\mu\nu}(x)=f^{abc}\omega^b(x)G^c_{\mu\nu}(x)=i[G_{\mu\nu}(x),
\omega(x)]\label{Ginfinitesimal}\, ,\eneq as required.

Up to here, we have been acting the covariant derivative $D_\mu$
on fields $\Psi$ which transform as
$\delta\Psi_k=-i\omega^aT^a_{kl}\Psi_l$. Now, we want to apply the
covariant derivative to Lie Algebra valued objects $V=V^aT^a$.
Before proving the form how the covariant derivative acts on the
Lie Algebra valued objects, it is useful to introduce a
generalized cross product for gauge group vectors $V=V^aT^a$,
$W=W^aT^a$ and $X=X^aT^a$, where the dummy index $a=1, 2,.\; .\;
., n$:\beeq \vec{V}(x)\equiv(V^1(x),.\; .\; ., V^n(x))\, ,\,\,
\vec{W}(x)\equiv(W^1(x),.\; .\; ., W^n(x))\, , \,\,
\vec{X}(x)\equiv(X^1(x),.\; .\; ., X^n(x))\; .\nonumber\eneq We
define \beeq (\vec{V}(x)\times\vec{W}(x))^a\equiv f^{abc} V^b(x)
W^c(x)\, ,\eneq which implies\be (\vec{V}(x)\times\vec{W}(x))^a&=&
f^{abc} V^b(x) W^c(x)=-i([T^b , T^c])^a V^b(x)
W^c(x)\nonumber\\
&=&-i([V(x) , W(x)])^a\, .\label{crosscommutator}\ee Consequently,
we can make use of the usual cross product relations like
\beeq\vec{V}\times \vec{W}=-\vec{W}\times \vec{V}\,
,\nonumber\eneq \beeq \vec{V}\times(\vec{W}\times\vec{X})
+\vec{W}\times(\vec{X}\times\vec{V})+\vec{X}\times(\vec{V}\times\vec{W})=0\,
,\label{idiesJacandanticom}\eneq where the later is known as the
Jacobi identity. In this new notation Eqs. \ref{infgaugA} and
\ref{Ginfinitesimal} can be written, respectively, as\beeq
\delta\vec{A}_{\mu}(x)=\vec{\omega}(x)\times\vec{A}_{\mu}(x)
-\frac{1}{g}\partial_\mu \vec{\omega}(x)\, ,\label{deltaA}\eneq
\beeq \delta\vec{G}_{\mu\nu}(x)=\vec{\omega}(x)
\times\vec{G}_{\mu\nu}(x)\,\label{deltaG}\eneq Now, we want to
show that the covariant derivative of $V=V^aT^a$ must be of the
form \beeq D_\mu V=\partial_\mu V-ig[A_\mu , V]\, ,\eneq to have
the transformation property $(D_\mu V)'=U(D_\mu V)U^{-1}$. In
other words, expressing the covariant derivative as $D_\mu V=T^a
(D_\mu V)^a$ and $\vec{D_\mu V}=\{(D_\mu V)^a: a=1,.\; .\; .,
n\}$, we will show that\beeq \vec{D_\mu
V}=\partial_\mu\vec{V}+g\vec{A}_\mu\times\vec{V}\,
,\label{covderV}\eneq has the required transformation property, in
infinitesimal form, \beeq \delta (\vec{D_\mu
V})=\vec{\omega}\times (\vec{D_\mu V})=\vec{\omega}\times
(\partial_\mu\vec{V}+g\vec{A}\times \vec{V})\;
,\label{formwewant}\eneq where \beeq \delta\vec{V}=
\vec{\omega}\times\vec{V}\; .\label{deltaV}\eneq In particular,
when $V$ is taken as the field strength tensor, we will have
$D_\mu G_{\nu\rho}=\partial_\mu G_{\nu\rho}-ig[A_\mu ,
G_{\nu\rho}]$, where $\delta
\vec{G}_{\mu\nu}=\vec{\omega}\times\vec{G}_{\mu\nu}$ as in Eq.
\ref{deltaG}. Now, to prove the claim Eq. \ref{covderV} let us
calculate the transformation\beeq \delta(\vec{D_\mu
V})=\delta(\partial_\mu \vec{V}+g \vec{A}_\mu \times
\vec{V})=\partial_\mu \delta \vec{V}+g \delta\vec{A}_\mu\times
\vec{V}+g \vec{A}_\mu \times \delta\vec{V}\, .\label{A28}\eneq
Using Eqs. \ref{deltaA} and \ref{deltaV} in Eq. \ref{A28} we
obtain \be \delta(\vec{D_\mu V})&=&\partial_\mu
(\vec{\omega}\times\vec{V})-(\partial_\mu
\vec{\omega})\times\vec{V}+g(\vec{\omega}\times
\vec{A}_\mu)\times\vec{V}+g\vec{A}_\mu\times
(\vec{\omega}\times\vec{V})\nonumber\\
&=&\vec{\omega}\times\partial_\mu\vec{V}+g\left(
(\vec{\omega}\times \vec{A}_\mu)\times\vec{V}+\vec{A}_\mu\times
(\vec{\omega}\times\vec{V})\right)\, .\label{deltaDV}\ee Finally,
the application of identities \ref{idiesJacandanticom} in Eq.
\ref{deltaDV} yield \beeq \delta(\vec{D_\mu
V})=\vec{\omega}\times\partial_\mu\vec{V}+g\vec{\omega}\times(\vec{A}_\mu\times\vec{V})
=\vec{\omega}\times
(\partial_\mu\vec{V}+g\vec{A}_\mu\times\vec{V})\; ,\eneq which is
the transformation we wanted in Eq. \ref{formwewant}. Thus, we
define $D_\mu\equiv\partial_\mu -ig[A_\mu ,\; ]$ or
$D_\mu\equiv\partial_\mu+g(\vec{A}_\mu\times\;)$. Another relation
we want to derive is \be [D_\mu ,
D_\nu]\vec{V}&=&\partial_\mu(\partial_\nu \vec{V}+g
\vec{A}_\mu\times \vec{V})+g\vec{A}_\mu(\partial_\nu \vec{V}+g
\vec{A}_\mu\times \vec{V})-(\mu\leftrightarrow\nu)\nonumber\\
&=&g(\partial_\mu \vec{A}_\nu - \partial_\nu
\vec{A}_\mu)\times\vec{V} + g^2 (\vec{A}_\mu\times
(\vec{A}_\nu\times\vec{V})-\vec{A}_\nu\times
(\vec{A}_\mu\times\vec{V}))\nonumber\\
&=&g(\partial_\mu \vec{A}_\nu - \partial_\nu \vec{A}_\mu+g
\vec{A}_\mu\times\vec{A}_\nu)\times\vec{V}\nonumber\\&=&g\;
\vec{G}_{\mu\nu}\times\vec{V}\, ,\label{DDDD}\ee where we have
used the identities \ref{idiesJacandanticom}. Using result
\ref{crosscommutator} in Eq. \ref{DDDD} we obtain \beeq [D_\mu ,
D_\nu]\vec{V}= -ig [\vec{G}_{\mu\nu} , \vec{V}]\, .\eneq

All the pure gauge types of vector potential, defined in
\ref{pure}, are vacuum solutions. This can be shown as follows.
For the pure gauge fields \be G_{\mu\nu}(x)&=&(\partial_\mu
A_\nu(x)-igA_\mu(x)A_\nu(x) )-(\mu\leftrightarrow \nu)
=-\frac{i}{g}\Big\{\big[\partial_\mu\left( U(x)\partial_\nu
U^{-1}(x)\right)\nonumber\\
&&+ U(x)\left(\partial_\mu U^{-1}(x)\right) U(x)\left(
\partial_\nu U^{-1}(x)\right)\big]-
\big[\mu\leftrightarrow\nu\big]\Big\}\label{pureU} \, .\ee On the
other hand, by taking the derivative of the identity
$U(x)U^{-1}(x)=I$, we obtain $(\partial_\mu U)U^{-1}+U\partial_\mu
U^{-1}=0$ which gives \beeq
\partial_\mu U(x)=-U(x)\left(\partial_\mu U^{-1}(x)\right)U(x)\; .
\label{therelation} \eneq Replacing $\partial_\nu U(x)$ in Eq.
\ref{pureU} using Eq. \ref{therelation} one finds \beeq
G_{\mu\nu}(x)=0\; . \label{purefield}\eneq

Recall Eq. \ref{Ginfinitesimal} that, unlike the Abelian case
where $f^{abc}=0$, $G^a_{\mu\nu}$ transforms nontrivially (under
gauge transformation Eq. \ref{A} the field tensor is gauge
covariant in the non-Abelian case whereas gauge invariant in the
Abelian case). However the trace ${\rm Tr}(G_{\mu\nu}
G^{\mu\nu})=\frac{1}{2}G^a_{\mu\nu}G^{a\mu\nu}$ is gauge
invariant: \beeq \delta[G^a_{\mu\nu}(x)G^{a\mu\nu}(x)]=2\delta
G^a_{\mu\nu}(x)
G^{a\mu\nu}(x)=2f^{abc}\omega^b(x)G^c_{\mu\nu}(x)G^{a\mu\nu}(x)=0
\;. \eneq This is because the indices $a$ and $c$ are
antisymmetric in $f^{abc}$ whereas symmetric in
$G^c_{\mu\nu}G^{a\mu\nu}$. Therefore, \beeq {{\mathcal{L}}}_{\rm
gauge}=-\frac{1}{2}{\rm
Tr}G_{\mu\nu}(x)G^{\mu\nu}(x)=-\frac{1}{4}G^a_{\mu\nu}(x)G^{a\mu\nu}(x)\,
, \eneq can be used as the kinetic term for the gauge vector field
$A^a_{\mu}(x)$. This is, in fact, the Yang Mills Lagrangian
density and in our the discussion we consider only the Yang Mills
field theory. The equations of motion are obtained from the
condition that the action $S=\int dx^4 {{\mathcal{L}}}_{\rm
gauge}$ be stationary against arbitrary variations of the gauge
fields. Suppressing the explicit spacetime dependence of the
fields, \beeq \delta S=-\int d^4x {\rm Tr}(G_{\mu\nu} \delta
G^{\mu\nu}) \, ,\eneq where $\delta G^{\mu\nu}=(\partial^\mu\delta
A^\nu-ig\delta A^\mu A^\nu -ig A^\mu \delta A^\nu) -
(\mu\leftrightarrow\nu)$. Thus, using $G_{\mu\nu}=-G_{\nu\mu}$, we
obtain \beeq \delta S=-2\int d^4x {\rm Tr}\left[
G_{\mu\nu}(\partial^\mu\delta A^\nu-ig\delta A^\mu A^\nu -ig A^\mu
\delta A^\nu)\right] \; .\eneq Integration of the first term by
parts yields \beeq \delta S= 2\int dx^4 {\rm Tr}
\left[\partial^\mu G_{\mu\nu}+ig G_{\mu\nu} \left(\delta A^\mu
A^\nu + A^\mu\delta A^\nu\right)\right]\; . \eneq Since ${\rm
Tr}\left[ G_{\mu\nu}\delta A^\mu A^\nu\right]=-{\rm Tr}\left[
A^\mu G_{\mu\nu}\delta A^\nu\right]$ we have \beeq \delta S= 2\int
dx^4 {\rm Tr} \left[(\partial^\mu G_{\mu\nu}-ig[A^\mu ,
G_{\mu\nu}])\delta A^\nu\right]\label{varS}\; . \eneq Thus the
field equations are \beeq D^\mu G_{\mu\nu}=\partial^\mu
G_{\mu\nu}-ig[A^\mu , G_{\mu\nu}]=0\label{covEqmot}\; ,\eneq or
equivalently, \beeq D^\mu G^a_{\mu\nu}=\partial^\mu
G^a_{\mu\nu}+gf^{abc}A^{b\mu}G^c_{\mu\nu}=0\, .\eneq Using the
Jacobi identity, \beeq \left[ D_\mu [D_\nu , D_\rho]\right]+\left[
D_\nu [D_\rho , D_\mu]\right]+\left[ D_\rho [D_\mu ,
D_\nu]\right]=0 \, ,\eneq and Eq. \ref{cov}, $[D_\mu ,
D_\nu]=-igG_{\mu\nu}$ we obtain the Bianchi identity:\beeq [D_\mu
, G_{\nu\rho}]+[D_\nu , G_{\rho\mu}]+[D_{\rho} , G_{\mu\nu}]=0\,
.\label{Bianchicommutators}\eneq Let's calculate the above
commutators:\be [D_\mu ,
G_{\nu\rho}]\Psi&=&\Big[\partial_\mu-ig[A_\mu , ] ,
G_{\nu\rho}\Big]\Psi=[\partial_\mu ,
G_{\nu\rho}]\Psi-ig\Big[[A_\mu ,\;
],G_{\nu\rho}\Big]\Psi\nonumber\\
&=&\partial_\mu(G_{\nu\rho}\Psi)-G_{\nu\rho}\partial_\mu\Psi-ig\left([A_\mu
G_{\nu\rho}\Psi]-G_{\nu\rho}[A_\mu ,
\Psi]\right)\nonumber\\
&=&\left(\partial_\mu G_{\nu\rho}-ig [A_\mu ,
G_{\nu\rho}]\right)\Psi=(D_\mu G_{\nu\rho})\Psi\, .\ee Likewise
$[D_\nu , G_{\rho\mu}]=D_\nu G_{\rho\mu}$ and $[D_{\rho} ,
G_{\mu\nu}]=D_{\rho} G_{\mu\nu}$. Hence the Bianchi identity given
in Eq. \ref{Bianchicommutators} reduces to \beeq D_\mu G_{\nu\rho}
+D_\nu G_{\rho\mu}+D_{\rho} G_{\mu\nu}=0\, .\eneq By introducing
the dual field strength tensor \beeq {}^\ast
G^{\mu\nu}=\frac{1}{2}\epsilon^{\mu\nu\rho\sigma} G_{\rho\sigma}
\label{dual} \eneq we can also write the Bianchi identity as \beeq
D_\mu {}^\ast G^{\mu\nu}=0\; . \label{BianchiDstarG=0}\eneq Here
we immediately observe that the (anti)self dual field
configurations, where $G_{\mu\nu}=\mp {}^\ast G_{\mu\nu}$
respectively, automatically satisfy the field equations $D_\mu
G^{\mu\nu}=0$ due to the Bianchi identity. This result is
appealing because it is easier to solve the first order self
duality equations ${}^\ast G^{\mu\nu}=G^{\mu\nu}$ rather than
second order field equations $D_\mu G^{\mu\nu}=0$.

The symmetric stress energy tensor is obtained easily by coupling
the theory to a background metric $g_{\mu\nu}$: \beeq
L=-\frac{1}{4} G^a_{\mu\nu} G^a_{\rho\sigma} g^{\mu\rho}
g^{\nu\sigma}\sqrt{-g}\; , \eneq and varying the action with
respect to $g_{\mu\nu}$: \beeq
T^{\mu\nu}\equiv-\frac{2}{\sqrt{-g}}\left(\frac{\partial
L}{\partial g_{\mu\nu}}-\partial_\rho \frac{\partial L}{\partial
g_{\mu\nu,\, \rho}}+\partial_\rho
\partial_\sigma \frac{\partial L}{\partial g_{\mu\nu,\,
\rho\sigma}}\right)\; .\eneq Using the identities $\delta g=g
g^{\mu\nu}\delta g_{\mu\nu}$ and $\partial g^{\rho\sigma}/\partial
g_{\mu\nu} =-(g^{\rho\mu} g^{\nu\sigma}+g^{\rho\nu}
g^{\mu\sigma})/2$ we find \beeq T^{\mu\nu}=-G^{a\mu\sigma}
{G^{a\nu}}
_\sigma+\frac{1}{4}g^{\mu\nu}G^a_{\rho\sigma}G^{a\rho\sigma}\; .
\eneq We define the electric field $\vec{E}\equiv{\vec{E}}^aT^a$
and the magnetic field $\vec{B}\equiv{\vec{B}}^aT^a$, which are
elements of the Lie Algebra:
\be E^{ai}&=&-G^{a0i}\; ,\nonumber\\
B^{ai}&=&-\frac{1}{2}\epsilon^{ijk}G^{ajk} \leftrightarrow
G^{aij}=-\epsilon^{ijk}B^{ak}\; . \ee In terms of $A^\mu=(A^0,
\vec{A})$ \be E^i&=&\partial^i A^0-\partial^0 A^i+ig [A^{0} ,
A^{i}]\, ,\\
B^i&=&-\epsilon^{ijk}(\partial^j A^k-ig A^j A^k )\, .\ee These
matrices are not gauge invariant, they transform according to Eq.
\ref{Ginfinitesimal}. The equations for non-Abelian electric and
magnetic fields can also be written as: \be
\vec{E}^a&=&-\vec{\nabla} A^{a0}-\frac{\partial\vec{A}^a}{\partial
t}+g f^{abc}{\vec{A}}^{b}
A^{c0}\label{vEa}\; ,\\
{\vec{B}}^a&=&\vec{\nabla}\times{\vec{A}}^a
-\frac{g}{2}f^{abc}{\vec{A}}^b\times{\vec{A}}^c\label{vBa}\; . \ee
The above equation implies that non-Abelian gauge theories allow
magnetic monopoles with magnetic charge density: \beeq
\rho\equiv\vec{\nabla}\cdot
\vec{B}^a=-\frac{g}{2}f^{abc}\vec{\nabla}\cdot({\vec{A}}^b\times{\vec{A}}^c)\;
. \eneq Total magnetic charge is the surface integral at spatial
infinity and there are monopole solutions for which the integral
does not vanish.

The components of the dual field tensor \ref{dual} are \be
{}^\ast G^{a0i}&=&\frac{1}{2}\epsilon^{ijk}G^{ajk}=-B^{ai}\; ,\nonumber\\
{}^\ast G^{aij}&=&-\epsilon^{ijk}G^{a0k}=\epsilon^{ijk}E^{ak}\;
.\label{dualfields} \ee Duality replaces $E^{ai}\rightarrow
B^{ai}$ and $B^{ai}\rightarrow -E^{ai}$. The Lagrangian density
and energy can be computed in terms of $E^{ai}$ and $B^{ai}$: \be
{{\mathcal{L}}}_{\rm
gauge}&=&-\frac{1}{4}(-2G^{a0i}G^{a0i}+G^{aij}G^{aij})=\frac{1}{2}({\vec{E}}^a
\cdot{\vec{E}}^a-{\vec{B}}^a\cdot{\vec{B}}^a)\label{canLag}\; ,\\
T^{00}&=&G^{a0i}G^{a0i}-g^{00}{{\mathcal{L}}}=\frac{1}{2}({\vec{E}}^a
\cdot{\vec{E}}^a+{\vec{B}}^a\cdot{\vec{B}}^a)\equiv{\mathcal{E}}\label{energy}\;
,\ee where ${\mathcal{E}}$ denotes the energy density. Using Eq.
\ref{dualfields} we find\beeq {}^\ast G^{a\mu\nu}
G^{a}_{\mu\nu}=-4{\vec{E}}^{a}\cdot {\vec{B}}^{a}\;
.\label{pseudoMin}\eneq The equations of motion \ref{covEqmot}, in
three vector form, reads \be
\vec{\nabla}\times{\vec{B}}^a&=&\frac{\partial{\vec{E}}^a}{\partial
t}+gf^{abc}(A^b_0{\vec{E}}^c+{\vec{A}}^b\times{\vec{B}}^c)\label{eqdyn}\; ,\\
\vec\nabla\cdot{\vec{E}}^a&=&gf^{abc}{\vec{A}}^b\cdot{\vec{E}}^c\label{eqconst}\;
. \ee They are the generalizations of Maxwell Equations.
\subsection{Classical Instanton Solutions of Non-Abelian Gauge
Theories}

A pure gauge theory can have ``static'' soliton solutions only in
four dimensional Euclidean space \cite{Deser} (here static means
that the solutions are independent of one of the four
coordinates). In this section, we will study such a solution,
called the instanton, which is characterized by a topological
charge and finite
action\cite{Huang,Dittrich,Rajaraman,Rubakov,Shifman}. Instantons
are important for the quantum theory, because they describe
tunneling between degenerate vacua. Instanton solutions play a
crucial role in understanding the quantum theory of gauge
theories. We will use most of the results obtained in this section
when we discuss the nontrivial vacuum structure of non Abelian
gauge theories, later in this Appendix.

We will concentrate on the Euclidean $SU(2)$ gauge theory which is
obtained by making the following replacements in the Minkowski
space version of the theory: \beeq x^0\rightarrow -ix^4 \; ,\;\;
g_{\mu\nu}=(+,-,-,-)\rightarrow g_{\mu\nu}=(+,+,+,+)\; ,\eneq
where the indices run from $1$ to $4$ in Euclidean space and
$\epsilon^{1234}=1$ ($\epsilon^{\mu\nu\rho\sigma}$ is a tensor
density; it has to be multiplied by the determinant of the
transformation when it is transformed. That is why it does not
pick up an $i$ under $x^0\rightarrow -ix^4$). Hence
$\partial_0\rightarrow i\partial_4$, $D_0\rightarrow i D_4$, and
$A_0(x)\rightarrow iA_4(x_E)$. Note that $A_0$ and $\partial_0$
are components of vectors hence transform similarly, whereas $x^0$
is a component of a co-vector and transforms oppositely. Then the
fields transform as\beeq {E}_j^{M}\rightarrow i{E}_j^E\; ,\;\;\;
{B}_j^{M}\rightarrow {B}_j^E\; .\label{Euclideanfields}\eneq
Defining $x^4\equiv T$ (hence $t=x^0\rightarrow -ix^4=-iT$), where
$T$ is a real parameter, the transformation of the action is\beeq
iS_M\!=\! i\!\!\!\int\!\!\! d^3x
dt{\mathcal{L}}_M(t,\vec{x})\!\rightarrow\!i\!\!\!\int\!\!\! d^3x
(-idT){\mathcal{L}}_M(-iT,\vec{x})\!=\! -\!\!\!\int\!\!\!
d^4x_E{\mathcal{L}}_E({x}_E)\!\equiv\! -S_E\;
,\label{Wickaction}\eneq where we define ${\mathcal{L}}_E
({x}_E)=-{\mathcal{L}}_M(-iT,\vec{x})$. Continuation to Euclidean
spacetime means that we consider the dynamical evolution of the
system in imaginary time. In principle, we must solve the
equations of motion in which the time $x^0$ is replaced by
$-ix^4$. The Lorentz invariance $SO(3,1)$ of the Lagrangian
density is replaced by invariance with respect to $SO(4)$
rotations in Euclidean space. The equations of motion determines
how the {\it fields} are to be continued into Euclidean space. For
example, for a real scalar field $\phi (x)$, the action in
Minkowski spacetime is \beeq S_M=\int dt d^3x
\left[\frac{1}{2}\left((\partial_0
\phi(x_M))^2-\nabla^2\phi(x_M)\right)
-U(\phi(x_M))\right]\nonumber\; ,\label{actionscalarMin}\eneq
where $U$ is the potential energy. Let's consider the equations of
motion\beeq
\partial_0^2\phi(x_M)-\partial_i^2\phi(x_M)+U'(\phi(x_M))=0\,
,\label{scalareqmotMin}\eneq where prime denotes the derivative
with respect to the field. If we Euclideanize the spacetime so
that $\partial_0\rightarrow i\partial_4$ and
$\partial_i\rightarrow\partial_i$, we obtain \beeq
-\partial_4^2\phi(x_E)-\partial_i^2\phi(x_E)+\tilde{U'}(\phi(x_E))=0\,
.\eneq To have the required $SO(4)$ invariance we must have\beeq
\phi(x_M)=\phi(x^0=t, \vec{x})\rightarrow\phi(x_E)=\phi(x^4=T,
\vec{x})\, ,\eneq \beeq U(\phi(x_M))\rightarrow
\tilde{U}(\phi(x_E))=-U(\phi(x_E))\, .\eneq Therefore, a real
scalar field $\phi(x)$ defined in Minkowski space is replaced by a
real scalar field $\phi(x)$ invariant under $SO(4)$ and the
potential energy picks up a minus sign. In Euclidean space Eq.
\ref{scalareqmotMin} is replaced by\beeq -\partial_E^2\phi_E -
U'(\phi(x_E))=0\; .\eneq The action in Eq. \ref{actionscalarMin}
therefore becomes\be S_M&\rightarrow& \int (-idx^4) d^3x
\Big[\frac{1}{2}\Big(\left(i\partial_{x^4}
\phi(x_E)\right)^2-\nabla^2\phi(x_E)\Big)
+U(\phi(x_E))\Big]\nonumber\\&\rightarrow&i\int d^4x_E \left[
\frac{1}{2}(\partial_E\phi(x_E))^2-U(\phi(x_E))\right]=iS_E\; ,\ee
which confirms Eq. \ref{Wickaction}. The relation between the path
integral formulation of the two regimes is therefore\beeq \int
[dA] e^{iS_M}\rightarrow\int [dA] e^{-S_E}\label{intSMSE}\; .\eneq
In the rest of the section, we will drop the index $E$ and adopt
the summation convention when the repeated indices are both lower
or upper. Using Eqs. \ref{canLag} and \ref{Euclideanfields} in Eq.
\ref{intSMSE}, we find that the Lagrangian in Euclidean space
reads \beeq {\mathcal{L}}=\frac{1}{4}G^a_{\mu\nu}
G^a_{\mu\nu}=\frac{1}{2}(E^a_i E^a_i + B^a_i B^a_i)\; .\eneq Non
Abelian electric and magnetic fields are assigned to the strength
tensor exactly the same way as in Minkowski spacetime:
$G^a_{4i}=-E^a_i$, and $G^a_{ij}=-\epsilon_{ijk}B^a_k$. These
definitions are consistently yield the above action. The dual
field strength tensor ${}^\ast
G^a_{\mu\nu}=\frac{1}{2}\epsilon_{\mu\nu\rho\sigma}G^a_{\rho\sigma}$
has the following components:\beeq {}^\ast G^a_{4i}=B^a_i\;
,\;\;\; {}^\ast G^a_{ij}=\epsilon_{ijk}E^a_k\; .\eneq Duality
replaces $E^a_i\leftrightarrow -B^a_i$, therefore self dual and
anti-self dual field configurations satisfy\beeq G^a_{\mu\nu}=\pm
{}^\ast G^a_{\mu\nu}\Leftrightarrow E^a_i=\mp B^a_i\;
.\label{selfdual}\eneq Due to the Bianchi identity (Eq.
\ref{BianchiDstarG=0}), selfdual and anti-selfdual solutions
automatically satisfy the equations of motion $D_\mu
G^a_{\mu\nu}=0$. Hence, it is sufficient to find gauge potentials
yielding $E^a_i=\mp B^a_i$ to find a solution for the equations of
motion.

In Euclidean space the stress energy tensor
$T_{\mu\nu}=G^a_{\mu\sigma}G^a_{\nu\sigma}-\frac{1}{4}g_{\mu\nu}G^a_{\rho\sigma}G^a_{\rho\sigma}$
has the following components:\be
T_{44}&=&\frac{1}{2}(E^a_iE^a_i-B^a_iB^a_i)\;
,\nonumber\\
T_{ij}&=&E^a_iE^a_j-B^a_iB^a_j-\frac{1}{2}\delta_{ij}(E^a_kE^a_k-B^a_kB^a_k)\;
,\nonumber\\
T_{4i}&=&\epsilon_{ijk} E^a_j B^a_k\, .
\label{Euclideanenenergymom}\ee All of the above components vanish
for the self dual or anti self dual fields (as
$E^a_i\rightarrow\mp B^a_i$). Thus, in Euclidean space, the energy
momentum tensor for (anti)self dual solutions is identical to
zero.

The other $SO(4)$ ($SO(3,1)$
 in Minkowski spacetime) and gauge
invariant quantity is given by the pseudoscalar density:\beeq
{{\mathcal{T}}}\equiv\frac{1}{2}{\rm Tr}\, G_{\mu\nu} {}^\ast
G_{\mu\nu}=-E^a_iB^a_i\label{pseudoEuc}\;.\eneq This term is not
considered as a kinetic term because it is a pure divergence, it
can be written as \beeq {\mathcal{T}}\equiv\partial_\mu
{K}_\mu\equiv\frac{1}{2g^2}\partial_\mu {J}_\mu\label{totaldiv}\,
,\eneq with \beeq {K}_\mu\equiv\epsilon_{\mu\nu\rho\sigma}{\rm
Tr}[A_\nu
\partial_\rho A_\sigma -i \frac{2g}{3}A_\nu A_\rho A_\sigma]\,
.\eneq (Here, let us note that for the group $SU(2)$, one can
easily evaluate the traces by introducing Pauli spin matrices
$\{\sigma^a; a=1, 2, 3\}$ as the generators
$T^a\equiv\frac{1}{2}\sigma^a$, hence
$A_\mu=A^a_\mu\frac{1}{2}\sigma^a$. The identity $\sigma^a\sigma^b
= \delta^{ab}+i\epsilon^{abc}\sigma^c$ yields ${\rm Tr}
(\sigma^a\sigma^b)=2\delta^{ab}$, ${\rm Tr}
(\sigma^a\sigma^b\sigma^c)=2i\epsilon^{abc}$ and one obtains
$K_\mu=\frac{1}{2}\epsilon_{\mu\nu\rho\sigma}[A^a_\nu
\partial_\rho A^a_\sigma + \frac{g}{3}\epsilon^{abc}A^a_\nu A^b_\rho
A^c_\sigma]$.)

Let us prove Eq. \ref{totaldiv}:\be
{{\mathcal{T}}}&=&\frac{1}{4}\epsilon_{\mu\nu\rho\sigma}{\rm
Tr}\left\{\left[(\partial_\mu A_\nu -igA_\mu
A_\nu)-(\mu\leftrightarrow\nu)\right]\left[(\partial_\rho A_\sigma
-igA_\rho
A_\sigma)-(\rho\leftrightarrow\sigma)\right]\right\}\nonumber\\
&=&\epsilon_{\mu\nu\rho\sigma}{\rm Tr}\left[(\partial_\mu
A_\nu)(\partial_\rho A_\sigma) -ig(\partial_\mu A_\nu) A_\rho
A_\sigma - ig A_\mu A_\nu \partial_\rho A_\sigma -g^2 A_\mu A_\nu
A_\rho A_\sigma \right]\; .\nonumber\ee Using the cyclic property
of trace, the last term in the above equation can be eliminated.
Renaming $\mu\leftrightarrow \rho$ and $\nu\leftrightarrow \sigma$
in the third term, we obtain \beeq
{\mathcal{T}}=\epsilon_{\mu\nu\rho\sigma}{\rm
Tr}\left[(\partial_\mu A_\nu)(\partial_\rho A_\sigma)
-2ig(\partial_\mu A_\nu) A_\rho A_\sigma\right]\;.\eneq It is
trivial to show, again using the cyclic property and renaming the
necessary indices, that \beeq\epsilon_{\mu\nu\rho\sigma}{\rm
Tr}\left[(\partial_\mu A_\nu)A_\rho
A_\sigma\right]=\frac{1}{3}\partial_\mu {\rm Tr}\left[A_\nu A_\rho
A_\sigma\right]\; .\eneq Moreover, \beeq
\epsilon_{\mu\nu\rho\sigma}{\rm Tr}\left[(\partial_\mu
A_\nu)(\partial_\rho
A_\sigma)\right]=\epsilon_{\mu\nu\rho\sigma}\partial_\mu{\rm
Tr}\left[A_\nu\partial_\rho A_\sigma\right]\; ,\eneq because
$\epsilon_{\mu\nu\rho\sigma}{\rm
Tr}\left[A_\nu\partial_\mu\partial_\rho A_\sigma\right]=0$ due to
the symmetry of the partial derivatives and the anti symmetry of
the Levi Civita tensor density. Thus we have\beeq
{{\mathcal{T}}}=\partial_\mu\left[\epsilon_{\mu\nu\rho\sigma}{\rm
Tr}( A_\nu\partial_\rho A_\sigma -i\frac{2g}{3}A_\nu A_\rho
A_\sigma)\right]=\partial_\mu K_\mu\; ,\label{totder}\eneq as
claimed.

Integrating ${\mathcal{T}}$ over all four space we obtain the
topological charge (Pontryagin index) $q[{A}]\equiv
({g^2}/{8\pi^2}) \int d^4x {\mathcal{T}}$ of the Euclidean field
configuration $A^a(x)_\mu$:\be q[A]&=& \frac{g^2}{16\pi^2} \int
d^4x {\rm Tr} G_{\mu\nu}{}^\ast G_{\mu\nu}
=\frac{g^2}{32\pi^2}\int d^4x
G^a_{\mu\nu}{}^\ast G^a_{\mu\nu}\label{divvolint}\; ,\\
&=&\frac{g^2}{8\pi^2} \oint_{S^3_\infty} d\Sigma_\mu K_\mu +q_{\rm
sing}=\frac{1}{16\pi^2} \oint_{S^3_\infty} d\Sigma_\mu J_\mu
+q_{\rm sing}\; ,\label{surfaceintegralSING}\ee where the surface
integral is on the boundary of four space, $S^3$, with the measure
$d\Sigma$, and $q_{\rm sing}$ is included because of the possible
solutions that might contribute to the integral independent of
their behavior at the boundary (e.g., Dirac-delta like solutions
or solutions which yield divergent volume integral Eq.
\ref{divvolint} because of their singularities away from the
boundary). For nonsingular solutions $q_{\rm sing}=0$. Let us note
that $q[A]$ is gauge invariant:\beeq q'=\frac{g^2}{16\pi^2}\int
d^4x {\rm Tr} [G^{a'}_{\mu\nu}{}^\ast
G^{a'}_{\mu\nu}]=\frac{g^2}{16\pi^2}\int d^4x {\rm Tr}
[UG^{a}_{\mu\nu}U^{-1}U\;{}^\ast
G^{a}_{\mu\nu}U^{-1}]=q\;.\label{gaugeinvq}\eneq The importance of
the nonsingular solutions becomes clear in quantization of gauge
fields via path integral methods where the integration is over all
classical field configurations $A^a_\mu(x)$ in imaginary time that
interpolate between the vacua at $x_0=\pm \infty$. The dominant
contributions to this integration comes from the field
configurations for which $S^E[A^a_\mu]$ is stationary. Here lies
the particular relevance of the (anti)selfdual field
configurations which satisfy $G_{\mu\nu}^a=\mp {}^\ast
G_{\mu\nu}^a$. They are solutions of the classical field equations
$D_\mu G_{\mu \nu}=0$ due to the Bianchi identity $D_\mu {}^\ast
G_{\mu\nu}=0$ which is always satisfied.

Before trying to find solutions of the field equations, let us
show that the action for arbitrary Euclidean solutions is bounded
below by $|(8\pi^2/g^2)q[A^a_\mu]|$. In Euclidean space we
have\beeq {\rm Tr} [(G_{\mu\nu} \mp {}^\ast G_{\mu\nu})(G_{\mu\nu}
\mp {}^\ast G_{\mu\nu})]\geq 0\; ,\eneq since it is the sum of
squares. The above equation implies\beeq {\rm Tr} [G_{\mu\nu}
G_{\mu\nu}+{}^\ast G_{\mu\nu} {}^\ast G_{\mu\nu}]\geq \pm 2{\rm
Tr} [G_{\mu\nu} {}^\ast G_{\mu\nu}]\; .\label{ineqbound}\eneq In
Euclidean space
$\epsilon_{\rho\sigma\mu\nu}\epsilon_{\mu\nu\phi\theta}
=2[\delta_{\rho\phi}\delta_{\sigma\theta}-\delta_{\rho\theta}\delta_{\sigma\phi}]$,
thus ${\rm Tr} [{}^\ast G_{\mu\nu}\,{}^\ast G_{\mu\nu}]={\rm Tr}
[G_{\mu\nu} G_{\mu\nu}]$. Then, Eq. \ref{ineqbound} gives\beeq
{\rm Tr}[G_{\mu\nu} G_{\mu\nu}]\geq \pm {\rm Tr} [G_{\mu\nu}\,
{}^\ast G_{\mu\nu}]\; .\eneq This means for the action \beeq
S^E=\frac{1}{2}\int d^4x {\rm Tr} [G_{\mu\nu}
G_{\mu\nu}]\geq\pm\frac{1}{2}\int d^4x {\rm Tr} [G_{\mu\nu}
{}^\ast G_{\mu\nu}]\; .\eneq However, since ${\rm Tr} [G_{\mu\nu}
G_{\mu\nu}]\geq 0$, the action $S^E\geq 0$. Therefore we obtain
the result we want \beeq S^E\geq \left|\frac{1}{2}\int d^4x {\rm
Tr} [G_{\mu\nu} {}^\ast
G_{\mu\nu}]\right|=\frac{8\pi^2}{g^2}|q[A]|\; .\eneq The lower
bound is reached (i.e., the equality is satisfied), when
$G_{\mu\nu}=\pm {}^\ast G_{\mu\nu}$. Thus the action has its
minimum value $S^E=(8\pi^2/g^2)|q[A]|$ for (anti)self dual
solutions. We had already encountered another property of these
solutions. They also had zero energy momentum tensor (Eq.
\ref{Euclideanenenergymom}). We are now ready for the topological
interpretation of $q[A]$.

Recalling  Eqs. \ref{surfaceintegralSING} and \ref{ineqbound}, for
nonsingular field configurations, the Euclidean action: \beeq
S^E=\frac{1}{2}\int d^4x {\rm Tr} [G_{\mu\nu}
G_{\mu\nu}]\geq\frac{8\pi^2}{g^2}|q[A]|=\frac{1}{2g^2}\oint_{S^3_\infty}
d\Sigma_\mu J_\mu\; .\eneq Thus the minimum value of the action
depends on the behavior of the gauge fields at infinity. For $S^E$
to be finite, $G_{\mu\nu}$ has to decrease sufficiently fast to
zero at Euclidean infinity: \beeq \lim_{x^2\rightarrow\infty}
G_{\mu\nu} (x)=0\; .\eneq Therefore, the potential must go to a
pure gauge configuration at infinity:\beeq A_\mu\rightarrow
-\frac{i}{g}(\partial_\mu U)U^{-1} \hskip 1cm {\rm for}\hskip
0.5cm x^2\rightarrow\infty \; ,\label{boundcondA}\eneq which is
obtained from $A_\mu=0$ by a ``gauge transformation.'' Here the
term gauge transformation must be used with precaution. There are
so called ``large gauge transformations'' that yield physically
distinct vacua which are gauge inequivalent to the vacuum
$A_\mu=0$. We will discuss this issue, in detail, later in this
Appendix. The minimum value of the action is reached by
asymptotically pure gauge configurations that yield asymptotically
vanishing fields (Eq. \ref{purefield}). The boundary condition Eq.
\ref{boundcondA} defines a map at infinity. At the infinity of
four dimensional Euclidean space $E^4$ we have boundary
$S^3_\infty$. Equation \ref{boundcondA} associates an $SU(2)$
gauge group element $U(x)$ with each point on the boundary
3-sphere $S^3_\infty$ of $E^4$. Hence any solution with the
behavior of Eq. \ref{boundcondA} defines a map:\beeq x\rightarrow
U(x):\hskip 1cm S^3_\infty \rightarrow SU(2)\; .\eneq In fact,
since the $SU(2)$ group manifold is topologically the same as
$S^3$, the solution $A_\mu$ with condition Eq. \ref{boundcondA}
determines a map \beeq S^3_\infty\rightarrow S^3\; .\eneq All maps
of the $S^3$ onto itself are decomposed into homotopy classes by
an index $n\in Z$, called winding number. The integer $n$ is
simply the number of times $S^3$ gets covered by the map from
$S^3_\infty$. The trivial vacuum gauge field $A_\mu=0$ maps all
the points of $S^3_\infty$ to a single point of the gauge group
space $S^3$ (which can be identified as the (minus)identity of the
group). The trivial map has zero winding number since it does not
cover the $S^3$.

\subsubsection{The Winding Number}
\label{subsec:windingnumber} Now, we will show that for gauge
fields $A_\mu(x)$ with the asymptotic behavior of Eq.
\ref{boundcondA}, the winding number $n$ is given by
\cite{Huang,Dittrich,Rajaraman}\beeq q[A]=\frac{g^2}{32\pi^2} \int
d^4x G^a_{\mu\nu} {}^\ast G^a_{\mu\nu}=n \;. \eneq Let us start by
calculating the current $J_\mu$ for a pure gauge
$A_\mu=-(i/g)(\partial_\mu U)U^{-1}$:\be
J_\mu&=&2g^2\epsilon_{\mu\nu\rho\sigma}{\rm Tr}\left[A_\nu
\partial_\rho A_\sigma -i \frac{2g}{3}A_\nu A_\rho
A_\sigma\right]\nonumber\\
&=&-2\epsilon_{\mu\nu\rho\sigma}{\rm Tr}\left[(\partial_\nu U)
U^{-1}\partial_{\rho}\left[(\partial_\sigma
U)U^{-1}\right]-\frac{2}{3}(\partial_\nu U) U^{-1}(\partial_\rho
U) U^{-1}(\partial_\sigma U) U^{-1} \right]\nonumber \; .\ee
Evaluating the term\beeq
\partial_\rho\left[(\partial_\sigma U)U^{-1}\right]
=(\partial_\rho\partial_\sigma U)U^{-1}+(\partial_\sigma
U)\partial_\rho U^{-1}=(\partial_\rho\partial_\sigma
U)U^{-1}-(\partial_\sigma U)U^{-1}(\partial_\rho U)U^{-1}\;
,\label{J1}\eneq where we used identity Eq. \ref{therelation} to
write $\partial_\rho U^{-1} =-U^{-1}(\partial_\rho U)U^{-1}$, and
inserting back in Eq. \ref{J1} we find \beeq
J_\mu=-2\epsilon_{\mu\nu\rho\sigma}{\rm Tr}\left[-(\partial_\nu U)
U^{-1}(\partial_{\sigma} U )U^{-1}(\partial_\rho
U)U^{-1}-\frac{2}{3}(\partial_\nu U) U^{-1}(\partial_\rho U)
U^{-1}(\partial_\sigma U) U^{-1} \right]\nonumber \; .\eneq
Moreover, if we rename $\sigma\leftrightarrow\rho$ in the first
term above and sum up the terms, then \beeq
J_\mu=-\frac{2}{3}\epsilon_{\mu\nu\rho\sigma}{\rm
Tr}\left[(\partial_\nu U) U^{-1}(\partial_{\rho} U
)U^{-1}(\partial_\sigma U)U^{-1}\right]\; .\label{current}\eneq
Thus, for nonsingular solutions satisfying Eq. \ref{boundcondA}
\beeq q[A]= \frac{1}{16\pi^2}\oint _{S^3_\infty}\!\!\! d\Sigma_\mu
J_\mu=\frac{-1}{24\pi^2}\oint _{S^3_\infty}\!\!\!
d\Sigma_\mu\epsilon_{\mu\nu\rho\sigma}{\rm Tr}\left[(\partial_\nu
U) U^{-1}(\partial_{\rho} U )U^{-1}(\partial_\sigma
U)U^{-1}\right]\; ,\nonumber\eneq depends only on the group
element $U(x)$. It is interesting that the minimum value of an
arbitrary Euclidean action depends only on the properties of
$U(x)$ at infinity, and not on the details of the field
configurations at finite $x$. The boundary, $S^3_\infty$ is
parametrized by three angles $\psi_1(x)$, $\psi_2(x)$,
$\psi_3(x)$, while the group elements $U$ are characterized by
three angles $\theta_1(x)$, $\theta_2(x)$, $\theta_3(x)$: $U(x)=U(
\theta_a(x))$. Then, \beeq
\partial_\mu U\equiv\frac{\partial U}{\partial x^\mu}=\frac{\partial \theta_a}{\partial
x^\mu} \frac{\partial U}{\partial
\theta_a}\equiv(\partial_\mu\theta_a)\partial_a U\; .\eneq Hence,
\beeq q[A]=\frac{-1}{24\pi^2}\oint _{S^3_\infty}\!\!
d\Sigma_\mu\epsilon_{\mu\nu\rho\sigma} (\partial_\nu\theta_a)
(\partial_\rho\theta_b) (\partial_\sigma\theta_c) {\rm
Tr}\left[(\partial_a U) U^{-1}(\partial_{b} U )U^{-1}(\partial_c
U)U^{-1}\right]\; .\nonumber\eneq Since
$\epsilon_{\mu\nu\rho\sigma}$ is cyclic in $(\nu\rho\sigma)$ and
${\rm Tr}\left[(\partial_a U) U^{-1}(\partial_{b} U
)U^{-1}(\partial_c U)U^{-1}\right]$ is cyclic in (abc), we obtain
\beeq q[A]=\frac{-3!}{24\pi^2}\oint _{S^3_\infty}\!\!
d\Sigma_\mu\epsilon_{\mu\nu\rho\sigma} (\partial_\nu\theta_1)
(\partial_\rho\theta_2) (\partial_\sigma\theta_3) {\rm
Tr}\left[(\partial_1 U) U^{-1}(\partial_{2} U )U^{-1}(\partial_3
U)U^{-1}\right]\; .\nonumber\eneq The measure transform between
the coordinate systems as \beeq d\theta_1 d\theta_2
d\theta_3={(\rm Jacobian)}d\psi_1 d\psi_2 d\psi_3 \Rightarrow
d\theta_1 d\theta_2
d\theta_3=d\Sigma_\mu\epsilon_{\mu\nu\rho\sigma}
(\partial_\nu\theta_1) (\partial_\rho\theta_2)
(\partial_\sigma\theta_3) \; .\nonumber\eneq Therefore, we can
write $q[A]$ in the following form: \beeq
q[A]=\frac{-1}{4\pi^2}\;n\;\oint _{S^3_{SU(2)}}\!\! d\theta_1
d\theta_2 d\theta_3 {\rm Tr}\left[(\partial_1 U)
U^{-1}(\partial_{2} U )U^{-1}(\partial_3 U)U^{-1}\right]\;
.\label{qA}\eneq Now, here the integration is over the group
manifold of the gauge group $S^3_{SU(2)}$ instead of the boundary
manifold $S^3_\infty$. The integer $n\in Z$ is there because it
denotes how many points of $S^3_\infty$ is mapped to one point of
$S^3_{SU(2)}$. Choosing for $\theta_a$, the Euler angles, we can
always parametrize an $SU(2)$ group element  \beeq U(\theta_a)=
e^{\frac{i}{2}\theta_1\sigma_3}e^{\frac{i}{2}\theta_2\sigma_2}
e^{\frac{i}{2}\theta_3\sigma_3}\Rightarrow
U^{-1}(\theta_a)=e^{-\frac{i}{2}\theta_3\sigma_3}e^{-\frac{i}{2}\theta_2\sigma_2}
e^{-\frac{i}{2}\theta_1\sigma_3} \nonumber\; ,\eneq where
$0\leq\theta_1\leq 2\pi$, $0\leq\theta_2\leq\pi$,
$0\leq\theta_3\leq 4\pi$.  Then, after trivial cancellations, we
find \be{\mathcal{U}}&\equiv&{\rm Tr}\left[(\partial_1 U)
U^{-1}(\partial_{2} U )U^{-1}(\partial_3
U)U^{-1}\right]\nonumber\\ &=&-\frac{i}{8}{\rm Tr}\left[\sigma_3
e^{\frac{i}{2}\theta_1\sigma_3}\sigma_2
e^{\frac{i}{2}\theta_2\sigma_2}\sigma_3
e^{-\frac{i}{2}\theta_2\sigma_2}e^{-\frac{i}{2}\theta_1\sigma_3}\right]\;.\label{Trace}\ee
In Eq. \ref{Trace}, we can rewrite the term $I\equiv
e^{\frac{i}{2}\theta_2\sigma_2}\sigma_3
e^{-\frac{i}{2}\theta_2\sigma_2}$, by power expanding the
exponentials, as \be
{\mathcal{I}}&=&\sigma_3+(\frac{i}{2}\theta_2)[\sigma_2 ,
\sigma_3]+\frac{1}{2!}(\frac{i}{2}\theta_2)^2\left[\sigma_2 , [
\sigma_2 , \sigma_3]\right]
+\frac{1}{3!}(\frac{i}{2}\theta_2)^3\left[\sigma_2 ,\left[
\sigma_2, [ \sigma_2 ,
\sigma_3]\right]\right]+.\; .\; .\nonumber\\
&=&\sigma^3(1-\frac{1}{2!}\theta_2^2+.\; .\; .)-\sigma_1(\theta_2
-\frac{1}{3!}\theta_2^3+.\; .\; .)\nonumber\\
&=&\sigma_3 \cos{(\theta_2)}-\sigma_1\sin{(\theta_2)}\;
.\label{power}\ee Then Eq. \ref{Trace} becomes \be
{\mathcal{U}}&=&-\frac{i}{8}{\rm Tr}\left[\sigma_3
e^{\frac{i}{2}\theta_1\sigma_3}\sigma_2\left(\sigma_3
\cos{(\theta_2)}-\sigma_1\sin{(\theta_2)}\right)
e^{-\frac{i}{2}\theta_1\sigma_3}\right]\nonumber\\
&=&-\frac{i}{8}{\rm Tr}\left[
e^{\frac{i}{2}\theta_1\sigma_3}(-i\sigma_1)\left(\sigma_3
\cos{(\theta_2)}-\sigma_1\sin{(\theta_2)}\right)
e^{-\frac{i}{2}\theta_1\sigma_3}\right]\nonumber\\
&=&-\frac{1}{8}{\rm Tr}\left[-i\cos{(\theta_2)}
e^{\frac{i}{2}\theta_1\sigma_3}\sigma_2
e^{-\frac{i}{2}\theta_1\sigma_3}
+I\sin{(\theta_2)}\right]\nonumber\; .\ee Next, we use the
identity \beeq e^{\frac{i}{2}\theta_1\sigma_3}\sigma_2
e^{-\frac{i}{2}\theta_1\sigma_3}=\sigma_2\cos{(\theta_1)}+\sigma_1\sin{(\theta_1)}\;
,\eneq which can be verified the same way as in Eq. \ref{power}.
Thus \be {\mathcal{U}} &=&-\frac{1}{8}{\rm
Tr}\left[-i\cos{(\theta_2)}\cos{(\theta_1)} \sigma_2
-i\cos{(\theta_2)}\sin{(\theta_1)}
\sigma_1+I\sin{(\theta_2)}\right]\nonumber\\
&=&-\frac{1}{8}\sin{(\theta_2)}{\rm
Tr}\left[I\right]=-\frac{1}{4}\sin{(\theta_2)}\; .\label{14st2}\ee
Finally, we can calculate $q[A]$ using Eqs. \ref{qA} and
\ref{14st2}: \beeq q[A]= -\frac{1}{4\pi^2}\; n\;
\int_0^{2\pi}d\theta_1\int_0^{4\pi}d\theta_3\int_0^{\pi}d\theta_2
(-\frac{1}{4})\sin{(\theta_2)}=n\; \label{thiscanbeex}.\eneq
Equation \ref{thiscanbeex} can be expressed in Minkowski
spacetime. Since $(d^4x)_E=id^4x$; $G^a_{i4}=-iG^a_{i0}$; and
$\epsilon^{1230}=-1$ (whereas $\epsilon^{1234}=1$), we have \beeq
q_M[A]=\frac{g^2}{32\pi^2} \int d^4x G^a_{\mu\nu} {}^\ast
G^a_{\mu\nu}=-n \;.\label{EucMinn} \eneq The winding number $n$ is
a measure of how many times the group manifold $SU(2)$ is covered
by the map $U(\theta(x))$, when $x$ spans the whole boundary of
the four dimensional Euclidean space, $S^3_\infty$, once.

We now consider the simplest nontrivial case, $n=1$, the
instanton. Instantons are localized, nonsingular self dual
solutions of the classical Euclidean Yang-Mills field equations
$D_\mu G_{\mu\nu}=0$ with one unit of topological charge. They
have vanishing Euclidean energy momentum tensor: $T_{\mu\nu}=0$.
In particular the Hamiltonian density ${{\mathcal{H}}}=T_{00}=0$.
For instantons the associated action is\beeq
S^E=\frac{8\pi^2}{g^2}|q[A]|=\frac{8\pi^2}{g^2}\; .\eneq Having
defined the instantons, we construct an explicit solution in the
next section. \subsubsection{An Explicit Instanton Solution by
Construction} We start with making some useful
definitions\cite{Huang,Rubakov}: \beeq
\sigma_\mu\equiv(\sigma_a,\pm i),
\sigma_\mu^\dagger\equiv(\sigma_a,\mp i), \sigma_{\mu\nu}\equiv
i(\delta_{\mu\nu}-\sigma_\mu^\dagger\sigma_\nu),
{}^\ast\sigma_{\mu\nu}\equiv\frac{1}{2}\epsilon_{\mu\nu\rho\phi}\sigma_{\rho\phi}
=\pm\sigma_{\mu\nu}\; ,\eneq where $\mu$ and $\nu$ runs from one
to four, whereas $a$ runs from one to three. Notice that
$\sigma_{\mu\nu}$ is self dual if $\sigma_4=+i$, or anti self dual
if $\sigma_4=-i$. Let's continue with the following ansatz:\beeq
U\equiv
\frac{x_\mu\sigma_\mu^\dagger}{x}=\frac{\vec{x}\cdot\vec{\sigma}\mp
ix_4}{\sqrt{\vec{x}^2+x_4^2}}\;\;\;\Rightarrow\;\;\; U^{-1}=
\frac{x_\mu\sigma_\mu}{x}=\frac{\vec{x}\cdot\vec{\sigma}\pm
ix_4}{\sqrt{\vec{x}^2+x_4^2}}\;,\eneq so that \beeq A_\mu^{\rm
pure}=-\frac{i}{g}(\partial_\mu
U)U^{-1}=\frac{i}{g}\frac{(x_\mu-\sigma^\dagger_\mu(x_\nu\sigma_\nu))}{x^2}
=\frac{1}{g}\frac{\sigma_{\mu\nu}x_\nu}{x^2}\; ,\eneq where we
have used the identity
$(x_\mu\sigma_\mu)(x_\nu\sigma_\nu^\dagger)=x^2$. For a finite
energy solution we need an $A_\mu$ which is not a pure gauge over
the whole volume. We make the following ansatz for the potential
\beeq A_\mu=\frac{1}{g}\frac{\sigma_{\mu\nu}x_\nu}{x^2}f(x^2)\; .
\label{Asigma}\eneq We require $f(0)=0$ for regularity at $x=0$,
and $f(\infty)=1$ for finite energy. To make this a solution, we
only need to choose $f$ such that the field strength tensor is
self(anti-self) dual: $G_{\mu\nu}=\pm{}^\ast G_{\mu\nu}$. From Eq.
\ref{Asigma} we obtain \beeq G_{\mu\nu}=\frac{2}{g}\left\{
\frac{f(1-f)}{x^2}\sigma_{\mu\nu}+\left[f'-\frac{f(1-f)}{x^2}\right](\sigma_{\mu\rho}x_\rho
x_\nu-\sigma_{\nu\rho}x_\rho x_\mu)\right\}\; ,\eneq \beeq {}^\ast
G_{\mu\nu}=\frac{2}{g}\left\{
\frac{f(1-f)}{x^2}{}^\ast\sigma_{\mu\nu}+\left[f'-\frac{f(1-f)}{x^2}\right]
\epsilon_{\mu\nu\rho\phi}(\sigma_{\rho\gamma}x_\gamma
x_\phi-\sigma_{\phi\gamma}x_\gamma x_\rho)\right\}\; ,\eneq where
prime denotes derivative with respect to the argument. Thus,
$G_{\mu\nu}=\pm {}^\ast G_{\mu\nu}$, if the second term vanishes
\beeq f'-\frac{f(1-f)}{x^2}=0.\eneq The general solution
satisfying the required boundary conditions is\beeq
f(x^2)=\frac{cx^2}{cx^2+\lambda^2}\; .\eneq Note that as
$x^2\rightarrow\infty$, $f(x^2)\rightarrow 1$, hence
$A_\mu\rightarrow A^{\rm pure}_\mu$ as required. The solution with
topological charge $q=\mp 1$, has $c=1$, as we now show. Setting
$c=1$, we find \beeq
A_\mu=\frac{1}{g}\frac{\sigma_{\mu\nu}x_\nu}{x^2+\lambda^2}\;.\eneq
We can write the vector potential more explicitly: \beeq
\vec{A}=\frac{1}{g}\frac{(\vec{x}\times \vec{\sigma})\pm
x_4\vec{\sigma}}{x^2+\lambda^2}\Rightarrow
A^a_i=\frac{2}{g}\frac{\left(\epsilon_{ija}x_j\pm
\delta_{ai}x_4\right)}{{x^2+\lambda^2}}\, ,\nonumber\eneq \beeq
A_4=\mp\frac{1}{g}\frac{\vec{\sigma}\cdot\vec{x}}{x^2+\lambda^2}\Rightarrow
A^a_4=\mp \frac{2}{g}\frac{x^a}{x^2+\lambda^2}\;
.\label{explicitinstanton}\eneq The field strength tensor becomes
\beeq G_{\mu\nu}=\pm{}^\ast
G_{\mu\nu}=\pm\frac{2}{g}\frac{f(1-f)}{x^2}\sigma_{\mu\nu}
=\pm\frac{2}{g}\frac{\lambda^2}{(x^2+\lambda^2)}\sigma_{\mu\nu}\;
,\eneq from which we obtain \beeq E^a_i=G^a_{i4}=\pm
\frac{4}{g}\frac{\lambda^2}{(x^2+\lambda^2)^2}\delta^a_i\; .\eneq
Let us confirm that the topological charge $q[A]=\mp 1$. For
self(anti-self) dual fields: $E^a_i=\pm B^a_i$. Then, since
$(1/4)G^a_{\mu\nu}{}^\ast G^a_{\mu\nu}=-E^a_i B^a_i$, we have:
$(1/4)G^a_{\mu\nu}{}^\ast G^a_{\mu\nu}=\mp E^a_i E^a_i$. Hence,
\beeq q[A]=\frac{g^2}{8\pi^2}\int d^4x \frac{1}{4}G^a_{\mu\nu}
{}^\ast G^a_{\mu\nu}=\mp \frac{6\lambda^4}{\pi^2}\int
d^4x\frac{1}{(x^2+\lambda^2)^4} \; .\eneq In four dimensional
Euclidean space $d^4x= d\Omega_3r^3dr=\pi^2r^2d(r^2)$, thus \be
q[A]=\mp 6\lambda^4\!\!\int_0^\infty\!\!\!\!\! \frac{r^2
d(r^2)}{(r^2+\lambda^2)^4}=\mp 6\!\!\int_0^\infty\!\!\!\!\!
\frac{z dz}{(1+z)^4}=\mp\left[\frac{1}{3}\frac{1}{(1+z)^3}
-\frac{1}{2}\frac{1}{(1+z)^2}\right]\Big|_0^\infty=\mp 1\;
,\nonumber\ee where we changed the integration variable $r^2\equiv
\lambda^2 z$. Later we will show how instantons are related to
tunneling events between physically distinct gauge field vacua.
But, before that, we want to study the quantum theory of non
Abelian gauge fields in the next section.

\subsection{Canonical Formalism and Quantization}

Let us turn back and formulate the classical gauge theory using
the canonical procedure\cite{Huang,JackivVacuum}. The Hamiltonian
density and Hamiltonian are obtained using the standard recipe:
\beeq{\mathcal{H}}= \Pi^{a}_\mu
\partial_0
A^{a\mu} -{\mathcal{L}}_{\rm gauge} \; ,\eneq \beeq H=\int d^3 x
{\mathcal{H}}\; ,\eneq where $\Pi_\mu$ is the canonical conjugate
of the gauge field. From Lagrange density Eq. \ref{canLag}, we
find\beeq\Pi^a_0(x)=\frac{\partial{\mathcal{L}}}{\partial(\partial_0
A^{a0})}=0\, ,\eneq \beeq
\Pi^a_i(x)=\frac{\partial{\mathcal{L}}}{\partial(\partial_0
A^{ai})}=-G^{a0}\, _i=-E^{ai}\label{canmom}\; .\eneq Thus
\beeq{\mathcal{H}}=-{\vec{E}}^a\cdot
\frac{\partial{\vec{A}}^a}{\partial t} -\frac{1}{2}({\vec{E}}^a
\cdot{\vec{E}}^a-{\vec{B}}^a\cdot{\vec{B}}^a)\label{Hamden}\;
.\eneq Using Eq. \ref{vEa} we can re-express the first term of Eq.
\ref{Hamden}\be -{\vec{E}}^a\cdot
\frac{\partial{\vec{A}}^a}{\partial
t}&=&{\vec{E}}^a\cdot({\vec{E}}^a+\vec{\nabla}A^{a0}-gf^{abc}{\vec{A}}^bA^{c0})\nonumber\\
&=&{\vec{E}}^a\cdot{\vec{E}}^a
+\vec{\nabla}(A^{a0}{\vec{E}}^a)-A^{a0}(\vec{\nabla}\cdot{\vec{E}}^a
-gf^{abc}{\vec{A}}^b\cdot{\vec{E}}^c)\, .\ee Then, Eq.
\ref{Hamden} becomes\beeq
{\mathcal{H}}=\frac{1}{2}({\vec{E}}^a\cdot{\vec{E}}^a+{\vec{B}}^a\cdot{\vec{B}}^a)+\vec{\nabla}
(A^{a0} {\vec{E}}^a)-A^{a0}\vec{D}\cdot {\vec{E}}^a\, ,\eneq where
\beeq \vec{D}\cdot {\vec{E}}^a\equiv\vec{\nabla}\cdot
{\vec{E}}^a-gf^{abc}{\vec{A}}^b\cdot{\vec{E}}^c\, .\eneq If we try
to impose the canonical equal time commutators (Poisson bracket
relations in the classical theory), we immediately encounter
trouble since $\Pi_0=0$ because ${{\mathcal{L}}}_{\rm gauge}$ does
not depend on $\partial_0 A^{a0}$. Since $A^{a0}$ is not a
dynamical variable, we may eliminate it from the Hamiltonian by
solving for it in terms of the dynamical variables through the
classical constraint equation Eq. \ref{eqconst}. This constraint
equation $(\vec{D}\cdot {\vec{E}}^a=0)$ is nothing but the
variation of the action $\delta S$ given in Eq. \ref{varS} with
respect to $\delta A^0$. The component $A^{a0}$ may be expressed
in terms of the dynamical variables by inserting ${\vec{E}}^a$
given in Eq. \ref{vEa} into the constraint Eq. \ref{eqconst}:
\beeq \nabla^2 A^{a0}-f^{abc}{\vec{A}}^b\cdot (
2\vec{\nabla}A^{c0} +g\frac{\partial{\vec{A}}^c}{\partial t}
+g^2f^{cpq}A^{p0}{\vec{A}}^q)
+gf^{abc}A^{b0}\vec{\nabla}\cdot{\vec{A}}^c=0 \, .\eneq The usual
procedure however, is to use the gauge invariance and set $A^{a0}$
to zero. In this case Hamiltonian is the energy with density Eq.
\ref{energy}: \be H=\frac{1}{2}\int d^3x \left(
({\vec{E}}^a)^2+({\vec{B}}^b)^2\right)=\int d^3x T^{00}\; .\ee
When the canonical commutation relations are imposed \be \left[
A^{ia}(\vec{x}, t), A^{jb}(\vec{x}', t)\right]&=&0\; ,\nonumber\\
\left[E^{ia}(\vec{x}, t), E^{jb}(\vec{x}', t)\right]&=&0\; ,\nonumber\\
\left[ E^{ia}(\vec{x}, t), A^{jb}(\vec{x}',
t)\right]&=&i\delta^{ij}\delta^{ab}\delta(\vec{x}-\vec{x}')
\label{comEA}\; , \ee Hamilton equations $\frac{\partial
{\vec{A}^a}}{\partial t}=i[H,{\vec{A}}^a]$ and $\frac{\partial
{\vec{E}^a}}{\partial t}=i[H,{\vec{E}}^a]$ yield, in the temporal
gauge, the definition of the electric field \ref{vEa} and Ampere's
law Eq. \ref{eqdyn} (spatial components of Eq. \ref{covEqmot})
respectively. Let us verify these claims. The Hamilton equation
for ${\vec{A}}^a(x)$ yields \beeq \frac{\partial
{\vec{A}}^a}{\partial t}(x)=i[H,{\vec{A}^a(x)}]=\frac{i}{2}\int
d^3y\left\{ \left[(\vec{E}^b({y}))^2
,\vec{A}^a({x})\right]+\left[(\vec{B}^b({y}))^2
,\vec{A}^a({x})\right]\right\}\nonumber\ , \eneq however, since\be
\left[\vec{B}^b({y}),\vec{A}^a({x})\right]
\!\!\!&=&\!\!\!\left[\vec{\nabla}_y\times{\vec{A}}^b({y})
-\frac{g}{2}f^{bcd}{\vec{A}}^c({y})\times{\vec{A}}^d({y}),\vec{A}^a({x})\right]
=\left[\vec{\nabla}_y\times{\vec{A}}^b({y}),\vec{A}^a({x})\right]\nonumber\\
\!\!\!&=&\!\!\!\vec{\nabla}_y\times\left[{\vec{A}}^b({y})
,\vec{A}^a({x})\right]=0\ ,\ee using the distribution law and  Eq.
\ref{comEA} we recover the definition of the electric field:\be
\frac{\partial {\vec{A}}^a}{\partial t}(x)=\frac{i}{2}\int
d^3y\left[(\vec{E}^b({y}))^2 ,\vec{A}^a({x})\right]=-{\vec{E}}^a
(x)\, .\ee The Hamilton equation for ${E}^{aj}(x)$ yields \be
\frac{\partial {E}^{aj}}{\partial
t}(x)\!\!\!&=&\!\!\!i[H,{E}^{aj}(x)]\!=\!\frac{i}{2}\!\int\! d^3y
\left[(\vec{B}^b({y}))^2 , {E}^{aj}({x})\right] \!=\!{i}\!\int\!
d^3y {B}^{bi}({y})\!\!\left[{B}^{bi}({y}) , {E}^{aj}({x})\right]\
,\nonumber\\
\!\!\!&=&\!\!\!{i}\!\int\! d^3y
{B}^{bi}({y})\!\!\left[\epsilon^{ipq}\partial_{y^p}A^{bq}(y)
-\frac{g}{2}f^{bcd}\epsilon^{ipq}A^{cp}(y)A^{dq}(y) ,
{E}^{aj}({x})\right]\ ,\nonumber\\&\equiv&{\mathcal{I}}_a +
{\mathcal{I}}_b\label{HEcommutator}\ ,\ee where
\be{\mathcal{I}}_a &=& {i}\epsilon^{ipq}\!\int\! d^3y
{B}^{bi}({y})\partial_{y^p}\left[A^{bq}(y)
,{E}^{aj}({x})\right]=\epsilon^{ipj}\!\int\! d^3y
{B}^{ai}({y})\partial_{y^p}\delta(\vec{x}-\vec{y})\nonumber\\
&=&\epsilon^{ipj} \!\int\! d^3y\left\{
\partial_{y^p}\left({B}^{ai}({y})
\delta(\vec{x}-\vec{y})\right)-\left(\partial_{y^p}{B}^{ai}({y})\right)
\delta(\vec{x}-\vec{y})\right\}\nonumber\\
&=&\epsilon^{ipj}\left\{ \oint_\infty d^2\Sigma{B}^{ai}({x}) -
\partial_{x^p}{B}^{ai}({x})\right\}
=\epsilon^{jpi}\partial_{x^p}{B}^{ai}({x})\nonumber\\
&=&\left(\nabla\times \vec{B}^a(x)\right)^j\, ,\ee and \beeq
{\mathcal{I}}_b=-\frac{i}{2}gf^ {bcd}\epsilon^{ipq}\!\int\!
d^3y{B}^{bi}({y})\left[A^{cp}(y)A^{dq}(y) ,
E^{aj}(x)\right]\nonumber\eneq \beeq \hskip 1cm
=-\frac{1}{2}g\left\{f^{bca}\epsilon^{ipj}{B}^{bi}({x})A^{cp}(x)
+f^{bad}\epsilon^{ijq}{B}^{bi}({x})A^{dq}(x)\right\}\,
,\nonumber\eneq adding the two terms up we find \beeq
{\mathcal{I}}_b=-gf^{abc}\left(\vec{A}^b(x)\times\vec{B}^c(x)\right)^j\,
.\eneq Hence \beeq
{\mathcal{I}}_a+{\mathcal{I}}_b=\left(\nabla\times
\vec{B}^a(x)-gf^{abc}\vec{A}^b(x)\times\vec{B}^c(x)\right)^j\
,\eneq which implies \beeq \frac{\partial {\vec{E}}^{a}}{\partial
t}(x)=\vec{\nabla}\times{\vec{B}}^a(x)-gf^{abc}{\vec{A}}^b\times{\vec{B}}^c(x)
\equiv\vec{D}\times\vec{B}^a \; .\eneq However, the Gauss' law
(the time component of Eq. \ref{covEqmot}), $\vec{D}\cdot
{\vec{E}}^a=0$, is not found. Note that ${\vec{B}}^a(x)$, given by
Eq. \ref{vBa}, is not a fundamental variable and its definition
does not appear in the Hamilton equations either. ${\vec{B}}^a(x)$
is given by Eq. \ref{vBa}.

In Eq. \ref{A}, we have seen that, classically, the theory has a
symmetry which leaves the equations of motion invariant. We choose
the gauge $A^0(x)=0$. But this does not completely fix the
invariance. The infinitesimal form (Eq. \ref{infgaugA}) of the
residual symmetry transformation is \beeq \delta
{\vec{A}}^a(x)=-g^{-1}\vec{D} \omega^a(\vec{x})\equiv
-\frac{1}{g}\vec{\nabla}\omega^a(\vec{x})+f^{abc}{\vec{A}}^b(x)\omega(\vec{x})\,
. \eneq Here, unlike Eq. \ref{infgaugA}, $\omega(\vec{x})$ is time
independent. The time dependent gauge transformations would
generate nonzero $A^{a0}(x)$ in view of Eq. \ref{A}. Quantum
mechanically this residual gauge invariance is expressed through
the operator equation\cite{Huang,JackivVacuum}: \beeq
[\vec{D}\cdot{\vec{E}}^a(x), H]=i\frac{\partial}{\partial
t}\vec{D}\cdot\vec{E}^a=i\vec{D}\cdot\dot{\vec{E}}^a=0\;
.\label{comDEH}\eneq To see that $[\vec{D}\cdot{\vec{E}}^a(x),
H]=0$, let us first simplify Eq. \ref{comDEH}: \beeq
[\vec{D}\cdot{\vec{E}}^a(x), H]=i(\partial_i {\dot{E}}_i^a
+gf^{abc}\dot{A}_i^bE_i^c-gf^{abc}A_i^b\dot{E}_i^c)\; .\eneq Using
$\dot{A}_i^b=-{E}_i^b$, we obtain \beeq
[\vec{D}\cdot{\vec{E}}^a(x), H]=i(\partial_i
{\dot{E}}_i^a-gf^{abc}A_i^b\dot{E}_i^c)\; .\eneq On the other
hand, from the equations of motion ($D_\nu^{ab}F^{b\nu\mu}=0$), we
have \beeq D_i^{ab}F^{bi0}=0\eneq\beeq
D_0^{ab}F^{b0i}+D_j^{ab}F^{bji}
=-(\partial_0\delta^{ab}-gf^{abd}A^d_0)E^b+D_j^{ab}F^{bji}=0\label{eqmotikinci}\;
.\eneq In the temporal gauge, Eq. \ref{eqmotikinci} becomes \beeq
\dot{E}^a_i=D_j^{ab}F^b_{ji} \; .\eneq Then the commutator $
[\vec{D}\cdot{\vec{E}}^a(x), H]=i\vec{D}\cdot\dot{\vec{E}}^a$ can
be expressed as\be [\vec{D}\cdot{\vec{E}}^a(x),
H]&=&i(\delta^{ab}\partial_i
\dot{E}^b_i-gf^{abc}A^b_i\dot{E}^c_i)=iD^{ab}_i\dot{E}^b_i=iD^{ab}_iD^{bc}_jF^c_{ji}\nonumber\\
&=&\frac{i}{2}(D^{ab}_iD^{bc}_j-D^{ab}_jD^{bc}_i)F^c_{ji}=\frac{i}{2}[D_i
, D_j]^{ac}F^c_{ji}\; ,\label{EqA149}\ee where \be [D_i ,
D_j]^{ac}\!\!\!&=&\!\!\!D^{ab}_iD^{bc}_j-D^{ab}_jD^{bc}_i\nonumber\\
\!\!\!&=&\!\!\!(\delta^{ab}\partial_i-gf^{abe}A^e_i)
(\delta^{bc}\partial_j-gf^{bcd}A^d_j)-(\delta^{ab}\partial_j-gf^{abe}A^e_j)
(\delta^{bc}\partial_i-gf^{bcd}A^d_i)\nonumber\\
\!\!\!&=&\!\!\!-gf^{acb}(\partial_i A_j^b-\partial_j
A_i^b)+g^2f^{abe}f^{bcd}(A_i^e A_j^d - A_j^e A_i^d)\nonumber\\
\!\!\!&=&\!\!\!-gf^{acb}(\partial_i A_j^b-\partial_j
A_i^b)+g^2(f^{abe}f^{bcd}-f^{abd}f^{bce})A_i^e A_j^d \;
.\label{EqA150}\ee Using the Jacobi identity,
$f^{abe}f^{bcd}-f^{abd}f^{bce}=f^{acb}f^{bde}$, we recast Eq.
(\ref{EqA150}) as\be [D_i , D_j]^{ac}&=&-gf^{acb}(\partial_i
A_j^b-\partial_j A_i^b)+g^2f^{acb}f^{bde}A_i^e A_j^d\nonumber\\
&=&-gf^{acb}(\partial_i A_j^b-\partial_j A_i^b+f^{bpq}A^p_i
A^q_j)=-gf^{acb}F^b_{ij}\;.\label{EqA151}\ee Hence, Eqs.
(\ref{EqA149}) and (\ref{EqA151}) give \beeq
[\vec{D}\cdot{\vec{E}}^a(x), H]=\frac{i}{2}[D_i ,
D_j]^{ac}F^c_{ji}=-\frac{i}{2}gf^{abc}F^b_{ij} F^c_{ij}=0\; .\eneq
The commutator vanishes because the indices $a$ and $b$ are
antisymmetric in $f^{abc}$ whereas they are symmetric in $F^b_{ij}
F^c_{ij}$.

Therefore, an eigenfunction $\Psi[\vec{A}]$ of H can be chosen as
a simultaneous eigenfunction of $\vec{D}\cdot{\vec{E}}^a(x)$.
Thus, Gauss' law is imposed on the Hilbert space by choosing the
eigenvalue of $\vec{D}\cdot{\vec{E}}^a(x)$ to be zero \beeq
\vec{D}\cdot{\vec{E}}^a(x)\Psi[\vec{A}(x)]=0 \label{Gauss}\,
.\eneq We are particularly interested in vacuum configurations.
Classically the ground state must correspond to time independent
minimum energy solution. We have, therefore, $G_{\mu\nu}=0$. To
prove this, consider the action for a time independent solution
(i.e., the space integral of the Lagrangian
density)\cite{Itzykson}: \beeq S=S_1-S_2=\frac{1}{2}\int d^3x
\left( \vec{E}^a\cdot\vec{E}^a -\vec{B}^a\cdot\vec{B}^a\right)\;
.\eneq Since the total energy ${\mathcal{E}}$ is the sum
$S_1+S_2$, its finiteness implies the finiteness of $S_1$, $S_2$
and $S$. Any solution, if exists, is unstable (not a solution)
under the scale transformations:\beeq A^{a0}(\vec{x})\rightarrow
\rho\lambda A^{a0}(\lambda \vec{x})\; ,
\;\;A^{ai}(\vec{x})\rightarrow \lambda A^{ai}(\lambda \vec{x})\;
.\eneq Hence\beeq \vec{E}^{ai}(\vec{x})\rightarrow \rho\lambda^2
\vec{E}^{ai}(\vec{y})\; ,
\;\;\vec{B}^{ai}(\vec{x})\rightarrow\lambda^4
\vec{B}^{ai}(\vec{y})\; ,\eneq where $\vec{y}\equiv \lambda
\vec{x}$. Therefore, \be S_1&=&(1/2)\int d^3x
E^{ai}(\vec{x})E^{ai}(\vec{x})\rightarrow
S_1=(1/2)\lambda\rho^2\int d^3y
E^{ai}(\vec{y})E^{ai}(\vec{y})\nonumber\; ,\\
S_2&=&\frac{1}{2}\int d^3x
B^{ai}(\vec{x})B^{ai}(\vec{x})\rightarrow
S_2=\frac{1}{2}\lambda^4\int d^3y
B^{ai}(\vec{y})B^{ai}(\vec{y})\nonumber\; ,\\
S&=&S_1-S_2\rightarrow \lambda\rho^2S_1-\lambda S_2\; , \ee which
should be stationary at $\rho=\lambda=1$. Thus, the variations of
$\lambda$ and $\rho$ at $\lambda=\rho=1$ yield
${\mathcal{L}}_1-{\mathcal{L}}_2=0$ and
$2{\mathcal{L}}_1-{\mathcal{L}}_2=0$, respectively. Together they
give ${\mathcal{L}}_1={\mathcal{L}}_2=0$, hence $G_{\mu\nu}=0$ (as
expected in the ground state $\vec{E}^a=\vec{B}^a=0$). This means
that (in the temporal gauge) the field $\vec{A}(\vec{x})$ is a
pure gauge: $\vec{A}(\vec{x})=(-i/g)(\nabla U(\vec{x}))
U^{-1}(\vec{x})$. Now, to make progress, we make a very important
hypothesis regarding gauge transformations. We assume that the
allowed gauge transformation matrices $U(\vec{x})$ have the same
definite space-time independent limit $U_\infty$ at spatial
infinity: \beeq\lim_{|x|\rightarrow\infty}
U(\vec{x})=U_\infty={\rm const.}\label{Ulimit}\eneq Physically,
this assumption is necessary at least to have a well defined
non-Abelian electric charge. The total electric charge $Q^a\equiv
g\int dr f^{abc} {\vec{A}}^b\cdot{\vec{E}}^c =\int dr \nabla\cdot
{\vec{E}}^a=\int d\vec{S}\cdot \vec{E}^a$. Under a local gauge
transformation charge transforms as \beeq Q^a\rightarrow \int
d\vec{S} \cdot U^{-1}{\vec{E}}^a U\; .\eneq If $U$ has a constant
limit $U_\infty$ at spatial infinity, $Q^a$ transforms by the
global gauge transformation $U^{-1}_\infty {Q}^a U_\infty$.
Otherwise, the transformed charge has no simple relation to the
original charge and we must conclude that the non-Abelian electric
charge is not well defined. Let us also note that, because of
condition that all pure gauge potentials have the same limit at
infinity, the possibility of transitions between different the
vacuum configurations can not be ruled out by kinetic
considerations. Such transitions are only possible if the
different vacuum field configurations do not vary at spatial
infinity. Otherwise a transition would require an infinite kinetic
energy (the integral $\int d^3x (\partial_0 A_i)^2$ would diverge)
and an infinite action.

We can regard $U(\vec{x})$ as a continuous mapping of three
dimensional space to the gauge group $G$. Equation \ref{Ulimit}
identifies spatial infinity as one point. This is equivalent to
compactifying the spatial manifold $R^3$ to $S^3$. Thus, we only
need to consider the maps of $S^3\rightarrow G$. A theorem due to
Bott \cite{Bott} states that any continuous mapping of $S^3$ into
a simple Lie group $G$ can be continuously deformed to a mapping
into an $SU(2)$ subgroup of $G$. Therefore, for a gauge theory
with simple group structure, it is sufficient to consider
$S^3\rightarrow SU(2)$. The manifold of the $SU(2)$ gauge group is
also $S^3$. Hence the matrix functions $U$ are mappings of
$S^3\rightarrow S^3$. These can be categorized into homotopy
classes labelled by an integer, called the ``winding number'' (the
number of times the spatial $S^3$ is covered by the group manifold
$S^3)$ of the mapping \cite{JackiwRebbi}. Class $n=0$ represents
the set of gauge transformations which do not wind the target
$S^3$ at all, and hence are homotopic (continuously deformable) to
$U=I$ (the mappings of all points in $S^3$ to the identity of
$S^3$); transformations of class $n=1$ are those that wind the
target $S^3$ only once. Remember that the most general element of
$SU(2)$ is of the form: \beeq U=\exp{(\pm
i\frac{\phi}{2}\hat{x}\cdot\vec{\sigma})}=v_0 \pm
i\vec{v}\cdot\vec{\sigma}\label{mapU}\eneq where $\vec{\sigma}$
are the Pauli matrices, $v_0\equiv\cos{(\phi /2)}$,
$\vec{v}\equiv\hat{x}\sin{(\phi/2)}$. As $\phi$ runs from $0$ to
$2\pi$, map Eq. \ref{mapU} covers the $SU(2)$ manifold once.
Therefore, for the maps of class $n=1$, replacing $\phi$ by a
continuous monotonic function $\omega(|\vec{x}|)$ with appropriate
boundary conditions, we can write the gauge transformations of the
form\beeq U_1(x)=\exp{(\pm
i\vec{{\sigma}}\cdot{\hat{x}}\omega(|\vec{x}|))} \label{uabst}\;
.\eneq To wind $S^3$ once, as $|\vec{x}|$ runs from $0$ to
$\infty$, $\omega(|\vec{x}|)$ should satisfy\beeq
\omega(\infty)=\pi, \hskip 1cm \omega(0)=0\label{bound}\; .\eneq
As a representative of $U_1$, choosing\beeq
\omega(|\vec{x}|)=\frac{\pi|\vec{x}|}{\sqrt{|\vec{x}|^2
+\lambda^2}}\; ,\label{omega(x)}\eneq we have\beeq
U_1(\vec{x})=\exp{(\pm \frac{i\pi\sigma^i {x}^i}{\sqrt{|\vec{x}|^2
+\lambda^2}})}=\cos{(\frac{\pi|\vec{x}|}{\sqrt{|\vec{x}|^2
+\lambda^2}})}\pm
i\vec{\sigma}\cdot\hat{x}\sin{(\frac{\pi|\vec{x}|}{\sqrt{|\vec{x}|^2
+\lambda^2}})}\label{map1}\; ,\eneq where $\lambda$ is an
arbitrary number. $U_1$ can also be written, by making the
deformation $\sin({\pi|\vec{x}|}/{\sqrt{|\vec{x}|^2
+\lambda^2}})\rightarrow
2\lambda|\vec{x}|/(|\vec{x}|^2+\lambda^2)$, as \beeq
U_1(\vec{x})=\frac{\lambda^2-|\vec{x}|^2}{|\vec{x}|^2+\lambda^2}
\pm
i2\lambda\frac{\vec{\sigma}\cdot\vec{x}}{|\vec{x}|^2+\lambda^2}=\sigma^\mu
{\hat{n}}_\mu\label{Uratio}\, ,\eneq where $\sigma^\mu=(I,\mp
i\vec{\sigma})$ and the unit
vector\beeq{\hat{n}}_\mu=\left(\frac{\lambda^2-|\vec{x}|^2}{|\vec{x}|^2+\lambda^2},
-2\lambda\frac{\vec{x}}{|\vec{x}|^2+\lambda^2}\right)\, . \eneq As
$\vec{x}$ ranges over all of $3$ dimensional space, $n_\mu$ traces
out a sphere $S^3$ in four dimensional spacetime. Thus, indeed,
$U_1$ maps a $3$-space onto $S^3\simeq SU(2)$, covering it once.
Equation \ref{Ulimit} is satisfied since $U_\infty=-1$. Map Eq.
\ref{map1} can not be continuously deformed to identity (or to any
other $n=0$ transformations) without violating Eq. \ref{Ulimit}.
Indeed, Eq. \ref{map1} only yields a unique limit: \beeq U_\infty
=\cos{(\phi)}\pm i\vec{\sigma}\cdot\hat{x} \sin{(\phi)}\; ,\eneq
as $|\vec{x}|\rightarrow\infty$ independently of the direction of
approach if $\phi$ is an integer multiple of $\pi$. Gauge
transformations $U_n(\vec{x})$ of class $n$, can be taken as
$[U_1(\vec{x})]^n$ and $A^{ai}$ is calculated by \ref{pure}. If
$n=0$, then $U=I$ and we have the trivial vacuum $A^{ai}=0$. $U_1$
given in Eq. \ref{Uratio} yields \be
\vec{A}(\vec{x})&=&-\frac{i}{g}\left(\nabla U_1(\vec{x})\right)
U^{-1}_1(\vec{x})\nonumber\\
&=&\frac{1}{g}\frac{2\lambda}{(|\vec{x}|^2+\lambda^2)^2}\left[ \mp
\vec{\sigma}(|\vec{x}|^2-\lambda^2)\pm
2\vec{x}(\vec{\sigma}\cdot\vec{x})
+2\lambda(\vec{x}\times\vec{\sigma})\right] \label{vecA}\; ,\ee
where we used
$U_1^{-1}(\vec{x})=\frac{\lambda^2-|\vec{x}|^2}{|\vec{x}|^2+\lambda^2}
\mp
i2\lambda\frac{\vec{\sigma}\cdot\vec{x}}{|\vec{x}|^2+\lambda^2}$
and the identity
$\sigma^i(\vec{\sigma}\cdot\vec{x})=\sigma^i\sigma^jx^j=x^i+i\epsilon^{ijk}x^j\sigma^k$.
In the quantum considerations we will show that only $n=0$
homotopy class leaves the wave function of the system gauge
invariant whereas the higher classes change the wave function by a
phase. Therefore, homotopically equivalent transformations are
also gauge equivalent. $U_1$ transforms any classical vacuum
configuration in class $n$ to one in class $(n+1)$. Hence, any
gauge transformation of class $(n+m)$ is a product of a member of
class $n$ with a member of class $m$. The existence of gauge
transformations which are not homotopic to the identitity is what
distinguishes the non-Abelian theory from the Abelian one. In the
latter, all gauge transformations fall in the homotopically
trivial $n=0$ class. Thus the vacuum is physically unique in that
case.

An analytic expression for the winding number of a gauge
transformation $U(x)$ can be given by introducing a second
non-conserved topological charge\cite{Dittrich}:\beeq
Q_T(t)=\frac{1}{16\pi^2}\int d^3x J_0(\vec{x}, t)\; ,\eneq where
$J_{\mu}$, defined in Eq. \ref{totaldiv}, satisfying
${\mathcal{T}}\equiv ({1}/{2g^2})\partial_\mu J_\mu\equiv
{({1}/{4})}G^a_{\mu\nu}{}^\ast G^{a}_{\mu\nu}$. Recall that first
topological charge, defined in Eq. \ref{surfaceintegralSING}, was
absolutely gauge invariant winding number $q[A]$ (or the
Pontryagin index) of the Euclidean field configuration
$A_\mu^a(x)$. The new topological number is related to the issue
of how to connect different Minkowski vacua via tunneling events
(instantons in Euclidean space). Physically $Q_T[A^a_\mu, t]$ is a
number which characterizes a field configuration
$A_\mu(t,\vec{x})$ in three dimensional space at a fixed time $t$.
For an arbitrary field, $Q_T(t)$ is neither an integer nor has
topological meaning. But for pure gauges
$A_\mu(x)=-\frac{i}{g}(\partial_\mu U(x))U^{-1}(x)$, as shown
below, $Q_T(t)\in Z$ gives the homotopy class index (or winding
number) of the gauge transformation $U(x)$. Recall Eq.
\ref{current} that for a pure gauge\beeq
J_0=-\frac{2}{3}\epsilon_{ijk}{\rm Tr} [ (\partial_i U) U^{-1}
(\partial_j U) U^{-1} (\partial_k U) U^{-1}]\; .\eneq Let $x^0=t$
be fixed and assume that $U(t, \vec{x}) \rightarrow 1$ as
$|\vec{x}|\rightarrow \infty$. Using the Euler angles $\theta^a$
and going through the same procedure we used for $q[A]$ in section
\ref{subsec:windingnumber}, we obtain \be
Q_T(t)&=&\frac{-1}{12\pi^2}\int d^3x \epsilon_{ijk} (\partial_i
\theta_a)(\partial_j \theta_b)(\partial_k \theta_c)\cdot {\rm Tr}
[(\partial_a U) U^{-1} (\partial_b U)
U^{-1} (\partial_c U) U^{-1}]\nonumber\\
&=&\frac{-3!}{12\pi^2}\int d^3x \epsilon_{ijk} (\partial_i
\theta_1)(\partial_j \theta_2)(\partial_k \theta_3)\cdot {\rm Tr}
[(\partial_1 U) U^{-1} (\partial_2 U) U^{-1} (\partial_3 U)
U^{-1}]\nonumber\\
&=&\frac{-n}{2\pi^2}\int d\theta_1 d\theta_2 d\theta_3
(-\frac{1}{4} \sin{(\theta_2)})=n\; .\ee Note that $Q_T$ can be
changed by large gauge transformations whereas $q$ is gauge
invariant (Eq. \ref{gaugeinvq}). Integrating the four divergence
$\partial_\mu J^\mu(x)$ over space and assuming that $\vec{J}$
vanishes faster than $O(1/r^2)$ as $r\rightarrow\infty$ one
obtains \beeq
\partial_0 Q_T(t)=\partial_0\int d^3x J_0=\int d^3x
\partial_\mu J^\mu\, .
\eneq Integrating the above equation over time we find \beeq
Q_T(\infty)-Q_T(-\infty)=q \, .\eneq The interpretation of this
formula is as follows. Euclidean solution $A_\mu(\vec{x}, x_4)$
interpolates between the real time Minkowski potentials
$A_\mu(t=\pm\infty, \vec{x})$ in the distant past and future. The
imaginary time $x_4=iT$ is the interpolating parameter. In
general, this is not a tunnelling process since $A_\mu(t=-\infty,
\vec{x})$ and $A_\mu(t=\infty, \vec{x})$ are not separated by a
barrier. However, when the fields $A_\mu(t=\pm\infty, \vec{x})$
are pure gauges (i.e., vacua) so that the $Q_T(t=\pm\infty)$ are
their respective homotopy class labels, then the
$A_\mu(t=\pm\infty, \vec{x})$ are inequivalent if
$Q_T(t=-\infty)\not= Q_T(t=\infty)$ (i.e., they are separated by a
barrier in Minkowski spacetime. Hence, solutions with $q\not= 0$
describe tunneling events between topologically inequivalent vacua
and later it will be shown that in Euclidean space the instanton
in the $A_4^a=0$ gauge has exactly this behavior). To see this,
observe that the field configuration of pure gauge $\vec{A}$ in
Eq. \ref{vecA} has zero potential energy (the term in the energy
functional not containing derivatives of the field $A_i$ with
respect to time): \beeq \frac{1}{4}\int dx^3 G^a_{ij}G^{aij}=0\,
,\label{potentialeneryij}\eneq and that there is no zero energy
evolution of the system which adiabatically connects configuration
Eq. \ref{vecA}, where $Q_T(t=-\infty)=1$, with the configuration
$\vec{A}=0$, where $Q_T(t=\infty)=0$. To examplify why all the
paths joining the two field configurations in real time must go
over an energy barrier, consider the class of field
configurations\cite{Rajaraman} \beeq
A_i^{(\gamma)}(\vec{x})=\gamma A^{(1)}_i(\vec{x})\; ,\eneq where
$\gamma\in [0, 1]$ is a real time independent parameter and
$A^{(1)}_i(\vec{x})$ is the pure gauge classical vacuum in the
$n=1$ class given in Eq. \ref{vecA}. We obtain the pure gauges
$A_i=0$ and $A_i=A_i^{1}$ for $\gamma=0$ and $\gamma=1$,
respectively. They both give $G_{ij}=0$ and hence zero energy for
$\gamma=0, 1$. For $\gamma\in (0, 1)$, however,
$A_i^{(\gamma)}(\vec{x})$ is not a pure gauge.
 Although the electric field $G^{0i}$ still vanishes because
$A_0^{(\gamma)}=0$ and $A_i^{(\gamma)}$ is time independent, the
magnetic field $B_i=(1/2)\epsilon_{ijk}G^{jk}$ does not\beeq
G_{jk}=\gamma(\partial_j A_k^{(1)}-\partial_k
A_j^{(1)})-ig\gamma^2[A_j^{(1)} , A_k^{(1)}]=
-ig(\gamma^2-\gamma)[A_j^{(1)} , A_k^{(1)}]\not= 0\; ,\eneq for
$\gamma\in (0, 1)$ . The energy, $(-1/8)\int d^3x {\rm Tr}
(G_{ij}G^{ij})
>0$ for $0<\gamma<1$, and proportional to
$(g^2/8)(\gamma^2-\gamma)^2$ (note that $\int d^3x {\rm Tr}
(G_{ij}G^{ij})$ is finite because $A_i^{(1)}$ falls as
$|\vec{x}|^{-2}$ as $|\vec{x}|\rightarrow\infty$ as shown in Eq.
\ref{vecA}). As $\gamma$ varies in the interval $[0 , 1]$, in
field space $A^{(\gamma)}_i$ plots a path which connects the two
classical vacua from $n=0$ and $n=1$ sectors. The energy of the
configuration can be interpreted as an energy barrier between the
two vacua. However, (due to the nonlinear nature of the Yang-Mills
theory) $A^{(\gamma)}_i$ does not solve the field equations, hence
the path is classically forbidden; the barrier is impenetrable. In
the quantum theory, tunneling will occur across this barrier. A
semiclassical description of tunneling can be given by solving the
classical equations of motion with imaginary time, thus achieving
an evolution which is classically forbidden. The solutions with
$q\not= 0$ describe tunnelling between topologically inequivalent
vacua. As we will show next, the instanton in the temporal gauge
$(A^{\rm inst-temp}_i, A_{4}=0)$, has exactly this behavior. For
$x_4=it\rightarrow-\infty$, $A_{i}^{\rm inst-temp}\rightarrow
(-i/g)(\partial_i U(\vec{x}))U^{-1}(\vec{x})$, whereas for
$x_4=it\rightarrow\infty$, $A_{i}^{\rm inst-temp}\rightarrow 0$.
The exact solution of the imaginary time equations of motion
interpolates between the $n=1$ and $n=0$ vacuum field
configuration. Hence the instanton solutions with $q\not= 0$
describe tunnelling between topologically inequivalent vacua. This
is another way of saying that we are going under an energy
barrier. To be more practical\cite{Huang}, we will now use the
explicit instanton solution given in Eq.
\ref{explicitinstanton}:\beeq \vec{A}^{\rm
inst}=\frac{1}{g}\frac{(\vec{x}\times \vec{\sigma})\pm
x_4\vec{\sigma}}{\vec{x}^2+x_4^2+\lambda^2}\; ,\;\;\; A^{\rm
inst}_4=\mp\frac{\vec{\sigma}\cdot\vec{x}}{\vec{x}^2+x_4^2+\lambda^2}\;
,\eneq and demonstrate the relation between vacuum tunneling and
instantons. To bring the instanton field to the temporal gauge we
make a gauge transformation\beeq A^{\rm inst}_\mu(x)\rightarrow
A^{\rm inst-temp}_\mu(x)=U(x)A^{\rm
inst}_\mu(x)U^{-1}(x)-\frac{i}{g}(\partial_\mu U(x))U^{-1}(x)\;
,\eneq such that $A^{\rm inst-temp}_4(x)=0$. This yields\beeq
\partial_4 U(x)=-igU(x)A_4^{\rm inst-temp}=\pm i U(x)
\frac{\vec{\sigma}\cdot\vec{x}}{|\vec{x}|^2+x_4^2+\lambda^2}\;
,\eneq which can be integrated to give \beeq U(\vec{x},
x_4)=\exp{\left(\pm
i\frac{\vec{\sigma}\cdot\vec{x}}{|\vec{x}|}f(\vec{x},
x_4)\right)}\; ,\eneq with \beeq
f(\vec{x},x_4)=\frac{|\vec{x}|}{\sqrt{|\vec{x}|^2+\lambda^2}}\left[\theta_0
+\arctan{\left(\frac{x_4}{\sqrt{|\vec{x}|^2+\lambda^2}}\right)}\right]
\; .\eneq By setting the integration constant
$\theta_0=\frac{\pi}{2}$ we find that \beeq \lim_{x_4\rightarrow
-\infty} f(\vec{x},x_4)=0\, ,\nonumber\eneq \beeq
\lim_{x_4\rightarrow \infty}
f(\vec{x},x_4)=\omega(|\vec{x}|)=\frac{\pi
|\vec{x}|}{\sqrt{|\vec{x}|^2+\lambda^2}}\; . \eneq Hence \beeq
\lim_{x_4\rightarrow -\infty} U(\vec{x},x_4)=1\, ,\nonumber\eneq
\beeq \lim_{x_4\rightarrow \infty}
U(\vec{x},x_4)=U_1(\vec{x})=\exp{(\pm i\frac{\pi
\vec{\sigma}\cdot\vec{x}}{\sqrt{|\vec{x}|^2+\lambda^2}})}\; ,
\eneq where $U_1(\vec{x})$ is given in Eq. \ref{map1}. Since
$A^{\rm inst-temp}_4=0$, the instanton field in the temporal gauge
is \beeq \vec{A}^{\rm inst-temp}=\frac{1}{g} U(\vec{x},
x_4)\frac{(\vec{x}\times \vec{\sigma})\pm
x_4\vec{\sigma}}{\vec{x}^2+x_4^2+\lambda^2} U^{-1}(\vec{x},
x_4)-\frac{i}{g}(\nabla U(\vec{x}, x_4))U^{-1}(\vec{x}, x_4)\;
.\eneq The first term of the above equation vanishes as
$|x_4|\rightarrow\infty$. As $x_4\rightarrow \mp\infty$, the
second term yields the following limits: \beeq
\lim_{x_4\rightarrow -\infty} \vec{A}^{\rm inst-temp}=0\,
,\nonumber\eneq \beeq \lim_{x_4\rightarrow \infty} \vec{A}^{\rm
inst-temp}(\vec{x})=-\frac{i}{g}(\nabla
U_1(\vec{x}))U^{-1}_1(\vec{x})\; \eneq which is given explicitly
in Eq. \ref{vecA}. Thus, the instanton field in the limit
$x_4\rightarrow -\infty$, is the trivial vacuum, whereas in the
limit $x_4\rightarrow \infty$, it has the form of the pure gauge
obtained from the large gauge transformations of class $n=1$. This
means that the instanton interpolates, in Euclidean time, between
two field configurations differing by a large gauge
transformation.

\subsection{The Vacuum Angle}

Let us now study how quantum theory of the Yang Mills theory
responds to gauge transformations\cite{Huang,JackivVacuum}. We
work in the temporal gauge $A^a_0(x)=0$. Recall that in this gauge
the Hamiltonian is \beeq H=\frac{1}{2}\int d^3x \left( E^a_i(x)
E^a_i(x)+B^a_i(x) B^a_i(x)\right)\, ,\nonumber\eneq where
$\vec{E}^a(x)=-{\partial A^a_i(x)}/{\partial t}$ and
$\vec{B}^a(x)=\vec{\nabla}\vec{A}^a(x)-\frac{g}{2}f^{abc}\vec{A}^b\times\vec{A}^c$.
The canonical equal time commutators are \beeq\left[
A^{ia}(\vec{x}, t), A^{jb}(\vec{x}', t)\right]=
\left[E^{ia}(\vec{x}, t), E^{jb}(\vec{x}', t)\right]=0\,
,\nonumber\eneq \beeq\left[ E^{ia}(\vec{x}, t), A^{jb}(\vec{x}',
t)\right]=i\delta^{ij}\delta^{ab}\delta(\vec{x}-\vec{x}')\,
.\nonumber\eneq In the functional Schrodinger representation of
the quantum theory\cite{Hatfield}, the operator ${A}^a_i(\vec{x})$
is diagonal. Let $|a\rangle$ be an eigenstate of $A^a_i(\vec{x})$
with eigenvalue (eigenfunction) $a^a_i(\vec{x})$: \beeq
A^a_i(\vec{x})|a\rangle=a^a_i(\vec{x})|a\rangle\, .\eneq $A^a_i$
is an operator, $a_i$ is a function. In this representation, a
functional differential representation of the conjugate momentum
\beeq {{E}_i}^a(\vec{x})=i\frac{\delta}{\delta
{{a}_i}^a(\vec{x})}\, ,\eneq is also a differential representation
of the above commutators. The state $|\Psi\rangle$ is represented
by the wave functional $\Psi[a]= \langle a|\Psi\rangle$.
$|\Psi\rangle$ is an eigenstate of the above Hamiltonian, if the
wave functional $\Psi[a]$ is a solution of the time independent
functional Shrodinger equation: \beeq \frac{1}{2} \int d^3x \left(
-\frac{\delta}{\delta {{a}_i}^a(\vec{x})}\frac{\delta}{\delta
{{a}_i}^a(\vec{x})}+{{b}_i}^a(\vec{x}){{b}_i}^a(\vec{x})\right)
\Psi[{a}]=E_0\Psi[a] \, ,\label{functionalshrodinger}\eneq where
${\vec{b}}^a(\vec{x})=\nabla\times{\vec{a}}^a(\vec{x})-\frac{g}{2}f^{abc}
{\vec{a}}^b(\vec{x}){\vec{a}}^c(\vec{x})$ is the eigenvalue of the
diagonal and time independent magnetic field operator. Recall also
that in the temporal gauge there is no Hamiltonian equation of
motion corresponding to Gauss' law, $\vec{D}\cdot\vec{E}^a=0$.
Therefore, we must add this equation as a constraint on the
states, $\vec{D}\cdot\vec{E}^a|\Psi\rangle=0$. The wave
functionals $\Psi[a]$ that satisfy Eq. \ref{functionalshrodinger},
must also satisfy the functional Gauss' law constraint \beeq
\vec{D}\cdot\frac{\delta}{\delta \vec{a}(\vec{x})}\Psi[a]=0\,
.\eneq To interpret this constrained let's consider what happens
to $\Psi[a]$ under a gauge transformation. When we gauge transform
$\vec{a}^a$, we cause a change in the functional $\Psi[a]$. Any
change in the wave functional due to a change in $\vec{a}^a$ is
given by\beeq \delta \Psi[a]=\int d^3x \frac{\delta\Psi}{\delta
a^{a}_i(\vec{x})} \delta a^{a}_i(\vec{x})\, .\eneq When the change
in $\vec{a}^a(\vec{x})$ is due to a gauge transformation, then
(using Eq. \ref{deltaa_i}) \beeq \delta
a^{a}_i(\vec{x})=-\frac{1}{g} (\partial_i\omega^{a}(\vec{x})
-gf^{abc}\omega^b(\vec{x})a^{c}_i(\vec{x}))\,
.\label{deltaa_i}\eneq Therefore, a change in $\Psi[a]$ due to
$\delta\vec{a}^a(\vec{x})$ of Eq. \ref{deltaa_i}, is \beeq \delta
\Psi[\vec{a}]=\frac{i}{g}\left\{ \int d^3x
\left[{{E}_i}^a(\vec{x})\Psi[a]\right]\left(\partial_i\omega^a(\vec{x})
-gf^{abc}\omega^b(\vec{x})a^{c}_i(\vec{x})\right)\right\} \,
,\label{182}\eneq where we used $\frac{\delta}{\delta
{{a}_i}^a(\vec{x})}=-i{{E}_i}^a(\vec{x})$. Equation \ref{182} can
also be written as \beeq\delta \Psi[\vec{a}]=\frac{i}{g} \int
d^3x\left\{
\partial_i\left[\omega^a(\vec{x}){{E}_i}^a(\vec{x})\right]-\omega^a(\vec{x})\partial_i{\vec{E}}^a(\vec{x})
-gf^{abc}\omega^b(\vec{x})a^{c}_i(\vec{x}){{E}_i}^a(\vec{x})\right\}
\Psi[a]\, ,\label{183}\nonumber\eneq which gives \beeq \delta
\Psi[\vec{a}]=\frac{i}{g}\left( \int
dS\cdot{\vec{E}}^a(\vec{x})\omega^a(\vec{x})-\int dx^3\,
\omega^a(\vec{x}) (\vec{D}\cdot
\vec{E})^a\right)\Psi[\vec{a}]\label{varPsi}\; .\eneq Notice that
the Gauss' law operator $(\vec{D}\cdot \vec{E})^a$ acting on the
wave functional $\Psi[a]$ has appeared in Eq. \ref{varPsi}. If we
assume \beeq \lim_{|\vec{x}|\rightarrow\infty} \omega^a(\vec{x})=
0\label{0class}\, ,\eneq the surface term vanishes, and if the
wave functional satisfies the Gauss' law constraint, then the
right hand side of Eq. \ref{varPsi} vanishes completely, and hence
the wave functional is ``invariant under a gauge transformation,''
$\delta \Psi[\vec{a}]=0$. On the other hand. if we want the wave
functional to be gauge invariant, then the functional must satisfy
the Gauss' law constraint and the gauge transformation must be of
class $n=0$ which satisfies Eq. \ref{0class}. Remember that the
transformations which satisfy Eq. \ref{0class} are only class
$n=0$ gauge transformations since:\beeq
\lim_{|\vec{x}|\rightarrow\infty}\omega(\vec{x})= n\pi\; . \eneq
Thus, the Gauss' law constraint implies the physical wave
functional to be ``class $n=0$ gauge invariant.'' Thus,
$(\vec{D}\cdot \vec{E})^a$ is the generator of class $n=0$ gauge
transformations under which $\Psi[\vec{a}]$ is invariant.
Therefore, we will call the wave functionals obtained by gauge
transformations satisfying Eq. \ref{0class} ``gauge equivalent.''
Let us also note that, when we say ``gauge invariant,'' we also
mean invariant under time independent gauge transformations (i.e.,
$\omega^a(\vec{x})$ is independent of time. We noted earlier that
the temporal gauge condition does not fully fix the gauge. Any
transformation that is time independent leaves $A^a_0=0$). Calling
${\mathcal{G}}_n$ the unitary operator which implements the gauge
transformation in the $n$th homotopy class (operator
representation of $U_n(\vec{x})=[U_1(\vec{x})]^n$), we have
${\mathcal{G}}_0 \Psi[\vec{a}]=\Psi[\vec{a}]$. Next we consider
the action of ${\mathcal{G}}_1$ which takes the $n$th sector of
vacua to the $(n+1)$th sector (hence
${\mathcal{G}}_n\equiv({\mathcal{G}}_1)^n$). For this class
$\omega\rightarrow \pi$ as $|\vec{x}|\rightarrow\infty$, hence
gauge invariance of $\Psi[\vec{a}]$ can not be forced by Gauss'
law (Eq. \ref{varPsi}). However, since ${\mathcal{G}}_1$ commutes
with all observables (they are gauge invariant), the only effect
it can make on physical states is to change them by a phase: \beeq
{\mathcal{G}}_1\Psi[\vec{a}]=e^{-i\Theta}\Psi[\vec{a}]\hskip 0.5cm
{\rm and \;\;hence}\hskip
0.5cm{\mathcal{G}}_n\Psi[\vec{a}]=e^{-in\Theta}\Psi[\vec{a}]
\label{opG}\; ,\eneq where $\Theta$ is a free, real parameter. In
other words, since under a gauge transformation of class
$n\not=0$, $\Psi[a]$ is not necessarily invariant, but since the
Hamiltonian is locally gauge invariant, $[H,
(\vec{D}\cdot\vec{E})^a]=0$, all energy eigenfunctions can be
chosen so that they change at most by a constant phase, which is
the same for all eigenfunctions:
$\Psi[a_n]={\mathcal{G}}_n\Psi[a]=e^{-in\Theta}\Psi[a]$, where
$a_n$ is the transform of $a$ by a class $n$ gauge transformation
$a_n=(U_1)^n a (U_1)^{-n} -\frac{i}{g}(U_1)^n\nabla(U_1)^{-n}$;
$(n=0, \pm 1, \pm 2,.\; .\; .)$ This is the origin of the famous
angle in Yang-Mills theory. We will denote the vacuum state wave
functional by $\Psi_\Theta[a]$ and the vacuum state characterized
by $\Theta$ is called the ``$\Theta$-vacuum.''

Now, we can ask what the physical vacuum looks like. We have seen
that states in different homotopy sectors are not forced to be
gauge equivalent by Gauss' law.  We have an infinite number of
distinct topological vacuum $|n\rangle$ in each sector. Non of
these is the true vacuum, since they will tunnel into one another.
This is expected since, as we have seen, the Yang-Mills theory has
finite action instanton solutions. Therefore, assuming that the
tunneling does occur, one can anticipate that the correct vacua
will be linear superposition of the topological vacua and are
given by $|\Theta\rangle=\sum_{-\infty}^\infty
e^{-in\Theta}|n\rangle$. In the next paragraph, we make this claim
more rigourously, and show that it is plausible.

In Hilbert space, we represent the vacuum associated with the
homotopy class index $n$ by the state vector $|n\rangle$, with
$\langle n|m\rangle=\delta_{nm}$. The topological charge operator
$Q_T=({1}/{16\pi^2})\int d^3x J_0$ has the property
$Q_T|n\rangle=n|n\rangle$. Large gauge transformations change $n$
so that: \beeq {\mathcal{G}}_1 |n\rangle=|n+1\rangle ,\hskip 1cm
{\rm hence}\hskip 1cm |n\rangle={\mathcal{G}}_n|n=0\rangle\;
.\eneq Therefore $[Q_T, {\mathcal{G}}_1]={\mathcal{G}}_1$. The
state vector $|n\rangle$ is not invariant under ${\mathcal{G}}_n$.
However, the physical vacuum should be invariant, up to a phase
$\Theta$, under gauge transformations (i.e., since
$[{\mathcal{G}}_1 , H]=0$, their eigenstates are common. Then the
unitarity of ${\mathcal{G}}_1$ implies that the eigenstates
$|\Theta\rangle$ of $H$ must satisfy ${\mathcal{G}}_1
|\Theta\rangle =e^{-i\Theta}|\Theta\rangle$). Thus, we superimpose
the n-vacua to form the so called $\Theta$ vacuum: \beeq
|\Theta\rangle=\sum_{n=-\infty}^{\infty} e^{in\Theta}|n\rangle\; .
\eneq Indeed,\beeq
{\mathcal{G}}_1|\Theta\rangle=\sum_{n=-\infty}^{\infty}
e^{in\Theta}|n+1\rangle=e^{-i\Theta}\sum_{n=-\infty}^{\infty}
e^{i(n+1)\Theta}|n+1\rangle=e^{-i\Theta}|\Theta\rangle\; , \eneq
justifying Eq. \ref{opG}. Different $\Theta$-vacua are
orthogonal:\beeq\langle\Theta'|\Theta\rangle=\sum_{n,m}
e^{i(n\Theta-m\Theta')}\langle m|n\rangle=\sum_n
e^{in(\Theta-\Theta')}=2\pi\delta(\Theta-\Theta')\eneq Thus, there
exists a continuum of $\Theta$-vacua and each one is a suitable
ground state for the physical gauge theory. No gauge invariant
operator $B$ $([B,{\mathcal{G}}_1]=0)$, can generate transitions
between different $\Theta$-vacua\cite{Rajaraman}: \beeq
\langle\Theta'|B|\Theta\rangle=\sum_{m,n}e^{in\Theta-im\Theta'}\langle
m|B|n\rangle=\sum_{m,n}e^{i(n+p)\Theta-i(m+p)\Theta'}\langle
m+p|B|n+p\rangle\eneq where $p$ is an arbitrary integer. Bracket
$\langle m+p|B|n+p\rangle=\langle
m|({\mathcal{G}}_1^\dagger)^pB({\mathcal{G}}_1)^p|n\rangle=\langle
m|B|n\rangle$, since $[B,{\mathcal{G}}_1]=0$ and ${\mathcal{G}}_1$
is unitary. Therefore, we have \be
\langle\Theta'|B|\Theta\rangle=e^{ip(\Theta-\Theta')}\sum_{m,n}e^{in\Theta-im\Theta'}\langle
m|B|n\rangle=e^{ip(\Theta-\Theta')}\langle\Theta'|B|\Theta\rangle\;
, \ee which implies, for all $p\in Z$,\be
(1-e^{ip(\Theta-\Theta')})\langle\Theta'|B|\Theta\rangle=0\;
,\nonumber\ee or
$\langle\Theta'|B|\Theta\rangle=\langle\Theta|B|\Theta\rangle\delta(\Theta-\Theta')$.
Hence, each $|\Theta\rangle$ is the vacuum of a separate sector of
states.

Here, let us note that we claimed all physical observables commute
with the large gauge transformation ${\mathcal{G}}_1$ but we
consider the wave functions $|n\rangle$ and $|n+1\rangle$ which
are related exactly by ${\mathcal{G}}_1$ as distinct. Although
physics is the same (the observables have the same value) in
$|n\rangle$ and in $|n+1\rangle$, as we show below, the existence
of distinct vacua changes the Lagrangian density by \beeq
{\mathcal{L}}_\Theta=-\frac{g^2}{32\pi^2}\,\Theta\; {}^\ast
G^{a\mu\nu} G^a_{\mu\nu}=\frac{g^2}{8\pi^2}\,\Theta\,
E^{ai}B^{ai}\; \label{thetaction}.\eneq The new $\Theta$-term,
${\mathcal{L}}_\Theta$, is a total divergence; it does not effect
the equations of motion. However, it changes the quantum theory
nonperturbatively  in a $\Theta$ dependent manner. The
relationship between the canonical momentum conjugate to $A^{a}_i$
and $E^{a}_i$ (Eq. \ref{canmom}), also changes \beeq
\Pi^a_i=-E^{ai}-\frac{g^2}{8\pi^2}\,\Theta\, B^{ai}\; .\eneq The
angle in the Yang Mills theory, therefore, can not be avoided by
postulating the (large) gauge invariance of the states. It
reappears as an ambiguity in the definition of the Lagrangian, and
hence in the canonical variables. Another difference introduced by
the $\Theta$-term is the violation of time reversal $T$ and parity
$P$ symmetries. Under $T$, $E_i\rightarrow E_i$, while
$B_i\rightarrow -B_i$. Under $P$, $E_i\rightarrow -E_i$, while
$B_i\rightarrow B_i$. Thus, under either $T$ or $P$,
${\mathcal{L}}_\Theta$ changes sign, while the original Lagrangian
density, proportional to $E^a_iE^a_i- B^a_iB^a_i$, is invariant.

To prove the claim Eq. \ref{thetaction}, we recall the Feynman-Kac
formula which gives the transition amplitude between $n$-vacua:
\beeq \lim_{T\rightarrow\infty} \langle m|e^{-TH}|n\rangle=\int
[dA]_{n-m} e^{-S_E}\; ,\label{FeyKac}\eneq where $H$ is the
Hamiltonian of the system and, \beeq S_E=\frac{1}{4}
\int_{-T/2}^{T/2} d\tau\int d^3x
F^a_{\mu\nu}F^a_{\mu\nu}=\frac{1}{2}\int_{-T/2}^{T/2} d\tau\int
d^3x (\vec{E}^a\cdot\vec{E}^a+\vec{B}^a\cdot\vec{B}^a) \, \eneq is
the Euclidean action between times $-T/2$ and $T/2$. Notice that
in Eq. \ref{FeyKac} we integrate over Euclidean gauge field
configurations which interpolate between the vacua $|n\rangle$ and
$|m\rangle$. Thus, the functional integral in Eq. \ref{FeyKac} is
restricted by the boundary conditions on the winding number
$Q_T[A]$: \beeq Q_T[A(\vec{x}, \tau=-T/2)]=n \hskip 0.5cm {\rm
and} \hskip 0.5cm Q_T[A(\vec{x}, \tau=T/2)]=m\; ,\eneq with
$q=m-n$. In other words, in Feynman-Kac formula, for
$|T|\rightarrow \infty$, we consider configurations which yield
finite contributions. Since their Euclidean action is finite,
$G_{\mu\nu}$ must vanish at infinity in all Euclidean directions.
This means that $A_\mu$ is a pure gauge field and yields a mapping
from $S^3$ to the group classified by the winding numbers. Then,
we consider the transition amplitude ${\mathcal{A}}$ between
$\Theta$-vacua: \beeq
{\mathcal{A}}=\lim_{T\rightarrow\infty}\langle\Theta'|e^{-TH}|\Theta\rangle
=\sum_{n,m}e^{i(n\Theta-m\Theta')}\lim_{T\rightarrow\infty}\langle
m|e^{-TH}| n\rangle\; , \eneq replacing the summation index $m$ by
$p=(m-n)$, we have\beeq
{\mathcal{A}}=\sum_{n,p}e^{i(n\Theta-p\Theta'-n\Theta')}\lim_{T\rightarrow\infty}\langle
p+n|e^{-TH}| n\rangle\;.\eneq The amplitude $\langle p+n|e^{-TH}|
n\rangle= \langle p|e^{-TH}| 0\rangle$ since $[H, G]=0$.
Therefore, \be
{\mathcal{A}}&=&\sum_p\sum_ne^{in(\Theta-\Theta')}e^{-ip\Theta}\lim_{T\rightarrow\infty}\langle
p|e^{-TH}|0\rangle\nonumber\\
&=&\delta(\Theta-\Theta')\sum_pe^{-ip\Theta}\int [dA]_m
e^{-S_E}\;,\ee where the $\delta$-function is obtained by summing
over $n$, and Eq. \ref{FeyKac} is used. Now we use the formula for
topological charge  $q={(g^2/32\pi^2)}\int d^4x
G^a_{\mu\nu}{}^\ast G^a_{\mu\nu}$, which is an integer, to replace
$p$ with $q$. Hence \be
{\mathcal{A}}&=&\delta(\Theta-\Theta')\sum_q\int[dA]_q
\exp{\left(-S^E-\frac{ig^2}{32\pi^2}\,\Theta\int d^4x
G^a_{\mu\nu}{}^\ast
G^a_{\mu\nu}\right)}\nonumber\\
&=&\delta(\Theta-\Theta')\int e^{-S^E_{\rm
 eff}}\;,\ee where the integration is over all fields, and \be
 S^E_{\rm eff}&=&\int
 d^4x_E {\mathcal{L}}^E_{\rm eff}=\int
 d^4x_E \left( \frac{1}{4}G^a_{\mu\nu} G^a_{\mu\nu} +\frac{ig^2}{32\pi^2}\Theta G^a_{\mu\nu}{}^\ast
 G^a_{\mu\nu}\right)\\&=&\int
 d^4x_E \left(\frac{1}{2}(E^a_iE^a_i +B^a_iB^a_i)+\frac{ig^2}{32\pi^2}\Theta (-4E^a_iB^a_i)\right)\, .\ee
To obtain the action in the Minkowski spacetime, we recall the
necessary transformations: $d^4x_E\rightarrow id^4x_M$,
$E^E_i\rightarrow -iE^E_i$. Then, the effective action transforms
as \be S^E_{\rm eff}&\rightarrow&\int
 id^4x \left(\frac{1}{2}(B^a_iB^a_i-E^a_iE^a_i)
 +\frac{ig^2}{32\pi^2}\Theta (4iE^a_iB^a_i)\right)\nonumber\; ,\\
&\rightarrow& i\int
 d^4x \left(\frac{1}{4} G^a_{\mu\nu} G^{a\mu\nu}
 +\frac{g^2}{32\pi^2}\Theta G^a_{\mu\nu} {}^\ast
 G^{a\mu\nu}\right)
 \; ,\ee where we dropped the index $M$ and used Eqs. \ref{canLag},
 \ref{pseudoMin}, and \ref{pseudoEuc}. Since $S^E\rightarrow -i\int d^4x_M
{\mathcal{L}}^M_{\rm
 eff}$ (Eq. \ref{Wickaction}), we find the effective action as\beeq {\mathcal{L}}^M_{\rm
 eff}= -\frac{1}{4} G^a_{\mu\nu} G^{a\mu\nu}
 -\frac{g^2}{32\pi^2}\Theta G^a_{\mu\nu} {}^\ast G^{a\mu\nu}\;
 .\eneq This proves claim Eq.
\ref{thetaction}. Since the $\Theta$-angle is an arbitrary real
parameter, whose absolute value can only be measured, the
$\Theta$-term in Minkowski spacetime can also be chosen with a
positive sign.

\myend